\documentclass[superscriptaddress, showpacs, nofootinbib, amsmath, amssymb, aps, prd, notitlepage]{revtex4-1}
\usepackage{amsmath,amssymb,amsfonts,mathrsfs,bbold}
\usepackage{yfonts}

\usepackage[dvipsnames]{xcolor}

\usepackage{float}
\usepackage{graphics,graphicx}
\usepackage{dcolumn}
\usepackage{bm}
\usepackage{hyperref}
\usepackage{comment}
\usepackage{xcolor}
\usepackage{slashed}
\usepackage{pstricks}
\usepackage{color}
\usepackage{simplewick}
\usepackage{caption}
\usepackage{subcaption}

\usepackage{ragged2e} 
\DeclareCaptionJustification{justified}{\justifying}

\usepackage{multirow}
\usepackage{mathbbol}
\usepackage{mathtools}

\usepackage{slashed}
\usepackage{simpler-wick}

\usepackage[normalem]{ulem}

\usepackage{tikz}
\usetikzlibrary{decorations.pathmorphing, decorations.markings, shapes.geometric, shapes.misc, calc, math}
\tikzstyle{gluon}=[decorate, decoration={coil,aspect=0.8, amplitude=1.5pt,  segment length=3pt}]

\usepackage{stackrel}
\usepackage[most]{tcolorbox}

\allowdisplaybreaks[1]

\def\eq#1{{Eq.~(\ref{#1})}}
\def\fig#1{{Fig.~\ref{#1}}}
\newcommand{\ben}{\begin{eqnarray*}}
\newcommand{\een}{\end{eqnarray*}}
\newcommand{\un}[1]{\underline{#1}}

\newcommand{\pd}{\partial}

\newcommand{\ee}{\underline{\varepsilon}}
\newcommand{\ul}[1]{\underline{#1}}

\newcommand{\tr}{\mbox{tr}}
\newcommand{\Tr}{\mbox{Tr}}
\newcommand{\thalf}{\tfrac{1}{2}}
\newcommand{\llangle}{\Big\langle \!\! \Big\langle}
\newcommand{\rrangle}{\Big\rangle \!\! \Big\rangle}
\newcommand{\half}{{1\over 2}}

\newcommand{\as}{\alpha_s}

\newcommand{\bra}[1]{\left\langle #1 \right|}
\newcommand{\ket}[1]{\left| #1 \right\rangle}

\makeatletter
\DeclareRobustCommand{\cev}[1]{%
  {\mathpalette\do@cev{#1}}%
}
\newcommand{\do@cev}[2]{%
  \vbox{\offinterlineskip
    \sbox\z@{$\m@th#1 x$}%
    \ialign{##\cr
      \hidewidth\reflectbox{$\m@th#1\vec{}\mkern4mu$}\hidewidth\cr
      \noalign{\kern-\ht\z@}
      $\m@th#1#2$\cr
    }%
  }%
}
\makeatother

\begin{document}

\title{Quark and Gluon Helicity Evolution at Small $x$: Revised and Updated}

\author{Florian Cougoulic} 
         \email[Email: ]{florian.f.cougoulic@jyu.fi}
         \affiliation{Department of Physics, P.O. Box 35, 40014 University of Jyv\"{a}skyl\"{a}, Finland}

\author{Yuri V. Kovchegov} 
         \email[Email: ]{kovchegov.1@osu.edu}
         \affiliation{Department of Physics, The Ohio State
           University, Columbus, OH 43210, USA}
           
\author{Andrey Tarasov}
         \email[Email: ]{tarasov.3@osu.edu}
         \affiliation{Department of Physics, The Ohio State
           University, Columbus, OH 43210, USA}
         \affiliation{Joint BNL-SBU Center for Frontiers in Nuclear Science (CFNS) at Stony Brook University, Stony Brook, NY 11794, USA}

\author{Yossathorn Tawabutr}
         \email[Email: ]{tawabutr.1@osu.edu}
         \affiliation{Department of Physics, The Ohio State
           University, Columbus, OH 43210, USA}

\begin{abstract}
We revisit the problem of small Bjorken-$x$ evolution of the gluon and flavor-singlet quark helicity distributions in the shock wave ($s$-channel) formalism. Earlier works on the subject in the same framework \cite{Kovchegov:2015pbl,Kovchegov:2016zex,Kovchegov:2018znm} resulted in an evolution equation for the gluon field-strength $F^{12}$ and quark ``axial current" ${\bar \psi} \gamma^+ \gamma^5 \psi$ operators (sandwiched between the appropriate light-cone Wilson lines) in the double-logarithmic approximation (summing powers of $\as \, \ln^2 (1/x)$ with $\as$ the strong coupling constant). In this work, we observe that an important mixing of the above operators with another gluon operator, $\cev{D}^i \, {D}^i$, also sandwiched between the light-cone Wilson lines (with the repeated transverse index $i=1,2$ summed over), was missing in the previous works \cite{Kovchegov:2015pbl,Kovchegov:2016zex,Kovchegov:2018znm}. This operator has the physical meaning of the sub-eikonal (covariant) phase: its contribution to helicity evolution is shown to be proportional to another sub-eikonal operator, $D^i - \cev{D}^i$, which is related to the Jaffe-Manohar polarized gluon distribution \cite{Jaffe:1989jz}. In this work we include this new operator into small-$x$ helicity evolution, and construct novel evolution equations mixing all three operators ($D^i - \cev{D}^i$, $F^{12}$, and ${\bar \psi} \gamma^+ \gamma^5 \psi$), generalizing the results of \cite{Kovchegov:2015pbl,Kovchegov:2016zex,Kovchegov:2018znm}. We also construct closed double-logarithmic evolution equations in the large-$N_c$ and large-$N_c \& N_f$ limits, with $N_c$ and $N_f$ the numbers of quark colors and flavors, respectively. Solving the large-$N_c$ equations numerically we obtain the following small-$x$ asymptotics of the quark and gluon helicity distributions $\Delta \Sigma$ and $\Delta G$, along with the $g_1$ structure function,
\begin{align}
    \Delta \Sigma (x, Q^2) \sim \Delta G (x, Q^2) 
    \sim g_1 (x, Q^2) \sim \left( \frac{1}{x} \right)^{3.66 \, \sqrt{\frac{\as \, N_c}{2 \pi}}} ,  \notag
\end{align}
in complete agreement with the earlier work by Bartels, Ermolaev and Ryskin \cite{Bartels:1996wc}.
\end{abstract}

\maketitle
\tableofcontents


\section{Introduction}
\label{sec:intro}

Understanding the proton spin puzzle \cite{Aidala:2012mv,Accardi:2012qut,Leader:2013jra,Aschenauer:2013woa,Aschenauer:2015eha,Boer:2011fh,Proceedings:2020eah,Ji:2020ena,AbdulKhalek:2021gbh} is one of the main goals of contemporary hadronic physics. Apart from being a question of general scientific interest testing our understanding of the proton internal structure, the proton spin puzzle is one of the central topics to be addressed by the experimental program at the future Electron-Ion Collider (EIC) \cite{Accardi:2012qut,Boer:2011fh,Proceedings:2020eah,AbdulKhalek:2021gbh}.

The main question of the proton spin puzzle is how the proton spin (1/2) is made up of the contributions of quark and gluon helicities and their orbital angular momenta (OAM) (see \cite{Aidala:2012mv,Leader:2013jra,Ji:2020ena} and references therein for reviews). It is usually formulated in terms of either Jaffe-Manohar \cite{Jaffe:1989jz} or Ji \cite{Ji:1996ek} spin sum rules. The Jaffe-Manohar sum rule \cite{Jaffe:1989jz} reads
\begin{equation}
S_q+L_q+S_G+L_G=\frac{1}{2},
\label{eqn:JM}
\end{equation}
where $S_q$ and $S_G$ are the contributions to the spin of the proton carried by the quarks and gluons, respectively, and $L_q$ and $L_G$ are their OAM. All four terms on the left-hand side of \eq{eqn:JM} can be written as integrals over the Bjorken $x$ variable. For the quark and gluon spin contributions, $S_q$ and $S_G$, the decomposition is (see \cite{Bashinsky:1998if,Hagler:1998kg,Harindranath:1998ve,Hatta:2012cs,Ji:2012ba} for decompositions for the OAM terms)
\begin{align}
S_q(Q^2) = \frac{1}{2} \int\limits_0^1 dx \; \Delta\Sigma(x,Q^2), \ \ \  \ \
S_G(Q^2) = \int\limits_0^1 dx \; \Delta G(x,Q^2),
\label{eqn:SqSG}
\end{align}
where
\begin{equation}
\Delta\Sigma(x,Q^2) = \sum_{f=u,d,s,\ldots} \Delta q_f^+ (x, Q^2) 
\label{eqn:DeltaSigma}
\end{equation}
is the flavor-singlet quark helicity distribution function with $\Delta q_f^+ = \Delta q_f + \Delta {\bar q}_f$ \cite{Ethier:2017zbq,Adamiak:2021ppq}. Here $\Delta q_f$ and $\Delta {\bar q}_f$ are the quark and anti-quark helicity distributions for each quark flavor $f$, while $\Delta G$ is the gluon helicity distribution. As usual, the distributions also depend on the momentum scale $Q^2$. The current values of the proton spin carried by the quarks and gluons, as extracted from the experimental data, are $S_q (Q^2 = 10\, \mbox{GeV}^2) \approx 0.15 \div 0.20$ for $x \in [0.001, 1]$, and $S_G (Q^2 = 10 \, \mbox{GeV}^2) \approx 0.13 \div 0.26$, for $x \in [0.05, 1]$ (see \cite{Accardi:2012qut,Leader:2013jra,Aschenauer:2013woa,Aschenauer:2015eha,Proceedings:2020eah,Ji:2020ena} for reviews). The fact that the sum of these two numbers comes up short of $1/2$, especially if one takes into account the error bars, is the proton spin puzzle: we do not know where the rest of the proton spin is. The remaining missing spin of the proton may be found in the quark and gluon OAM and/or at smaller values of $x$.  

The latter possibility received a lot of attention in the literature, starting with the groundbreaking work by Bartels, Ermolaev and Ryskin (BER) \cite{Bartels:1995iu,Bartels:1996wc}, which studied the small-$x$ asymptotics of the $g_1$ structure function along with $\Delta \Sigma$ and $\Delta G$ employing the infrared evolution equations (IREE) approach pioneered in \cite{Kirschner:1983di,Kirschner:1994rq,Kirschner:1994vc,Griffiths:1999dj}. The phenomenology based on BER work was developed in \cite{Blumlein:1995jp,Blumlein:1996hb}. The BER approach resummed double logarithms in $x$, that is, powers of $\as \, \ln^2 (1/x)$. This is known as the double-logarithmic approximation (DLA). In the pure-glue case the BER approach resulted in the asymptotics given by
\begin{align}\label{asymptotics}
    \Delta \Sigma (x, Q^2) \sim \Delta G (x, Q^2) 
    \sim g_1 (x, Q^2) \sim \left( \frac{1}{x} \right)^{3.66 \, \sqrt{\frac{\as \, N_c}{2 \pi}}} .  
\end{align}

More recently, an effort has been under way \cite{Kovchegov:2015pbl, Hatta:2016aoc, Kovchegov:2016zex, Kovchegov:2016weo, Kovchegov:2017jxc, Kovchegov:2017lsr, Kovchegov:2018znm, Kovchegov:2019rrz, Boussarie:2019icw, Cougoulic:2019aja, Kovchegov:2020hgb, Cougoulic:2020tbc, Chirilli:2021lif, Adamiak:2021ppq, Kovchegov:2021lvz} to reproduce BER results \cite{Bartels:1995iu,Bartels:1996wc} and, possibly, expand on them using the $s$-channel/shock wave approach to small-$x$ evolution from \cite{Mueller:1994rr,Mueller:1994jq,Mueller:1995gb,Balitsky:1995ub,Balitsky:1998ya,Kovchegov:1999yj,Kovchegov:1999ua,Jalilian-Marian:1997dw,Jalilian-Marian:1997gr,Weigert:2000gi,Iancu:2001ad,Iancu:2000hn,Ferreiro:2001qy} (see \cite{Gribov:1984tu,Iancu:2003xm,Weigert:2005us,JalilianMarian:2005jf,Gelis:2010nm,Albacete:2014fwa,Kovchegov:2012mbw,Morreale:2021pnn} for reviews) modified to work at the sub-eikonal level and beyond \cite{Altinoluk:2014oxa, Chirilli:2018kkw, Altinoluk:2020oyd, Altinoluk:2021lvu}. (Small-$x$ asymptotics of parton distribution functions (PDFs) and transverse momentum-dependent PDFs (TMDs) can be classified by the leading power of $x$: our notation is such that, neglecting the quantum-evolution order-$\as$ or $\sqrt{\as}$ corrections to the power of $x$, the eikonal distributions scale as $\sim 1/x$, the sub-eikonal ones scale as $\sim x^0$, the sub-sub-eikonal ones scale as $\sim x$, etc.) The approach to helicity evolution in the $s$-channel formalism developed in \cite{Kovchegov:2015pbl, Kovchegov:2016zex, Kovchegov:2016weo, Kovchegov:2017jxc, Kovchegov:2017lsr, Kovchegov:2018znm, Cougoulic:2019aja, Kovchegov:2020hgb} resulted in the small-$x$ asymptotics of $\Delta \Sigma$ and $\Delta G$ different from that found by BER.\footnote{In the flavor non-singlet channel, the two approaches were in complete agreement \cite{Bartels:1995iu, Kovchegov:2016zex} at large $N_c$.} Despite the cross-check in  \cite{Kovchegov:2016zex} and an alternative calculation in \cite{Chirilli:2021lif} the origin of the difference remained unknown. In this work we identify the sub-eikonal operator which was not included in the approach of \cite{Kovchegov:2015pbl, Kovchegov:2016zex, Kovchegov:2016weo, Kovchegov:2017jxc, Kovchegov:2017lsr, Kovchegov:2018znm, Cougoulic:2019aja, Kovchegov:2020hgb}: after including it, we obtain a new set of small-$x$ evolution equations for helicity, whose solution gives the asymptotics \eqref{asymptotics} consistent with BER.

When going beyond the eikonal approximation, the degrees of freedom are no longer the light-cone Wilson lines: instead one has to modify the Wilson lines by inserting one or more sub-eikonal operators between segments of Wilson lines \cite{Altinoluk:2014oxa,Balitsky:2015qba,Balitsky:2016dgz, Kovchegov:2017lsr, Kovchegov:2018znm, Chirilli:2018kkw, Jalilian-Marian:2018iui, Jalilian-Marian:2019kaf, Altinoluk:2020oyd, Kovchegov:2021iyc, Altinoluk:2021lvu}. The sub-eikonal operators entering the helicity evolution of \cite{Kovchegov:2015pbl, Kovchegov:2016zex, Kovchegov:2016weo, Kovchegov:2017jxc, Kovchegov:2017lsr, Kovchegov:2018znm, Cougoulic:2019aja, Kovchegov:2020hgb} are the gluon field strength operator $F^{12}$ or the bi-local quark operator ${\bar \psi} (x_2) \gamma^+ \gamma^5 \psi (x_1)$. When wrapped around by light-cone Wilson lines they lead to the operators in Eqs.~\eqref{VG1} and \eqref{Vq1} (or Eqs.~\eqref{UG1} and \eqref{Uq1}) below. (Our calculations here are carried out in $A^- =0$ light-cone gauge of the projectile, while the expressions for the operators are valid in any gauge where the gluon field $A_\mu$ vanishes at $x^- \to \pm \infty$.) The operators $F^{12}$ and ${\bar \psi} (x_2) \gamma^+ \gamma^5 \psi (x_1)$ enter the calculation with the helicity-dependent prefactor, e.g., with $\sigma \, \delta_{\sigma, \sigma'}$ in the quark helicity basis, as defined in light-cone perturbation theory (LCPT) \cite{Lepage:1980fj,Brodsky:1997de}. This is what makes them natural operators for helicity evolution. The helicity evolution of \cite{Kovchegov:2015pbl, Kovchegov:2016zex, Kovchegov:2017lsr, Kovchegov:2018znm} mixes these two operators with each other. However, since $F^{12}$ is a local operator, it cannot be used to construct a PDF: hence the mapping of evolution from \cite{Kovchegov:2015pbl, Kovchegov:2016zex, Kovchegov:2017lsr, Kovchegov:2018znm} onto the spin-dependent Dokshitzer-Gribov-Lipatov-Altarelli-Parisi (DGLAP) evolution equation \cite{Gribov:1972ri,Altarelli:1977zs,Dokshitzer:1977sg} in the gluon sector has been problematic \cite{Kovchegov:2016zex}. At the same time, $F^{12}$ is not the only gluon operator at the sub-eikonal order: there exists another sub-eikonal operator, $\cev{D}^i \, D^i$, as derived in  \cite{Hatta:2016aoc,Balitsky:2015qba,Altinoluk:2020oyd,Kovchegov:2021iyc}. Here $D_\mu = \pd_\mu - i g A_\mu$ is the right-acting covariant derivative, $\cev{D}_\mu = \cev{\pd}_\mu + i g A_\mu$ is the left-acting covariant derivative, and $i=1,2$ is the transverse index. The operator $\cev{D}^i \, D^i$, whose contribution is simplified in our helicity-evolution calculations to $D^i - \cev{D}^i$, is related to the Jaffe-Manohar gluon helicity PDF \cite{Jaffe:1989jz}, as we show below. The $\cev{D}^i \, D^i$ operator enters the calculations with a helicity-independent prefactor $\delta_{\sigma, \sigma'}$; an expression for this operator, sandwiched between the light-cone Wilson lines, is given below in \eq{VxyG2} (or \eqref{UG2}). In the background field method \cite{Abbott:1980hw,Abbott:1981ke} this operator arises naturally due to the canonical momentum squared term,  $(P^i)^2$, present even in a scalar particle propagator \cite{Balitsky:2015qba,Balitsky:2016dgz}. In this work we show that small-$x$ helicity evolution mixes the $D^i - \cev{D}^i$ operator with $F^{12}$ and ${\bar \psi} (x_2) \gamma^+ \gamma^5 \psi (x_1)$. This mixing was neglected in \cite{Kovchegov:2015pbl, Kovchegov:2016zex, Kovchegov:2017lsr, Kovchegov:2018znm} due to the apparent helicity-independence of the $\cev{D}^i \, D^i$ term, which gives rise to the $D^i - \cev{D}^i$ operator (i.e., due to the fact that it comes in with $\delta_{\sigma, \sigma'}$). The physical origin of the mixing still requires further understanding. Here we note that the mixing probably happens because the quark and gluon polarization indices $\sigma$ and $\lambda$ in LCPT are not true helicities: they are projections of the particle's spin on the fixed $z$-axis, instead of being spin projections on the direction of the particle's 3-momentum.  

In this paper we derive a new small-$x$ evolution equations for helicity distributions mixing the three operators  $D^i - \cev{D}^i$, $F^{12}$, and ${\bar \psi} (x_2) \gamma^+ \gamma^5 \psi (x_1)$. The equations resum longitudinal logarithms, keeping the accompanying transverse integrals exactly. We, therefore, can and do extract the DLA evolution equations from them, obtaining two closed systems of integral evolution equations in the 't Hooft's large-$N_c$ \cite{tHooft:1973alw} and Veneziano's large-$N_c \& N_f$ \cite{Veneziano:1976wm} limits. Performing a numerical solution of the large-$N_c$ helicity evolution equation we arrive at the asymptotics given in \eq{small_x_asymp}, thus reproducing BER results.

The paper is structured as follows. In Section~\ref{sec:S-matrix} we summarize the results of the earlier calculation \cite{Kovchegov:2021iyc} for high-energy $S$-matrices of massless quarks and gluons scattering on the background quark and gluon fields at the sub-eikonal accuracy. As we mentioned, these results are consistent with the earlier calculation \cite{Altinoluk:2020oyd}. The relevant sub-eikonal operators are given in Eqs.~\eqref{VqG} and \eqref{UqG} for the quarks and gluons, respectively. 

To identify which sub-eikonal operators are relevant for helicity distributions and for the $g_1$ structure function, we re-analyse these quantities at small $x$ in Sec.~\ref{sec:PDFs+g1}.\footnote{Note that there is a possible role of the topological effects in $g_1$ due to the chiral anomaly, which has been debated in the literature \cite{Altarelli:1988nr,Carlitz:1988ab,Jaffe:1989jz,Shore:1990zu,Shore:1991dv}, see also the recent works \cite{Tarasov:2020cwl,Tarasov:2021yll,Hatta:2020ltd,Radyushkin:2022qvt}. This can be inferred from the first moment of the structure function and is related to the $U_A(1)$ problem in QCD. However, we leave the question about the relation of our results to the chiral anomaly for future publications.} In Section~\ref{sec:glue_helicity} we reconstruct the known result \cite{Hatta:2016aoc,Kovchegov:2017lsr} that the gluon helicity TMD and PDF at small $x$ are related to the dipole amplitude $G^j$ from \eq{Gj}, dependent on the novel sub-eikonal operator ${D}^i - \cev{D}^i$ entering \eq{Vi}. This is summarized in Eqs.~\eqref{glue_hel_TMD57} and \eqref{JM_DeltaG} with the amplitude $G_2$ entering those equations defined in \eq{decomp}. Small-$x$ quark helicity distributions are studied in Sec.~\ref{sec:quark_helicity} with the result given by Eqs.~\eqref{TMD19} and \eqref{DeltaSigma}. The dipole amplitude $Q$ in those equations, defined in \eq{Q_def}, contains the operators $F^{12}$ and ${\bar \psi} (x_2) \gamma^+ \gamma^5 \psi (x_1)$: this part of the results for quark helicity distributions was known before \cite{Kovchegov:2015pbl, Kovchegov:2016zex, Kovchegov:2018znm}. The amplitude $G_2$ in the same expressions contains the new operator ${D}^i - \cev{D}^i$, which, in turn, originated in the sub-eikonal $\cev{D}^i \, D^i$ operator. (Comparing this with \eq{JM_DeltaG} we conclude that the $\cev{D}^i \, D^i$ operator is related to the Jaffe-Manohar distribution.) This $G_2$ contribution in Eqs.~\eqref{TMD19} and \eqref{DeltaSigma} is new compared to \cite{Kovchegov:2015pbl, Kovchegov:2016zex, Kovchegov:2017lsr, Kovchegov:2018znm}. Finally, in Sec.~\ref{sec:g1} we re-analyse the $g_1$ structure function at small $x$, arriving at Eqs.~\eqref{g1_final} and \eqref{g1_DLA}. Again, the contribution of the dipole $Q$ has been known before \cite{Kovchegov:2015pbl, Kovchegov:2016zex, Kovchegov:2018znm, Kovchegov:2021lvz}, while the contribution of $G_2$ is new. 

The small-$x$ evolution of the sub-eikonal operators ${D}^i - \cev{D}^i$, $F^{12}$ and ${\bar \psi} (x_2) \gamma^+ \gamma^5 \psi (x_1)$ with the appropriate light-cone Wilson lines is studied in Sections~\ref{sec:hel_evo} and \ref{sec:hel_evo_bfm}. The calculation in Sec.~\ref{sec:hel_evo_bfm} is done using the background field method \cite{Abbott:1981ke,Balitsky:1995ub}, while in Sec.~\ref{sec:hel_evo} the calculation employs a hybrid formalism developed in \cite{Kovchegov:2017lsr, Kovchegov:2018znm, Kovchegov:2021iyc} which combines the elements of LCPT \cite{Lepage:1980fj,Brodsky:1997de} with the background field method: we refer to it as the light-cone operator treatment (LCOT) method. In Sec. \ref{sec:prbfm} we derive the structure of sub-eikonal operators from the analysis of quark and gluon propagators in the background field. At the level of sub-eikonal operators the main evolution equations we obtain are \eqref{Q_evol_main}, \eqref{G_adj_evol}, \eqref{Gi_evol_main}, and \eqref{Gi_adj_evol_main} in Sec.~\ref{sec:evo_op}. The same equations are derived again in Sec.~\ref{sec:hel_evo_bfm} using the background field method (see Eqs.~\eqref{V10evol} and \eqref{V10i_evol} there). These equations contain leading logarithms in the longitudinal integral in their kernels, along with the exact transverse integrations, similar to the unpolarized small-$x$ evolution \cite{Kuraev:1977fs,Balitsky:1978ic,Mueller:1994rr,Mueller:1994jq,Mueller:1995gb,Balitsky:1995ub,Balitsky:1998ya,Kovchegov:1999yj,Kovchegov:1999ua,Jalilian-Marian:1997dw,Jalilian-Marian:1997gr,Weigert:2000gi,Iancu:2001ad,Iancu:2000hn,Ferreiro:2001qy}. Using the technique of \cite{Kovchegov:2015pbl, Kovchegov:2018znm} we take the DLA limit of those equations obtaining closed large-$N_c$ evolution equations \eqref{eq_LargeNc} in Sec.~\ref{sec:large-Nc}. Similarly, the large-$N_c \& N_f$ evolution equations are studied in Sec.~\ref{sec:large-NcNf}, resulting in the closed system of equations \eqref{eq_LargeNcNf}. These large-$N_c$ and large-$N_c \& N_f$ equations extend and generalize the results of \cite{Kovchegov:2015pbl, Kovchegov:2018znm}. We cross-check our large-$N_c$ evolution equations \eqref{eq_LargeNc} against the small-$x$ limit of the pure-glue spin-dependent DGLAP evolution in Sec.~\ref{sec:cross-check} and find an agreement up to and including three loops \cite{Mertig:1995ny,Moch:2014sna}, the highest-known order for the spin-dependent DGLAP splitting functions.

The large-$N_c$ evolution equations \eqref{eq_LargeNc} are solved numerically in Sec.~\ref{sec:numerics}, following the technique of \cite{Kovchegov:2016weo, Kovchegov:2020hgb, Adamiak:2021ppq}. The resulting numerical solution for the amplitudes $G$ (defined in \eq{Ggdef} with $Q \approx G$ at large $N_c$) and $G_2$ is plotted in Fig. \ref{fig:ln_3d}. The extracted intercepts are given in \eq{intercept_results}; within the uncertainty, they are the same for both amplitudes. This leads to the asymptotics of \eq{small_x_asymp}, in complete agreement with BER, \eq{asymptotics}. 

We conclude in Sec.~\ref{sec:conclusion} by outlining future directions of this research program.


\section{Sub-eikonal Quark and Gluon $S$-Matrices in the Background Field}
\label{sec:S-matrix}

We define our light-cone coordinates by $x^\pm = (t \pm z)/\sqrt{2}$, while the transverse vectors are denoted by ${\un x} = (x^1, x^2)$ with ${\un x}_{ij} = {\un x}_i - {\un x}_j$ and $x_{ij} = |{\un x}_{ij}|$ for $i,j = 0,1,2, \ldots$ labeling the partons. Our proton is always moving in the light-cone plus direction, while the projectile quarks and gluons are moving in the light-cone minus direction. The gluon field is denoted by $A_\mu^a$, while the quark and anti-quark fields are $\psi$ and $\bar \psi$. The calculations for small-$x$ evolution will be carried out in $A^{a \, -} =0$ gauge. However, the expressions for the operators in this Section are also valid in the Lorenz gauge $\pd_\mu A^{a \, \mu} =0$, and in any gauge where the gluon field vanishes at $x^- \to \pm \infty$ (cf. \cite{Altinoluk:2020oyd, Kovchegov:2021iyc}).  

We denote the fundamental light-cone Wilson lines by
\begin{align}\label{Vline}
V_{\un{x}} [x^-_f,x^-_i] = \mathcal{P} \exp \left[ ig \int\limits_{x^-_i}^{x^-_f} d{x}^- A^+ (0^+, x^-, \un{x}) \right]
\end{align}
with the abbreviation $V_{\un{x}} = V_{\un{x}} [\infty, - \infty]$ for infinite lines. Here $\mathcal{P}$ is the path ordering operator, $A^{\mu} = \sum_a A^{a \, \mu} \, t^a$ is the background gluon field with $t^a$ the fundamental SU($N_c$) generators, and $g$ is the strong coupling constant. The adjoint light-cone Wilson line is defined similarly as
\begin{align}\label{Uline}
U_{\un{x}} [x^-_f,x^-_i] = \mathcal{P} \exp \left[ ig \int\limits_{x^-_i}^{x^-_f} d{x}^- {\cal A}^+ (0^+, x^-, \un{x}) \right]
\end{align}
with ${\cal A}^{\mu} = \sum_a A^{a \, \mu} \, T^a$, where $T^a$ are the adjoint SU($N_c$) generators, $(T^a)_{bc} = - i f^{abc}$. Again, $U_{\un{x}} = U_{\un{x}} [\infty, - \infty]$.

Define an $S$-matrix for the quark--target scattering in the helicity basis\footnote{Helicity basis refers to the $(\pm)$-interchanged Brodsky-Lepage spinor basis defined below in \eq{anti-BLspinors}, which is commonly used in LCPT.
In LCPT, the $e_z$ spatial direction is used for spin quantization: we refer to the projection of particle's spin onto the $z$-axis as \textit{``helicity''}. We note that the proper helicity is the projection of spin onto the momentum of the particle. This difference between the true helicity and LCPT ``helicity" requires us to keep both $\delta_{\sigma,\sigma'}$ and $\sigma\delta_{\sigma,\sigma'}$ structures in, e.g., \eq{V_sub-eikonal}, even when only helicity-dependent quantities are considered. In the rest of the manuscript, both helicity basis and $(\pm)$-interchanged Brodsky-Lepage spinor basis are used interchangeably.} by 
\begin{align}\label{Vxy}
V_{\un{x}, \un{y}; \sigma', \sigma}  \equiv \int \frac{d^2 p_{in}}{(2\pi)^2} \, \frac{d^2 p_{out}}{(2\pi)^2} \, e^{i \un{p}_{out} \cdot \un{x} - i \un{p}_{in} \cdot \un{y}} \ \left[ \delta_{\sigma, \sigma'} \, (2 \pi)^2 \, \delta^2 \left( \un{p}_{out} - \un{p}_{in} \right) + i \, A^q_{\sigma', \sigma} (\un{p}_{out}, \un{p}_{in}) \right],
\end{align}
where $A^q (\un{p}_{out}, \un{p}_{in})$ is the scattering amplitude for a quark on a target with $\un{p}_{in}$ and $\un{p}_{out}$ the incoming and outgoing quark transverse momenta, respectively, while $\sigma'$ and $\sigma$ are the outgoing and incoming quark helicities. The amplitude $A$ is normalized such that $A = M/(2 s)$ \cite{Kovchegov:2012mbw}, where $M$ is the conventional textbook scattering amplitude and $s$ is the center-of-mass energy squared. 

Neglecting the quark mass, which does not affect small-$x$ evolution, at the sub-eikonal order the quark $S$-matrix is \cite{Balitsky:2015qba,Altinoluk:2020oyd,Kovchegov:2021iyc,Kovchegov:2018znm,Chirilli:2018kkw}\footnote{Note that the sign in front of the $\gamma^+ \gamma^5$ term in \eq{V_sub-eikonal} is different from that in \cite{Kovchegov:2018znm,Kovchegov:2021iyc} while in agreement with \cite{Kovchegov:2018zeq}. In \cite{Kovchegov:2018znm} one has to correct Eq.~(45) by replacing $\rho(\sigma) \to \rho (-\sigma)$ on its right-hand side. Similarly, one should replace $\rho^T (\sigma') \to \rho^T (- \sigma')$ in Eq.~(48) of \cite{Kovchegov:2018znm}. This would modify Eq.~(51) of \cite{Kovchegov:2018znm} to agree with our \eq{V_sub-eikonal}.}
\begin{align}
\label{V_sub-eikonal}
& V_{\un{x}, \un{y}; \sigma', \sigma} = V_{\un{x}} \, \delta^2 (\un{x} - \un{y}) \, \delta_{\sigma,\sigma'} \\
&  + \frac{i \, P^+}{s} \int\limits_{-\infty}^{\infty} d{z}^- d^2 z \ V_{\un{x}} [ \infty, z^-] \, \delta^2 (\un{x} - \un{z}) \, \left[ -   \delta_{\sigma,\sigma'}  \cev{D}^i \, D^i  + g \, \sigma \, \delta_{\sigma, \sigma'} \, F^{12} \right] \! (z^-, {\un z}) \, V_{\un{y}} [ z^-, -\infty] \, \delta^2 (\un{y} - \un{z}) \notag \\
& - \frac{g^2 P^+}{2 \, s} \delta^2 (\un{x} - \un{y}) \!\! \int\limits_{-\infty}^{\infty} \!\! d{z}_1^- \! \int\limits_{z_1^-}^\infty d z_2^- V_{\un{x}} [ \infty, z_2^-] \, t^b \, \psi_{\beta} (z_2^-,\un{x}) \, U_{\un{x}}^{ba} [z_2^-,z_1^-] \left[ \delta_{\sigma,\sigma'} \, \gamma^+  - \sigma \, \delta_{\sigma, \sigma'} \, \gamma^+ \gamma^5 \right]_{\alpha \beta} \bar{\psi}_\alpha (z_1^-,\un{x}) \, t^a \, V_{\un{x}} [ z_1^-, -\infty] , \notag
\end{align}
where $D^i = \pd^i - i g A^i$, and $\cev{D}^i = \cev{\pd}^i + i g A^i$. 

The $S$-matrix in \eq{V_sub-eikonal} has two distinct polarization-dependent structures, $\sigma \, \delta_{\sigma, \sigma'}$ and $\delta_{\sigma, \sigma'}$. At the sub-eikonal level (that is, for everything except for the first term on the right-hand side of \eq{V_sub-eikonal}), we define the ``polarized Wilson lines" $V_{\un x}^{\textrm{pol} [1]}$ and $V_{{\ul x}, {\un y}}^{\textrm{pol} [2]}$ by \cite{Kovchegov:2018znm,Kovchegov:2021iyc}
\begin{align}\label{Vxy_sub-eikonal}
V_{\un{x}, \un{y}; \sigma', \sigma} \bigg|_{\textrm{sub-eikonal}}  \equiv \sigma \, \delta_{\sigma, \sigma'} \, V_{\un x}^{\textrm{pol} [1]} \, \delta^2 ({\un x} - {\un y}) + \delta_{\sigma, \sigma'} \, V_{{\ul x}, {\un y}}^{\textrm{pol} [2]}.
\end{align}
$V_{\un x}^{\textrm{pol} [1]}$ and $V_{{\ul x}, {\un y}}^{\textrm{pol} [2]}$ can be read off \eq{V_sub-eikonal} using their definition in \eq{Vxy_sub-eikonal}. In the following it will be helpful to separate the gluon and quark contributions to $V_{\un x}^{\textrm{pol} [1]}$ and $V_{{\ul x}, {\un y}}^{\textrm{pol} [2]}$. Therefore, we define
\begin{align}\label{VqG_decomp}
V_{\un x}^{\textrm{pol} [1]} = V_{\un x}^{\textrm{G} [1]} + V_{\un x}^{\textrm{q} [1]}, \ \ \  V_{{\ul x}, {\un y}}^{\textrm{pol} [2]} = V_{{\ul x}, {\un y}}^{\textrm{G} [2]} + V_{{\ul x}}^{\textrm{q} [2]} \, \delta^2 ({\un x} - {\un y}) ,
\end{align}
such that
\begin{subequations}\label{VqG}
\begin{align}
& V_{\un x}^{\textrm{G} [1]}  = \frac{i \, g \, P^+}{s} \int\limits_{-\infty}^{\infty} d{x}^- V_{\un{x}} [ \infty, x^-] \, F^{12} (x^-, {\un x}) \, \, V_{\un{x}} [ x^-, -\infty]  , \label{VG1} \\
& V_{\un x}^{\textrm{q} [1]}  = \frac{g^2 P^+}{2 \, s} \int\limits_{-\infty}^{\infty} \!\! d{x}_1^- \! \int\limits_{x_1^-}^\infty d x_2^- V_{\un{x}} [ \infty, x_2^-] \, t^b \, \psi_{\beta} (x_2^-,\un{x}) \, U_{\un{x}}^{ba} [x_2^-, x_1^-] \, \left[ \gamma^+ \gamma^5 \right]_{\alpha \beta} \, \bar{\psi}_\alpha (x_1^-,\un{x}) \, t^a \, V_{\un{x}} [ x_1^-, -\infty] , \label{Vq1} \\
& V_{{\ul x}, {\un y}}^{\textrm{G} [2]}  = - \frac{i \, P^+}{s} \int\limits_{-\infty}^{\infty} d{z}^- d^2 z \ V_{\un{x}} [ \infty, z^-] \, \delta^2 (\un{x} - \un{z}) \, \cev{D}^i (z^-, {\un z}) \, D^i  (z^-, {\un z}) \, V_{\un{y}} [ z^-, -\infty] \, \delta^2 (\un{y} - \un{z}) , \label{VxyG2} \\
& V_{{\ul x}}^{\textrm{q} [2]} = - \frac{g^2 P^+}{2 \, s} \int\limits_{-\infty}^{\infty} \!\! d{x}_1^- \! \int\limits_{x_1^-}^\infty d x_2^- V_{\un{x}} [ \infty, x_2^-] \, t^b \, \psi_{\beta} (x_2^-,\un{x}) \, U_{\un{x}}^{ba} [x_2^-, x_1^-] \, \left[ \gamma^+ \right]_{\alpha \beta} \, \bar{\psi}_\alpha (x_1^-,\un{x}) \, t^a \, V_{\un{x}} [ x_1^-, -\infty] \label{Vq2}.
\end{align}
\end{subequations}
Curiously, only $V_{{\ul x}, {\un y}}^{\textrm{G} [2]}$ is truly a non-local operator in the transverse plane. 

Similar to \eq{Vxy} we define the $S$-matrix for the gluon-target scattering by
\begin{align}\label{Uxy}
U_{\un{x}, \un{y}; \lambda', \lambda}  \equiv \int \frac{d^2 p_{in}}{(2\pi)^2} \, \frac{d^2 p_{out}}{(2\pi)^2} \, e^{i \un{p}_{out} \cdot \un{x} - i \un{p}_{in} \cdot \un{y}} \ \left[ \delta_{\lambda, \lambda'} \, (2 \pi)^2 \, \delta^2 \left( \un{p}_{out} - \un{p}_{in} \right) + i \, A^G_{\lambda', \lambda} (\un{p}_{out}, \un{p}_{in}) \right],
\end{align}
with the gluon scattering amplitude $A^G_{\lambda', \lambda} (\un{p}_{out}, \un{p}_{in})$ on the background-field target normalized in the same way as the quark one above. At the sub-eikonal level the $S$-matrix is \cite{Balitsky:2015qba,Altinoluk:2020oyd,Kovchegov:2021iyc,Kovchegov:2018znm,Chirilli:2021lif}\footnote{Similar to \eq{V_sub-eikonal}, the sign of the $\gamma^+ \gamma^5$ term in \eq{U_sub-eikonal} is different from that in Eq.~(64) of \cite{Kovchegov:2018znm}: correcting $\rho^T (\sigma) \to \rho^T (-\sigma)$ in Eq.~(58) and $\rho(\sigma) \to \rho (-\sigma)$ in Eq. (60), both in \cite{Kovchegov:2018znm}, would change the sign of the $\gamma^+ \gamma^5$ term in Eq.~(64) of \cite{Kovchegov:2018znm}, making it agree with our \eq{U_sub-eikonal}.}
\begin{align}
\label{U_sub-eikonal}
& (U_{\un{x}, \un{y}; \lambda', \lambda})^{ba} = (U_{\un{x}})^{ba} \, \delta^2 (\un{x} - \un{y}) \, \delta_{\lambda, \lambda'} \\
& + \frac{i P^+}{s} \int\limits_{-\infty}^\infty d z^- d^2 z \, (U_{\un{x}} [\infty, z^-])^{bb'} \delta^2 ({\un x} - {\un z}) \left[  2 g \lambda \, \delta_{\lambda, \lambda'} ({\cal F}^{12})^{b'a'} - \delta_{\lambda, \lambda'} \, \cev{\underline{\mathscr{D}}}^{b'c} \cdot  \underline{\mathscr{D}}^{ca'} \right] \!\! (z^-, {\un z}) \, (U_{\un{y}} [z^-, -\infty])^{a'a} \delta^2 ({\un y} - {\un z}) \notag \\
& - \frac{g^2 P^+}{2 \, s} \delta^2 (\un{x} - \un{y}) \!\! \int\limits_{-\infty}^{\infty} \!\! d{z}_1^- \int\limits_{z_1^-}^\infty d z_2^- \notag \\ 
& \hspace*{2cm} \times \, (U_{\un{x}} [ \infty, z_2^-])^{bb'} \, \bar{\psi} (z_2^-,\un{x}) \, t^{b'} \, V_{\un{x}} [z_2^-,z_1^-] \, \left[ \delta_{\lambda, \lambda'} \, \gamma^+  - \lambda \, \delta_{\lambda, \lambda'} \, \gamma^+ \gamma^5 \right] \, t^{a'} \, \psi (z_1^-,\un{x}) \,  (U_{\un{x}} [ z_1^-, -\infty])^{a'a} - \mbox{c.c.} . \notag
\end{align}
Here ${\cal F}^{12} = \sum_a { F}^{a \, 12}  \, T^a$ is the adjoint gluon field strength tensor, while the adjoint covariant derivatives are $\cev{\underline{\mathscr{D}}}^{ab} = \cev{\un{\nabla}} \, \delta^{ab} + g f^{acb} \un{A}^c$  and  $\underline{\mathscr{D}}^{ab} = \un{\nabla} \, \delta^{ab} - g f^{acb} \un{A}^c$ (or, simply, $\mathscr{D}^{ab}_i = \partial_i\delta^{ab} - ig (T^c)_{ab} A_i^c$ and $\cev{\mathscr{D}}^{ab}_i = \cev{\partial}_i\delta^{ab} + ig (T^c)_{ab} A_i^c$, using $(T^c)_{ab} = -if^{abc}$ with ${\un A} = (A^1, A^2) = - (A_1, A_2)$).

Just as in the fundamental representation, for the adjoint $S$-matrix at hand we can identify two polarization structures, $\lambda \, \delta_{\lambda, \lambda'}$ and $\delta_{\lambda, \lambda'}$, and define $U_{{\ul x}}^{\textrm{pol} [1]}$ and $U_{{\ul x}, {\un y}}^{\textrm{pol} [2]}$ by
\begin{align}\label{Uxy_sub-eikonal}
(U_{{\ul x}, {\un y}; \lambda', \lambda})^{b a} \bigg|_{\textrm{sub-eikonal}}  \equiv \lambda \, \delta_{\lambda, \lambda'} \, (U_{{\ul x}}^{\textrm{pol} [1]})^{b a} \, \delta^2 ({\un x} - {\un y}) + \delta_{\lambda, \lambda'} \, (U_{{\ul x}, {\un y}}^{\textrm{pol} [2]})^{b a} 
\end{align}
at the sub-eikonal order, excluding the eikonal term in \eq{U_sub-eikonal}. Again, we separate the quark and gluon operator contributions by writing
\begin{align}\label{UqG_decomp}
U_{\un x}^{\textrm{pol} [1]} = U_{\un x}^{\textrm{G} [1]} + U_{\un x}^{\textrm{q} [1]}, \ \ \  U_{{\ul x}, {\un y}}^{\textrm{pol} [2]} = U_{{\ul x}, {\un y}}^{\textrm{G} [2]} + U_{{\ul x}}^{\textrm{q} [2]} \, \delta^2 ({\un x} - {\un y}) ,
\end{align}
with
\begin{subequations}\label{UqG}
\begin{align}
& (U_{\un x}^{\textrm{G} [1]})^{ba} = \frac{2 \, i \, g \, P^+}{s} \int\limits_{-\infty}^{\infty} d{x}^- (U_{\un{x}} [ \infty, x^-])^{bb'} \, ({\cal F}^{12})^{b'a'} (x^-, {\un x}) \, (U_{\un{x}} [ x^-, -\infty])^{a'a}  , \label{UG1} \\
& (U_{\un x}^{\textrm{q} [1]})^{ba} = \frac{g^2 P^+}{2 \, s} \!\! \int\limits_{-\infty}^{\infty} \!\! d{x}_1^- \! \int\limits_{x_1^-}^\infty d x_2^- (U_{\un{x}} [ \infty, x_2^-])^{bb'} \bar{\psi} (x_2^-,\un{x}) \, t^{b'} V_{\un{x}} [x_2^-,x_1^-] \, \gamma^+ \gamma^5 \, t^{a'} \psi (x_1^-,\un{x})  (U_{\un{x}} [ x_1^-, -\infty])^{a'a} + \mbox{c.c.}  ,  \label{Uq1} \\
& (U_{{\ul x}, {\un y}}^{\textrm{G} [2]})^{ba}  = - \frac{i \, P^+}{s} \int\limits_{-\infty}^{\infty} d{z}^- d^2 z \ (U_{\un{x}} [ \infty, z^-])^{bb'} \, \delta^2 (\un{x} - \un{z}) \,\cev{\underline{\mathscr{D}}}^{b'c} (z^-, {\un z}) \, \underline{\mathscr{D}}^{ca'}  (z^-, {\un z}) \, (U_{\un{y}} [ z^-, -\infty])^{a'a} \, \delta^2 (\un{y} - \un{z}) , \label{UG2}  \\
& (U_{{\ul x}}^{\textrm{q} [2]} )^{ba} = - \frac{g^2 P^+}{2 \, s} \int\limits_{-\infty}^{\infty} \!\! d{x}_1^- \! \int\limits_{x_1^-}^\infty d x_2^- (U_{\un{x}} [ \infty, x_2^-])^{bb'} \, \bar{\psi} (x_2^-,\un{x}) \, t^{b'} \, V_{\un{x}} [x_2^-,x_1^-] \, \gamma^+ \, t^{a'} \, \psi (x_1^-,\un{x}) \,  (U_{\un{x}} [ x_1^-, -\infty])^{a'a} - \mbox{c.c.} . \label{Uq2}
\end{align}
\end{subequations}
Once more, only $U_{{\ul x}, {\un y}}^{\textrm{G} [2]}$ is non-local in the transverse plane.


\section{Quark and Gluon Helicity Distributions and $g_1$ Structure Function at Small $x$}
\label{sec:PDFs+g1}


\subsection{Gluon Helicity Distribution}
\label{sec:glue_helicity}

We begin with the dipole gluon helicity TMD, defined as \cite{Bomhof:2006dp}
\begin{align}\label{glue_hel_TMD}
  g_{1L}^{G \, dip} (x, k_T^2) &= \frac{-2 i}{x \, P^+}
  \frac{1}{(2\pi)^3} \, \half\sum_{S_L} S_L \, \int d \xi^- \, d^2\xi
  \: \: e^{i x P^+ \, \xi^- } \: e^{
    - i \un{k} \cdot \un{\xi} }
 \\ & \hspace{1cm} \times
\bra{P, S_L} \epsilon^{ij} \, \tr \left[ F^{+i} (0) \: {\cal
    U}^{[+]}[0,\xi] \: F^{+j} (\xi) \: {\cal U}^{[-]} [\xi, 0]
\right] \ket{P, S_L}_{\xi^+ = 0}, \notag
\end{align}
where $\mathcal{U}^{[+]}$ and $\mathcal{U}^{[-]}$ are the future- and past-pointing Wilson line staples, $k_T = |{\un k}|$, and $\epsilon^{ij}$ is the transverse Levi-Civita symbol with $\epsilon^{12} = +1$.
The Jaffe-Manohar (JM) gluon helicity PDF is then \cite{Jaffe:1989jz}
\begin{align} 
  \label{e:coll}
  \Delta G (x, Q^2) & = \int\limits^{Q^2} d^2 k \, g_{1L}^{G \, dip} (x, k_T^2) = \frac{-2 i}{x \, P^+}
  \frac{1}{4\pi} \, \half\sum_{S_L} S_L \, \int\limits_{-\infty}^\infty d \xi^- \, e^{i x P^+ \, \xi^- } 
  \\ 
%
%
& \hspace{1cm} \times \bra{P, S_L} \epsilon^{ij} \, F^{a+i} (0^+, 0^- , {\un 0}) \: 
    U^{ab}_{\un 0} [0,\xi^-] \: F^{b+j} (0^+, \xi^- , {\un 0}) 
 \ket{P, S_L} , \notag
\end{align}
where $U^{ab}_{\un 0}$ now is a regular adjoint light-cone Wilson line \eqref{Uline} connecting the two points in the correlator. 

We rewrite \eq{glue_hel_TMD} as
\begin{align}\label{glue_hel_TMD2}
  g_{1L}^{G \, dip} (x, k_T^2) &= \frac{-2 i}{x \, P^+ \, V^-}
  \frac{1}{(2\pi)^3} \, \half\sum_{S_L} S_L \, \int d \xi^- \, d^2\xi \, d \zeta^- \, d^2\zeta \:
   e^{i x P^+ \, (\xi^- - \zeta^-) } \: e^{
    - i \un{k} \cdot ( \un{\xi} - \un{\zeta}) }
 \\ & \hspace{1cm} \times
\bra{P, S_L} \epsilon^{ij} \, \tr \left[ F^{+i} (\zeta) \: {\cal
    U}^{[+]}[\zeta,\xi] \: F^{+j} (\xi) \: {\cal U}^{[-]} [\xi, \zeta]
\right] \ket{P, S_L}_{\xi^+ = \zeta^+ = 0}, \notag
\end{align}
with the (infinite) volume factor $V^- = \int d x^- d^2 x$. The JM gluon helicity PDF is now given by 
\begin{align}\label{glue_hel_PDF}
  \Delta G (x, Q^2) &= \int\limits^{Q^2} d^2 k \, g_{1L}^{G \, dip} (x, k_T^2) = \frac{-2 i}{x \, P^+ \, L^-}
  \frac{1}{4\pi} \, \half\sum_{S_L} S_L \, \int d \xi^- \, d \zeta^- \, e^{i x P^+ \, (\xi^- - \zeta^-) } 
 \\ & \hspace{1cm} \times
\bra{P, S_L} \epsilon^{ij} \, F^{a+i} (0^+, \zeta^-, {\un 0}) \: 
    U^{ab}_{\un 0} [\zeta^- , \xi^-] \: F^{b+j} (0^+, \xi^- , {\un 0})
 \ket{P, S_L} ,\notag
\end{align}
where $L^- = \int d x^-$.

In any gauge where the field $A^\perp$ is zero at $x^- \to \pm \infty$ we can rewrite \eq{glue_hel_PDF} as
\begin{align}\label{glue_hel_PDF200}
  \Delta G (x, Q^2) &= \frac{-2 i}{x \, P^+ \, L^-}
  \frac{1}{2\pi} \, \half\sum_{S_L} S_L \, \int\limits_{-\infty}^\infty d \xi^- \, d \zeta^- \, e^{i x P^+ \, (\xi^- - \zeta^-) } 
\\ &  \times
\bra{P, S_L} \epsilon^{ij} \, \tr \Big[ V_{\un 0} [-\infty, \zeta^-] \, F^{+i} (0^+, \zeta^-, {\un 0}) \: V_{\un 0} [\zeta^-, \infty] \, V_{\un 0} [\infty, \xi^-] \,
    F^{+j} (0^+, \xi^- , {\un 0}) \, V_{\un 0} [\xi^-, - \infty] \Big] \notag 
 \ket{P, S_L} .
\end{align}
We further note that in a gauge where the field $A^\perp$ is zero at $x^- \to \pm \infty$ we have \cite{Kovchegov:2017lsr,Hatta:2016aoc}
\begin{align}\label{JMexp1}
& \int\limits_{-\infty}^\infty d \xi^- e^{i x P^+ \, \xi^-} V_{\un 0} [\infty, \xi^-] \,
    F^{+j} (0^+, \xi^- , {\un 0}) \, V_{\un 0} [\xi^-, - \infty]   \\ 
& = \int\limits_{-\infty}^\infty d \xi^- e^{i x P^+ \, \xi^-} \left\{  \pd^+ \left( V_{\un 0} [\infty, \xi^-] \,
    A^{j} (0^+, \xi^- , {\un 0}) \, V_{\un 0} [\xi^-, - \infty] \right)  - V_{\un 0} [\infty, \xi^-] \,
    (\pd^j A^+) \, V_{\un 0} [\xi^-, - \infty] \right\} \notag \\ 
& = - \int\limits_{-\infty}^\infty d \xi^- e^{i x P^+ \, \xi^-} V_{\un 0} [\infty, \xi^-] \,
    (\pd^j A^+ + i x P^+ \, A^j) \, V_{\un 0} [\xi^-, - \infty], \notag
\end{align}
such that 
\begin{align}\label{glue_hel_PDF3}
  \Delta G (x, Q^2) &= \frac{-2 i}{x \, P^+ \, L^-}
  \frac{1}{2\pi} \, \half\sum_{S_L} S_L \, \int\limits_{-\infty}^\infty d \xi^- \, d \zeta^- \, e^{i x P^+ \, (\xi^- - \zeta^-) } 
\\ &  \times
\bra{P, S_L} \epsilon^{ij} \, \tr \Big[ (\pd^i A^+ - i x P^+ \, A^i) \: V_{\un 0} [\zeta^-, \xi^-] \,
    (\pd^j A^+ + i x P^+ \, A^j) \, V_{\un 0} [\xi^-, \zeta^-] \Big] \notag 
 \ket{P, S_L} .
\end{align}

Similarly, for the dipole gluon helicity TMD we write
\begin{align}\label{glue_hel_TMD3}
  g_{1L}^{G \, dip} & (x, k_T^2) = \frac{-2 i}{x \, P^+ \, V^-}
  \frac{1}{(2\pi)^3} \, \half\sum_{S_L} S_L \, \int\limits_{-\infty}^\infty d \xi^- \, d^2\xi \, d \zeta^- \, d^2\zeta \:
   e^{i x P^+ \, (\xi^- - \zeta^-) } \: e^{
    - i \un{k} \cdot ( \un{\xi} - \un{\zeta}) }
\\ &  \times
\bra{P, S_L} \epsilon^{ij} \, \tr \Big[ V_{\un \zeta} [-\infty, \zeta^-] \, (\pd^i A^+ - i x P^+ \, A^i) \: V_{\un \zeta} [\zeta^-, \infty] \, V_{\un \xi} [\infty, \xi^-] \,
    (\pd^j A^+ + i x P^+ \, A^j) \, V_{\un \xi} [\xi^-, -\infty] \Big] \notag 
 \ket{P, S_L} .
\end{align}

Let us simplify the gluon helicity TMD operator \eqref{glue_hel_TMD3} at small $x$, expanding it down to sub-eikonal order. Start by defining a ``Lipatov vertex"
\begin{align}\label{Lj}
L^j (x, {\un k}) \equiv \int\limits_{-\infty}^\infty d \xi^- \, d^2\xi \, e^{i x P^+ \, \xi^- - i \un{k} \cdot \un{\xi} } \ V_{\un \xi} [\infty, \xi^-] \,
    (\pd^j A^+ + i x P^+ \, A^j) \, V_{\un \xi} [\xi^-, -\infty]
\end{align}
and rewriting the gluon dipole helicity TMD as
\begin{align}\label{glue_hel_TMD44}
  g_{1L}^{G \, dip} (x, k_T^2) = \frac{-2 i}{x \, P^+ \, V^-}
  \frac{1}{(2\pi)^3} \, \half\sum_{S_L} S_L \, \bra{P, S_L} \epsilon^{ij} \, \tr \Big[  L^{i \, \dagger} (x, {\un k}) \, L^j (x, {\un k}) \Big] 
 \ket{P, S_L} .
\end{align}

Next let us expand the Lipatov vertex \eqref{Lj} in powers of $x$, that is, in {\sl eikonality}. We get
\begin{align}\label{Lj2}
L^j (x, {\un k}) = \int\limits_{-\infty}^\infty d \xi^- \, d^2\xi \, e^{- i \un{k} \cdot \un{\xi} } \ V_{\un \xi} [\infty, \xi^-] \,
    \left[ \pd^j A^+ + i x P^+ \left(\xi^- \, \pd^j A^+ + A^j \right) + {\cal O} (x^2) \right] \, V_{\un \xi} [\xi^-, -\infty] .
\end{align}
At order-$x^0$ we get the standard result, 
\begin{align}\label{x0ord}
\int\limits_{-\infty}^\infty d \xi^- \, d^2\xi \, e^{- i \un{k} \cdot \un{\xi} } \, V_{\un \xi} [\infty, \xi^-] \, \left(
    \pd^j A^+  \right)\, V_{\un \xi} [\xi^-, -\infty] = \int d^2\xi \, e^{- i \un{k} \cdot \un{\xi} } \, \frac{1}{i g} \, \pd^j V_{\un \xi} = - \frac{k^j}{g} \, \int d^2\xi \, e^{- i \un{k} \cdot \un{\xi} } \, V_{\un \xi} .
\end{align}
At order-$x$, let us simplify the $\xi^- \, \pd^j A^+$ term. Writing
\begin{align}
\label{Ditrick}
    \xi^- = \lim_{L^- \to +\infty} \frac{1}{2} \, \left[ - \int\limits_{\xi^-}^{L^-/2} d z^- + \int\limits^{\xi^-}_{-L^-/2} d z^- \right]
\end{align}
we obtain
\begin{align}\label{Dishiftderiv}
    & i x P^+  \int\limits_{-\infty}^\infty d \xi^- \, d^2\xi \, e^{- i \un{k} \cdot \un{\xi} } \ V_{\un \xi} [\infty, \xi^-] \,
     \left[\xi^- \,  \pd^j A^+ (\xi) \right]\, V_{\un \xi} [\xi^-, -\infty]  \\ 
& = - \frac{x P^+}{2 g} \int d^2\xi \, e^{- i \un{k} \cdot \un{\xi} } \, \int\limits_{-\infty}^\infty d z^- \, V_{\un \xi} [\infty, z^-] \, \left[ \pd^j  - \cev{\pd}^j \right] \,  V_{\un \xi} [z^-, -\infty].  \nonumber
\end{align}

The entire Lipatov vertex becomes 
\begin{align}\label{Lj3}
L^j (x, {\un k}) = - \frac{k^j}{g} \, \int d^2\xi \, e^{- i \un{k} \cdot \un{\xi} } \, V_{\un \xi} - \frac{x P^+}{2 g} \int d^2\xi \, e^{- i \un{k} \cdot \un{\xi} } \, \int\limits_{-\infty}^\infty d z^- \, V_{\un \xi} [\infty, z^-] \, \left[ D^j - {\cev D}^j \right] \,  V_{\un \xi} [z^-, -\infty] + {\cal O} (x^2),
\end{align}
where we have employed the right-acting covariant derivative $D^j = \pd^j - i g A^j$ and the left-acting covariant derivative ${\cev D}^j = {\cev \pd}^j + i g A^j$ (see \cite{Lorce:2012ce} for the $D^j - {\cev D}^j$ operator arising in the definitions of quark OAM in the proton). Substituting \eq{Lj3} into \eq{glue_hel_TMD44} and expanding the latter to order-$x$, we see that only the cross-talk between the leading-order term and the $D^j  - {\cev D}^j$ term in \eq{Lj3} survives, yielding
\begin{align}\label{glue_hel_TMD444}
  g_{1L}^{G \, dip} & (x, k_T^2) = \frac{- 2 i s}{P^+ \, V^- \, g^2}
  \frac{1}{(2\pi)^3} \, \half\sum_{S_L} S_L \, \epsilon^{ij} \, k^i \, \int d^2 \zeta \, d^2 \xi \, e^{
    - i \un{k} \cdot ( \un{\xi} - \un{\zeta}) } \, \bra{P, S_L}  \tr \Big[  V^\dagger_{\un \zeta} \, V_{\un{\xi}}^{j \, \textrm{G} [2]}  - \left( V_{\un{\zeta}}^{j \, \textrm{G} [2]} \right)^\dagger  V_{\un \xi} \Big]  \ket{P, S_L} ,
\end{align}
where we have defined the fundamental polarized Wilson line of a different type from those in Eqs.~\eqref{VqG} above, by
\begin{align}\label{Vi}
V_{\un{z}}^{i \, \textrm{G} [2]} \equiv \frac{P^+}{2 s} \, \int\limits_{-\infty}^{\infty} d {z}^- \, V_{\un{z}} [ \infty, z^-] \, \left[ {D}^i (z^-, \un{z}) - \cev{D}^i (z^-, \un{z}) \right] \, V_{\un{z}} [ z^-, -\infty]  .
\end{align}

Defining the standard (but polarization-dependent) ``CGC averaging" by
\begin{align}
\Big\langle \ldots \Big\rangle \equiv \half\sum_{S_L} S_L \, \frac{1}{2 P^+ V^-} \, \bra{P, S_L} \ldots \ket{P, S_L}
\end{align}
and the sub-eikonal one by \cite{Kovchegov:2015pbl}
\begin{align}\label{double_angle_brackets}
\llangle \ldots \rrangle \equiv s \,  \Big\langle \ldots \Big\rangle
\end{align}
we recast \eq{glue_hel_TMD444} as
\begin{align}\label{glue_hel_TMD445}
  g_{1L}^{G \, dip} & (x, k_T^2) = \frac{- 4 i}{g^2 \, (2\pi)^3}
   \, \epsilon^{ij} \, k^i \, \int d^2 \zeta \, d^2 \xi \, e^{
    - i \un{k} \cdot ( \un{\xi} - \un{\zeta}) } \, \llangle  \tr \Big[  V^\dagger_{\un \zeta} \, V_{\un{\xi}}^{j \, \textrm{G} [2]}  - \left( V_{\un{\zeta}}^{j \, \textrm{G} [2]} \right)^\dagger  V_{\un \xi} \Big]  \rrangle .
\end{align}
This result should be compared to Eq.~(35) in \cite{Kovchegov:2017lsr}. The definition of the polarized Wilson line in Eq.~(34) of \cite{Kovchegov:2017lsr} is different from our \eq{Vi} by keeping only 
$ 2 ig \, A^i$ instead of the covariant derivative difference, $D^i - \cev{D}^i$ 
and excluding the normalization factor of $\frac{1}{s}$. The former explains the sign difference between our \eq{glue_hel_TMD445} and Eq.~(35) in \cite{Kovchegov:2017lsr}.

Finally, interchanging ${\un \zeta} \leftrightarrow {\un \xi}$ in the second term of \eq{glue_hel_TMD445} and replacing ${\un k} \to - {\un k}$ in the same term (which we can do since each term in \eq{glue_hel_TMD445} depends on $k_T^2$ and does not depend on the direction of $\un k$), we arrive at
\begin{align}\label{glue_hel_TMD446}
  g_{1L}^{G \, dip} & (x, k_T^2) = \frac{-4 i}{g^2 \, (2\pi)^3}
   \, \epsilon^{ij} \, k^i \, \int d^2 \zeta \, d^2 \xi \, e^{
    - i \un{k} \cdot ( \un{\xi} - \un{\zeta}) } \, \llangle  \tr \Big[  V^\dagger_{\un \zeta} \, V_{\un{\xi}}^{j \, \textrm{G} [2]} + \left( V_{\un{\xi}}^{j \, \textrm{G} [2]} \right)^\dagger  V_{\un \zeta} \Big]  \rrangle .
\end{align}
Defining the polarized dipole amplitude of the second kind
\begin{align}\label{Gj}
G^j_{10} (zs) \equiv \frac{1}{2 N_c} \, \llangle  \tr \Big[  V^\dagger_{\un 0} \, V_{\un{1}}^{j \, \textrm{G} [2]} + \left( V_{\un{1}}^{j \, \textrm{G} [2]} \right)^\dagger  V_{\un 0} \Big]  \rrangle
\end{align}
we obtain (cf. Eq.~(38) in \cite{Kovchegov:2017lsr})
\begin{align}\label{glue_hel_TMD447}
  g_{1L}^{G \, dip} & (x, k_T^2) = \frac{- 8 i N_c}{g^2 \, (2\pi)^3}
   \, \epsilon^{ij} \, k^i \, \int d^2 x_0 \, d^2 x_1 \, e^{
    - i \un{k} \cdot \un{x}_{10} } \, G^j_{10} \left( zs = \frac{Q^2}{x} \right) .
\end{align}
Here ${\un x}_{10} = {\un x}_1 - {\un x}_0$ for the transverse-plane position vectors ${\un x}_1$ and ${\un x}_0$, with $x_{10} = |{\un x}_{10}|$ to be used later on.

Similar to \cite{Kovchegov:2017lsr} we can introduce the following decomposition of the impact-parameter integrated amplitude $G^j$:
\begin{align}\label{decomp}
  \int d^2 \left( \frac{x_{1} + x_0}{2} \right) \, G^{i}_{10} (z s) = (x_{10})_\bot^i \, G_1
  (x_{10}^2, z s) + \epsilon^{ij} \, (x_{10})_\bot^j \, G_2
  (x_{10}^2, z s) .
\end{align} 
Substituting this into \eq{glue_hel_TMD446} we see that $G_1$ does not contribute. We get (cf. Eqs.~(40) and (41) in \cite{Kovchegov:2017lsr})
\begin{align}\label{glue_hel_TMD57}
g_{1L}^{G \, dip} (x, k_T^2) & = \frac{8 i N_c}{g^2 \, (2\pi)^3}
   \, \int d^2 x_{10} \, e^{- i \un{k} \cdot \un{x}_{10} } \, {\un k} \cdot {\un x}_{10} \, G_2
  \left(  x_{10}^2,  zs = \frac{Q^2}{x} \right) \\ 
  & = \frac{N_c}{\as 2 \pi^4} \, \int d^2 x_{10} \, e^{- i \un{k} \cdot \un{x}_{10} } \, \left[ 1 + x_{10}^2 \frac{\pd}{\pd x_{10}^2 } \right] \, G_2 \left(  x_{10}^2,  zs = \frac{Q^2}{x} \right). \notag
\end{align}

The gluon helicity PDF is obtained by integrating over ${\un k}$, which yields (cf. Eq.~(124) in \cite{Kovchegov:2017lsr})
\begin{align}\label{JM_DeltaG}
\Delta G (x, Q^2) = \frac{2 N_c}{\as \pi^2} \, \left[ \left( 1 + x_{10}^2 \frac{\pd}{\pd x_{10}^2 } \right) \, G_2 \left(  x_{10}^2,  zs = \frac{Q^2}{x} \right) \right]_{x_{10}^2 = \frac{1}{Q^2}}.
\end{align}

We conclude, just as in \cite{Kovchegov:2017lsr}, that the amplitude $G_2$ gives us both the gluon dipole helicity TMD \eqref{glue_hel_TMD57} and the gluon helicity PDF \eqref{JM_DeltaG} at small $x$. The difference here is in the definition of the operator in \eq{Vi}, which is different here from that employed in \cite{Kovchegov:2017lsr}, where the partial-derivative part of the full covariant derivative was discarded as a term independent of helicity.


\subsection{Quark Helicity Distribution}
\label{sec:quark_helicity}

To include both sub-eikonal terms from \eq{V_sub-eikonal} into quark helicity distribution we can employ the analysis carried out in \cite{Kovchegov:2018znm}, which applies here as well, with the diagram B from \cite{Kovchegov:2018znm} (see \fig{FIG:diagbdet} below) again giving the only contribution we need to keep. Just as in  \cite{Kovchegov:2018znm,Kovchegov:2018zeq}, we will work with the $(\pm)$-interchanged Brodsky-Lepage spinors \cite{Lepage:1980fj}  (referred there as the anti-BL spinors)
\begin{align}\label{anti-BLspinors}
u_\sigma (p) = \frac{1}{\sqrt{\sqrt{2} \, p^-}} \, [\sqrt{2} \, p^- + m \, \gamma^0 +  \gamma^0 \, {\un \gamma} \cdot {\un p} ] \,  \rho (\sigma), \ \ \ v_\sigma (p) = \frac{1}{\sqrt{\sqrt{2} \, p^-}} \, [\sqrt{2} \, p^- - m \, \gamma^0 +  \gamma^0 \, {\un \gamma} \cdot {\un p} ] \,  \rho (-\sigma),
\end{align}
with $p^\mu = \left( \frac{{\un p}^2+ m^2}{2 p^-}, p^-, {\un p} \right)$ and
\begin{align}
  \rho (+1) \, = \, \frac{1}{\sqrt{2}} \, \left(
  \begin{array}{c}
      1 \\ 0 \\ -1 \\ 0
  \end{array}
\right), \ \ \ \rho (-1) \, = \, \frac{1}{\sqrt{2}} \, \left(
  \begin{array}{c}
        0 \\ 1 \\ 0 \\ 1
  \end{array}
\right) .
\end{align}

We begin with Eq.~(15) in \cite{Kovchegov:2018znm}, which we modify by replacing (for the massless quarks we will consider from now on) \cite{Kovchegov:2021iyc}
\begin{align}\label{V_replacement}
{\bar v}_{\sigma_1} (k_1) \left( \hat{V}_{{\un w}}^\dagger \right)^{ji} v_{\sigma_2} (k_2)  \to 2\, \sqrt{k_1^- k_2^-} \, \int d^2 z \, \left( V_{{\un z}, {\un w}; -\sigma_2, -\sigma_1}^\dagger \right)^{ji} ,
\end{align} 
which accounts for both the notation change (to the quark $S$-matrix from \eq{V_sub-eikonal}) and the fact that the anti-quark position may be different on the two sides of the shock wave, as depicted in \fig{FIG:diagbdet}. (Here $i,j$ are the anti-quark color indices. The shock wave, representing the proton target, is shown by the shaded rectangle in \fig{FIG:diagbdet}.) Additionally, we need to replace $e^{i \un{k} \cdot (\un{w} - \un{\zeta})} \to e^{i \un{k} \cdot (\un{z} - \un{\zeta})}$ in the same Eq.~(15) of \cite{Kovchegov:2018znm}. We end up with
\footnote{The overall sign difference between our \eq{TMD15} and Eq.~(15) in \cite{Kovchegov:2018znm} is due to the need to correct the on-shell anti-quark factor in the latter such that  $- \slashed{k}_2 \, 2\pi\delta(k_2^2) \to \slashed{k}_2 \, 2\pi\delta(k_2^2)$.}
\begin{align}\label{TMD15}
g_{1L}^q (x, k_T^2) & =  \frac{2 P^+}{(2\pi)^3}  \:  \int d^{2} \zeta 
\, d^2 w \, d^2 z \, \frac{d^2 k_1 \, d k_1^-}{(2\pi)^3} \, e^{i \ul{k}_1  \cdot (\ul{w} - \ul{\zeta}) + i {\un k} \cdot (\ul{z} - \ul{\zeta}) } \, \theta (k_1^-) \, \sum_{\sigma_1, \, \sigma_2} {\bar v}_{\sigma_2} (k_2)  \thalf \gamma^+ \gamma^5 v_{\sigma_1} (k_1)  \, 2 \, \sqrt{k_1^- k_2^-}
\\ & \times \, \Bigg\langle \mbox{T} \, \tr \left[ V_{\ul \zeta} \, V_{{\un z}, {\un w}; -\sigma_2, -\sigma_1}^\dagger \right]  \Bigg\rangle \, \frac{1}{\left[ 2 k_1^- x P^+ + {\un k}_1^2 - i \epsilon k_1^-  \right] \, \left[ 2 k_1^- x P^+ + {\un k}^2 + i \epsilon k_1^- \right]}  \Bigg|_{k_2^- = k_1^-, k_1^2 =0, k_2^2 =0, {\un k}_2 = - {\un k}}  + \mbox{c.c.} \notag
\end{align}
for the quark helicity TMD with a future-pointing (semi-inclusive Deep Inelastic Scattering, or SIDIS) Wilson-line staple.
\begin{figure}[ht]
\centering
\includegraphics[width=0.5\linewidth]{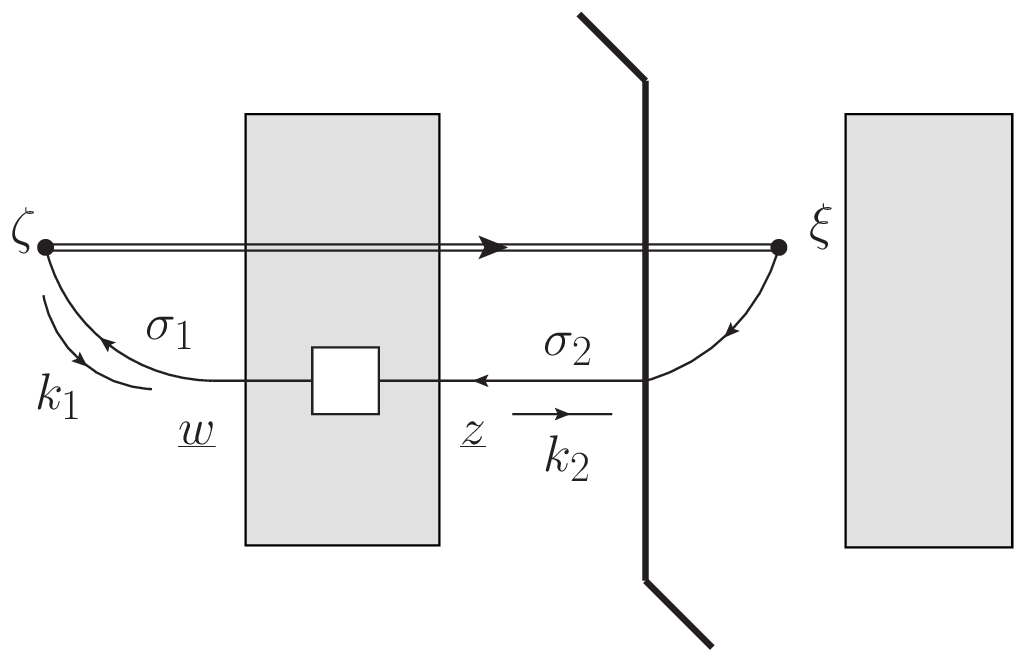}  
\caption{Diagram of class B with kinematics specified. The antiquark propagates from $\zeta$ to $\un{w}$ with momentum $k_1$, undergoes a sub-eikonal interaction with the proton which changes its transverse position from $\un{w}$ on the left of the shock wave (the left shaded rectangle) to $\un{z}$ on the right of the shock wave, and then propagates from $\un{z}$ to $\xi$ with momentum $k_2$. The sub-eikonal interaction with the proton shock wave (shaded rectangle) is denoted by the white box.}
\label{FIG:diagbdet}
\end{figure}

Using
\begin{align}\label{spinors2}
2\, \sqrt{k_1^- k_2^-} \, {\bar v}_{\sigma_2} (k_2) \thalf \gamma^+ \gamma^5 v_{\sigma_1} (k_1)  = 
\sigma_1 \, \delta_{\sigma_2 \sigma_1} \, (\ul{k}_2 \cdot \ul{k}_1) - i \, \delta_{\sigma_2 \sigma_1} \,  (\ul{k}_2 \times \ul{k}_1),
\end{align}
in \eq{TMD15}, along with \eq{Vxy_sub-eikonal}, and assuming that $2 k_1^- x P^+ \ll {\un k}^2, {\un k}_1^2$ to simplify the denominators at small $x$, we obtain  
\begin{align}\label{TMD16}
g_{1L}^q (x, k_T^2) & =  \frac{4 P^+}{(2\pi)^3}  \:  \int d^{2} \zeta 
\, d^2 w \, d^2 z \, \frac{d^2 k_1 \, d k_1^-}{(2\pi)^3} \, e^{i \ul{k}_1  \cdot (\ul{w} - \ul{\zeta}) + i {\un k} \cdot (\ul{z} - \ul{\zeta}) } \, \theta (k_1^-) \,  \frac{1}{{\un k}_1^2  \, {\un k}^2 }
\\ & \times \, \left[ \,  {\un k} \cdot {\un k}_1 \, \delta^2 ({\un z} - {\un w} ) \, \Bigg\langle \mbox{T} \, \tr \left[ V_{\ul \zeta} \, V_{{\un w}}^{\textrm{pol} [1] \, \dagger} \right]  \Bigg\rangle  + i  \, {\un k} \times {\un k}_1 \, \Bigg\langle \mbox{T} \, \tr \left[ V_{\ul \zeta} \, V_{{\un z}, {\un w}}^{\textrm{pol} [2] \, \dagger} \right] \Bigg\rangle \right] + \mbox{c.c.} . \notag
\end{align}
Note that the contribution of the eikonal term in \eq{V_sub-eikonal} to \eq{TMD15} is zero, as was shown in \cite{Kovchegov:2018znm}.

Performing the ${\un k}_1$ integration and adding the complex conjugate terms explicitly in \eq{TMD16} we arrive at
\begin{align}\label{TMD17}
g_{1L}^q (x, k_T^2) & = - \frac{4 i P^+}{(2\pi)^5}  \:  \int d^{2} \zeta 
\, d^2 w \,  \int\limits_0^{p_2^-} d k_1^- \, \Bigg\{  e^{i {\un k} \cdot (\ul{w} - \ul{\zeta}) } 
\, \frac{{\un k}}{{\un k}^2} \cdot \frac{\ul{\zeta} - \ul{w}}{|\ul{\zeta} - \ul{w}|^2} \, \Bigg\langle \mbox{T} \, \tr \left[ V_{\ul \zeta} \, V_{{\un w}}^{\textrm{pol} [1] \, \dagger} \right] + \bar{\mbox{T}} \, \tr \left[ V_{{\un \zeta}}^{\textrm{pol} [1]} \, V_{\ul w}^\dagger \right]  \Bigg\rangle  \\
& + i  \, \frac{{\un k}}{{\un k}^2} \times \frac{\ul{\zeta} - \ul{w}}{|\ul{\zeta} - \ul{w}|^2} \, \int d^2 z \, \Bigg\langle e^{i {\un k} \cdot (\ul{z} - \ul{\zeta}) }  \, \mbox{T} \, \tr \left[ V_{\ul \zeta} \, V_{{\un z}, {\un w}}^{\textrm{pol} [2] \, \dagger} \right] + e^{-i {\un k} \cdot (\ul{z} - \ul{\zeta}) }  \, \bar{\mbox{T}} \, \tr \left[ V_{{\un z}, {\un w}}^{\textrm{pol} [2]} \, V_{\ul \zeta}^\dagger \right]  \Bigg\rangle \Bigg\}  . \notag
\end{align}
We have also integrated over $\un z$ in the first term in the curly brackets of \eq{TMD17} and replaced ${\un \zeta} \leftrightarrow {\un w}$ in the term containing the second trace from the first angle brackets. 

We concentrate on the second term on the right of \eq{TMD17}. Employing \eq{VqG_decomp}, we see that the quark operator contribution to that term is proportional to
\begin{align}\label{unpol_q1}
\propto \int d^{2} \zeta 
\, d^2 w \, \frac{{\un k}}{{\un k}^2} \times \frac{\ul{\zeta} - \ul{w}}{|\ul{\zeta} - \ul{w}|^2} \, \Bigg\langle e^{i {\un k} \cdot (\ul{w} - \ul{\zeta}) }  \, \mbox{T} \, \tr \left[ V_{\ul \zeta} \, V_{{\un w}}^{\textrm{q} [2] \, \dagger} \right] + e^{-i {\un k} \cdot (\ul{w} - \ul{\zeta}) }  \, \bar{\mbox{T}} \, \tr \left[ V_{{\un w}}^{\textrm{q} [2]} \, V_{\ul \zeta}^\dagger \right]  \Bigg\rangle .
\end{align}
For a longitudinally polarized target proton, the expectation values of the impact-parameter integrated traces in \eq{unpol_q1} are functions of the dipole size only. Let us illustrate this with the first such trace: the absence of any preferred transverse direction in the longitudinally polarized target means
\begin{align}
\int d^{2} \left( \frac{\zeta + w}{2} \right) \,  \Bigg\langle \mbox{T} \, \tr \left[ V_{\ul \zeta} \, V_{{\un w}}^{\textrm{q} [2] \, \dagger} \right] \Bigg\rangle = f (|{\un \zeta} - {\un w}|^2),
\end{align}
such that 
\begin{align}\label{unpol_q2}
& \int d^{2} \zeta \, d^2 w \, \frac{{\un k}}{{\un k}^2} \times \frac{\ul{\zeta} - \ul{w}}{|\ul{\zeta} - \ul{w}|^2} \, \Bigg\langle e^{i {\un k} \cdot (\ul{w} - \ul{\zeta}) }  \, \mbox{T} \, \tr \left[ V_{\ul \zeta} \, V_{{\un w}}^{\textrm{q} [2] \, \dagger} \right] \Bigg\rangle \\
& = \frac{{\un k}}{{\un k}^2} \times \int d^2 (\zeta - w) \, e^{i {\un k} \cdot (\ul{w} - \ul{\zeta}) }  \, \frac{\ul{\zeta} - \ul{w}}{|\ul{\zeta} - \ul{w}|^2} \, f (|{\un \zeta} - {\un w}|^2)  \propto  {\un k} \times {\un k} =0.  \notag
\end{align}
Applying a similar argument to the second term in \eq{unpol_q1}, we see that the quark operator $V_{{\un w}}^{\textrm{q} [2]}$ does not contribute to the quark dipole TMD at small $x$ in \eq{TMD17}.

We next consider the gluon contribution to the second term on the right of \eq{TMD17}. To evaluate this term, it is easier to go back to \eq{TMD16}, the second term of which can be written as 
\begin{align}\label{unpol_G1}
&- \frac{4  \, (P^+)^2}{s \, (2\pi)^6}  \: \int\limits_0^{p_2^-} d k_1^- \,  \int d^{2} \zeta 
\, d^2 w \, d^2 k_1  \, e^{i (\ul{k}_1 + {\un k})  \cdot (\ul{w} - \ul{\zeta}) } \, \frac{ {\un k} \times {\un k}_1}{{\un k}^2  \, {\un k}_1^2} \, \int\limits_{-\infty}^\infty d y^- \\ 
& \times  \, \Bigg\langle \mbox{T} \, \tr \left[ V_{\ul \zeta} \, V_{{\un w}} [-\infty, y^-] \, \left( \cev{D}^i_{\un w} - i k_1^i \right) \, \left( D^i_{\un w} - i k^i \right) V_{{\un w}} [y^-, \infty] \right] \Bigg\rangle + \mbox{c.c.}  \notag
\end{align}
with the help of \eq{VxyG2}. Further, writing $D^i_{\un w} = (D^i_{\un w}/2) + (D^i_{\un w}/2)$ and integrating one of these terms by parts, while performing the same operation for $\cev{D}^i_{\un w}$, we arrive at
\begin{align}\label{unpol_G2}
& - \frac{(P^+)^2}{s \, (2\pi)^6}  \: \int\limits_0^{p_2^-} d k_1^- \,  \int d^{2} \zeta 
\, d^2 w \, d^2 k_1  \, e^{i (\ul{k}_1 + {\un k})  \cdot (\ul{w} - \ul{\zeta})  } \, \frac{ {\un k} \times {\un k}_1}{{\un k}^2  \, {\un k}_1^2} \, \int\limits_{-\infty}^\infty d y^- \\ 
& \times  \, \Bigg\langle \mbox{T} \, \tr \left[ V_{\ul \zeta} \, V_{{\un w}} [-\infty, y^-] \, \left( \cev{D}^i_{\un w} - D^i_{\un w} + i (k^i - k_1^i) \right) \, \left( D^i_{\un w} - \cev{D}^i_{\un w} + i (k_1^i - k^i) \right) V_{{\un w}} [y^-, \infty] \right] \Bigg\rangle + \mbox{c.c.} . \notag
\end{align}
The arguments similar to those used to show that the quark operator contribution to this term vanishes apply here to the $(D^i_{\un w} - \cev{D}^i_{\un w})^2$ and $(k_1^i - k^i)^2$ terms as well, leaving only the ``cross-talk" between the $D^i_{\un w} - \cev{D}^i_{\un w}$ and $k_1^i - k^i$ in \eq{unpol_G2}. Employing the definition \eqref{Vi}, we recast those remaining non-zero terms in \eq{unpol_G2} as
\begin{align}\label{unpol_G3}
\frac{4 \, i \, P^+}{(2\pi)^6}  \: \int\limits_0^{p_2^-} d k_1^- \,  \int d^{2} \zeta 
\, d^2 w \, d^2 k_1  \, e^{i (\ul{k}_1 + {\un k})  \cdot (\ul{w} - \ul{\zeta})  } \, \frac{ {\un k} \times {\un k}_1}{{\un k}^2  \, {\un k}_1^2} \, (k^i - k_1^i) \, \Bigg\langle \mbox{T} \, \tr \left[ V_{\ul \zeta} \, V_{{\un w}}^{i \, \textrm{G} \, [2] \, \dagger} \right] - \bar{\mbox{T}} \, \tr \left[ V_{{\un \zeta}}^{i \, \textrm{G} \, [2]}  V_{\ul w}^\dagger \, \right]  \Bigg\rangle . 
\end{align}

Further, employing 
\begin{align}\label{derivative}
- \pd_2^j \left( \frac{x_{20}^i}{x_{20}^2} \right) = \frac{\delta^{ij} \, x_{20}^2 - 2 x_{20}^i x_{20}^j}{x_{20}^4} + \delta^{ij} \, \pi \, \delta^2 ({\un x}_{20})
\end{align}
we perform the Fourier transform over ${\un k}_1$, obtaining 
\begin{align}\label{unpol_G4}
- \frac{4 \, P^+}{(2\pi)^5}  \: \int\limits_0^{p_2^-} d k_1^- \,  \int d^{2} \zeta 
\, d^2 w \, e^{i {\un k} \cdot (\ul{w} - \ul{\zeta})  } & \ \frac{ \epsilon^{mj} k^m }{{\un k}^2 } \,\left[ k^i \frac{(\ul{w} - \ul{\zeta})^j}{|\ul{w} - \ul{\zeta}|^2} + i \, \frac{\delta^{ij} \, |\ul{w} - \ul{\zeta}|^2 - 2 (\ul{w} - \ul{\zeta})^i (\ul{w} - \ul{\zeta})^j}{|\ul{w} - \ul{\zeta}|^4} \right] \\ 
& \times \, \Bigg\langle \mbox{T} \, \tr \left[ V_{\ul \zeta} \, V_{{\un w}}^{i \, \textrm{G} \, [2] \, \dagger} \right] - \bar{\mbox{T}} \, \tr \left[ V_{{\un \zeta}}^{i \, \textrm{G} \, [2]}  V_{\ul w}^\dagger \, \right]  \Bigg\rangle , \notag
\end{align}
where we have also used the fact that $\tr \left[ V_{\ul w} \, V_{{\un w}}^{i \, \textrm{G} \, [2] \, \dagger} \right] =0$. Interchanging the integration variables ${\un \zeta} \leftrightarrow {\un w}$ along with flipping the sign ${\un k} \to - {\un k}$ of the transverse momentum in the second term of  \eq{unpol_G4} (which is allowed since each term in \eq{unpol_G4} is a function of $k_T^2$ only), we obtain
\begin{align}\label{unpol_G5}
- \frac{4 \, P^+}{(2\pi)^5}  \: \int\limits_0^{p_2^-} d k_1^- \,  \int d^{2} \zeta 
\, d^2 w \, e^{i {\un k} \cdot (\ul{w} - \ul{\zeta})  } & \ \frac{ \epsilon^{mj} k^m }{{\un k}^2 } \,\left[ k^i \frac{(\ul{w} - \ul{\zeta})^j}{|\ul{w} - \ul{\zeta}|^2} + i \, \frac{\delta^{ij} \, |\ul{w} - \ul{\zeta}|^2 - 2 (\ul{w} - \ul{\zeta})^i (\ul{w} - \ul{\zeta})^j}{|\ul{w} - \ul{\zeta}|^4} \right] \\ 
& \times \, \Bigg\langle \mbox{T} \, \tr \left[ V_{\ul \zeta} \, V_{{\un w}}^{i \, \textrm{G} \, [2] \, \dagger} \right] + \bar{\mbox{T}} \, \tr \left[ V_{{\un w}}^{i \, \textrm{G} \, [2]}  V_{\ul \zeta}^\dagger \, \right]  \Bigg\rangle . \notag
\end{align}

To further simplify the matrix elements of the traces in \eq{unpol_G5} we can employ the relations given by Eqs.~(22) of \cite{Kovchegov:2018znm} (see also \cite{Mueller:2012bn}),
\begin{subequations}\label{Wilson_rels}
\begin{align}
\left\langle \mbox{T} \, \mbox{tr} \left[ V_{\ul x} \, V_{{\un y}}^{\textrm{pol} \, \dagger} \right]  \right\rangle = \left\langle \mbox{tr} \left[ V_{\ul x} \, V_{{\un y}}^{\textrm{pol} \, \dagger} \right]  \right\rangle ,  \label{Wilson_rels_a} \\
\left\langle \bar{\mbox{T}} \, \mbox{tr} \left[ V_{\ul x} \, V_{{\un y}}^{\textrm{pol} \, \dagger} \right]  \right\rangle = \left\langle \mbox{tr} \left[ V_{{\un y}}^{\textrm{pol} \, \dagger} \, V_{\ul x} \right]  \right\rangle , \label{Wilson_rels_b}
\end{align}
\end{subequations}
where the ordering of the operators on the right is important, since the right Wilson line belongs to the amplitude, while the left one is in the complex conjugate amplitude. Application of Eqs.~\eqref{Wilson_rels} yields
\begin{align}\label{Tdrop}
\Bigg\langle \mbox{T} \, \tr \left[ V_{\ul \zeta} \, V_{{\un w}}^{i \, \textrm{G} \, [2] \, \dagger} \right] + \bar{\mbox{T}} \, \tr \left[ V_{{\un w}}^{i \, \textrm{G} \, [2]}  V_{\ul \zeta}^\dagger \, \right]  \Bigg\rangle = \Bigg\langle \tr \left[ V_{\ul \zeta} \, V_{{\un w}}^{i \, \textrm{G} \, [2] \, \dagger} \right] + \tr \left[ V_{{\un w}}^{i \, \textrm{G} \, [2]}  V_{\ul \zeta}^\dagger \, \right]  \Bigg\rangle .
\end{align}
Comparing this with \eq{Gj}, we see that the objects in the angle brackets in the two equations are similar, but not quite the same: the order of the Wilson lines is different in the trace. As we noted above, the order of Wilson lines matters for the operators here. To remedy this issue, we note that the quark helicity TMD is PT-even: hence, we can substitute \eq{Tdrop} back into \eq{unpol_G5} and apply the PT-transformation to the latter, leaving it invariant (while, in the process, changing the SIDIS Wilson-line staple to the Drell-Yan (DY) one for the TMD). For infinite Wilson lines in question we have
\begin{align}
V_{\ul \zeta} \ \xrightarrow[]{\textrm{PT}} \ V_{- \ul \zeta}^\dagger , \ \ \ V_{{\un w}}^{i \, \textrm{G} \, [2]}  \ \xrightarrow[]{\textrm{PT}} \ V_{-{\un w}}^{i \, \textrm{G} \, [2] \, \dagger} . 
\end{align}
This means that, under PT, the expression in \eq{Tdrop} becomes $(2 N_c / s) \, G^i_{-{\un w}, -{\un \zeta}}$ (cf. \eq{Gj}), where the sign change in front of $\un w$ and $\un \zeta$ is not important, since these are integration variables. Due to the PT-invariance of the quark helicity TMD, we obtain for \eq{unpol_G5}
\begin{align}\label{unpol_G6}
- \frac{4 \, N_c}{(2\pi)^5}  \: \int\limits_{\Lambda^2/s}^{1} \frac{d z}{z} \,  \int d^{2} \zeta 
\, d^2 w \, e^{i {\un k} \cdot (\ul{w} - \ul{\zeta})  } & \ \frac{ \epsilon^{mj} k^m }{{\un k}^2 } \,\left[ k^i \frac{(\ul{w} - \ul{\zeta})^j}{|\ul{w} - \ul{\zeta}|^2} + i \, \frac{\delta^{ij} \, |\ul{w} - \ul{\zeta}|^2 - 2 (\ul{w} - \ul{\zeta})^i (\ul{w} - \ul{\zeta})^j}{|\ul{w} - \ul{\zeta}|^4} \right] \, G^i_{{\un w}, {\un \zeta}} (zs) ,
\end{align}
where $z = k_1^-/p_2^-$ and $\Lambda$ is an infrared (IR) cutoff. (Note that the PT-symmetry argument would not have been needed if we had started with the quark TMD with the DY Wilson-line staple or interchanged the past- and forward-pointing staples in \eq{glue_hel_TMD}.)

Replacing the second term in \eq{TMD17} by the expression \eqref{unpol_G6}, and adding the contribution of the anti-quark helicity TMD as it was done in \cite{Kovchegov:2018znm} to obtain the flavor-singlet quark helicity TMD, we arrive at
\begin{align}\label{TMD18}
g_{1L}^S (x, k_T^2) = & \ - \frac{8 \, N_c \, N_f}{(2\pi)^5} \:  \int\limits_{\Lambda^2/s}^1 \frac{d z}{z} \,  \int d^{2} \zeta 
\, d^2 w \, e^{ i {\un k} \cdot (\ul{w} - \ul{\zeta})}  \,  \Bigg\{ i \, \frac{\un{\zeta} - \un{w}}{|\un{\zeta} - \un{w}|^2} \cdot  \frac{\un{k} }{\un{k}^2} \, Q_{{\un w}, {\un \zeta}} (zs) \\
& + \frac{ \epsilon^{mj} k^m }{{\un k}^2 } \,\left[ k^i \frac{(\ul{w} - \ul{\zeta})^j}{|\ul{w} - \ul{\zeta}|^2} + i \, \frac{\delta^{ij} \, |\ul{w} - \ul{\zeta}|^2 - 2 (\ul{w} - \ul{\zeta})^i (\ul{w} - \ul{\zeta})^j}{|\ul{w} - \ul{\zeta}|^4} \right] \, G^i_{{\un w}, {\un \zeta}} (zs) \Bigg\} , \notag 
\end{align}
where we have also summed over quark flavors, generating a factor of $N_f$ by assuming, for simplicity, that all flavors give equal contributions. As in \cite{Kovchegov:2018znm}, we have defined the ``original" polarized dipole amplitude
\begin{align}\label{Q_def}
Q_{{\un w}, {\un \zeta}} (zs) \equiv \frac{1}{2 N_c} \, \textrm{Re} \, \llangle \mbox{T} \, \tr \left[ V_{\ul \zeta} \, V_{{\un w}}^{\textrm{pol} [1] \, \dagger} \right] + \mbox{T} \, \tr \left[ V_{{\un w}}^{\textrm{pol} [1]} \, V_{\ul \zeta}^\dagger \right]  \rrangle .
\end{align}
While the helicity evolution we will derive below is independent of quark flavor, the initial conditions for $Q_{{\un w}, {\un \zeta}} (zs)$ may be flavor-dependent \cite{Adamiak:2021ppq}, meaning that our simplified assumption of flavor symmetry may need to be generalized by replacing $N_f \to \sum_f$ and $Q_{{\un w}, {\un \zeta}} (zs) \to Q^f_{{\un w}, {\un \zeta}} (zs)$ in \eq{TMD18} to include the potential flavor-dependence of the amplitudes $Q^f_{{\un w}, {\un \zeta}} (zs)$.

The flavor-singlet quark helicity PDF \eqref{eqn:DeltaSigma} is
\begin{align}\label{DeltaSigma1}
\Delta \Sigma (x, Q^2) = \int\limits^{Q^2} d^2 k_T \, g_{1L}^S (x, k_T^2) .
\end{align}
Using \eq{TMD18} in \eq{DeltaSigma1} while imposing the $\tfrac{1}{x} > z s x_{10}^2$ lifetime ordering yields
\begin{align}\label{DeltaSigma}
\Delta \Sigma (x, Q^2) = - \frac{N_c \, N_f}{2 \pi^3} \:  \int\limits_{\Lambda^2/s}^1 \frac{d z}{z} \,  \int\limits_{\frac{1}{zs}}^{\min \left\{ \frac{1}{z Q^2} , \frac{1}{\Lambda^2} \right\}} \frac{d x^{2}_{10}}{x_{10}^2}  \, \left[  Q (x^2_{10} , zs) + 2 \, G_2 (x^2_{10} , zs) \right]
\end{align}
where we have employed the decomposition \eqref{decomp} and
\begin{align}\label{Q_int}
Q (x^2_{10} , zs) \equiv \int d^2 \left( \frac{x_0 + x_1}{2} \right) \, Q_{10} (zs). 
\end{align}
Equation \eqref{DeltaSigma} is to be compared to Eq.~(8b) in \cite{Kovchegov:2016zex} or, equivalently, Eq.~(5) in \cite{Kovchegov:2016weo}, which contain only the first term in the square brackets of \eq{DeltaSigma}. 

Using the decomposition \eqref{decomp} and \eq{Q_int}, \eq{TMD18} can be rewritten as
\begin{align}\label{TMD19a}
g_{1L}^S (x, k_T^2) = \frac{8 \, N_c \, N_f}{(2\pi)^5}  \int\limits_{\Lambda^2/s}^1 \frac{d z}{z} \int d^{2} x_{10} 
\, e^{ i {\un k} \cdot \ul{x}_{10} }  \,  \left[ i \, \frac{\ul{x}_{10}}{{x}_{10}^2} \cdot  \frac{\un{k} }{\un{k}^2} \, \left[ Q (x_{10}^2 , zs) + G_2 (x_{10}^2 , zs)  \right] - \frac{({\un k} \times {\un x}_{10})^2}{{\un k}^2 \, x_{10}^2} \, G_2 (x_{10}^2 , zs) \right] . 
\end{align}
The integral over the angles of ${\un x}_{10}$ in the last term on the right of \eq{TMD19a} can be cast into the same form as in the first term \cite{Kovchegov:2019rrz}. This yields
\begin{align}\label{TMD19}
g_{1L}^S (x, k_T^2) = \frac{8 \, i \, N_c \, N_f}{(2\pi)^5}  \int\limits_{\Lambda^2/s}^1 \frac{d z}{z} \int d^{2} x_{10} 
\, e^{ i {\un k} \cdot \ul{x}_{10} }   \, \frac{\ul{x}_{10}}{{x}_{10}^2} \cdot  \frac{\un{k} }{\un{k}^2} \, \left[ Q (x_{10}^2 , zs) + 2 \, G_2 (x_{10}^2 , zs)  \right] . 
\end{align}

We see that both the quark and gluon helicity TMDs and PDFs at small $x$ can be expressed in terms of the polarized dipole amplitudes $Q (x^2_{10} , zs)$ and $G_2 (x^2_{10} , zs)$. These dipole amplitudes enter the expressions \eqref{TMD19} and \eqref{DeltaSigma} for the quark helicity TMD and PDF in a specific linear combination, $Q + 2 \, G_2$.


\subsection{$g_1$ Structure Function}
\label{sec:g1}

Next we consider DIS on a longitudinally polarized proton. The hadronic tensor can be written as (see \cite{Zyla:2020zbs,Lampe:1998eu} for a systematic exposition)
\begin{align}\label{Wmn}
W_{\mu\nu} & \ \equiv \frac{1}{4 \pi M_p} \, \int d^4 x \, e^{i q \cdot x} \, \bra{P, S_L} j_\mu (x) \, j_\nu (0) \ket{P, S_L} \\
& = W^\textrm{sym}_{\mu\nu} + i \, \epsilon_{\mu\nu\rho\sigma} \frac{q^\rho}{M_p \,  P \cdot q} \left[ S^\sigma g_1 (x, Q^2) + \left( S^\sigma - \frac{S \cdot q}{P \cdot q} \, P^\sigma \right) \, g_2 (x, Q^2) \right], \notag
\end{align}
where $M_p$ is the proton mass and $W^{\textrm{sym}}_{\mu\nu}$ denotes the spin-independent ($\mu \leftrightarrow \nu$ symmetric) part of the hadronic tensor, dependent on the $F_1$, $F_2$ structure functions. As usual, $j_\mu$ is the quark electromagnetic current operator and the 4-dimensional Levi-Civita symbol is defined with $\epsilon_{0123} = +1$ \cite{bjorken1965relativistic}. We will work in the proton rest frame where $P^\mu = (M_p, \vec{0})$ and the spin 4-vector is $S^\mu = (0,0,0, \Sigma M_p)$ for the longitudinally polarized proton with polarization $\Sigma = \pm 1$. Adjusting the frame further such that the virtual photon momentum is $q^\mu = (- Q^2/(2 q^-),  q^-, {\un 0})$ in the $(+, -, \perp)$ light-cone notation, we have the photon polarizations vectors $\epsilon_{T \lambda}^\mu = (0, 0, {\un \epsilon}_\lambda)$ for transverse polarizations (with ${\un \epsilon}_\lambda = (-1/\sqrt{2}) (\lambda , i)$ and $\lambda = \pm 1$) and $\epsilon_L^\mu = (Q/(2 q^-), q^-/Q , {\un 0} )$ for the longitudinal polarization (see e.g. \cite{Kovchegov:2012mbw}). 

Consider the $\gamma^* + p$ scattering cross section, 
\begin{align}\label{sigma_DIS1}
\sigma^{\gamma^* p} = \frac{4 \pi^2 \alpha_{EM}}{q^0} \ W_{\mu\nu} \, \epsilon^{* \, \mu} \, \epsilon^\nu 
\end{align}
with $\alpha_{EM}$ the fine structure constant. We are interested in the spin-dependent part of this cross section, which we obtain by using the spin-dependent part of $W_{\mu\nu}$ from \eq{Wmn} in \eq{sigma_DIS1}. In the frame we are working in, one can see that only transverse values of $\mu, \nu$ contribute to the spin-dependent part of $\sigma^{\gamma^* p}$: this means only transverse photon polarizations contribute. Assuming that the virtual photon is transversely polarized with polarization $\lambda$, after some algebra we obtain the spin-dependent cross section
\begin{align}\label{sigma_DIS2}
\sigma^{\gamma^* p} (\lambda, \Sigma) = \frac{4 \pi^2 \alpha_{EM}}{q^0} \ W_{\mu\nu} \, \epsilon_{T \lambda}^{* \, \mu} \, \epsilon_{T \lambda}^\nu = - \frac{8 \pi^2 \alpha_{EM} \, x}{Q^2} \, \lambda \, \Sigma \, \left[ g_1 (x, Q^2) - \frac{4 x^2 M_p^2}{Q^2} \, g_2 (x, Q^2) \right]. 
\end{align}
The object in the square brackets of \eq{sigma_DIS2} is equal to the virtual photon spin asymmetry $A_1$ multiplied by the spin-independent  structure function $F_1 (x, Q^2)$ \cite{Lampe:1998eu}. The factor of $x$ in the prefactor of \eq{sigma_DIS2}, which is absent in the analogue of this equation for the spin-independent case \cite{Kovchegov:2012mbw}, indicates that the spin-dependent cross section is indeed sub-eikonal at small $x$. Furthermore, at small $x$ we have $4 x^2 M_p^2/Q^2 \ll 1$ (which is also true in the standard perturbative approaches which assume large $Q^2$): this allows us to neglect the second term in the square brackets of \eq{sigma_DIS2}, since it is a sub-sub-sub-eikonal contribution (that is, a contribution suppressed by $x^3$ compared to the eikonal scattering). We thus write
\begin{align}\label{g1_0}
g_1 (x, Q^2)  = - \frac{Q^2}{16 \pi^2 \alpha_{EM} \, x} \, \left[ \sigma^{\gamma^* p} (+, +) - \sigma^{\gamma^* p} (-, +) \right] .
\end{align}

\begin{figure}[ht]
\centering
\includegraphics[width= 0.9 \textwidth]{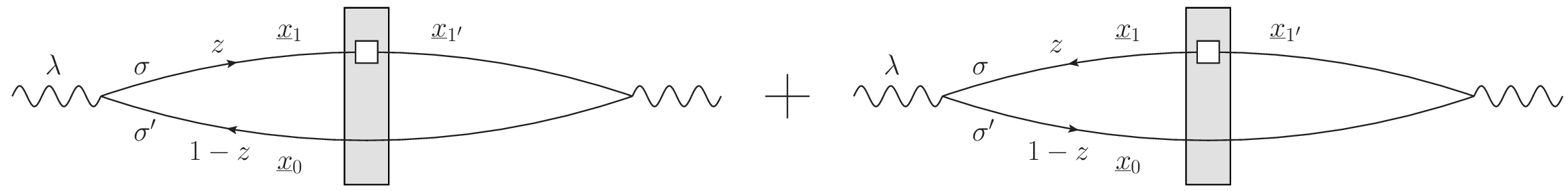}  
\caption{Diagrams needed for the calculation of the $g_1$ structure function in the dipole picture of DIS. The proton shock wave is denoted by the shaded rectangle, while the white box denotes the sub-eikonal interaction with the target.}
\label{FIG:g1}
\end{figure}

We see that to obtain the $g_1$ structure function, we need to find the polarization-dependent part of the $\gamma^* + p$ scattering cross section, $\sigma^{\gamma^* p}$, with the transversely polarized photon. Working in the dipole picture of DIS, appropriate at small $x$, we write (cf. \cite{Kovchegov:2012mbw}), keeping in mind the eikonal and sub-eikonal terms,
\begin{align}\label{g1_1}
 \sigma^{\gamma^* p} (\lambda, \Sigma) = & \, - \int \frac{d^2 x_1 \, d^2 x_{1'} \, d^2 x_0}{4 \pi}  \int\limits_0^1 \frac{dz}{z \, (1-z)}  \, \sum_{\sigma, \sigma', f} 2 \, \textrm{Re} \left\{ \Psi_{\sigma, \sigma'; \lambda}^{\gamma^*\to q\overline{q}} ({\un x}_{10}, z) \, \left[ \Psi_{\sigma, \sigma'; \lambda}^{\gamma^*\to q\overline{q}} ({\un x}_{1'0}, z) \right]^* \left\langle \textrm{T} \, \tr \left[ V^\textrm{pol}_{\un{1}', {\un 1}; \sigma, \sigma} \, V_{\un 0}^\dagger \right] \right\rangle (z) \right. \notag \\ 
 & \left.  - \, \Psi_{\sigma', \sigma; \lambda}^{\gamma^*\to q\overline{q}} ({\un x}_{01}, 1-z) \, \left[ \Psi_{\sigma', \sigma; \lambda}^{\gamma^*\to q\overline{q}} ({\un x}_{01'}, 1-z) \right]^*  \left\langle \textrm{T} \, \tr \left[  V_{\un 0} \, V^{\textrm{pol} \, \dagger}_{\un{1}', {\un 1}; -\sigma, -\sigma} \right] \right\rangle (z) \right\}, 
\end{align}
where the light-cone wave function of a transversely polarized virtual photon in the conventions of \cite{Kovchegov:2012mbw} is
\begin{align}\label{Qtog3a}  
\Psi_{\sigma, \sigma'; \lambda}^{\gamma^*\to q\overline{q}}({\un x}_{10},z) &= \frac{eZ_f}{2\pi}\sqrt{z(1-z)}\left[\delta_{\sigma,-\sigma'}\left(1-2z-\sigma\lambda\right)ia_f\frac{\ee_{\lambda}\cdot {\un x}_{10}}{x_{10}}K_1\left(x_{10} \, a_f\right) + \delta_{\sigma\sigma'}\frac{m_f}{\sqrt{2}}\left(1+\sigma\lambda\right)K_0\left(x_{10} \, a_f\right)  \right] . 
\end{align}
Equation \eqref{g1_1} is illustrated in \fig{FIG:g1}. For each quark flavor $f$, $m_f$ is the quark mass, $Z_f$ is the fractional charge of the quark, and $a^2_f = z(1-z)Q^2 + m^2_f$ with $z$ the fraction of the photon's light-cone ($-$) momentum carried by the quark or by the antiquark, as labeled in the diagrams in \fig{FIG:g1}. The overall minus sign in \eq{g1_1} reflects the sign difference between the real part of (the interaction term in)  the $S$-matrix and the imaginary part of the forward scattering amplitude.

The light-cone wave function in \eq{Qtog3a} is defined in such a way that the quark is located at ${\un x}_1$ in the transverse plane, while the anti-quark is at ${\un x}_{0}$. As before, ${\un x}_{ij} = {\un x}_i - {\un x}_j$ with $x_{ij} = |{\un x}_{ij}|$. The dipole sizes before and after scattering on the shock wave in \eq{g1_1} are $x_{10}$ and $x_{1'0}$, respectively. In \eq{Qtog3a}, the quark and the anti-quark carry polarizations $\sigma$ and $\sigma'$, respectively, while the photon carries polarization $\lambda$. 

One can easily show that the eikonal part of the $S$-matrix $V^\textrm{pol}_{\un{1}', {\un 1}; \sigma, \sigma}$ does not contribute a $\lambda$-dependent term in \eq{g1_1} that would contribute to \eq{g1_0}. Therefore, concentrating on the sub-eikonal terms, we substitute Eqs.~\eqref{Vxy_sub-eikonal} and \eqref{Qtog3a} into \eq{g1_1} and sum over $\sigma, \sigma'$. This gives
\begin{align}\label{g1_2}
& \sigma^{\gamma^* p} (+, +) - \sigma^{\gamma^* p} (-, +) = - \sum_f \frac{2 \, \alpha_{EM} \, Z^2_f}{\pi^2} \int d^2 x_1 \, d^2 x_{1'} \, d^2 x_0 \, \int\limits_0^1 dz \\
& \times \, \textrm{Re} \bigg\{ - i \, [z^2 + (1-z)^2] \, a_f^2 \, \frac{{\un x}_{10} \times {\un x}_{1'0}}{x_{10} \, x_{1'0}} \, K_1 (x_{10} \, a_f) \, K_1 (x_{1'0} \, a_f) \, \left\langle \textrm{T} \, \tr \left[ V^{\textrm{G} [2]}_{\un{1}', {\un 1}} \, V_{\un 0}^\dagger \right] - \textrm{T} \, \tr \left[  V_{\un 0} \, V^{\textrm{G} [2] \, \dagger}_{\un{1}', {\un 1}} \right] \right\rangle (z) \notag \\
&  + \delta^2 ({\un x}_{11'}) \, \left[ (2 z -1) \, a_f^2 \, \left[ K_1(x_{10} \, a_f) \right]^2 + m_f^2 \, \left[ K_0 (x_{10} \, a_f) \right]^2 \right] \left\langle \textrm{T} \, \tr \left[ V^{\textrm{pol} [1]}_{\un 1} \, V_{\un 0}^\dagger \right] + \textrm{T} \, \tr \left[  V_{\un 0} \, V^{\textrm{pol} [1] \, \dagger}_{\un 1} \right] \right\rangle (z)  \bigg\}. \notag
\end{align}
Note that the quark operator $V^{\textrm{q} [2]}$ does not contribute. 

Finally, we employ the definition of $V^{\textrm{G} [2]}_{\un{1}', {\un 1}}$ from \eq{VxyG2}, along with the polarized Wilson line \eqref{Vi} and the dipole amplitude definitions \eqref{Gj}, \eqref{decomp}, \eqref{Q_def}, and \eqref{Q_int}, in \eq{g1_2}. Inserting the result into \eq{g1_0}, after some algebra and after invoking the PT-symmetry argument we employed earlier on, we obtain our final expression for the small-$x$ structure function $g_1$ in terms of the polarized dipole amplitudes:
\begin{align}\label{g1_final}
g_1 (x, Q^2)  = - \sum_f \frac{N_c \, Z^2_f}{4 \pi^4} \int d^2 x_{10} \, \int\limits_{\Lambda^2/s}^1 \frac{dz}{z} \, \Bigg\{ & \, 2 \, [z^2 + (1-z)^2] \, a_f^2 \,\left[ K_1(x_{10} \, a_f) \right]^2 \, G_2 (x_{10}^2, zs) \\
& + \left[ (1 - 2 z) \, a_f^2 \, \left[ K_1(x_{10} \, a_f) \right]^2 - m_f^2 \, \left[ K_0 (x_{10} \, a_f) \right]^2 \right] \, Q (x_{10}^2, zs) \Bigg\}. \notag
\end{align}

We can cross-check the result \eqref{g1_final} by considering the double-logarithmic limit of its integrals. Expanding the integrand of \eq{g1_final} for $z \ll 1$ and $x_{10} \, a_f \ll 1$ and keeping only the double-logarithmic terms yields
 \begin{align}\label{g1_DLA}
g_1 (x, Q^2)  = - \sum_f \frac{N_c \, Z^2_f}{4 \pi^3} \int\limits_{\Lambda^2/s}^1 \frac{dz}{z} \,  \int\limits^{\min \left\{ \frac{1}{z Q^2} , \frac{1}{\Lambda^2} \right\}}_\frac{1}{zs} \frac{d x^2_{10}}{x_{10}^2} \, \left[ Q (x_{10}^2, zs) + 2 \, G_2 (x_{10}^2, zs) \right], 
\end{align}
where the lower limit of the $x_{10}^2$-integral arises from the $z s x_{10}^2 \gg 1$ conditions, which, in turn, follows from the validity of the shock wave (dipole picture of DIS) approximation (see e.g. \cite{Cougoulic:2019aja}), and is also implicitly applied to the full \eq{g1_final}. 

Equation \eqref{g1_DLA} should be compared to \eq{DeltaSigma}, also written in the double-logarithmic approximation. One can rewrite \eq{DeltaSigma} as \eq{eqn:DeltaSigma}
with 
\begin{align}\label{q+}
\Delta q_f^+ (x, Q^2) = - \frac{N_c}{2 \pi^3} \:  \int\limits_{\Lambda^2/s}^1 \frac{d z}{z} \,  \int\limits_{\frac{1}{zs}}^{\min \left\{ \frac{1}{z Q^2} , \frac{1}{\Lambda^2} \right\}} \frac{d x^{2}_{10}}{x_{10}^2}  \, \left[  Q (x^2_{10} , zs) + 2 \, G_2 (x^2_{10} , zs) \right].
\end{align}
Comparing Eqs.~\eqref{q+} and \eqref{g1_DLA} we arrive at the well-known relation \cite{Lampe:1998eu}
\begin{align}
g_1 (x, Q^2)  = \frac{1}{2} \sum_f \, Z^2_f \, \Delta q_f^+ (x, Q^2),
\end{align}
thus confirming consistency of our Eqs.~\eqref{g1_DLA} and \eqref{DeltaSigma}. This completes the cross-check of \eq{g1_final}.

~\\~~\\

We conclude this Section by summarizing its main results: at small $x$, the flavor-singlet quark and gluon helicity PDFs and TMDs ($\Delta \Sigma (x, Q^2) , g_{1L}^S (x, k_T^2), \Delta G (x, Q^2), g_{1L}^{G \, dip} (x, k_T^2)$) along with the $g_1$ structure function can all be expressed in terms of the polarized dipole amplitudes $Q (x^2_{10} , zs)$ and $G_2 (x^2_{10} , zs)$. Therefore, to describe these observables we need to construct evolution equations for these two polarized dipole amplitudes. In the earlier literature \cite{Kovchegov:2016zex,Kovchegov:2016weo,Kovchegov:2018znm}, the contributions of the amplitude $G_2$ to the quark helicity TMD and PDF, and to the $g_1$ structure function, have been omitted.


\section{Helicity Evolution at Small $x$}
\label{sec:hel_evo}

Our next step is to derive small-$x$ evolution equations for the polarized dipole amplitudes in Eqs.~\eqref{Q_def} and \eqref{Gj}, which we summarize here again for convenience:
\begin{align}\label{Q_def2}
Q_{10} (zs) \equiv \frac{1}{2 N_c} \, \textrm{Re} \, \llangle \mbox{T} \, \tr \left[ V_{\ul 0} \, V_{\un 1}^{\textrm{pol} [1] \, \dagger} \right] + \mbox{T} \, \tr \left[ V_{\ul 1}^{\textrm{pol} [1]} \, V_{\ul 0}^\dagger \right]  \rrangle (zs),
\end{align}
\begin{align}\label{Gj2}
G^i_{10} (zs) \equiv \frac{1}{2 N_c} \, \llangle  \tr \Big[  V^\dagger_{\un 0} \, V_{\un{1}}^{i \, \textrm{G} [2]} + \left( V_{\un{1}}^{i \, \textrm{G} [2]} \right)^\dagger  V_{\un 0} \Big]  \rrangle (zs).
\end{align}
Ultimately, in the evolution equations we would replace $G^i$ by $G_2$, defined in the decomposition \eqref{decomp}. 

The evolution equations will be derived in the double-logarithmic approximation (DLA), which is defined as resumming powers of $\as \, \ln^2 (1/x)$. We will then compare our results to those obtained earlier in \cite{Kovchegov:2015pbl,Kovchegov:2016zex,Kovchegov:2016weo,Kovchegov:2017lsr,Kovchegov:2018znm}.


\subsection{Evolution Equations in the Operator Form}
\label{sec:evo_op}


\subsubsection{Evolution Equations for Fundamental and Adjoint $Q_{10} (zs)$}
\label{sec:Qevol}

Following the procedure outlined in \cite{Kovchegov:2017lsr,Kovchegov:2018znm}, we construct the evolution in the operator language using the shock wave approximation for the polarized target. We suggest that the procedure we employ, which uses the operator language in light cone time-ordered Feynman diagrams (cf. also \cite{Mueller:2012bn}), could be called the light-cone operator treatment (LCOT). We will again work in the frame where the target proton has a large $P^+$ momentum, while the projectile Wilson lines are oriented along the $x^-$-axis. To construct the evolution we will need gluon and quark propagators in the shock wave background. The operators in the polarized dipole amplitudes $Q_{10}$ and $G_{10}^i$ depend on the gluon field components $A^+, {\un A}$ and on the quark fields $\psi$, $\bar \psi$: we will need propagators connecting those fields. (We are working in $A^-=0$ light-cone gauge.)

\begin{figure}[ht]
\centering
\includegraphics[width= \textwidth]{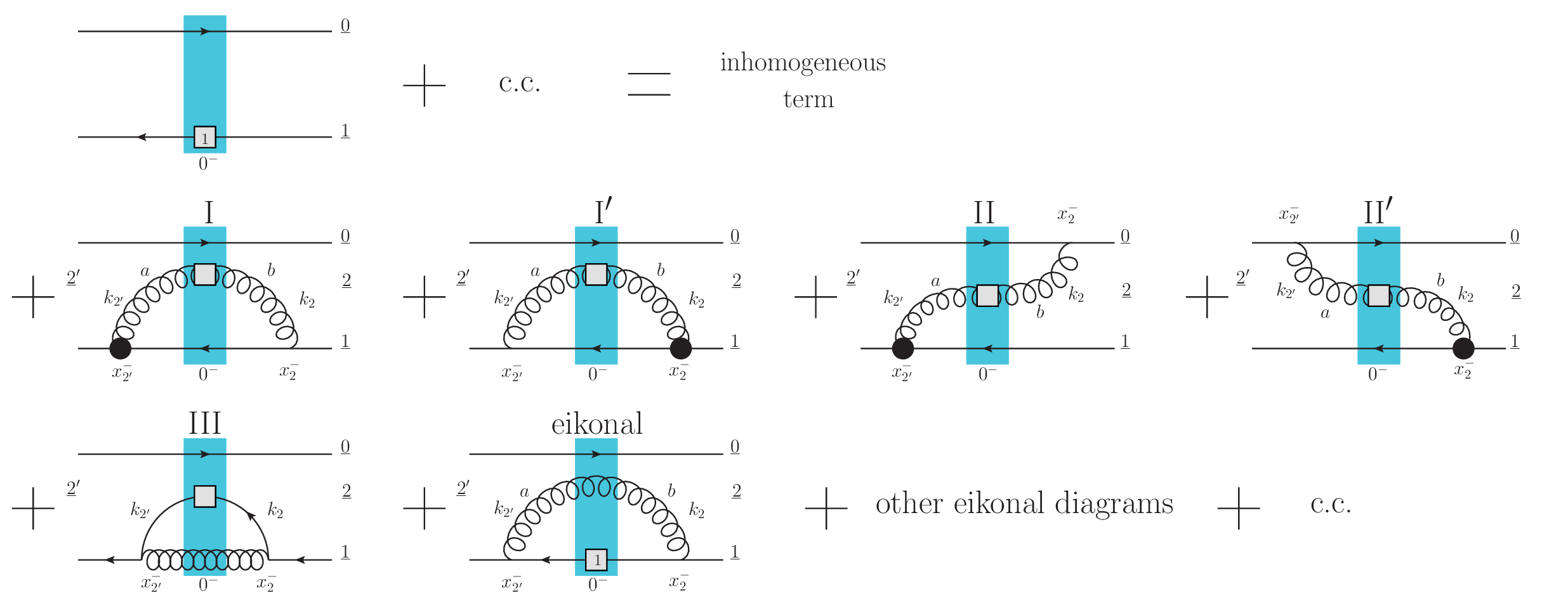}
\caption{Diagrams representing the evolution of the fundamental polarized dipole amplitude $Q_{10}$. The vertical shaded rectangle represents the shock wave. The square box on the gluon and quark lines represents the sub-eikonal interaction with the target given by \eq{Uxy_sub-eikonal} for gluons and \eq{Vxy_sub-eikonal} for quarks. The same square box, but with number 1 in it, on the quark line denotes the interaction described by $V_{\ul 1}^{\textrm{pol} [1]}$ only. The black circle denotes the sub-eikonal quark-gluon vertex generated by the $F^{12}$ operator in \eq{VG1}, that is, by the $F^{12}$ part of $V_{\ul 1}^{\textrm{pol} [1]}$, which, in turn, contributes to $Q_{10}$ through \eq{Q_def2}. All momenta flow to the right.}
\label{FIG:Q_evol}
\end{figure}

We begin with the amplitude $Q_{10} (zs)$. Diagrams contributing to its evolution are shown in \fig{FIG:Q_evol}. These are the same diagrams as in the earlier works on the subject \cite{Kovchegov:2015pbl,Kovchegov:2017lsr,Kovchegov:2018znm}, except now the square box on the line going through the shock wave indicates both terms in \eq{Uxy_sub-eikonal} for the gluon line and both terms in \eq{Vxy_sub-eikonal} for the quark line. In the past works \cite{Kovchegov:2015pbl,Kovchegov:2017lsr,Kovchegov:2018znm}, only the first term in each of those equations was included. 

In the gluon sector, the sub-eikonal propagator contributing to the evolution of $Q_{10} (zs)$ is $\contraction[2ex]{}{a}{^\perp}{a} a^\perp a^+$. It contributes to diagrams I, I$'$, II, II$'$ in \fig{FIG:Q_evol}. 
Following the steps detailed in \cite{Kovchegov:2017lsr,Kovchegov:2018znm} while including both polarization structures from \eq{Uxy_sub-eikonal} gives (for the propagator in the diagram II)
\begin{align} \label{+perp_sub_eik_1}
&\: \int\limits_{-\infty}^0 dx_{2'}^- 
\int\limits_0^\infty dx_2^- \, 
\contraction[2ex]
{}
{a}{_{\bot}^{i \, a}
(x_{2'}^- , \ul{x}_1) \:}
{a}
a_{\bot}^{i \, a} (x_{2'}^- , \ul{x}_1) \:
a^{+ \, b} (x_2^- , \ul{x}_0) =
 \sum_{\lambda, \lambda'}  \int d^2 x_2 \, d^2 x_{2'}  \left[
  \int\limits_{-\infty}^0 dx_{2'}^- \int \frac{d^4 k_{2'}}{(2\pi)^4} e^{i
    k_{2'}^+ x_{2'}^-} e^{i \ul{k}_{2'} \cdot \ul{x}_{2'1}} \frac{-i}{k_{2'}^2 +
    i\epsilon} \, \epsilon_{\lambda'}^{i \, *} \right]
 \\ & \times
\Bigg[ (U_{{\ul 2}, {\un 2}'; \lambda, \lambda'}^\textrm{pol})^{b a} \: 2\pi (2 k_2^-) \: \delta(k_2^- -
k_{2'}^-) \Bigg]
%
%
\left[ \int\limits_0^\infty dx_2^- \int \frac{d^4 k_2}{(2\pi)^4} e^{-i
    k_2^+ x_2^-} e^{-i \ul{k}_2 \cdot \ul{x}_{20}} \frac{-i}{k_2^2 +
    i\epsilon} \frac{{\un \epsilon}_\lambda \cdot {\un k}_2}{k_2^-} \right].\notag
\end{align}
The propagator \eqref{+perp_sub_eik_1} is separated by the square brackets into the interaction with the shock wave and two free-gluon propagators on either side of the shock wave. It neglects the instantaneous terms in the free-gluon propagators in the light-cone perturbation theory (LCPT) terminology \cite{Lepage:1980fj,Brodsky:1997de}, which is justified since such terms do not generate longitudinal logarithms, and, hence, do not contribute to the DLA evolution. 

Substituting \eq{Uxy_sub-eikonal} into \eq{+perp_sub_eik_1}, summing over polarizations and integrating over $k_2$ and $k_{2'}$ (except for $k^- = k_2^- = k_{2'}^-$) yields
\begin{align}\label{+perp_sub_eik_2}
\int\limits_{-\infty}^0 dx_{2'}^- \, & 
\int\limits_0^\infty dx_2^- \, 
\contraction[2ex]
{}
{a}{_{\bot}^{i \, a}
(x_{2'}^- , \ul{x}_1) \:}
{a}
\: 
a_{\bot}^{i \, a} (x_{2'}^- , \ul{x}_1) \:
a^{+ \, b} (x_2^- , \ul{x}_0) =
\\ & =
- \frac{1}{4 \pi^3} \, \int\limits_0^{p_2^-} d k^- \left[ \int d^2 x_2 \, \ln \left( \frac{1}{x_{21} \Lambda} \right) \, \frac{\epsilon^{ij} x_{20}^j}{x_{20}^2} \, (U_{{\ul 2}}^{\textrm{pol} [1]})^{b a} - i  \int d^2 x_2 \, d^2 x_{2'} \, \ln \left( \frac{1}{x_{2'1} \Lambda} \right) \, \frac{x_{20}^i}{x_{20}^2} \, (U_{{\ul 2}, {\un 2}'}^{\textrm{pol} [2]})^{b a}  \,   \right]. \notag 
\end{align}
The first term on the right of \eq{+perp_sub_eik_2} was obtained before in \cite{Kovchegov:2017lsr}.

As can be seen from the diagrams I and I$'$, or II and II$'$ in \fig{FIG:Q_evol}, the propagator \eqref{+perp_sub_eik_2} enters the evolution of  $Q_{10} (zs)$ together with the similar propagator, with the $x^-$-ordering of the endpoints reversed, along with the color indices $a, b$ of the gluon fields interchanged,
\begin{align}\label{+perp_sub_eik_3}
\int\limits_{-\infty}^0 dx_{2'}^- \, & 
\int\limits_0^\infty dx_2^- \, 
\contraction[2ex]
{}
{a}{_{\bot}^{i \, b}
(x_2^- , \ul{x}_1) \:}
{a}
\: 
a_{\bot}^{i \, b} (x_2^- , \ul{x}_1) \:
a^{+ \, a} (x_{2'}^- , \ul{x}_0) =
\\ & =
- \frac{1}{4 \pi^3} \, \int\limits_0^{p_2^-} d k^- \left[ \int d^2 x_2 \, \ln \left( \frac{1}{x_{21} \Lambda} \right) \, \frac{\epsilon^{ij} x_{20}^j}{x_{20}^2} \, (U_{{\ul 2}}^{\textrm{pol} [1]})^{b a} + i  \int d^2 x_2 \, d^2 x_{2'} \, \ln \left( \frac{1}{x_{21} \Lambda} \right) \, \frac{x_{2'0}^i}{x_{2'0}^2} \, (U_{{\ul 2}, {\un 2}'}^{\textrm{pol} [2]})^{b a}  \,   \right]. \notag 
\end{align}
One can clearly see that the eikonal Wilson line contribution $(U_{\un{2}})^{ba} \, \delta^2 (\un{x}_{22'})$ from \eq{U_sub-eikonal}, which is neglected here as a non-DLA contribution, would have entered Eqs.~\eqref{+perp_sub_eik_2} and \eqref{+perp_sub_eik_3} in the same way as $(U_{{\ul 2}, {\un 2}'}^{\textrm{pol} [2]})^{b a}$ does: this eikonal contribution would exactly vanish in the sum of Eqs.~\eqref{+perp_sub_eik_2} and \eqref{+perp_sub_eik_3}, justifying our neglecting of this contribution. 

Adding Eqs.~\eqref{+perp_sub_eik_2} and \eqref{+perp_sub_eik_3}, and employing \eq{UqG_decomp}, after some algebra we arrive at
\begin{align}\label{+perp_sub_eik_4}
& \int\limits_{-\infty}^0 dx_{2'}^- \, 
\int\limits_0^\infty dx_2^- \, \left[
\contraction[2ex]
{}
{a}{_{\bot}^{i \, a}
(x_{2'}^- , \ul{x}_1) \:}
{a}
\: 
a_{\bot}^{i \, a} (x_{2'}^- , \ul{x}_1) \:
a^{+ \, b} (x_2^- , \ul{x}_0) +
\contraction[2ex]
{}
{a}{_{\bot}^{i \, b}
(x_2^- , \ul{x}_1) \:}
{a}
\: 
a_{\bot}^{i \, b} (x_2^- , \ul{x}_1) \:
a^{+ \, a} (x_{2'}^- , \ul{x}_0) \right]
\\ & =
- \frac{1}{4 \pi^3} \, \int\limits_0^{p_2^-} d k^-  \int d^2 x_2 \, \Bigg\{ 2 \, \ln \left( \frac{1}{x_{21} \Lambda} \right) \, \frac{\epsilon^{ij} x_{20}^j}{x_{20}^2} \, (U_{{\ul 2}}^{pol [1]})^{b a} \notag \\ 
& + \frac{P^+}{s}  \int\limits_{-\infty}^\infty d z^- \, \left[ \frac{x_{20}^i}{x_{20}^2} \, \frac{x_{21}^j}{x_{21}^2} + \ln \left( \frac{1}{x_{21} \Lambda} \right) \, \left( \frac{\delta^{ij} \, x_{20}^2 - 2 x_{20}^i x_{20}^j}{x_{20}^4} + \delta^{ij} \, \pi \, \delta^2 ({\un x}_{20}) \right)  \right] \notag \\
& \times \, \left( U_{\ul 2}[\infty, z^-] \, \left[ \mathscr{D}^j - \cev{\mathscr{D}}^j \right] (z^-, {\un x}_2) \, U_{\ul 2}[z^-, - \infty]  \right)^{b a}  
\Bigg\}, \notag 
\end{align}
where we have used \eq{derivative}. Note that the contributions of $U^{\textrm{q} [2]}_{\un 2}$ from \eq{Uq2} also canceled in the sum \eqref{+perp_sub_eik_4}, similar to how the eikonal contributions disappeared earlier. 

Defining the gluon contribution to the adjoint polarized Wilson line of the second kind by (cf. \eq{Vi}) 
\begin{align}\label{Ui}
U_{\un{z}}^{i \, \textrm{G} [2]} \equiv \frac{P^+}{2 s} \, \int\limits_{-\infty}^{\infty} d {z}^- \, U_{\un{z}} [ \infty, z^-] \, \left[ \mathscr{D}^i (z^-, \un{z}) - \cev{\mathscr{D}}^i (z^-, \un{z}) \right]  \, U_{\un{z}} [ z^-, -\infty]  
\end{align}
we rewrite \eq{+perp_sub_eik_4} as
\begin{align}\label{+perp_sub_eik_5}
& \int\limits_{-\infty}^0 dx_{2'}^- \, 
\int\limits_0^\infty dx_2^- \, \left[
\contraction[2ex]
{}
{a}{_{\bot}^{i \, a}
(x_{2'}^- , \ul{x}_1) \:}
{a}
\: 
a_{\bot}^{i \, a} (x_{2'}^- , \ul{x}_1) \:
a^{+ \, b} (x_2^- , \ul{x}_0) +
\contraction[2ex]
{}
{a}{_{\bot}^{i \, b}
(x_2^- , \ul{x}_1) \:}
{a}
\: 
a_{\bot}^{i \, b} (x_2^- , \ul{x}_1) \:
a^{+ \, a} (x_{2'}^- , \ul{x}_0) \right]
\\ & 
=
- \frac{1}{2 \pi^3} \, \int\limits_0^{p_2^-} d k^-  \int d^2 x_2 \, 
\Bigg\{ \ln \left( \frac{1}{x_{21} \Lambda} \right) \, \frac{\epsilon^{ij} x_{20}^j}{x_{20}^2} \, \left( U_{{\ul 2}}^{\textrm{pol} [1]} \right)^{b a} \notag
\\ & +  \left[ \frac{x_{20}^i}{x_{20}^2} \, \frac{x_{21}^j}{x_{21}^2} + \ln \left( \frac{1}{x_{21} \Lambda} \right) \, \left( \frac{\delta^{ij} \, x_{20}^2 - 2 x_{20}^i x_{20}^j}{x_{20}^4} + \delta^{ij} \, \pi \, \delta^2 ({\un x}_{20}) \right) \right] \, \left( U_{\un{2}}^{j \, \textrm{G} [2]} \right)^{b a} \Bigg\}.\notag
\end{align}

Let us pose here to review the time-ordering arguments, previously detailed in \cite{Cougoulic:2019aja}. For the shock-wave picture to be valid, the light-cone lifetime of a gluon, which is $\sim 2 k^-/k_\perp^2$ for a gluon with momentum $k$, should be longer than the extent of the ``core" shock wave, the target proton, $\sim 1/P^+$. This gives $2 k^- P^+ \gg k_\perp^2$. Since $k^- = z p_2^-$ with $p_2^-$ the momentum of the original probe, $2 k^- P^+ = z s$. For a dipole, $k_\perp \sim 1/x_\perp$ with $x_\perp$ the dipole size. The lifetime ordering condition becomes $zs x_\perp^2 \gg 1$. The delta function in \eq{+perp_sub_eik_5}, which puts $x_{20}=0$, should be interpreted as putting $x^2_{20} = 1/zs$, since $1/zs$ is the shortest possible distance squared in the scattering system at hand. This means that $z s x^2_{20} =1$ and the gluon emission, corresponding to the delta-function term, is inside the ``core"  shock wave (and deep inside the shock wave made out of the subsequent emissions in the evolution). Therefore, the subsequent emissions cannot be outside the shock wave, and, hence, cannot generate longitudinal logarithms of energy. Hence, further evolution stops in the delta-function term in \eq{+perp_sub_eik_5}, and the delta function only contributes to the inhomogeneous term in the evolution equations. A similar observation has already been made in \cite{Kovchegov:2021iyc}. We will, therefore, discard the contribution of the delta function $\delta^2 ({\un x}_{20})$ to the evolution kernel in the following. It is possible that the delta-function term is canceled by the instantaneous term for the gluon propagator (in the LCPT terminology \cite{Lepage:1980fj,Brodsky:1997de}): we are not including such terms in the DLA calculation at hand and, hence, cannot verify that.

The contribution of the propagator \eqref{+perp_sub_eik_5} to the evolution of $F^{12} = \epsilon^{ij} \partial^i A^j$ (at the Abelian level) is proportional to
\begin{align}\label{+perp_sub_eik_6}
& \epsilon^{ji} \pd_1^j \left[ \int\limits_{-\infty}^0 dx_{2'}^- \, 
\int\limits_0^\infty dx_2^- \, \left(
\contraction[2ex]
{}
{a}{_{\bot}^{i \, a}
(x_{2'}^- , \ul{x}_1) \:}
{a}
\: 
a_{\bot}^{i \, a} (x_{2'}^- , \ul{x}_1) \:
a^{+ \, b} (x_2^- , \ul{x}_0) +
\contraction[2ex]
{}
{a}{_{\bot}^{i \, b}
(x_2^- , \ul{x}_1) \:}
{a}
\: 
a_{\bot}^{i \, b} (x_2^- , \ul{x}_1) \:
a^{+ \, a} (x_{2'}^- , \ul{x}_0) \right)
 \right] = \\ 
& 
- \frac{1}{2 \pi^3} \int\limits_0^{p_2^-} d k^- \int d^2 x_2 \, 
\Bigg\{  \frac{{\un x}_{21}}{x_{21}^2} \cdot \frac{{\un x}_{20}}{x_{20}^2} \left(U_{{\ul 2}}^{ \textrm{pol} [1]} \right)^{b a} + \left[ \frac{\epsilon^{ij} \, (x_{20}^j + x_{21}^j)}{x_{20}^2 \, x_{21}^2}  + \frac{2 \, {\un x}_{20} \times {\un x}_{21}}{x_{20}^2 \, x_{21}^2} \left( \frac{x_{21}^i}{x_{21}^2} - \frac{x_{20}^i}{x_{20}^2}\right) \right] \left( U_{\un{2}}^{i \, \textrm{G} [2]} \right)^{b a}   \Bigg\}. \notag 
\end{align}
The first term on the right agrees with (twice) the equation~(65) in \cite{Kovchegov:2017lsr}. Its contribution to the evolution of $Q_{10} (zs)$ in \fig{FIG:Q_evol} has been studied before \cite{Kovchegov:2017lsr,Kovchegov:2018znm}. Therefore, we need to concentrate on the contribution of the second term on the right of \eq{+perp_sub_eik_6}. 

Employing \eq{+perp_sub_eik_6} we see that the contribution of the diagrams I, I$'$, II and II$'$ from \fig{FIG:Q_evol} to the evolution of $Q_{10} (zs)$ is (cf.  \cite{Kovchegov:2015pbl,Kovchegov:2017lsr,Kovchegov:2018znm,Kovchegov:2021lvz})
\begin{align}\label{I+I'+II+II'}
& \textrm{I} + \textrm{I}' + \textrm{II} + \textrm{II}' = \frac{\as \, N_c}{2 \pi^2} \, \int\limits_{\frac{\Lambda^2}{s}}^z \frac{d z'}{z'} \, \int d^2 x_2 \Bigg\{ \left[ \frac{1}{x_{21}^2} -  \frac{{\un x}_{21}}{x_{21}^2} \cdot \frac{{\un x}_{20}}{x_{20}^2} \right] \, \frac{1}{N_c^2} \llangle  \mbox{T} \, \tr \left[ t^b V_{\ul 0} t^a V_{\un 1}^{\dagger} \right] \left(U_{{\ul 2}}^{\textrm{pol} [1]} \right)^{b a} + \textrm{c.c.} \rrangle (z' s) \\ 
& + \left[ 2 \frac{\epsilon^{ij} \, x_{21}^j}{x_{21}^4} - \frac{\epsilon^{ij} \, (x_{20}^j + x_{21}^j)}{x_{20}^2 \, x_{21}^2}  - \frac{2 \, {\un x}_{20} \times {\un x}_{21}}{x_{20}^2 \, x_{21}^2} \left( \frac{x_{21}^i}{x_{21}^2} - \frac{x_{20}^i}{x_{20}^2}\right) \right] \frac{1}{N_c^2} \llangle  \mbox{T} \, \tr \left[ t^b V_{\ul 0} t^a V_{\un 1}^{\dagger} \right]  \left( U_{\un{2}}^{i \, \textrm{G} [2]} \right)^{b a}  + \textrm{c.c.} \rrangle (z' s)  \Bigg\} . \notag 
\end{align}
Here the emitted gluon's longitudinal momentum is $k^- = z' \, p_2^-$, while the minimum minus momentum fraction in the parent dipole is labeled $z$  \cite{Kovchegov:2015pbl,Kovchegov:2017lsr,Kovchegov:2018znm,Kovchegov:2021lvz}. The second line of \eq{I+I'+II+II'} was not present in the earlier works \cite{Kovchegov:2015pbl,Kovchegov:2017lsr,Kovchegov:2018znm,Kovchegov:2021lvz}. 

While the eikonal diagrams in \fig{FIG:Q_evol} are evaluated the same way as usual \cite{Mueller:1994rr,Mueller:1994jq,Mueller:1995gb,Balitsky:1995ub,Balitsky:1998ya,Kovchegov:1999yj,Kovchegov:1999ua}, the contribution of the diagram III is evaluated using the anti-quark propagator through the shock wave \cite{Kovchegov:2018znm}
\begin{align}\label{q_propagator1}
& \int\limits_{-\infty}^0 dx_{2'}^- \, 
\int\limits_0^\infty dx_2^- \, 
\contraction
{}
{\bar\psi^i_\alpha}
{(x_2^- , {\un x}_1) \:}
{psi^j_\beta}
\bar\psi^i_\alpha (x_2^- , {\un x}_1) \: 
\psi^j_\beta (x_{2'}^- , {\un x}_1) 
= \sum_{\sigma, \sigma'}  \int d^2 x_2 \, d^2 x_{2'} \, \left[ \, \int\limits_{-\infty}^0 dx_{2'}^- \, \int
 \frac{d^4 k_{2'}}{(2\pi)^4} \, e^{i k_{2'}^+ x_{2'}^-} \, 
e^{i \ul{k}_{2'} \cdot \ul{x}_{2'1}} \,  \frac{i}{k_{2'}^2 + i\epsilon}  \, \left( v_{\sigma'} (k_{2'}) \right)_\beta \right] \notag
\\ & \times
\left[ \left( - V_{{\un 2}, {\un 2}' ; - \sigma, - \sigma'}^\dagger \right)^{ji} \, (2 k_2^-) \, (2\pi) \, \delta(k_2^- - k_{2'}^-) \right]  \, \left[ \int\limits_0^\infty dx_2^- \, \int \frac{d^4 k_2}{(2\pi)^4} \, e^{-i k_2^+ x_2^-} \, e^{- i \ul{k}_2 \cdot \ul{x}_{21} }  \, \frac{i}{k_2^2 + i\epsilon} \, \left( {\bar v}_\sigma (k_2) \right)_\alpha \right] ,
\end{align}
where $\alpha, \beta$ are the spinor indices, while $i, j$ are the color indices. The propagator \eqref{q_propagator1} is taken to be local in the transverse plane, since this is how it always enters our evolution in \fig{FIG:Q_evol}. Once again, we neglect the instantaneous terms, as being beyond the DLA we are constructing here. 

Simplifying the propagator \eqref{q_propagator1} to (again, $k^- = k_2^- = k_{2'}^-$)
\begin{align}\label{q_propagator2}
& \int\limits_{-\infty}^0 dx_{2'}^- \, 
\int\limits_0^\infty dx_2^- \, 
\contraction
{}
{\bar\psi^i_\alpha}
{(x_2^- , {\un x}_1) \:}
{psi^j_\beta}
\bar\psi^i_\alpha (x_2^- , {\un x}_1) \: 
\psi^j_\beta (x_{2'}^- , {\un x}_1) 
= - \frac{1}{\pi} \sum_{\sigma}  \int d k^- \, k^-  \int d^2 x_2 \, d^2 x_{2'} \, \left[ \int
 \frac{d^2 k_{2'}}{(2\pi)^2} \, 
e^{i \ul{k}_{2'} \cdot \ul{x}_{2'1}} \,  \frac{1}{{\un k}_{2'}^2 }  \, \left( v_{\sigma} (k_{2'}) \right)_\beta \right] 
\\ & \times
\left( \sigma \, V_{\un 2}^{\textrm{pol} [1] \, \dagger} \, \delta^2 ({\un x}_{22'}) - V_{{\ul 2}, {\un 2}'}^{\textrm{pol} [2] \, \dagger} \right)^{ji} \, \left[ \int \frac{d^2 k_2}{(2\pi)^2} \, e^{- i \ul{k}_2 \cdot \ul{x}_{21} }  \, \frac{1}{{\un k}_2^2} \, \left( {\bar v}_\sigma (k_2) \right)_\alpha \right] , \notag
\end{align}
we use it to contract the quark fields in the definition \eqref{Q_def2} of $Q_{10} (zs)$, where the quark field dependence enters through \eq{Vq1} (see \cite{Kovchegov:2018znm}). Employing \eq{spinors2}, we arrive at the following contribution of the diagram III to the evolution of $Q_{10} (zs)$:
\begin{align}
\textrm{III} = \frac{\as}{4 \pi^2 N_c} \, \int\limits_{\frac{\Lambda^2}{s}}^z \frac{d z'}{z'} \, \int d^2 x_2 \Bigg\{ & \, \frac{1}{x_{21}^2} \, \llangle  \mbox{T} \,\tr \left[ t^b \, V_{\un 0} \, t^a \, V_{\un 2}^{\textrm{pol} [1] \, \dagger} \right] \, U_{\un 1}^{ba} \rrangle (z' s) \\
&+ i \int d^2 x_{2'} \, \frac{{\un x}_{21}}{x_{21}^2} \times \frac{{\un x}_{2'1}}{x_{2'1}^2} \, \llangle  \mbox{T} \, \tr \left[ t^b \, V_{\un 0} \, t^a \, V_{{\ul 2}, {\un 2}'}^{\textrm{pol} [2] \, \dagger} \right] \, U_{\un 1}^{ba} \rrangle (z' s) + \textrm{c.c.} \Bigg\} \notag .
\end{align}
Employing Eqs.~\eqref{VqG_decomp} and \eqref{VxyG2} the integral over ${\un x}_{2'}$ can be carried out, yielding
\begin{align}\label{III}
\textrm{III} = \frac{\as}{4 \pi^2 N_c} \, \int\limits_{\frac{\Lambda^2}{s}}^z \frac{d z'}{z'} \, \int d^2 x_2 \Bigg\{ & \,  \frac{1}{x_{21}^2} \, \llangle  \mbox{T} \, \tr \left[ t^b \, V_{\un 0} \, t^a \, V_{\un 2}^{\textrm{pol} [1] \, \dagger} \right] \, U_{\un 1}^{ba} \rrangle (z' s) \\
&+ 2 \, \frac{\epsilon^{ij} \, {\un x}_{21}^j}{x_{21}^4} \, \llangle  \mbox{T} \, \tr \left[ t^b \, V_{\un 0} \, t^a \, V_{\ul 2}^{i \, \textrm{G} [2] \, \dagger} \right] \, U_{\un 1}^{ba} \rrangle (z' s) + \textrm{c.c.} \Bigg\}  . \notag
\end{align}
While the first term on the right-hand side of \eq{III} was obtained before \cite{Kovchegov:2015pbl,Kovchegov:2018znm}, the second term is new. 

Combining Eqs.~\eqref{I+I'+II+II'} and \eqref{III}, while adding the well-known contribution \cite{Mueller:1994rr,Mueller:1994jq,Mueller:1995gb,Balitsky:1995ub,Balitsky:1998ya,Kovchegov:1999yj,Kovchegov:1999ua} of the eikonal diagrams  from \fig{FIG:Q_evol}, and suppressing the time-ordering sign for brevity, we obtain our final evolution equation for the fundamental polarized dipole amplitude $Q_{10} (zs)$:
\begin{align}\label{Q_evol_main}
& \frac{1}{2 N_c} \, \llangle \tr \left[ V_{\ul 0} \, V_{\un 1}^{\textrm{pol} [1] \, \dagger} \right] + \textrm{c.c.}   \rrangle (zs) =   \frac{1}{2 N_c} \, \llangle \tr \left[ V_{\ul 0} \, V_{\un 1}^{\textrm{pol} [1] \, \dagger} \right] + \textrm{c.c.}  \rrangle_0 (zs) \\
& + \frac{\as \, N_c}{2 \pi^2} \, \int\limits_{\frac{\Lambda^2}{s}}^z \frac{d z'}{z'} \, \int d^2 x_2 \Bigg\{ \left[ \frac{1}{x_{21}^2} -  \frac{{\un x}_{21}}{x_{21}^2} \cdot \frac{{\un x}_{20}}{x_{20}^2} \right] \, \frac{1}{N_c^2} \llangle  \tr \left[ t^b V_{\ul 0} t^a V_{\un 1}^{\dagger} \right] \left(U_{{\ul 2}}^{\textrm{pol} [1]} \right)^{b a} + \textrm{c.c.} \rrangle (z' s) \notag \\ 
& + \left[ 2 \frac{\epsilon^{ij} \, x_{21}^j}{x_{21}^4} - \frac{\epsilon^{ij} \, (x_{20}^j + x_{21}^j)}{x_{20}^2 \, x_{21}^2}  - \frac{2 \, {\un x}_{20} \times {\un x}_{21}}{x_{20}^2 \, x_{21}^2} \left( \frac{x_{21}^i}{x_{21}^2} - \frac{x_{20}^i}{x_{20}^2}\right) \right] \frac{1}{N_c^2} \llangle  \tr \left[ t^b V_{\ul 0} t^a V_{\un 1}^{\dagger} \right]  \left( U_{\un{2}}^{i \, \textrm{G} [2]} \right)^{b a}  + \textrm{c.c.} \rrangle (z' s)  \Bigg\}  \notag \\
& + \frac{\as N_c}{4 \pi^2 } \, \int\limits_{\frac{\Lambda^2}{s}}^z \frac{d z'}{z'} \, \int \frac{d^2 x_2}{x_{21}^2} \Bigg\{ \frac{1}{N_c^2} \, \llangle \tr \left[ t^b \, V_{\un 0} \, t^a \, V_{\un 2}^{\textrm{pol} [1] \, \dagger} \right] \, U_{\un 1}^{ba} \rrangle (z' s) + 2 \, \frac{\epsilon^{ij} \, {\un x}_{21}^j}{x_{21}^2} \, \frac{1}{N_c^2} \, \llangle \tr \left[ t^b \, V_{\un 0} \, t^a \, V_{\ul 2}^{i \, \textrm{G} [2] \, \dagger} \right] \, U_{\un 1}^{ba} \rrangle (z' s) + \textrm{c.c.} \Bigg\} \notag \\
& + \frac{\as \, N_c}{2 \pi^2} \, \int\limits_{\frac{\Lambda^2}{s}}^z \frac{d z'}{z'} \, \int d^2 x_2 \, \frac{x_{10}^2}{x_{21}^2 \, x_{20}^2} \,  \Bigg\{ \frac{1}{N_c^2} \, \llangle \tr \left[ t^b \, V_{\un 0} \, t^a \, V_{\un 1}^{\textrm{pol} [1] \, \dagger} \right] \, U_{\un 2}^{ba} \rrangle (z' s)  - \frac{C_F}{N_c^2} \, \llangle \tr \left[ V_{\un 0} \, V_{\un 1}^{\textrm{pol} [1] \, \dagger} \right] \rrangle (z' s) + \textrm{c.c.}  \Bigg\}  .  \notag
\end{align}
Here the $0$ subscript on the angle brackets, $\llangle \ldots \rrangle_0$, denotes the inhomogeneous term \cite{Kovchegov:2015pbl,Kovchegov:2017lsr,Kovchegov:2018znm,Kovchegov:2021lvz}, which is the polarized dipole amplitude calculated in the quasi-classical approximation of the  Glauber--Mueller/McLerran--Venugopalan model  \cite{Mueller:1989st,McLerran:1993ni,McLerran:1993ka,McLerran:1994vd}, extended in \cite{Cougoulic:2020tbc} to include helicity dependence. 

For brevity reasons we did not include into \eq{Q_evol_main} the $\theta$-functions imposing the lifetime-ordering condition (of the daughter parton lifetime compared to the parent parton lifetime) 
\cite{Kovchegov:2015pbl,Kovchegov:2017lsr,Kovchegov:2018znm,Cougoulic:2019aja,Kovchegov:2021lvz}. We imply that every IR-divergent integral in \eq{Q_evol_main} is regulated via multiplication of the integrand by such a condition: e.g., by $\theta (z \, x_{10}^2 - z' \, x_{21}^2)$. Similarly, the ultraviolet (UV) divergences are regulated by the lifetime ordering condition discussed above, $\min \{ x_{21}^2, x_{20}^2 \} > 1/(z' s)$.

The equation \eqref{Q_evol_main} contains the DLA evolution of $Q_{10} (zs)$, resumming all powers of $\as \, \ln^2 (1/x)$ for this amplitude. In fact, it includes part of the single-logarithmic evolution too, by resumming all terms with the longitudinal logarithms of $x$: thus, it sums up some of the powers of $\as \, \ln (1/x)$. These terms were labeled SLA$_L$ in \cite{Kovchegov:2021lvz}, for the single-logarithmic approximation terms, coming from the longitudinal logarithms. 

Note also that the expressions in this Section have been written down by ignoring the nuances of  properly ordering the Wilson lines in the correlators discussed in detail in \cite{Kovchegov:2018znm}. In part, this is due to the PT-symmetry argument presented above, which shows that such issues are not relevant for the helicity operators at hand. Additionally, the ordering of operators is not important in the quasi-classical approximation, which is applicable to helicity observables as shown in \cite{Cougoulic:2020tbc}.

As is the case with Balitsky hierarchy \cite{Balitsky:1995ub,Balitsky:1998ya}, \eq{Q_evol_main} is not closed. It will become a closed equation only in the large-$N_c$ and large-$N_c \& N_f$ limits considered below (see also \cite{Kovchegov:2015pbl,Kovchegov:2017lsr,Kovchegov:2018znm,Kovchegov:2021lvz}). Additionally, different from the earlier works \cite{Kovchegov:2015pbl,Kovchegov:2017lsr,Kovchegov:2018znm,Kovchegov:2021lvz}, this evolution equation mixes polarized ``Wilson lines" of the first ``[1]" and second ``[2]" kind, in the notation of Eqs.~\eqref{Vxy_sub-eikonal} and \eqref{Uxy_sub-eikonal}. Hence, to close this equation, even in the large-$N_c$ and large-$N_c \& N_f$ limits, we will need to develop evolution equations for the polarized ``Wilson lines" of the second ``[2]" kind.

Before doing that, we need to construct the evolution of the adjoint analogue of the amplitude $Q_{10} (zs)$, defined by \cite{Kovchegov:2018znm,Kovchegov:2021lvz}
\begin{align}\label{G_adj_def}
G^\textrm{adj}_{10} (zs) \equiv \frac{1}{2 (N_c^2 -1)} \, \mbox{Re} \, \llangle \mbox{T} \, \mbox{Tr} \left[ U_{\ul 0} \, U_{{\un 1}}^{\textrm{pol} [1] \, \dagger} \right] + \mbox{T} \, \mbox{Tr} \left[ U_{{\un 1}}^{\textrm{pol} [1]} \, U_{\ul 0}^\dagger \right] \rrangle (zs) 
\end{align}
with Tr denoting an adjoint trace, delineating it from the fundamental one. The evolution of 
$G^\textrm{adj}_{10} (zs)$ is needed because unlike the large-$N_c$ limit, in the large-$N_c \& N_f$ limit there is no simple relationship between the two amplitudes, $Q_{10}$ and $G^\textrm{adj}_{10}$, and both of them enter the corresponding evolution equations \cite{Kovchegov:2015pbl,Kovchegov:2018znm}. 

\begin{figure}[ht]
\centering
\includegraphics[width=  \textwidth]{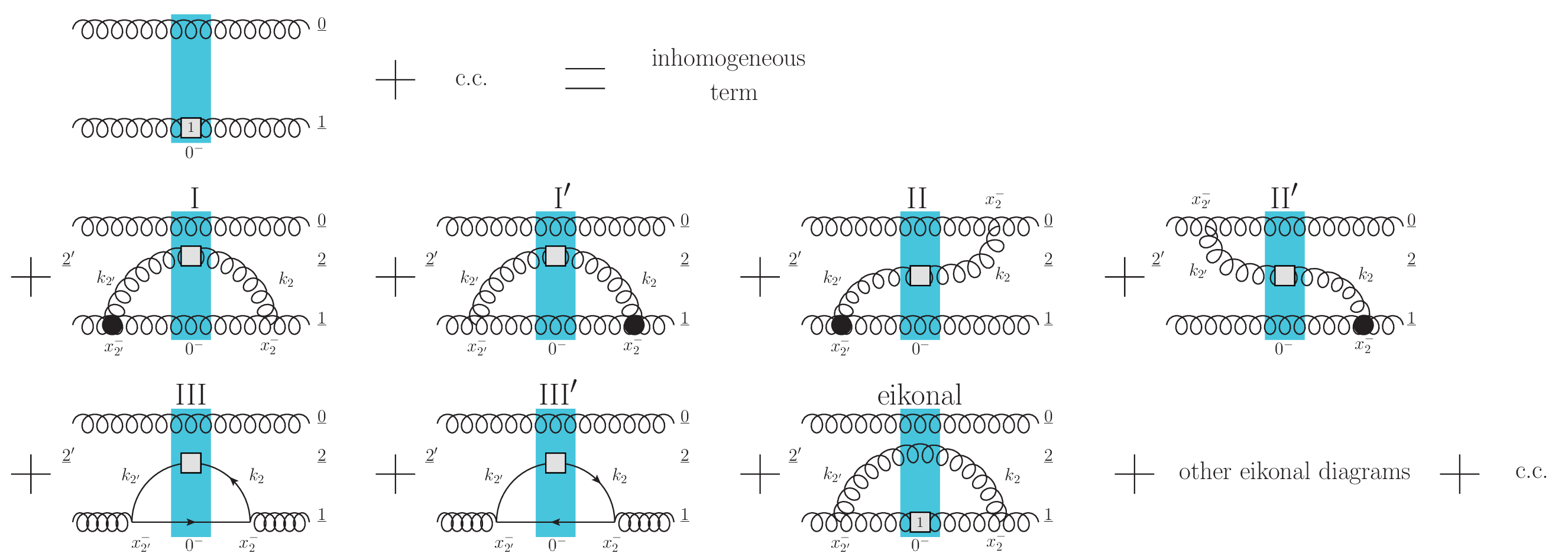}  
\caption{Diagrams representing the evolution of the adjoint polarized dipole amplitude $G^\textrm{adj}_{10}$.
Again, the square box on the gluon and quark lines represents the sub-eikonal interaction with the target given by \eq{Uxy_sub-eikonal} for gluons and \eq{Vxy_sub-eikonal} for quarks. The same square box, but with number 1 in it, on the gluon line denotes the interaction described by $U_{\ul 1}^{\textrm{pol} [1]}$ only. The black circle denotes the sub-eikonal triple-gluon vertex generated by the $F^{12}$ operator in \eq{UG1}, that is, by the $F^{12}$ part of $U_{\ul 1}^{\textrm{pol} [1]}$. All momenta flow to the right.}
\label{FIG:Gadj_evol}
\end{figure}

Our discussion of the evolution for $G^\textrm{adj}_{10}$ 
will be brief, since it mirrors the above derivation for the fundamental dipole; in addition, large parts of this calculation were done before in \cite{Kovchegov:2018znm}, albeit omitting the polarized Wilson lines of the second kind. The relevant diagrams are shown in \fig{FIG:Gadj_evol} and are similar to \fig{FIG:Q_evol}. The notation is the same as in \fig{FIG:Q_evol}, with the (minor) differences detailed in the caption of \fig{FIG:Gadj_evol}.

The contribution of diagrams I, I$'$, II, and II$'$ from \fig{FIG:Gadj_evol} is calculated in the same way as that for the same diagrams in \fig{FIG:Q_evol}, employing the propagator in \eq{+perp_sub_eik_6}, with the differences being $N_c^2 -1$ in the denominator of \eq{G_adj_def} as opposed to $N_c$ in the denominator of \eq{Q_def2}, the overall factor of 2 in \eq{UG1} absent in \eq{VG1}, and the adjoint representation versus fundamental. We get 
\begin{align}\label{adj_I+I'+II+II'}
& \textrm{I} + \textrm{I}' + \textrm{II} + \textrm{II}' = \frac{\as}{\pi^2} \, \int\limits_{\frac{\Lambda^2}{s}}^z \frac{d z'}{z'} \, \int d^2 x_2 \Bigg\{ \left[ \frac{1}{x_{21}^2} -  \frac{{\un x}_{21}}{x_{21}^2} \cdot \frac{{\un x}_{20}}{x_{20}^2} \right] \, \frac{1}{N_c^2 -1} \llangle  \mbox{T} \, \Tr \left[ T^b U_{\ul 0} T^a U_{\un 1}^{\dagger} \right] \left(U_{{\ul 2}}^{\textrm{pol} [1]} \right)^{b a} + \textrm{c.c.} \rrangle (z' s) \\ 
& + \left[ 2 \frac{\epsilon^{ij} \, x_{21}^j}{x_{21}^4} - \frac{\epsilon^{ij} \, (x_{20}^j + x_{21}^j)}{x_{20}^2 \, x_{21}^2}  - \frac{2 \, {\un x}_{20} \times {\un x}_{21}}{x_{20}^2 \, x_{21}^2} \left( \frac{x_{21}^i}{x_{21}^2} - \frac{x_{20}^i}{x_{20}^2}\right) \right] \frac{1}{N_c^2 -1} \llangle  \mbox{T} \, \Tr \left[ T^b U_{\ul 0} T^a U_{\un 1}^{\dagger} \right]  \left( U_{\un{2}}^{i \, \textrm{G} [2]} \right)^{b a}  + \textrm{c.c.} \rrangle (z' s)  \Bigg\} . \notag 
\end{align}

Diagrams III and III' in \fig{FIG:Gadj_evol} are calculated similar to the diagram III in \fig{FIG:Q_evol}, with the propagator \eqref{q_propagator2} and the operator \eqref{Uq1} coming in particularly handy. This gives
\begin{align}\label{adj_III}
\textrm{III} + \textrm{III}'  = - \frac{\as \, N_f}{2 \pi^2 (N_c^2 -1)} \, \int\limits_{\frac{\Lambda^2}{s}}^z \frac{d z'}{z'} \, \int d^2 x_2 \Bigg\{ & \, \frac{1}{x_{21}^2} \, \llangle \mbox{T} \, \tr \left[ t^b \, V_{\un 1} \, t^a \, V_{\un 2}^{\textrm{pol} [1] \, \dagger} \right] \, U_{\un 0}^{ba} \rrangle (z' s) \\
& + 2 \, \frac{\epsilon^{ij} \, {\un x}_{21}^j}{x_{21}^4} \, \llangle \mbox{T} \, \tr \left[ t^b \, V_{\un 1} \, t^a \, V_{\ul 2}^{i \, \textrm{G} [2] \, \dagger} \right] \, U_{\un 0}^{ba} \rrangle  (z' s) + \textrm{c.c.} \Bigg\}  , \notag 
\end{align}
where we multiplied everything by the number of quark flavors $N_f$ to account for the sum over flavors in the loop. 

Finally, combining Eqs.~\eqref{adj_I+I'+II+II'} and \eqref{adj_III}, adding the well-known contribution of the eikonal diagrams in \fig{FIG:Gadj_evol}, and again suppressing the time-ordering sign for brevity, we arrive at the evolution equation for the adjoint polarized dipole of the first kind, 
\begin{align}\label{G_adj_evol}
& \frac{1}{2 (N_c^2 -1)} \, \mbox{Re} \, \llangle \mbox{Tr} \left[ U_{\ul 0} \, U_{{\un 1}}^{\textrm{pol} [1] \, \dagger} \right] + \textrm{c.c.}  \rrangle (zs) = \frac{1}{2 (N_c^2 -1)} \, \mbox{Re} \, \llangle \mbox{Tr} \left[ U_{\ul 0} \, U_{{\un 1}}^{\textrm{pol} [1] \, \dagger} \right] + \textrm{c.c.}  \rrangle_0 (zs) \\
& + \frac{\as}{\pi^2} \, \int\limits_{\frac{\Lambda^2}{s}}^z \frac{d z'}{z'} \, \int d^2 x_2 \Bigg\{ \left[ \frac{1}{x_{21}^2} -  \frac{{\un x}_{21}}{x_{21}^2} \cdot \frac{{\un x}_{20}}{x_{20}^2} \right] \, \frac{1}{N_c^2 -1} \llangle \Tr \left[ T^b U_{\ul 0} T^a U_{\un 1}^{\dagger} \right] \left(U_{{\ul 2}}^{\textrm{pol} [1]} \right)^{b a} + \textrm{c.c.} \rrangle (z' s) \notag \\ 
& + \left[ 2 \frac{\epsilon^{ij} \, x_{21}^j}{x_{21}^4} - \frac{\epsilon^{ij} \, (x_{20}^j + x_{21}^j)}{x_{20}^2 \, x_{21}^2}  - \frac{2 \, {\un x}_{20} \times {\un x}_{21}}{x_{20}^2 \, x_{21}^2} \left( \frac{x_{21}^i}{x_{21}^2} - \frac{x_{20}^i}{x_{20}^2}\right) \right] \frac{1}{N_c^2 -1} \llangle  \Tr \left[ T^b U_{\ul 0} T^a U_{\un 1}^{\dagger} \right]  \left( U_{\un{2}}^{i \, \textrm{G} [2]} \right)^{b a}  + \textrm{c.c.} \rrangle (z' s)  \Bigg\} \notag \\
& - \frac{\as \, N_f}{2 \pi^2 (N_c^2 -1)} \int\limits_{\frac{\Lambda^2}{s}}^z \frac{d z'}{z'}  \int d^2 x_2 \, \Bigg\{ \frac{1}{x_{21}^2} \, \llangle \tr \left[ t^b \, V_{\un 1} \, t^a \, V_{\un 2}^{\textrm{pol} [1] \, \dagger} \right] \, U_{\un 0}^{ba} \rrangle (z' s) \notag \\ 
& \hspace*{5cm} + 2 \, \frac{\epsilon^{ij} \, {\un x}_{21}^j}{x_{21}^4} \, \llangle  \tr \left[ t^b \, V_{\un 1} \, t^a \, V_{\ul 2}^{i \, \textrm{G} [2] \, \dagger} \right] \, U_{\un 0}^{ba} \rrangle (z' s) + \textrm{c.c.} \Bigg\}  \notag \\
& + \frac{\as}{2 \pi^2} \, \int\limits_{\frac{\Lambda^2}{s}}^z \frac{d z'}{z'} \, \int d^2 x_2 \, \frac{x_{10}^2}{x_{21}^2 \, x_{20}^2} \, \frac{1}{N_c^2-1} \, \Bigg\{  \llangle \Tr \left[ T^b \, U_{\un 0} \, T^a \, U_{\un 1}^{\textrm{pol} [1] \, \dagger} \right] \, U_{\un 2}^{ba} \rrangle (z' s)  - N_c \, \llangle \Tr \left[ U_{\un 0} \, U_{\un 1}^{\textrm{pol} [1] \, \dagger} \right] \rrangle (z' s) + \textrm{c.c.}  \Bigg\}  .  \notag
\end{align}
Just like \eq{Q_evol_main}, \eq{G_adj_evol} resums both the DLA and SLA$_L$ terms. The regulator $\theta (z \, x_{10}^2 - z' \, x_{21}^2)$ is implied for the IR-divergent integrals while the $\min \{ x_{21}^2, x_{20}^2 \} > 1/(z' s)$ condition regulates the UV divergences. In the previous version of this evolution in the literature \cite{Kovchegov:2018znm}, the terms in the third and fifth lines were absent.


\subsubsection{Evolution Equations for Fundamental and Adjoint $G^i_{10} (zs)$}
\label{sec:evo_op_Gi}

\begin{figure}[ht]
\centering
\includegraphics[width=  \textwidth]{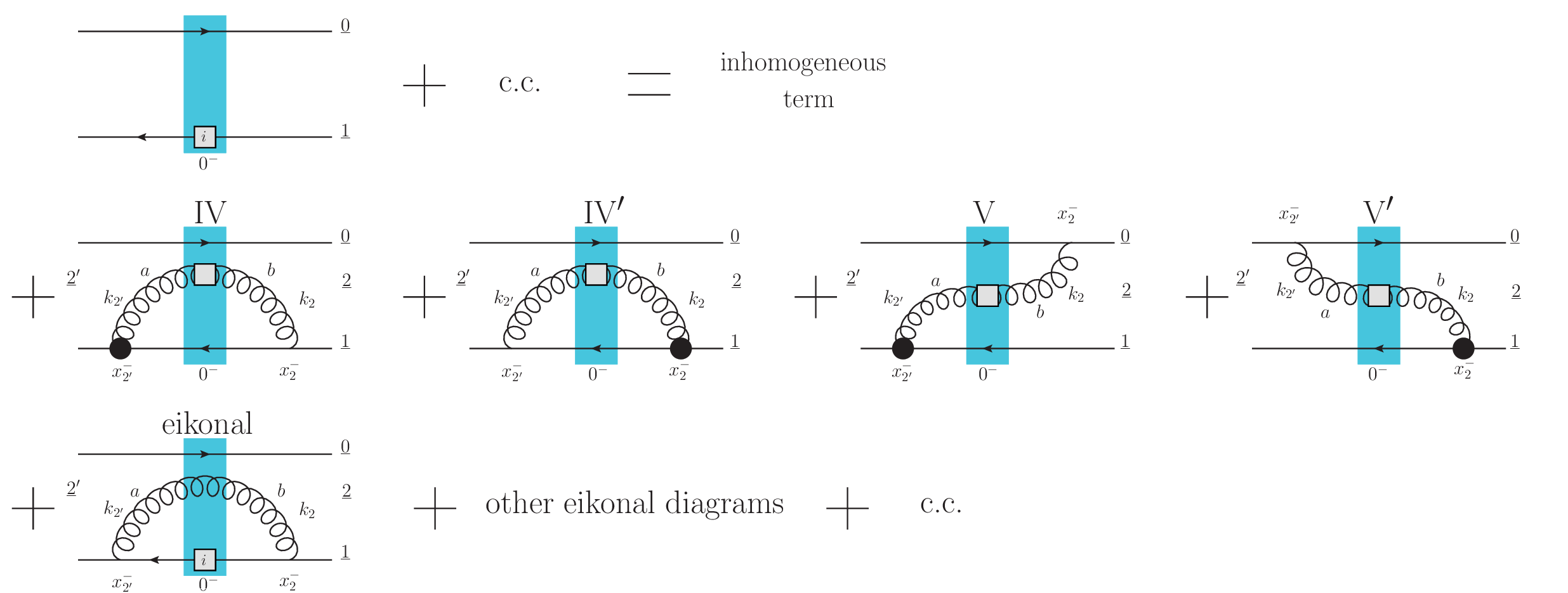}  
\caption{Diagrams representing the evolution of the polarized dipole amplitude of the second kind, $G^i_{10} (zs)$. Once again, the square box on the gluon and quark lines represents the sub-eikonal interaction with the target given by \eq{Uxy_sub-eikonal} for gluons and \eq{Vxy_sub-eikonal} for quarks. The same square box, but with an $i$ in it, on the quark line denotes the interaction described by $V_{\un{1}}^{i \, \textrm{G} [2]}$. The black circle denotes the sub-eikonal vertex generated by the $z^- {\pd}^i A^+ (z^-, \un{z}) + A^i (z^-, \un{z})$ operator in \eq{Vi2}, which contributes to $G^{i}_{10}$ through \eq{Gj2}. All momenta flow to the right.}
\label{FIG:Gi_evol}
\end{figure}

Our next step is to construct the evolution equation for the polarized amplitude of the second kind, $G^i_{10} (zs)$, defined in \eq{Gj2}. The process is very similar to the evolution equations for the polarized dipoles of the first kind constructed above in Sec.~\ref{sec:Qevol}. The diagrams contributing to the evolution of $G^i_{10} (zs)$ are shown in \fig{FIG:Gi_evol} (cf. Fig.~3 in \cite{Kovchegov:2017lsr}). Since the polarized Wilson line of the second kind $V_{\un{z}}^{i \, \textrm{G} [2]}$ from \eq{Vi} is a purely gluonic operator, the evolution of $G^i_{10} (zs)$ in \fig{FIG:Gi_evol} does not involve soft quark emissions, unlike the evolution of $Q_{10}$, which contains diagram III in \fig{FIG:Q_evol}. Since the eikonal diagrams' contribution is the same as above and in the literature \cite{Mueller:1994rr,Mueller:1994jq,Mueller:1995gb,Balitsky:1995ub,Balitsky:1998ya,Kovchegov:1999yj,Kovchegov:1999ua}, we only need to calculate diagrams IV, IV$'$, V, and V$'$ in \fig{FIG:Gi_evol}. 

Start with the operator in \eq{Vi}, which we rewrite as
\begin{align}\label{Vi2}
V_{\un{z}}^{i \, \textrm{G} [2]} & \equiv \frac{P^+}{2 s} \, \int\limits_{-\infty}^{\infty} d {z}^- \, V_{\un{z}} [ \infty, z^-] \, \left[ {D}^i (z^-, \un{z}) - \cev{D}^i (z^-, \un{z}) \right]  \, V_{\un{z}} [ z^-, -\infty]   \\ 
& = - i g \, \frac{P^+}{s} \, \int\limits_{-\infty}^{\infty} d {z}^- \, V_{\un{z}} [ \infty, z^-] \, \left[  z^- {\pd}^i A^+ (z^-, \un{z}) + A^i (z^-, \un{z}) \right] \, V_{\un{z}} [ z^-, -\infty]. \notag
\end{align}

Sub-eikonal evolution of the operator in \eq{Vi2} depicted in the diagram V of \fig{FIG:Gi_evol} includes the following propagator:
\begin{align} \label{Vi4}
& \int\limits_{-\infty}^0 dx_{2'}^- \,  
\int\limits_0^\infty dx_2^- \, \Big[ x_{2'}^- {\pd}^i a^{+ \, a} (x_{2'}^- , \ul{x}_1)
\contraction[2ex]
{}
{+}{a^{i \, a}
(x_{2'}^- , \ul{x}_1) \Big] \:}
{a}
\: 
+ a^{i \, a} (x_{2'}^- , \ul{x}_1) \Big] \:
a^{+ \, b} (x_1^- , \ul{x}_0) \\ & =
 \sum_{\lambda, \lambda'}  \int d^2 x_2 d^2 x_{2'}  \left[
  \int\limits_{-\infty}^0 dx_{2'}^- \int \frac{d^4 k_1}{(2\pi)^4} e^{i
    k_{2'}^+  x_{2'}^-} e^{i \ul{k}_1 \cdot \ul{x}_{2'1}} \frac{-i}{k_{2'}^2 +
    i\epsilon} \left( \epsilon_{\lambda'}^{*i} + i x_{2'}^- \, k_{2'}^i \, \frac{{\un \epsilon}_{\lambda'}^* \cdot {\un k}_{2'}}{k_{2'}^-} \right)  \right]
\notag \\ & \hspace{1cm} \times
\Bigg[ \left( U_{{\ul 2}, {\un 2}'; \lambda, \lambda'}^\textrm{pol} \right)^{b a} \: 2\pi (2 k_{2'}^-) \: \delta(k_2^- -
k_{2'}^-) \Bigg]
%
%
\left[ \int\limits_0^\infty dx_2^- \int \frac{d^4 k_2}{(2\pi)^4} e^{-i
    k_2^+ x_2^-} e^{-i \ul{k}_2 \cdot \ul{x}_{20}} \frac{-i}{k_2^2 +
    i\epsilon} \frac{{\un \epsilon}_\lambda \cdot {\un k}_2}{k_2^-} \right]. \notag 
\end{align}
The contraction sign over the square brackets in \eq{Vi4} is an abbreviated notation implying the sum of contractions $x_{2'}^- {\pd}^i  \contraction[2ex]{}{a}{^{+ \, a}}{a} a^{+ \, a} a^{+ \, b}$ and $\contraction[2ex]{}{a}{^{i \, a}}{a} a^{i \, a} a^{+ \, b}$.

We integrate \eq{Vi4} over $x_{2'}^-$ and $x_2^-$ first, then over $k_2^+, k_{2'}^+, k_{2'}^-$ (note the second-order pole in the $k_{2'}^+$ integral due to the extra power of $x_{2'}^-$), Fourier-transform over ${\un k}_2$ and ${\un k}_{2'}$, and use \eq{Uxy_sub-eikonal} to sum over polarizations. The following Fourier transform integral comes in handy:
\begin{align}\label{Fr1}
\int \frac{d^2 k}{(2 \pi)^2} \frac{e^{i {\un k} \cdot {\un x}}}{k_\perp^2} \left[ \delta^{ij} - \frac{2 k^i k^j}{k_\perp^2} \right] = - \frac{1}{4 \pi} \, \left[ \delta^{ij} - \frac{2 x^i x^j}{x_\perp^2} \right].
\end{align} 
In the end one arrives at 
\begin{align}\label{Vi5}
& \int\limits_{-\infty}^0 dx_{2'}^- \,  
\int\limits_0^\infty dx_2^- \, \Big[ x_{2'}^- {\pd}^i a^{+ \, a} (x_{2'}^- , \ul{x}_1)
\contraction[2ex]
{}
{+}{a^{i \, a}
(x_{2'}^- , \ul{x}_1) \Big] \:}
{a}
\: 
+ a^{i \, a} (x_{2'}^- , \ul{x}_1) \Big] \:
a^{+ \, b} (x_2^- , \ul{x}_0)  \\ & = \frac{1}{(2 \pi)^3} \int\limits_0^{p_2^-} d k^- \Bigg\{ \int d^2 x_2 \left[ \frac{\epsilon^{ij} x_{20}^j}{x_{20}^2} - 2 x_{21}^i \frac{{\un x}_{21} \times {\un x}_{20}}{x_{21}^2 \, x_{20}^2} \right] \left( U_{{\ul 2}}^{\textrm{pol} [1]} \right)^{b a}  - i  \int d^2 x_2 d^2 x_{2'} \left[ \frac{ x_{20}^i}{x_{20}^2} - 2 x_{2'1}^i \frac{{\un x}_{2'1} \cdot {\un x}_{20}}{x_{2'1}^2 \, x_{20}^2} \right] \left( U_{{\ul 2}, {\un 2}'}^{\textrm{pol} [2]} \right)^{b a} \Bigg\}, \notag
\end{align}
where, as usual, $k^- = k_2^- = k_{2'}^-$.

Similarly, for the other time ordering which enters in diagram V$'$ from \fig{FIG:Gi_evol} we obtain
\begin{align}\label{Vi6}
& \int\limits_{-\infty}^0 dx_{2'}^- \,  
\int\limits_0^\infty dx_2^- \, \Big[ x_2^- {\pd}^i a^{+ \, b} (x_2^- , \ul{x}_1)
\contraction[2ex]
{}
{+}{a^{i \, b}
(x_2^- , \ul{x}_1) \Big] \:}
{a}
\: 
+ a^{i \, b} (x_2^- , \ul{x}_1) \Big] \:
a^{+ \, a} (x_{2'}^- , \ul{x}_0) \\ & = \frac{1}{(2 \pi)^3} \int\limits_0^{p_2^-} d k^- \Bigg\{ \int d^2 x_2 \left[ \frac{\epsilon^{ij} x_{20}^j}{x_{20}^2} - 2 x_{21}^i \frac{{\un x}_{21} \times {\un x}_{20}}{x_{21}^2 \, x_{20}^2} \right] \left( U_{{\ul 2}}^{\textrm{pol} [1]} \right)^{b a} + i  \int d^2 x_2 d^2 x_{2'} \left[ \frac{ x_{2'0}^i}{x_{2'0}^2} - 2 x_{21}^i \frac{{\un x}_{21} \cdot {\un x}_{2'0}}{x_{21}^2 \, x_{2'0}^2} \right] \left( U_{{\ul 2}, {\un 2}'}^{\textrm{pol} [2]} \right)^{b a} \Bigg\}, \notag 
\end{align}
such that the sum of both time orderings \eqref{Vi5} and \eqref{Vi6} is
\begin{align}\label{aaa}
& \int\limits_{-\infty}^0 dx_{2'}^- \,  
\int\limits_0^\infty dx_2^- \left\{ \Big[ x_{2'}^- {\pd}^i a^{+ \, a} (x_{2'}^- , \ul{x}_1)
\contraction[2ex]
{}
{+}{a^{i \, a}
(x_{2'}^- , \ul{x}_1) \Big] \:}
{a}
\: 
+ a^{i \, a} (x_{2'}^- , \ul{x}_1) \Big] \:
a^{+ \, b} (x_2^- , \ul{x}_0) +
\Big[ x_2^- {\pd}^i a^{+ \, b} (x_2^- , \ul{x}_1)
\contraction[2ex]
{}
{+}{a^{i \, b}
(x_2^- , \ul{x}_1) \Big] \:}
{a}
\: 
+ a^{i \, b} (x_2^- , \ul{x}_1) \Big] \:
a^{+ \, a} (x_{2'}^- , \ul{x}_0) \right\}
\notag \\ & = \frac{1}{4 \pi^3} \, \int\limits_0^{p_2^-} d k^- \int d^2 x_2 \, \Bigg\{\left[ \frac{\epsilon^{ij} x_{20}^j}{x_{20}^2} - 2 x_{21}^i \frac{{\un x}_{21} \times {\un x}_{20}}{x_{21}^2 \, x_{20}^2} \right] \, \left( U_{{\ul 2}}^{\textrm{pol} [1]} \right)^{b a} \\ 
& +  \left[ \delta^{ij} \left( 2 \frac{{\un x}_{20} \cdot {\un x}_{21}}{x_{20}^2 \, x_{21}^2} + \frac{1}{x_{20}^2} \right) + 2 \frac{x_{21}^i \, x_{20}^j}{x_{21}^2 \, x_{20}^2} \left( 2 \frac{{\un x}_{20} \cdot {\un x}_{21}}{x_{20}^2} + 1 \right) - 2 \frac{x_{21}^i \, x_{21}^j}{x_{21}^2 \, x_{20}^2} \left( 2 \frac{{\un x}_{20} \cdot {\un x}_{21}}{x_{21}^2} + 1 \right) - 2 \frac{x_{20}^i \, x_{20}^j}{x_{20}^4}  \right] \, \left( U_{{\ul 2}}^{j \, \textrm{G} [2]} \right)^{b a} \Bigg\}. \notag 
\end{align}
Here we have neglected the delta-function terms, similar to those appearing in \eq{derivative}. 

Employing the propagator \eqref{aaa} to calculate diagrams IV, IV$'$, V, and V$'$ in \fig{FIG:Gi_evol}, and adding in the eikonal contribution, which is the same as in \eq{Q_evol_main}, we derive the evolution equation for $G^i_{10} (zs)$:
\begin{align}\label{Gi_evol_main}
& \frac{1}{2 N_c} \, \llangle  \tr \left[  V_{\un 0} \, V_{\un{1}}^{i \, \textrm{G} [2] \, \dagger}  \right] + \textrm{c.c.}   \rrangle (zs) =   \frac{1}{2 N_c} \, \llangle \tr \left[  V_{\un 0} \, V_{\un{1}}^{i \, \textrm{G} [2] \, \dagger}  \right] + \textrm{c.c.}  \rrangle_0 (zs) \\
& + \frac{\as \, N_c}{4 \pi^2} \, \int\limits_{\frac{\Lambda^2}{s}}^z \frac{d z'}{z'} \, \int d^2 x_2 \Bigg\{ \left[ \frac{\epsilon^{ij} x_{21}^j}{x_{21}^2} - \frac{\epsilon^{ij} x_{20}^j}{x_{20}^2} + 2 x_{21}^i \frac{{\un x}_{21} \times {\un x}_{20}}{x_{21}^2 \, x_{20}^2} \right] \, \frac{1}{N_c^2} \llangle  \tr \left[ t^b V_{\ul 0} t^a V_{\un 1}^{\dagger} \right] \left(U_{{\ul 2}}^{\textrm{pol} [1]} \right)^{b a} + \textrm{c.c.} \rrangle (z' s) \notag \\ 
& + \left[ \delta^{ij} \left( \frac{3}{x_{21}^2} -  2 \, \frac{{\un x}_{20} \cdot {\un x}_{21}}{x_{20}^2 \, x_{21}^2} - \frac{1}{x_{20}^2} \right)  - 2 \frac{x_{21}^i \, x_{20}^j}{x_{21}^2 \, x_{20}^2} \left( 2 \frac{{\un x}_{20} \cdot {\un x}_{21}}{x_{20}^2} + 1 \right) + 2 \frac{x_{21}^i \, x_{21}^j}{x_{21}^2 \, x_{20}^2} \left( 2 \frac{{\un x}_{20} \cdot {\un x}_{21}}{x_{21}^2} + 1 \right) + 2 \frac{x_{20}^i \, x_{20}^j}{x_{20}^4} - 2 \frac{x_{21}^i \, x_{21}^j}{x_{21}^4}   \right] \notag \\
& \times \, \frac{1}{N_c^2} \llangle  \tr \left[ t^b V_{\ul 0} t^a V_{\un 1}^{\dagger} \right]  \left( U_{\un{2}}^{j \, \textrm{G} [2]} \right)^{b a}  + \textrm{c.c.} \rrangle (z' s)  \Bigg\}  \notag \\
& + \frac{\as \, N_c}{2 \pi^2} \, \int\limits_{\frac{\Lambda^2}{s}}^z \frac{d z'}{z'} \, \int d^2 x_2 \, \frac{x_{10}^2}{x_{21}^2 \, x_{20}^2} \,  \Bigg\{ \frac{1}{N_c^2} \, \llangle \tr \left[ t^b \, V_{\un 0} \, t^a \, V_{\un 1}^{i \, \textrm{G} [2] \, \dagger} \right] \, \left( U_{\un 2} \right)^{ba} \rrangle (z' s)  - \frac{C_F}{N_c^2} \, \llangle \tr \left[ V_{\un 0} \, V_{\un 1}^{i \, \textrm{G} [2] \, \dagger} \right] \rrangle (z' s) + \textrm{c.c.}  \Bigg\}  .  \notag
\end{align}
In the version of \eq{Gi_evol_main} constructed in \cite{Kovchegov:2017lsr} (see Eq.~(80) there), the term in the third and fourth lines of \eq{Gi_evol_main} was missing, and the kernel of the term in the second line was different, since the contributions of the fields $a^{+ \, a} (x_{2'}^-, \ul{x}_1)$ and $a^{+ \, b} (x_2^-, \ul{x}_1)$ in \eq{aaa} was neglected.

The adjoint version of \eq{Gi_evol_main} can be constructed by analogy. One gets
\begin{align}\label{Gi_adj_evol_main}
& \frac{1}{2 (N_c^2 -1)} \, \llangle  \Tr \left[  U_{\un 0} \, U_{\un{1}}^{i \, \textrm{G} [2] \, \dagger}  \right] + \textrm{c.c.}   \rrangle (zs) =   \frac{1}{2 (N_c^2 -1)} \, \llangle \Tr \left[  U_{\un 0} \, U_{\un{1}}^{i \, \textrm{G} [2] \, \dagger}  \right] + \textrm{c.c.}  \rrangle_0 (zs) \\
& + \frac{\as}{4 \pi^2} \, \int\limits_{\frac{\Lambda^2}{s}}^z \frac{d z'}{z'} \, \int d^2 x_2 \Bigg\{ \left[ \frac{\epsilon^{ij} x_{21}^j}{x_{21}^2} - \frac{\epsilon^{ij} x_{20}^j}{x_{20}^2} + 2 x_{21}^i \frac{{\un x}_{21} \times {\un x}_{20}}{x_{21}^2 \, x_{20}^2} \right] \, \frac{1}{N_c^2-1} \llangle  \Tr \left[ T^b U_{\ul 0} T^a U_{\un 1}^{\dagger} \right] \left(U_{{\ul 2}}^{\textrm{pol} [1]} \right)^{b a} + \textrm{c.c.} \rrangle (z' s) \notag \\ 
& + \left[ \delta^{ij} \left( \frac{3}{x_{21}^2} -  2 \, \frac{{\un x}_{20} \cdot {\un x}_{21}}{x_{20}^2 \, x_{21}^2} - \frac{1}{x_{20}^2} \right)  - 2 \frac{x_{21}^i \, x_{20}^j}{x_{21}^2 \, x_{20}^2} \left( 2 \frac{{\un x}_{20} \cdot {\un x}_{21}}{x_{20}^2} + 1 \right) + 2 \frac{x_{21}^i \, x_{21}^j}{x_{21}^2 \, x_{20}^2} \left( 2 \frac{{\un x}_{20} \cdot {\un x}_{21}}{x_{21}^2} + 1 \right) + 2 \frac{x_{20}^i \, x_{20}^j}{x_{20}^4} - 2 \frac{x_{21}^i \, x_{21}^j}{x_{21}^4}   \right] \notag \\
& \times \, \frac{1}{N_c^2-1} \llangle  \Tr \left[ T^b U_{\ul 0} T^a U_{\un 1}^{\dagger} \right]  \left( U_{\un{2}}^{j \, \textrm{G} [2]} \right)^{b a}  + \textrm{c.c.} \rrangle (z' s)  \Bigg\}  \notag \\
& + \frac{\as}{2 \pi^2} \, \int\limits_{\frac{\Lambda^2}{s}}^z \frac{d z'}{z'} \, \int d^2 x_2 \, \frac{x_{10}^2}{x_{21}^2 \, x_{20}^2} \, \frac{1}{N_c^2 -1} \, \Bigg\{ \llangle \Tr \left[ T^b \, U_{\un 0} \, T^a \, U_{\un 1}^{i \, \textrm{G} [2] \, \dagger} \right] \left( U_{\un 2} \right)^{ba} \rrangle (z' s)  - N_c \, \llangle \Tr \left[ U_{\un 0} \, U_{\un 1}^{i \, \textrm{G} [2] \, \dagger} \right] \rrangle (z' s) + \textrm{c.c.}  \Bigg\}  .  \notag
\end{align}
Equations \eqref{Gi_evol_main} and \eqref{Gi_adj_evol_main} resum both the DLA and SLA$_L$ terms. The regulator $\theta (z \, x_{10}^2 - z' \, x_{21}^2)$ is implied again for all the IR-divergent integrals in these equations, while the $\min \{ x_{21}^2, x_{20}^2 \} > 1/(z' s)$ condition again regulates the UV divergences. 

Equations \eqref{Q_evol_main}, \eqref{G_adj_evol}, \eqref{Gi_evol_main}, and \eqref{Gi_adj_evol_main} form a closed set of equation at the level of (polarized) Wilson lines. To achieve a closed set of equations at the level of (polarized) dipole amplitude, we need to take the large-$N_c$ or the large-$N_c \& N_f$ limits \cite{Kovchegov:2015pbl,Kovchegov:2017lsr,Kovchegov:2018znm,Kovchegov:2021lvz}. This is what we will do next.


\subsection{Evolution Equations in the Large-$N_c$ Limit}
\label{sec:large-Nc}

To obtain the large-$N_c$ limit of the helicity evolution at hand, we will follow the standard approach \cite{Kovchegov:2018znm}. We start with \eq{G_adj_evol}, and drop the term proportional to $N_f$ on its right-hand side, as being due to the quark loop correction, suppressed at large $N_c$. Similarly neglecting all quark loop contribution, we replace
\begin{align}\label{pol->G}
U_{\un{x}}^{\textrm{pol} [1]} \to U_{\un{x}}^{\textrm{G} [1]}
\end{align}
everywhere in \eq{G_adj_evol}, thus discarding $U_{\un{x}}^{\textrm{q} [1]}$ in \eq{UqG_decomp}. In the same spirit, we define the large-$N_c$ analogue of $Q_{10} (zs)$ from \eq{Q_def2} by \cite{Kovchegov:2018znm}
\begin{align}\label{Ggdef}
 G_{10} (z s) \equiv \frac{1}{2 \, N_c} \, \mbox{Re} \, \llangle \mbox{T} \, \tr \left[ V_{\un 0} \,  V_{\un 1}^{\textrm{G} [1] \,\dagger} \right] + \mbox{T} \,  \tr \left[ V_{\un 1}^{\textrm{G} [1]} \, V_{\un 0}^\dagger \right]   \rrangle (zs) .
\end{align}

We employ the well-known relation between the adjoint and fundamental Wilson lines,
\begin{align}\label{Uba}
(U_{\ul x})^{ba} = 2 \, \tr [t^b V_{\ul x} t^a V_{\ul x}^\dagger ].
\end{align}
Using \eq{Uba} one can show that (see Eq.~(73) in \cite{Kovchegov:2018znm})
\begin{align}\label{Uba1}
\left( U_{\un{x}}^{\textrm{G} [1]} \right)^{b a} = 4 \, \tr \left[ t^b \, V_{\un{x}} \, t^a \, V_{\un{x}}^{\textrm{G} [1] \, \dagger} \right] + 4 \, \tr \left[ t^b \, V_{\un{x}}^{\textrm{G} [1] }  \, t^a \, V_{\un{x}}^\dagger \right]. 
\end{align}
This relation, in turn, gives
\begin{align}\label{eq:Gadj_Gfund_largeNc}
G^\textrm{adj}_{10} (z s) = 4 \, G_{10} (z s) \, S_{10} (zs)
\end{align}
at large $N_c$. Here we have defined the ``standard" unpolarized dipole $S$-matrix \cite{Mueller:1994rr,Mueller:1994jq,Mueller:1995gb,Balitsky:1995ub,Balitsky:1998ya,Kovchegov:1999yj,Kovchegov:1999ua,Jalilian-Marian:1997dw,Jalilian-Marian:1997gr,Weigert:2000gi,Iancu:2001ad,Iancu:2000hn,Ferreiro:2001qy}
\begin{align}
S_{10} (zs) = \frac{1}{N_c} \, \left\langle \mbox{T} \, \tr \left[ V_{\un 0} \,  V_{\un 1}^{\dagger} \right] \right\rangle (zs).
\end{align}
We assume that $S_{10} (zs)$ is real, neglecting the odderon contribution to the imaginary part of $S_{10} (zs)$, as suppressed by a power of $\as$ \cite{Bartels:1980pe,Kwiecinski:1980wb,Nicolescu:1990ii,Janik:1998xj,Bartels:1999yt,Korchemsky:2001nx,Kovchegov:2003dm,Ewerz:2003xi,Hatta:2005as,Kovner:2005qj,Jeon:2005cf,Hagiwara:2020mqb}.

Similarly, defining the adjoint version of $G^i_{10} (z s)$ from \eq{Gj2} by
\begin{align}\label{G_i_adj_def}
G^{i \, \textrm{adj}}_{10} (zs) \equiv \frac{1}{2 (N_c^2 -1)} \, \mbox{Re} \, \llangle \mbox{T} \, \mbox{Tr} \left[ U_{\ul 0} \, U_{{\un 1}}^{i \, \textrm{G} [2] \, \dagger} \right] + \mbox{T} \, \mbox{Tr} \left[ U_{{\un 1}}^{i \, \textrm{G} [2]} \, U_{\ul 0}^\dagger \right] \rrangle (zs) ,
\end{align}
one can use \eq{Uba} to show that 
\begin{align}\label{Uba2}
\left( U_{\un{x}}^{i \, \textrm{G} [2]} \right)^{b a} = 2 \, \tr \left[ t^b \, V_{\un{x}} \, t^a \, V_{\un{x}}^{i \, \textrm{G} [2] \, \dagger} \right] + 2 \, \tr \left[ t^b \, V_{\un{x}}^{i \, \textrm{G} [2] }  \, t^a \, V_{\un{x}}^\dagger \right]
\end{align}
such that 
\begin{align}\label{eq:Gadj_i_Gfund_i_largeNc}
G^{i \, \textrm{adj}}_{10} (z s) = 2 \, G^i_{10} (z s) \, S_{10} (zs) .
\end{align}

Using the above results, along with the Fierz identity, we can similarly simplify the operators on the right-hand side of \eq{G_adj_evol} at large $N_c$, obtaining (cf. \cite{Kovchegov:2018znm,Kovchegov:2021lvz})
\begin{subequations}\label{largeNc1}
\begin{align}
& \frac{1}{N_c^2 -1} \llangle \Tr \left[ T^b U_{\ul 0} T^a U_{\un 1}^{\dagger} \right] \left(U_{{\ul 2}}^{\textrm{G} [1]} \right)^{b a} + \textrm{c.c.} \rrangle (z s) = 4 \, N_c \, S_{10} (zs) \, \left[ S_{20} (zs) \, G_{21} (zs) + S_{21} (zs) \, \Gamma_{20,21} (zs)  \right], \\
& \frac{1}{N_c^2 -1} \llangle \Tr \left[ T^b U_{\ul 0} T^a U_{\un 1}^{\dagger} \right] \left(U_{{\ul 2}}^{i \, \textrm{G} [2]} \right)^{b a} + \textrm{c.c.} \rrangle (z s) = 2 \, N_c \, S_{10} (zs) \, \left[ S_{20} (zs) \, G^i_{21} (zs) + S_{21} (zs) \, \Gamma^i_{20,21} (zs)  \right], \\
& \frac{1}{N_c^2 -1} \llangle \Tr \left[ T^b U_{\ul 0} T^a U_{\un 1}^{\textrm{G} [1] \dagger} \right] \left(U_{\ul 2} \right)^{b a} + \textrm{c.c.} \rrangle (z s) = 4 \, N_c \, S_{20} (zs) \, \left[ S_{10} (zs) \, G_{21} (zs) + S_{21} (zs) \, \Gamma_{10,21} (zs)  \right].
\end{align}
\end{subequations}

Here $\Gamma_{20,21} (zs)$ is the ``neighbor" polarized dipole amplitude of the first kind \cite{Kovchegov:2015pbl,Kovchegov:2016zex,Kovchegov:2017lsr,Kovchegov:2018znm}: its operator definition is the same as for $G_{20} (zs)$, see \eq{Ggdef}. However, the evolution in $\Gamma_{20,21} (zs)$ is subject to the lifetime of subsequent emissions limited by $z \, x_{21}^2$ from above. Hence the evolution depends on the size of the neighbor dipole $21$, justifying the name of the amplitude. Similarly, the ``neighbor" polarized dipole amplitude of the second kind, $\Gamma^i_{20,21} (zs)$, is defined by \eq{Gj2} with the same lifetime constraint on the subsequent evolution. The choice of which amplitude in Eqs.~\eqref{largeNc1} becomes the ``neighbor" amplitude is made assuming that $x_{21} \ll x_{20}$ in the DLA, as is justified by the kernel in \eq{G_adj_evol}.

Employing Eqs.~\eqref{eq:Gadj_Gfund_largeNc} and \eqref{largeNc1} along with the trick detailed in the Appendix~D of \cite{Kovchegov:2021lvz}, we arrive at the large-$N_c$ version of \eq{G_adj_evol}, 
\begin{align}\label{G_largeNc_evol}
& G_{10} (zs) = G_{10}^{(0)} (zs) + \frac{\as \, N_c}{2 \pi^2} \, \int\limits_{\frac{\Lambda^2}{s}}^z \frac{d z'}{z'} \, \int d^2 x_2 \Bigg\{ 2 \left[ \frac{1}{x_{21}^2} -  \frac{{\un x}_{21}}{x_{21}^2} \cdot \frac{{\un x}_{20}}{x_{20}^2} \right]  \left[ S_{20} (z's) \, G_{21} (z's) + S_{21} (z's) \, \Gamma^\textrm{gen}_{20,21} (z's)  \right] \\ 
& + \left[ 2 \frac{\epsilon^{ij} \, x_{21}^j}{x_{21}^4} - \frac{\epsilon^{ij} \, (x_{20}^j + x_{21}^j)}{x_{20}^2 \, x_{21}^2}  - \frac{2 \, {\un x}_{20} \times {\un x}_{21}}{x_{20}^2 \, x_{21}^2} \left( \frac{x_{21}^i}{x_{21}^2} - \frac{x_{20}^i}{x_{20}^2}\right) \right] \left[ S_{20} (z's) \, G^i_{21} (z's) + S_{21} (z's) \, \Gamma^{i \, \textrm{gen}}_{20,21} (z's)  \right] \notag \\
& +  \frac{x_{10}^2}{x_{21}^2 \, x_{20}^2} \, \left[ S_{20} (z's) \, G_{12} (z's) - \Gamma^\textrm{gen}_{10,21} (z's)  \right]  \Bigg\}  .  \notag
\end{align}

In spirit with summing up the DLA and SLA$_L$ terms simultaneously, we replaced $\Gamma_{10,21}$ and $\Gamma^{i}_{20,21}$ in \eq{G_largeNc_evol} by the generalized polarized dipole amplitudes \cite{Kovchegov:2017lsr,Kovchegov:2021lvz}
\begin{subequations}\label{eqn:Nc7}
\begin{align}
& \Gamma^\textrm{gen}_{10, 32} (zs) \equiv G_{10} (zs ) \, \theta\left(x_{32}-x_{10}\right) + \Gamma_{10,32} (zs) \, \theta (x_{10}-x_{32} ), \\
& \Gamma^{i \, \textrm{gen}}_{10, 32} (zs) \equiv G^i_{10} (zs ) \, \theta\left(x_{32}-x_{10}\right) + \Gamma^i_{10,32} (zs) \, \theta (x_{10}-x_{32} ).
\end{align}
\end{subequations}
The amplitudes in Eqs.~\eqref{eqn:Nc7} become $\Gamma_{10,21}$ and $\Gamma^{i}_{20,21}$, respectively, in the DLA limit, and reduce back to $G_{10}$ and $G^i_{10}$ for the SLA$_L$ terms, in which the ordering between the dipole size and its neighbor dipole size is not important. Note that in Eqs.~\eqref{eqn:Nc7}, $x_{10}$ and $x_{32}$ can be any general transverse separations, that is, neither of them is necessarily the parent or daughter dipole size.

To extract the DLA contribution from \eq{G_largeNc_evol}, we put $S_{21} = S_{20} = 1$ in it, since the evolution of the unpolarized dipole $S$-matrix is SLA$_L$ \cite{Balitsky:1995ub,Balitsky:1998ya,Kovchegov:1999yj,Kovchegov:1999ua,Jalilian-Marian:1997dw,Jalilian-Marian:1997gr,Weigert:2000gi,Iancu:2001ad,Iancu:2000hn,Ferreiro:2001qy}. In addition, it appears to be more convenient to integrate \eq{G_largeNc_evol} over the impact parameters, while employing \eq{decomp}. The same decomposition applies to $\Gamma^{i}_{20,21}$, since it depends only on the size $x_{21}$ of the dipole $21$, and not on its orientation in the transverse plane,
\begin{align}\label{decomp2}
  \int d^2 \left( \frac{x_{1} + x_0}{2} \right) \, \Gamma^{i}_{20,21} (z s) = (x_{20})_\bot^i \, \Gamma_1
  (x_{20}^2, x_{21}^2, z s) + \epsilon^{ij} \, (x_{20})_\bot^j \, \Gamma_2 (x_{20}^2, x_{21}^2, z s) .
\end{align} 
Defining (cf. \eq{Q_int})
\begin{align}\label{G_int}
G (x^2_{10} , zs) \equiv \int d^2 \left( \frac{x_0 + x_1}{2} \right) \, G_{10} (zs), \ \ \  \Gamma (x_{20}^2, x^2_{21}, zs) \equiv \int d^2 \left( \frac{x_0 + x_2}{2} \right) \, \Gamma_{20,21} (zs)
\end{align}
we write the impact-parameter integrated part of \eq{G_largeNc_evol} as
\begin{align}\label{G_DLA_largeNc_evol1}
G (x^2_{10} , zs) = G^{(0)} (x^2_{10} , zs) & + \frac{\as \, N_c}{2 \pi^2} \, \int\limits_{\frac{\Lambda^2}{s}}^z \frac{d z'}{z'} \, \int d^2 x_2 \Bigg\{ \frac{1}{x_{21}^2}  \theta (x_{10} - x_{21}) \left[\Gamma (x_{10}^2, x_{21}^2,  z's) + 3 \, G (x^2_{21}, z's) \right] \\ 
& + \Bigg[ \frac{2}{x_{21}^2} - \frac{{\un x}_{21} \cdot ({\un x}_{20} + {\un x}_{21})}{x_{20}^2 \, x_{21}^2}  + \frac{2 \, ({\un x}_{21} \times {\un x}_{20})^2}{x_{20}^4 \, x_{21}^2} \Bigg] \, G_2 (x_{21}^2, z' s) \notag \\ 
& + \Bigg[ 2 \, \frac{ {\un x}_{20} \cdot {\un x}_{21}}{x_{21}^4} - \frac{{\un x}_{20} \cdot ({\un x}_{20} + {\un x}_{21})}{x_{20}^2 \, x_{21}^2}  + \frac{2 \, ({\un x}_{21} \times {\un x}_{20})^2}{x_{20}^2 \, x_{21}^4}  \Bigg] \, \Gamma_2^\textrm{gen} (x_{20}^2, x_{21}^2, z' s) \Bigg\}  .  \notag
\end{align}
Here we have applied the DLA simplifications to the parts of the integral kernel containing amplitudes $G$ and $\Gamma$. Note that the contributions of $G_1$ and $\Gamma_1$ defined in the decompositions \eqref{decomp} and \eqref{decomp2} vanish, due to a single Levi-Civita symbol $\epsilon^{ij}$ multiplying those functions in the $x_2$ integrals: it is impossible to make a non-zero scalar quantity out of a single transverse vector ${\un x}_{10}$ and one factor of $\epsilon^{ij}$.

We now need to extract the DLA part of the kernel containing amplitudes $G_2$ and $\Gamma_2^\textrm{gen}$ in \eq{G_DLA_largeNc_evol1}.\footnote{Here and below, when extracting DLA parts of various evolution equations, we will assume that the impact parameter-integrated amplitudes without transverse indices, $G, \Gamma, G_2, \Gamma_2$, etc., do not contain integer powers of the dipole sizes, $x_{10}$, $x_{21}$, $x_{20}$, etc., and the dependence on these distances enters the amplitudes only as perturbatively small ($\sim \sqrt{\alpha_s}$ or $\sim \alpha_s$) powers or logarithms of $x_{10}$, $x_{21}$, $x_{20}$, etc. This assumption is supported by the Born-level initial conditions (the inhomogeneous terms) shown below (see also \cite{Kovchegov:2016zex,Kovchegov:2017lsr}).} The $x_2$-integral in those terms appears to have no IR divergence and no UV divergence as $x_{20} \to 0$. However, there is a divergence at $x_{21} \to 0$, due to the first term in each square bracket: keeping those terms only we obtain
\begin{align}\label{G_DLA_largeNc_evol2}
G (x^2_{10} , zs) = G^{(0)} (x^2_{10} , zs) & + \frac{\as \, N_c}{2 \pi^2} \, \int\limits_{\frac{\Lambda^2}{s}}^z \frac{d z'}{z'} \, \int d^2 x_2 \, \theta (x_{10} - x_{21})  \, \Bigg\{ \frac{1}{x_{21}^2}  \left[\Gamma (x_{10}^2, x_{21}^2,  z's) + 3 \, G (x^2_{21}, z's) \right] \\ 
& + \frac{2}{x_{21}^2} \, G_2 (x_{21}^2, z' s) + 2 \, \frac{ {\un x}_{20} \cdot {\un x}_{21}}{x_{21}^4}  \, \Gamma_2 (x_{20}^2, x_{21}^2, z' s) \Bigg\}  .  \notag
\end{align}
The last term in \eq{G_DLA_largeNc_evol2} contains a power-law divergence as $x_{21} \to 0$: however, this divergence vanishes after angular averaging. Writing ${\un x}_{20} = {\un x}_{10} + {\un x}_{21}$ in that term, and expanding in the powers of $x_{21} \ll x_{10}$ while keeping only divergent terms as $x_{21} \to 0$, we get 
\begin{align}\label{expansion}
& \int d^2 x_2 \, \theta (x_{10} - x_{21}) \, 2 \, \frac{ {\un x}_{20} \cdot {\un x}_{21}}{x_{21}^4}  \, \Gamma_2 (x_{20}^2, x_{21}^2, z' s) \\ 
& \approx \int d^2 x_2 \, \theta (x_{10} - x_{21})  \,  \frac{2}{x_{21}^2} \, \left[ \Gamma_2 (x_{10}^2, x_{21}^2, z' s)  + x_{10}^2 \frac{\pd}{\pd x_{10}^2} \Gamma_2 (x_{10}^2, x_{21}^2, z' s) \right]. \notag
\end{align}
The second term on the right of \eq{expansion} contains a logarithmic derivative with respect to $x_{10}^2$. Such derivative removes one power of $\ln x_{10}^2$, and is, therefore, outside of the DLA. Therefore, we neglect this term here, keeping in mind that it will need to be reinstated in the single-logarithmic approximation (SLA). Inserting the first term from the right-hand side of \eq{expansion} into \eq{G_DLA_largeNc_evol2}, we arrive at the DLA version of \eq{G_largeNc_evol},
\begin{align}\label{G_DLA_largeNc_evol}
G (x^2_{10} , zs) = G^{(0)} (x^2_{10} , zs) + \frac{\as \, N_c}{2 \pi} \, \int_{\frac{1}{s x_{10}^2}}^z \frac{d z'}{z'} \, \int^{x_{10}^2}_\frac{1}{z's} \frac{d x^2_{21}}{x_{21}^2} \,  \Bigg[ &  \Gamma (x_{10}^2, x_{21}^2,  z's) + 3 \, G (x^2_{21}, z's)  \\ 
& + 2 \, G_2 (x_{21}^2, z' s) + 2 \, \Gamma_2 (x_{10}^2, x_{21}^2, z' s) \Bigg]  .  \notag
\end{align}
The $G$ and $\Gamma$ terms in \eq{G_DLA_largeNc_evol} agree with that found in the literature \cite{Kovchegov:2015pbl,Kovchegov:2016zex,Kovchegov:2017lsr,Kovchegov:2018znm}, while the $G_2$ and $\Gamma_2$ terms are new. 

The DLA large-$N_c$ evolution equation for the neighbor amplitude $\Gamma$ can be found by analogy, employing the existing techniques \cite{Kovchegov:2015pbl,Kovchegov:2016zex,Kovchegov:2017lsr,Kovchegov:2018znm,Cougoulic:2019aja,Kovchegov:2021lvz}. One gets
\begin{align}\label{Gamma_DLA_largeNc_evol}
\Gamma (x_{10}^2, x_{21}^2,  z's) = G^{(0)} (x^2_{10} , z's) + \frac{\as \, N_c}{2 \pi} \, \int\limits_{\frac{1}{s x_{10}^2}}^{z'} \frac{d z''}{z''} \, \int\limits^{\min \left[ x_{10}^2 , x_{21}^2 \frac{z'}{z''} \right]}_\frac{1}{z''s} \frac{d x^2_{32}}{x_{32}^2} \,  \Bigg[ &  \Gamma (x_{10}^2, x_{32}^2,  z'' s) + 3 \, G (x^2_{32}, z'' s)  \\ 
& + 2 \, G_2 (x_{32}^2, z'' s) + 2 \, \Gamma_2 (x_{10}^2, x_{32}^2, z'' s) \Bigg]  .  \notag
\end{align}
Again, the $G_2$ and $\Gamma_2$ terms are new. 

Equations \eqref{G_DLA_largeNc_evol} and \eqref{Gamma_DLA_largeNc_evol} have to be supplemented by the large-$N_c$ DLA evolution equations for $G_2$ and $\Gamma_2$. We begin with \eq{Gi_evol_main} and perform the replacement \eqref{pol->G} in it, to remove quark loops which are negligible at large $N_c$. Employing Eqs.~\eqref{Uba}, \eqref{Uba1}, and \eqref{Uba2}, along with the Fierz identity, one can readily show that at large $N_c$
\begin{subequations}\label{largeNc2}
\begin{align}
& \frac{1}{N_c^2} \llangle  \tr \left[ t^b V_{\ul 0} t^a V_{\un 1}^{\dagger} \right] \left(U_{{\ul 2}}^{\textrm{G} [1]} \right)^{b a} + \textrm{c.c.} \rrangle (z s) = 2 \, S_{20} (zs) \, G_{21} (zs) +  2 \, S_{21} (zs) \, \Gamma_{20,21} (zs) , \\
& \frac{1}{N_c^2} \llangle  \tr \left[ t^b V_{\ul 0} t^a V_{\un 1}^{\dagger} \right]  \left( U_{\un{2}}^{j \, \textrm{G} [2]} \right)^{b a}  + \textrm{c.c.} \rrangle (z s) = S_{20} (zs) \, G^j_{21} (zs) +  S_{21} (zs) \, \Gamma^j_{20,21} (zs) , \\
& \frac{1}{N_c^2} \, \llangle \tr \left[ t^b \, V_{\un 0} \, t^a \, V_{\un 1}^{i \, \textrm{G} [2] \, \dagger} \right] \, \left( U_{\un 2} \right)^{ba} + \textrm{c.c.} \rrangle (z s) = S_{20} (zs) \, G^i_{12} (zs) .
\end{align}
\end{subequations}
Again, in selecting which dipole amplitude are of the ``neighbor" type, we assume that the UV divergences in the DLA limit come only from the $x_{21} \ll x_{10} \approx x_{20}$ region, and do not arise from the $x_{20} \ll x_{10} \approx x_{21}$ region.

Employing Eqs.~\eqref{largeNc2} in \eq{Gi_evol_main} we arrive at
\begin{align}\label{Gi_largeNc_evol}
& G^i_{10} (zs) =  G^{i \, (0)}_{10} (zs) + \frac{\as \, N_c}{2 \pi^2} \, \int_{\frac{\Lambda^2}{s}}^z \frac{d z'}{z'} \, \int d^2 x_2 \, \frac{x_{10}^2}{x_{21}^2 \, x_{20}^2} \,  \left[ S_{20} (z' s) \, G^i_{12} (z' s)  - \Gamma^{i \, \textrm{gen}}_{10,21} (z' s)  \right] \\
& + \frac{\as \, N_c}{4 \pi^2} \, \int_{\frac{\Lambda^2}{s}}^z \frac{d z'}{z'} \, \int d^2 x_2 \Bigg\{ 2 \left[ \frac{\epsilon^{ij} x_{21}^j}{x_{21}^2} - \frac{\epsilon^{ij} x_{20}^j}{x_{20}^2} + 2 x_{21}^i \frac{{\un x}_{21} \times {\un x}_{20}}{x_{21}^2 \, x_{20}^2} \right] \, \left[ S_{20} (z' s) \, G_{21} (z' s) +  S_{21} (z' s) \, \Gamma^\textrm{gen}_{20,21} (z' s) \right] \notag \\ 
& + \left[ \delta^{ij} \left( \frac{3}{x_{21}^2} -  2 \, \frac{{\un x}_{20} \cdot {\un x}_{21}}{x_{20}^2 \, x_{21}^2} - \frac{1}{x_{20}^2} \right)  - 2 \frac{x_{21}^i \, x_{20}^j}{x_{21}^2 \, x_{20}^2} \left( 2 \frac{{\un x}_{20} \cdot {\un x}_{21}}{x_{20}^2} + 1 \right) + 2 \frac{x_{21}^i \, x_{21}^j}{x_{21}^2 \, x_{20}^2} \left( 2 \frac{{\un x}_{20} \cdot {\un x}_{21}}{x_{21}^2} + 1 \right) + 2 \frac{x_{20}^i \, x_{20}^j}{x_{20}^4} - 2 \frac{x_{21}^i \, x_{21}^j}{x_{21}^4}   \right] \notag \\
& \times \, \left[ S_{20} (z' s) \, G^j_{21} (z' s) +  S_{21} (z' s) \, \Gamma^{j \, \textrm{gen}}_{20,21} (z' s) \right]  \Bigg\}  \, .  \notag
\end{align}
Equation~\eqref{Gi_largeNc_evol}, just like \eq{G_largeNc_evol}, resums both the DLA and SLA$_L$ terms.

To extract the DLA contribution from \eq{Gi_largeNc_evol}, we put $S_{21} = S_{20} = 1$ and integrate it over the impact parameters. Since we are interested in the amplitude $G_2$, we invert \eq{decomp} to write
\begin{align}\label{decomp_inv}
  G_2 (x_{10}^2, z s)  =  \frac{\epsilon^{ij} \, x_{10}^j }{x_{10}^2} \int d^2 \left( \frac{x_{1} + x_0}{2} \right) \, G^{i}_{10} (z s) .
\end{align} 
Performing the projection \eqref{decomp_inv}, we arrive at
\begin{align}\label{Gi_largeNc_DLA_evol1}
& G_2 (x_{10}^2, z s)  =  G_2^{(0)} (x_{10}^2, z s)  - \frac{\as \, N_c}{2 \pi^2} \, \int\limits_{\frac{\Lambda^2}{s}}^z \frac{d z'}{z'} \, \int \frac{d^2 x_2}{x_{21}^2 \, x_{20}^2} \,  \left[ {\un x}_{10} \cdot {\un x}_{21} \, G_2 (x^2_{21} , z' s)  + x_{10}^2 \, \Gamma^{\textrm{gen}}_2 (x^2_{10} , x^2_{21} , z' s)  \right]  \\
& + \frac{\as \, N_c}{4 \pi^2} \, \int\limits_{\frac{\Lambda^2}{s}}^z \frac{d z'}{z'} \, \int \frac{d^2 x_2}{x_{10}^2} \Bigg\{ 2 \left[ {\un x}_{10} \cdot \left( \frac{{\un x}_{21}}{x_{21}^2} - \frac{{\un x}_{20}}{x_{20}^2} \right) + 2 \frac{({\un x}_{21} \times {\un x}_{20})^2}{x_{21}^2 \, x_{20}^2} \right] \, \left[ G (x^2_{21} , z' s) +  \Gamma^\textrm{gen} (x^2_{20} , x^2_{21} , z' s) \right] \notag \\ 
& + \left[ {\un x}_{10} \cdot {\un x}_{21} \left( \frac{3}{x_{21}^2} -  2 \, \frac{{\un x}_{20} \cdot {\un x}_{21}}{x_{20}^2 \, x_{21}^2} - \frac{1}{x_{20}^2} \right)  + 2 \frac{({\un x}_{21} \times {\un x}_{20})^2}{x_{20}^2} \left( \frac{1}{x_{21}^2} + 2 \, \frac{{\un x}_{20} \cdot {\un x}_{21}}{x_{20}^2 \, x_{21}^2} - \frac{1}{x_{20}^2} \right) \right] \, G_2 (x_{21}^2, z' s) \notag \\
& + \left[ {\un x}_{10} \cdot {\un x}_{20} \left( \frac{3}{x_{21}^2} -  2 \, \frac{{\un x}_{20} \cdot {\un x}_{21}}{x_{20}^2 \, x_{21}^2} - \frac{1}{x_{20}^2} \right)  + 2 \frac{({\un x}_{21} \times {\un x}_{20})^2}{x_{21}^2} \left( - \frac{1}{x_{21}^2} + 2 \, \frac{{\un x}_{20} \cdot {\un x}_{21}}{x_{20}^2 \, x_{21}^2} + \frac{1}{x_{20}^2} \right) \right] \, \Gamma_2^\textrm{gen} (x_{20}^2, x_{21}^2, z' s) \Bigg\}  \,  .  \notag
\end{align}
Next, we need to extract the DLA part of the integral kernels in \eq{Gi_largeNc_DLA_evol1}. We start with the first term in the kernel on the right-hand side of \eq{Gi_largeNc_DLA_evol1}, the one involving a linear combination of $G_2$ and $\Gamma_2^{\text{gen}}$. It contains a UV divergence as $x_{21}\to 0$ but no divergence as $x_{20}\to 0$. We proceed to the next term in the kernel, the one multiplying $G + \Gamma^\textrm{gen}$: it has no UV divergences, neither at $x_{21} \to 0$ nor at $x_{20} \to 0$. It does have an IR divergence when $x_{21} \approx x_{20} \gg x_{10}$. Next, the term multiplying $G_2$ has no UV divergences but contains an IR divergence. Finally, the term multiplying $\Gamma_2$ contains an IR divergence as well, along with the UV divergence at $x_{21} \to 0$. The two UV divergences cancel. Performing all the aforementioned DLA simplification yields the DLA large-$N_c$ evolution for $G_2$:
\begin{align}\label{Gi_largeNc_DLA_evol}
& G_2 (x_{10}^2, z s)  =  G_2^{(0)} (x_{10}^2, z s) + \frac{\as \, N_c}{\pi} \, \int\limits_{\frac{\Lambda^2}{s}}^z \frac{d z'}{z'} \, \int\limits_{\max \left[ x_{10}^2 , \frac{1}{z' s} \right]}^{\min \left[ \frac{z}{z'} x_{10}^2, \frac{1}{\Lambda^2} \right]} \frac{d x^2_{21}}{x_{21}^2} \left[ G (x^2_{21} , z' s) + 2 \, G_2 (x_{21}^2, z' s)  \right] .
\end{align}
In arriving at \eq{Gi_largeNc_DLA_evol} we have also employed the fact that for $x_{21} \approx x_{20} \gg x_{10}$ we have $\Gamma^\textrm{gen} (x_{20}^2, x_{21}^2, z' s) \approx G (x_{21}^2, z' s)$ and $\Gamma_2^\textrm{gen} (x_{20}^2, x_{21}^2, z' s) \approx G_2 (x_{21}^2, z' s)$, since the two daughter dipole sizes are comparable to each other. We have also imposed light-cone time ordering conditions, $z x_{10}^2 \gg z' x_{21}^2 \gg 1/s$, along with the $1/\Lambda^2$ IR cutoff on the $x_{21}^2$ integration. Equation \eqref{Gi_largeNc_DLA_evol} is different from the corresponding equation for $G_2$ derived earlier in \cite{Kovchegov:2017lsr}, for the reasons stated above. 

The analogue of \eq{Gi_largeNc_DLA_evol} for the neighbor dipole amplitude $\Gamma_2$ is constructed similarly. We get
\begin{align}\label{Gamma_i_largeNc_DLA_evol}
& \Gamma_2 (x_{10}^2, x_{21}^2, z' s)  =  G_2^{(0)} (x_{10}^2, z' s) + \frac{\as \, N_c}{\pi} \, \int\limits_{\frac{\Lambda^2}{s}}^{z' \frac{x_{21}^2}{x_{10}^2}} \frac{d z''}{z''} \, \int\limits_{\max \left[ x_{10}^2 , \frac{1}{z'' s} \right]}^{\min \left[  \frac{z'}{z''} x_{21}^2 , \frac{1}{\Lambda^2} \right]} \frac{d x^2_{32}}{x_{32}^2} \left[ G (x^2_{32} , z'' s) + 2 \, G_2 (x_{32}^2, z'' s)  \right] .
\end{align}

Equations \eqref{G_DLA_largeNc_evol}, \eqref{Gamma_DLA_largeNc_evol},  \eqref{Gi_largeNc_DLA_evol}, and \eqref{Gamma_i_largeNc_DLA_evol} form a closed system of DLA evolution equations for helicity at large $N_c$. For convenience, we list them all here,
\begin{subequations}\label{eq_LargeNc}
\begin{tcolorbox}[ams align]
& G (x^2_{10} , zs) = G^{(0)} (x^2_{10} , zs) + \frac{\as \, N_c}{2 \pi} \, \int\limits_{\frac{1}{s x_{10}^2}}^z \frac{d z'}{z'} \, \int\limits^{x_{10}^2}_\frac{1}{z's} \frac{d x^2_{21}}{x_{21}^2} \,  \Bigg[  \Gamma (x_{10}^2, x_{21}^2,  z's) + 3 \, G (x^2_{21}, z's)  \notag \\ 
& \hspace*{9cm} + 2 \, G_2 (x_{21}^2, z' s) + 2 \, \Gamma_2 (x_{10}^2, x_{21}^2, z' s) \Bigg] , \label{Geq} \\
& \Gamma (x_{10}^2, x_{21}^2,  z's) = G^{(0)} (x^2_{10} , z's) + \frac{\as \, N_c}{2 \pi} \, \int\limits_{\frac{1}{s x_{10}^2}}^{z'} \frac{d z''}{z''} \, \int\limits^{\min \left[ x_{10}^2 , x_{21}^2 \frac{z'}{z''} \right]}_\frac{1}{z''s} \frac{d x^2_{32}}{x_{32}^2} \,  \Bigg[  \Gamma (x_{10}^2, x_{32}^2,  z'' s) + 3 \, G (x^2_{32}, z'' s)  \notag \\ 
& \hspace*{9cm} + 2 \, G_2 (x_{32}^2, z'' s) + 2 \, \Gamma_2 (x_{10}^2, x_{32}^2, z'' s) \Bigg] , \label{Gamma_eq}  \\
& G_2 (x_{10}^2, z s)  =  G_2^{(0)} (x_{10}^2, z s) + \frac{\as \, N_c}{\pi} \, \int\limits_{\frac{\Lambda^2}{s}}^z \frac{d z'}{z'} \, \int\limits_{\max \left[ x_{10}^2 , \frac{1}{z' s} \right]}^{\min \left[ \frac{z}{z'} x_{10}^2, \frac{1}{\Lambda^2} \right]} \frac{d x^2_{21}}{x_{21}^2} \left[ G (x^2_{21} , z' s) + 2 \, G_2 (x_{21}^2, z' s)  \right] , \label{G2eq} \\
& \Gamma_2 (x_{10}^2, x_{21}^2, z' s)  =  G_2^{(0)} (x_{10}^2, z' s) + \frac{\as \, N_c}{\pi} \, \int\limits_{\frac{\Lambda^2}{s}}^{z' \frac{x_{21}^2}{x_{10}^2}} \frac{d z''}{z''} \, \int\limits_{\max \left[ x_{10}^2 , \frac{1}{z'' s} \right]}^{\min \left[ \frac{z'}{z''} x_{21}^2 , \frac{1}{\Lambda^2} \right]} \frac{d x^2_{32}}{x_{32}^2} \left[ G (x^2_{32} , z'' s) + 2 \, G_2 (x_{32}^2, z'' s)  \right] . \label{Gamma2_eq}
\end{tcolorbox}
\end{subequations}
Note that $\Gamma (x_{10}^2, x_{21}^2,  z's)$ and $\Gamma_2 (x_{10}^2, x_{21}^2, z' s)$ are only defined for $x_{10} \ge x_{21}$. Let us also stress here that $\Lambda$ is taken here to be the IR cutoff, such that Eqs.~\eqref{eq_LargeNc}, as written, are only valid for $x_{10} < 1/\Lambda$.

The equations \eqref{eq_LargeNc} have to be solved with the appropriate initial conditions (inhomogeneous terms). At Born level, these are \cite{Kovchegov:2016zex,Kovchegov:2017lsr}
\begin{align}\label{non-homo}
G^{(0)} (x_{10}^2, zs) = \frac{\as^2 C_F}{2 N_c} \pi \, \left[ C_F \, \ln \frac{zs}{\Lambda^2} - 2 \, \ln (z s x_{10}^2) \right],  \ \ \ G_2^{(0)} (x_{10}^2, zs) = \frac{\as^2 C_F}{N_c} \pi \, \ln \frac{1}{x_{10} \Lambda} .
\end{align}
(The sign difference in $G_2^{(0)}$ compared to that in \cite{Kovchegov:2017lsr} is due to the sign difference of the $A^i$ term in the definition of $G^i$ employed here and in that work.)

The solution of Eqs.~\eqref{eq_LargeNc} would give us the gluon and quark helicity TMD and PDF along with the $g_1$ structure function at small $x$ by using Eqs.~\eqref{glue_hel_TMD57}, \eqref{JM_DeltaG}, \eqref{DeltaSigma}, \eqref{TMD19}, and \eqref{g1_final} (or \eq{g1_DLA}). In using the latter formulas we have to assume that, at large $N_c$, $Q (x_{10}^2, zs) \approx G (x_{10}^2, zs)$ (see Sec.~VI of \cite{Kovchegov:2020hgb} for a brief discussion of the subtleties associated with taking the large-$N_c$ limit of small-$x$ helicity evolution).


\subsection{Evolution Equations in the Large-$N_c \& N_f$ Limit}
\label{sec:large-NcNf}

In this Section, we consider another limit under which equations \eqref{Q_evol_main}, \eqref{G_adj_evol}, \eqref{Gi_evol_main}, and \eqref{Gi_adj_evol_main} form a closed set of equations, following the standard approach described in \cite{Kovchegov:2018znm}. Since $N_f$ and $N_c$ are taken to be comparable in this limit, we include both gluon and quark loop contributions. We also notice the distinction between the fundamental and adjoint dipole amplitudes. The fundamental dipole amplitudes we consider in this Section are
\begin{subequations}
\begin{align}
Q_{10}(zs) &= \frac{1}{2 \, N_c} \, \mbox{Re} \, \llangle \mbox{T} \, \tr \left[ V_{\un 0} \,  V_{\un 1}^{\textrm{pol} [1] \,\dagger} \right] + \mbox{T} \,  \tr \left[ V_{\un 1}^{\textrm{pol} [1]} \, V_{\un 0}^\dagger \right]   \rrangle (zs)  , \\
G^i_{10}(zs) &= \frac{1}{2 \, N_c} \, \mbox{Re} \, \llangle \mbox{T} \, \tr \left[ V_{\un 0} \,  V_{\un 1}^{i \, \textrm{G} [2] \, \dagger} \right] + \mbox{T} \,  \tr \left[ V_{\un 1}^{i \, \textrm{G} [2] } \, V_{\un 0}^\dagger \right]   \rrangle (zs)  . \label{Gi_repeat}
\end{align}
\end{subequations}
Since the polarized Wilson line of the second kind, $V_{\un x}^{i \, \textrm{G} [2] }$, contains no sub-eikonal quark operator, the evolution equation for $G^i_{10}(zs)$ in the large-$N_c \& N_f$ limit will be the same as in the large-$N_c$ limit given above in \eq{Gi_largeNc_evol}. Furthermore, the relation \eqref{eq:Gadj_i_Gfund_i_largeNc} still holds in the large-$N_c\& N_f$ limit, allowing us to consider only the fundamental dipole amplitude, $G^{i}_{10} (zs)$. As for the dipole amplitudes of the first kind, the 
large-$N_c \& N_f$ analogue of $G_{10}(zs)$ from \eq{Ggdef} is defined as \cite{Kovchegov:2021lvz}
\begin{align}\label{G_dfn_NcNf}
& {\widetilde G}_{10} (zs) = \frac{1}{2 N_c} \, \mbox{Re} \, \llangle \mbox{T} \, \mbox{tr} \left[ V_{\ul 0} \, W_{{\un 1}}^{\textrm{pol} [1] \,\dagger} \right] + \mbox{T} \, \mbox{tr} \left[ W_{{\un 1}}^{\textrm{pol} [1] } \, V_{\ul 0}^\dagger \right] \rrangle (zs) \, ,
\end{align}
where 
\begin{align}\label{Wpol_dfn}
W_{{\un x}}^{\textrm{pol} [1] } &= V_{{\un x}}^{\textrm{G} [1]} + \frac{g^2p_1^+}{4s} \, \int\limits_{-\infty}^{\infty}dx_1^- \int\limits_{x_1^-}^{\infty}dx_2^- \; V_{{\un x}}[\infty,x_2^-] \, \psi_{\alpha}(x_2^-,{\un x}) \, \left(\frac{1}{2}\gamma^+\gamma_5\right)_{\beta\alpha}\bar{\psi}_{\beta}(x_1^-,{\un x}) \, V_{{\un x}}[x_1^-,-\infty] \, .
\end{align}
In the large-$N_c \& N_f$ limit the amplitude \eqref{G_dfn_NcNf} is related to that in \eq{G_adj_def} by $G^{\textrm{adj}}_{10} (zs) = 4 \, S_{10} (zs) \, {\widetilde G}_{10} (zs)$. Note that there is no simple relation between ${\widetilde G}_{10} (zs)$ and $Q_{10}(zs)$ even at the large $N_c \& N_f$ \cite{Kovchegov:2018znm}. The main argument in favor of the definitions \eqref{G_dfn_NcNf} and \eqref{Wpol_dfn} is that the following relation holds at large $N_c\& N_f$ (c.f. Eqs. (74) and (83) in \cite{Kovchegov:2018znm} along with \eq{Uba1} above),
\begin{align}\label{Upol_rel_NcNf}
\left(U_{{\un x}}^{\textrm{pol} [1] }\right)^{ba} &= 4\,\textrm{tr}\left[W_{{\un x}}^{\textrm{pol} [1] \,\dagger}t^bV_{{\un x}}t^a\right] + 4\,\textrm{tr}\left[V_{{\un x}}^{\dagger}t^bW_{{\un x}}^{\textrm{pol} [1] }t^a\right] \, ,
\end{align}
which will simplify our derivations below. For each amplitude of $Q_{10}(zs)$, ${\widetilde G}_{10} (zs)$ and $G^i_{10}(zs)$, we will derive below its DLA evolution equation in the large-$N_c\& N_f$ limit, together with the evolution equation for its neighbor dipole amplitude. 

The evolution of the fundamental dipole amplitude, $Q_{10}(zs)$, follows from the evolution equation \eqref{Q_evol_main}. At large-$N_c\& N_f$, by employing Fierz identity several times along with \eq{eq:Gadj_i_Gfund_i_largeNc}, the expectation values of the operators in \eq{Q_evol_main} can be written as
\begin{subequations}\label{Q_evol_rels}
\begin{align}
 \frac{1}{N_c^2} \llangle  \tr \left[ t^b V_{\ul 0} t^a V_{\un 1}^{\dagger} \right] \left(U_{{\ul 2}}^{\textrm{pol} [1]} \right)^{b a} + \textrm{c.c.} \rrangle (z s) &= 2 \left[S_{21}(z s) \, {\widetilde \Gamma}_{20,21}(z s) + S_{20}(z s)  \, {\widetilde G}_{21}(z s)\right]\, , \\
 \frac{1}{N_c^2} \llangle  \tr \left[ t^b V_{\ul 0} t^a V_{\un 1}^{\dagger} \right] \left( U_{\un{2}}^{i \, \textrm{G} [2]} \right)^{b a} + \textrm{c.c.} \rrangle (z s) &=  S_{21}(z s)\,\Gamma^{i}_{20,21}(z s) + S_{20}(z s)\, G^{i}_{21}(z s) \, ,  \\
 \frac{1}{N_c^2} \, \llangle \tr \left[ t^b \, V_{\un 0} \, t^a \, V_{\un 2}^{\textrm{pol} [1] \, \dagger} \right] \, U_{\un 1}^{ba} + \textrm{c.c.} \rrangle (z s) &= S_{10}(z s)\,Q_{21}(z s)\, ,   \\
 \frac{1}{N_c^2} \, \llangle \tr \left[ t^b \, V_{\un 0} \, t^a \, V_{\ul 2}^{i \, \textrm{G} [2] \, \dagger} \right] \, U_{\un 1}^{ba} + \textrm{c.c.} \rrangle (z s) &=  S_{10}(z s)  \, G^{i}_{21}(z s)\,.
\end{align}
\end{subequations}
Here, ${\widetilde \Gamma}_{20,21}$ is the neighbour counterpart of ${\widetilde G}_{10}$ defined in \eq{G_dfn_NcNf}, while $\Gamma^{i}_{20,21}$ is, again, the neighbour amplitude for $G^i_{10}$ from  \eq{Gi_repeat}. Below we will also employ ${\bar \Gamma}_{20,21}$, the neighbour counterpart of the amplitude $Q_{10}$.

Employing Eqs.~\eqref{Q_evol_rels} along with (at large $N_c\& N_f$) 
\begin{align}
    \frac{1}{N_c^2} \, \llangle \tr \left[ t^b \, V_{\un 0} \, t^a \, V_{\un 1}^{\textrm{pol} [1] \, \dagger} \right] \, U_{\un 2}^{ba} + \textrm{c.c.}  \rrangle (z s) = S_{20}(z s) \, Q_{12}(z s) ,
\end{align}
we rewrite \eq{Q_evol_main} as
\begin{align}\label{Q_evol_2}
& Q_{10}(zs) =  Q_{10}^{(0)}(zs) \\
& + \frac{\as \, N_c}{2 \pi^2} \, \int\limits_{\frac{\Lambda^2}{s}}^z \frac{d z'}{z'} \, \int d^2 x_2 \, \Bigg\{ 2 \, \left[ \frac{1}{x_{21}^2} -  \frac{{\un x}_{21}}{x_{21}^2} \cdot \frac{{\un x}_{20}}{x_{20}^2} \right] \, \left( S_{21}(z's) \, {\widetilde \Gamma}^{\text{gen}}_{20,21}(z's) +  S_{20}(z's) \, {\widetilde G}_{21}(z's)\right) \notag \\ 
& + \left[ 2 \frac{\epsilon^{ij} \, x_{21}^j}{x_{21}^4} - \frac{\epsilon^{ij} \, (x_{20}^j + x_{21}^j)}{x_{20}^2 \, x_{21}^2}  - \frac{2 \, {\un x}_{20} \times {\un x}_{21}}{x_{20}^2 \, x_{21}^2} \left( \frac{x_{21}^i}{x_{21}^2} - \frac{x_{20}^i}{x_{20}^2}\right) \right] \left( S_{21}(z's) \,\Gamma^{i \, \textrm{gen}}_{20,21}(z's) +  S_{20}(z's) \, G^{i }_{21}(z's)\right)  \Bigg\}  \notag \\
& + \frac{\as N_c}{4 \pi^2 } \, \int\limits_{\frac{\Lambda^2}{s}}^z \frac{d z'}{z'} \, \int \frac{d^2 x_2}{x_{21}^2} \, S_{10}(z's) \, \Bigg\{ Q_{21}(z's) +  \frac{2\epsilon^{ij} \, {x}_{21}^j}{x_{21}^2} \, G^{i}_{21}(z's) \Bigg\} \notag \\
& + \frac{\as \, N_c}{2 \pi^2} \, \int\limits_{\frac{\Lambda^2}{s}}^z \frac{d z'}{z'} \, \int d^2 x_2 \, \frac{x_{10}^2}{x_{21}^2 \, x_{20}^2} \,  \Bigg\{ S_{20}(z's)\,Q_{12}(z's)  - \overline{\Gamma}^{\text{\,gen}}_{10,21}(z's)  \Bigg\} \, .  \notag
\end{align}
In \eq{Q_evol_2}, similar to what we did in \eq{G_largeNc_evol}, we replaced neighbor dipole amplitudes by their generalized polarized dipole amplitude counterparts. For the purposes of this Section, the generalized dipole amplitudes in \eq{Q_evol_2} are defined as \cite{Kovchegov:2017lsr}
\begin{subequations}\label{Gamma_gen_NcNf}
\begin{align}
 & \overline{\Gamma}^{\text{\,gen}}_{10,32}(zs) \equiv Q_{10} (zs ) \, \theta\left(x_{32}-x_{10}\right) + \overline{\Gamma}_{10,32} (zs) \, \theta (x_{10}-x_{32} ), \\
 & {\widetilde \Gamma}^{\textrm{gen}}_{10, 32} (zs) \equiv {\widetilde G}_{10} (zs ) \, \theta\left(x_{32}-x_{10}\right) + {\widetilde \Gamma}_{10,32} (zs) \, \theta (x_{10}-x_{32} ), \\
 & \Gamma^{i \, \textrm{gen}}_{10, 32} (zs) \equiv G^{i}_{10} (zs ) \, \theta\left(x_{32}-x_{10}\right) + \Gamma^{i}_{10,32} (zs) \, \theta (x_{10}-x_{32} ).
\end{align}
\end{subequations}
Similar to Eqs.~\eqref{eqn:Nc7}, neither $x_{10}$ nor $x_{32}$ is necessarily the size of the parent or daughter dipole. Rather, they can be any general transverse separations. As one can infer from their definitions in Eqs.~\eqref{Gamma_gen_NcNf}, the generalized dipole amplitudes only reduce to the neighbor dipole amplitudes when $x_{32}\ll x_{10}$, as it is the only regime where the lifetime ordering needs to be expressed using a different transverse separation from the current dipole size. Otherwise, the generalized dipole amplitudes reduce to their ``regular" counterparts.

To further simplify the evolution equation \eqref{Q_evol_2} in preparation for rewriting it in the DLA form, we neglect the single-logarithmic unpolarized evolution \cite{Kuraev:1977fs,Balitsky:1978ic,Mueller:1994rr,Mueller:1994jq,Mueller:1995gb,Balitsky:1995ub,Balitsky:1998ya,Kovchegov:1999yj,Kovchegov:1999ua,Jalilian-Marian:1997dw,Jalilian-Marian:1997gr,Weigert:2000gi,Iancu:2001ad,Iancu:2000hn,Ferreiro:2001qy} and put all the unpolarized dipole $S$-matrices to 1. Subsequently, we integrate \eq{Q_evol_2} over the impact parameter, $\underline{b} = \frac{\underline{x}_1+\underline{x}_0}{2}$. Upon such integration, $G^{i}_{10}(zs)$ and $\Gamma^{i}_{20,21}(zs)$ decompose in a similar fashion to Eqs.~\eqref{decomp} and \eqref{decomp2}, that is,
\begin{subequations}\label{decomp3}
\begin{align}
&  \int d^2 \left( \frac{x_{1} + x_0}{2} \right) \, G^{i}_{10} (z s) = (x_{10})_\bot^i \, G_1
  (x_{10}^2, z s) + \epsilon^{ij} \, (x_{10})_\bot^j \, G_2
  (x_{10}^2, z s)\, , \\
&  \int d^2 \left( \frac{x_{1} + x_0}{2} \right) \, \Gamma^{i}_{20,21} (z s) = (x_{20})_\bot^i \, \Gamma_1
  (x_{20}^2, x_{21}^2, z s) + \epsilon^{ij} \, (x_{20})_\bot^j \, \Gamma_2 (x_{20}^2, x_{21}^2, z s)\, .
\end{align}
\end{subequations}
Performing all the mentioned steps in \eq{Q_evol_2}, we obtain (cf. \eq{G_DLA_largeNc_evol1})
\begin{align}\label{Q_evol_3}
& Q(x^2_{10}, zs) =  Q^{(0)}(x^2_{10},zs) \\
& + \frac{\as \, N_c}{2 \pi^2} \, \int\limits_{\frac{\Lambda^2}{s}}^z \frac{d z'}{z'} \, \int d^2 x_2 \, \Bigg\{ 2 \, \left[ \frac{1}{x_{21}^2} -  \frac{{\un x}_{21}}{x_{21}^2} \cdot \frac{{\un x}_{20}}{x_{20}^2} \right] \, \left[  {\widetilde \Gamma}^{\text{gen}}(x^2_{20},x^2_{21},z's) +  {\widetilde G} (x^2_{21},z's)\right] \notag \\ 
& + \left[ 2 \frac{{\un x}_{20} \cdot {\un x}_{21}}{x_{21}^4} - \frac{1}{x_{21}^2} - \frac{{\un x}_{20} \cdot {\un x}_{21}}{x_{20}^2 \, x_{21}^2}  + \frac{2 \, ({\un x}_{20} \times {\un x}_{21})^2}{x_{20}^2 \, x_{21}^4}   \right]  \Gamma^{\textrm{gen}}_{2}(x^2_{20},x^2_{21},z's)  \notag \\
& + \left[  \frac{2}{x_{21}^2} - \frac{{\un x}_{20} \cdot {\un x}_{21}}{x_{20}^2 \, x_{21}^2} - \frac{1}{x_{20}^2}  + \frac{2 \, ({\un x}_{20} \times {\un x}_{21})^2}{x_{20}^4 \, x_{21}^2}  \right]   G_{2}(x^2_{21},z's)  \Bigg\}  \notag \\
& + \frac{\as N_c}{4 \pi^2 } \, \int\limits_{\frac{\Lambda^2}{s}}^z \frac{d z'}{z'} \, \int \frac{d^2 x_2}{x_{21}^2} \left[  Q(x^2_{21},z's) + 2 \, G_{2}(x^2_{21},z's) \right] \notag \\
& + \frac{\as \, N_c}{2 \pi^2} \, \int\limits_{\frac{\Lambda^2}{s}}^z \frac{d z'}{z'} \, \int d^2 x_2 \, \frac{x_{10}^2}{x_{21}^2 \, x_{20}^2} \, \left[ Q(x^2_{21},z's)  - \overline{\Gamma}^{\text{\,gen}}(x^2_{10},x^2_{21},z's)  \right] \, ,  \notag
\end{align}
where we defined the impact-parameter integrated dipole amplitudes in a similar fashion to \eqref{Q_int} and \eqref{G_int}. In particular,
\begin{subequations}\label{GQ_int}
\begin{align}
 & \overline{\Gamma}(x_{20}^2, x^2_{21}, zs) \equiv \int d^2 \left( \frac{x_0 + x_2}{2} \right) \, \overline{\Gamma}_{20,21} (zs)\, , \\
& {\widetilde G} (x^2_{10} , zs) \equiv \int d^2 \left( \frac{x_0 + x_1}{2} \right) \, {\widetilde G}_{10} (zs)\, , \\ 
& {\widetilde \Gamma} (x_{20}^2, x^2_{21}, zs) \equiv \int d^2 \left( \frac{x_0 + x_2}{2} \right) \, {\widetilde \Gamma}_{20,21} (zs) \, .
\end{align}
\end{subequations}
Note that, similar to \eq{G_DLA_largeNc_evol1}, all the terms in \eq{Q_evol_3} involving $G_1$ or $\Gamma_1$ vanish upon integration over $\underline{x}_2$ because each of them contains a single Levi-Civita symbol, $\epsilon^{ij}$, along with a single transverse vector ${\un x}_{10}$: it is impossible to construct a non-zero scalar quantity out of such ingredients.

\eq{Q_evol_3} has no DLA term in the $x_{20} \ll x_{10}$ regime. However, there is at least one DLA term in both $x_{10} \ll x_{21} \approx x_{20}$ and $x_{21} \ll x_{10}$ regimes. Combining all the DLA terms together and taking lifetime ordering into account to specify the integration limits, we obtain the following DLA evolution equation for $Q(x^2_{10},zs)$ in the large-$N_c\& N_f$ limit,
\begin{align}\label{Q_evol_final}
Q(x^2_{10},zs) &= Q^{(0)}(x^2_{10},zs) + \frac{\alpha_sN_c}{2\pi} \int_{\max\{\Lambda^2,1/x^2_{10}\}/s}^{z} \frac{dz'}{z'}   \int_{1/z's}^{x^2_{10}}  \frac{dx^2_{21}}{x_{21}^2}    \left[ 2 \, {\widetilde \Gamma} (x^2_{10},x^2_{21},z's) + 2\, {\widetilde G} (x^2_{21},z's) \right. \\
&\;\;\;\;\;\;\;\;\left.+ \; Q(x^2_{21},z's) -  \overline{\Gamma}(x^2_{10},x^2_{21},z's) + 2 \, \Gamma_2(x^2_{10},x^2_{21},z's) + 2 \, G_2(x^2_{21},z's)   \right] \notag \\
&\;\;\;\;+ \frac{\alpha_sN_c}{4\pi} \int_{\Lambda^2/s}^{z} \frac{dz'}{z'}   \int_{1/z's}^{x^2_{10}z/z'}  \frac{dx^2_{21}}{x_{21}^2} \left[Q(x^2_{21},z's) + 2 \, G_2(x^2_{21},z's) \right]   ,\notag
\end{align}
where we changed the lower limit of the $z'$-integral in the first term of \eq{Q_evol_final} in order to ensure that $z's$ remains larger than $\Lambda^2$ for any value of $x^2_{10}$. A feature of \eq{Q_evol_final}, which is similar to previous treatments of the evolution equations at large-$N_c\& N_f$ \cite{Kovchegov:2015pbl,Kovchegov:2018znm}, is that the squared dipole size, $x_{10}^2$, can exceed the scale $\frac{1}{\Lambda^2}$ \cite{Kovchegov:2020hgb}. In contrast to the large-$N_c$ evolution \eqref{eq_LargeNc}, we no longer consider $\Lambda$ as the infrared cutoff in this regime. Rather, $1/\Lambda$ is understood as the typical transverse size of the target \cite{Kovchegov:2020hgb,Itakura:2003jp}, which may or may not be larger than the size $x_{10}$ of the projectile dipole.

Similar to what we did in the large-$N_c$ limit, we deduce the evolution equation for $\overline{\Gamma}(x^2_{10},x^2_{21},z's)$ by analogy to \eq{Q_evol_final}, obtaining
\begin{align}\label{Q_evol_nb}
\overline{\Gamma}&(x^2_{10},x^2_{21},z's) = Q^{(0)}(x^2_{10},z's) + \frac{\alpha_sN_c}{2\pi} \int_{\max\{\Lambda^2,1/x^2_{10}\}/s}^{z'} \frac{dz''}{z''}   \int_{1/z''s}^{\min\{x^2_{10}, x^2_{21}z'/z''\}}  \frac{dx^2_{32}}{x_{32}^2}    \left[ 2\, {\widetilde \Gamma}(x^2_{10},x^2_{32},z''s) \right. \\
&\;\;\;\;\;\;\;\;\left.+ \; 2\, {\widetilde G}(x^2_{32},z''s) +  Q(x^2_{32},z''s) -  \overline{\Gamma}(x^2_{10},x^2_{32},z''s) + 2 \, \Gamma_2(x^2_{10},x^2_{32},z''s) + 2 \, G_2(x^2_{32},z''s) \right] \notag \\
&\;\;\;\;+ \frac{\alpha_sN_c}{4\pi} \int_{\Lambda^2/s}^{z'} \frac{dz''}{z''}   \int_{1/z''s}^{x^2_{21}z'/z''}  \frac{dx^2_{32}}{x_{32}^2} \left[Q(x^2_{32},z''s) + 2 \, G_2(x^2_{32},z''s) \right]   .\notag
\end{align}

Now, we move on to consider the other polarized dipole amplitude of the first kind, ${\widetilde G}_{10} (zs)$. The general evolution equation we need for the large-$N_c \& N_f$ evolution of ${\widetilde G}_{10} (zs)$ has been derived in \eq{G_adj_evol} for the related $G_{10}^{\text{adj}}(zs)$. We simplify the equation in the large-$N_c\& N_f$ limit: we first apply the Fierz identity several times, together with \eq{Upol_rel_NcNf}, to obtain the following relations (where we, again, suppress the time-ordering sign for brevity):
\begin{subequations}\label{G_adj_evol_rels}
\begin{align}
& \frac{1}{N_c^2-1} \llangle \Tr \left[U_{\ul 0}  U_{{\un 1}}^{\textrm{pol} [1] \, \dagger}  \right]  + \textrm{c.c.} \rrangle (zs) = 8 \, S_{10}(zs) \, {\widetilde G}_{10} (zs)  \, , \label{G_adj_evol_rels_a} \\
 & \frac{1}{N_c^2 -1} \llangle \Tr \left[ T^b U_{\ul 0} T^a U_{\un 1}^{\dagger} \right] \left(U_{{\ul 2}}^{\textrm{pol} [1]} \right)^{b a} + \textrm{c.c.} \rrangle (z s) = 4 \, N_c \, S_{10}(z s) \left[S_{20}(z s)\, {\widetilde G}_{21} (z s) + S_{21}(z s) \, {\widetilde \Gamma}_{20,21}(z s)\right]  \, , \\
 & \frac{1}{N_c^2-1}  \llangle \Tr \left[ T^b \, U_{\un 0} \, T^a \, U_{\un 1}^{\textrm{pol} [1] \, \dagger} \right] \, U_{\un 2}^{ba} + \textrm{c.c.} \rrangle (z s) = 4 \, N_c \, S_{20}(z s) \left[S_{10}(z s) \, {\widetilde G}_{12}(z s) +  S_{21}(z s) \, {\widetilde \Gamma}_{10,21}(z s)\right]  \, .
\end{align}
\end{subequations}
Applying Eqs. \eqref{largeNc1}, \eqref{Q_evol_rels} and \eqref{G_adj_evol_rels} to \eq{G_adj_evol}, we obtain
\begin{align}\label{G_adj_evol_2}
& 4\, S_{10}(zs) \, {\widetilde G}_{10}(zs) = 4\, S^{(0)}_{10}(zs) \, {\widetilde G}_{10}^{(0)}(zs) \\
& + \frac{\as N_c}{\pi^2} \, \int\limits_{\frac{\Lambda^2}{s}}^z \frac{d z'}{z'} \, \int d^2 x_2 \, \Bigg\{ 4\, \left[ \frac{1}{x_{21}^2} -  \frac{{\un x}_{21}}{x_{21}^2} \cdot \frac{{\un x}_{20}}{x_{20}^2} \right] S_{10}(z's) \left[ S_{20}(z's) \, {\widetilde G}_{21} (z's) + S_{21}(z's) \, {\widetilde \Gamma}^{\textrm{gen}}_{20,21}(z's)  \right] \notag  \\
& + 2 \, \left[ 2 \frac{\epsilon^{ij} \, x_{21}^j}{x_{21}^4} - \frac{\epsilon^{ij} \, (x_{20}^j + x_{21}^j)}{x_{20}^2 \, x_{21}^2}  - \frac{2 \, {\un x}_{20} \times {\un x}_{21}}{x_{20}^2 \, x_{21}^2} \left( \frac{x_{21}^i}{x_{21}^2} - \frac{x_{20}^i}{x_{20}^2}\right) \right]    S_{10} (z's) \, \left[ S_{20} (z's) \, G^{i}_{21} (z's) + S_{21} (z's) \, \Gamma^{i\,\text{gen}}_{20,21} (z's)  \right]  \Bigg\} \notag \\
& - \frac{\as \, N_f}{2 \pi^2 } \int\limits_{\frac{\Lambda^2}{s}}^z \frac{d z'}{z'}  \int d^2 x_2 \, S_{10}(z's)\, \Bigg\{ \frac{1}{x_{21}^2} \, \overline{\Gamma}^{\text{gen}}_{20,21}(z's) +   \frac{2\epsilon^{ij} \, {\un x}_{21}^j}{x_{21}^4} \, \Gamma^{i \, \textrm{gen}}_{20,21}(z's)  \Bigg\}  \notag \\
& + \frac{2 \as N_c}{\pi^2} \, \int\limits_{\frac{\Lambda^2}{s}}^z \frac{d z'}{z'} \, \int d^2 x_2 \, \frac{x_{10}^2}{x_{21}^2 \, x_{20}^2}  \, \Bigg\{ S_{20}(z's) \left[ S_{10}(z's) \, {\widetilde G}_{12} (z's) +  S_{21}(z's) \, {\widetilde \Gamma}^{\text{gen}}_{10,21}(z's)\right]  - 2 \, S_{10}(z's) \, {\widetilde \Gamma}^{\text{gen}}_{10,21}(z's)  \Bigg\} \, . \notag
\end{align}
Once again, employing the trick from Appendix~D of \cite{Kovchegov:2021lvz}, we simplify \eq{G_adj_evol_2} to
\begin{align}\label{G_adj_evol_2.5}
& {\widetilde G}_{10} (zs) =  {\widetilde G}_{10}^{(0)}(zs) \\
& + \frac{\as N_c}{2\pi^2} \, \int\limits_{\frac{\Lambda^2}{s}}^z \frac{d z'}{z'} \, \int d^2 x_2 \, \Bigg\{ 2 \left[ \frac{1}{x_{21}^2} -  \frac{{\un x}_{21}}{x_{21}^2} \cdot \frac{{\un x}_{20}}{x_{20}^2} \right]  \left[ S_{20}(z's) \, {\widetilde G}_{21}(z's) + S_{21}(z's) \, {\widetilde \Gamma}^{\text{gen}}_{20,21}(z's)  \right] \notag  \\ 
& +  \left[ 2 \frac{\epsilon^{ij} \, x_{21}^j}{x_{21}^4} - \frac{\epsilon^{ij} \, (x_{20}^j + x_{21}^j)}{x_{20}^2 \, x_{21}^2}  - \frac{2 \, {\un x}_{20} \times {\un x}_{21}}{x_{20}^2 \, x_{21}^2} \left( \frac{x_{21}^i}{x_{21}^2} - \frac{x_{20}^i}{x_{20}^2}\right) \right] \, \left[ S_{20} (z's) \, G^{i}_{21} (z's) + S_{21} (z's) \, \Gamma^{i\,\text{gen}}_{20,21} (z's)  \right]  \Bigg\} \notag \\
& - \frac{\as \, N_f}{8 \pi^2 } \int\limits_{\frac{\Lambda^2}{s}}^z \frac{d z'}{z'}  \int d^2 x_2 \, \Bigg\{ \frac{1}{x_{21}^2} \, \overline{\Gamma}^{\text{gen}}_{20,21}(z's) +   \frac{2\epsilon^{ij} \, {\un x}_{21}^j}{x_{21}^4} \, \Gamma^{i \, \textrm{gen}}_{20,21}(z's)  \Bigg\}  \notag \\
& + \frac{\as N_c}{2 \pi^2} \, \int\limits_{\frac{\Lambda^2}{s}}^z \frac{d z'}{z'} \, \int d^2 x_2 \, \frac{x_{10}^2}{x_{21}^2 \, x_{20}^2}  \, \Bigg\{ S_{20}(z's) \, {\widetilde G}_{12}(z's)  -  {\widetilde  \Gamma}^{\text{gen}}_{10,21}(z's)  \Bigg\} \, . \notag
\end{align}
This is the DLA+SLA$_L$ large-$N_c \& N_f$ evolution equation for ${\widetilde G}_{10} (zs)$.

To extract the DLA limit, we put the unpolarized dipole $S$-matrices in \eq{G_adj_evol_2.5} to 1. Then, we integrate the resulting equation over the impact parameters, employing the definitions from Eqs. \eqref{Q_int}, \eqref{G_int}, \eqref{decomp3} and \eqref{GQ_int}. As a result, \eq{G_adj_evol_2.5} becomes
\begin{align}\label{G_adj_evol_3}
& {\widetilde G}(x^2_{10},zs) = {\widetilde G}^{(0)}(x^2_{10},zs) \\
& + \frac{\as N_c}{2  \pi^2} \, \int\limits_{\frac{\Lambda^2}{s}}^z \frac{d z'}{z'} \, \int d^2 x_2 \, \Bigg\{ 2 \left[ \frac{1}{x_{21}^2} -  \frac{{\un x}_{21}}{x_{21}^2} \cdot \frac{{\un x}_{20}}{x_{20}^2} \right]   \left[ {\widetilde G}(x^2_{21},z's) +  {\widetilde \Gamma}^{\text{gen}}(x^2_{20},x^2_{21},z's)  \right] \notag  \\ 
& + \left[  \frac{2}{x_{21}^2} - \frac{{\un x}_{21} \cdot {\un x}_{20}}{x_{20}^2 \, x_{21}^2} - \frac{1}{x_{20}^2}  + \frac{2 \, ({\un x}_{20} \times {\un x}_{21})^2}{x_{20}^4 \, x_{21}^2}  \right]   G_{2} (x^2_{21},z's)   \notag  \\ 
& +  \left[ 2 \frac{{\un x}_{21} \cdot {\un x}_{20}}{x_{21}^4} - \frac{1}{x_{21}^2} - \frac{{\un x}_{21} \cdot {\un x}_{20}}{x_{20}^2 \, x_{21}^2}  + \frac{2 \, ({\un x}_{20} \times {\un x}_{21})^2}{x_{20}^2 \, x_{21}^4}  \right]    \Gamma^{\text{gen}}_2(x^2_{20},x^2_{21},z's)     \Bigg\} \notag \\
& - \frac{\as \, N_f}{8 \pi^2 } \int\limits_{\frac{\Lambda^2}{s}}^z \frac{d z'}{z'}  \int d^2 x_2 \, \Bigg\{ \frac{1}{x_{21}^2} \,  \overline{\Gamma}^{\text{gen}}(x^2_{20},x^2_{21},z's) +   2 \frac{{\un x}_{21} \cdot {\un x}_{20}}{x_{21}^4} \,  \Gamma^{ \textrm{gen}}_2(x^2_{20},x^2_{21},z's)  \Bigg\}  \notag \\
& + \frac{\as N_c}{2 \pi^2} \, \int\limits_{\frac{\Lambda^2}{s}}^z \frac{d z'}{z'} \, \int d^2 x_2 \, \frac{x_{10}^2}{x_{21}^2 \, x_{20}^2}  \, \Bigg\{   {\widetilde G} (x^2_{21},z's) - {\widetilde \Gamma}^{\text{gen}}(x^2_{10},x^2_{21},z's)  \Bigg\} \, . \notag
\end{align}
For the same reason as in \eq{Q_evol_3}, all the terms involving $G_1$ and $\Gamma_1$ vanish.

In the $x_{20} \ll x_{10}$ regime, \eq{G_adj_evol_3} contains no DLA terms and is exclusively SLA$_L$. However, the equation contains DLA terms in both $x_{10} \ll x_{21} \approx x_{20}$ and $x_{21} \ll x_{10}$ regimes. Combining all the DLA terms with lifetime ordering taken into account to obtain the integration limits, we have
\begin{align}\label{G_adj_evol_final}
{\widetilde G}(x^2_{10},zs) &= {\widetilde G}^{(0)}(x^2_{10},zs) + \frac{\alpha_s N_c}{2\pi}\int_{\max\{\Lambda^2,1/x^2_{10}\}/s}^z\frac{dz'}{z'}\int_{1/z's}^{x^2_{10}} \frac{dx^2_{21}}{x^2_{21}} \\
&\;\;\;\;\;\;\;\;\;\;\times  \left[3 \, {\widetilde G}(x^2_{21},z's) + {\widetilde \Gamma}(x^2_{10},x^2_{21},z's) + 2 \, G_2(x^2_{21},z's)  +  2 \, \Gamma_2(x^2_{10},x^2_{21},z's)\right] \notag \\
&\;\;\;\;- \frac{\alpha_sN_f}{8\pi}  \int_{\Lambda^2/s}^z \frac{dz'}{z'}\int_{1/z's}^{x^2_{10}z/z'} \frac{dx^2_{21}}{x^2_{21}}  \left[   \overline{\Gamma}^{\text{gen}}(x^2_{20},x^2_{21},z's) +     2 \, \Gamma^{\text{gen}}_2(x^2_{20},x^2_{21},z's)  \right]  .\notag 
\end{align}
Notice that the lower limit of the longitudinal integral in the first term of \eq{G_adj_evol_final} is modified in a similar fashion to the first term of \eq{Q_evol_final}.

By analogy, the DLA evolution equation for the adjoint neighbor dipole amplitude of the first kind is
\begin{align}\label{G_adj_evol_nb}
{\widetilde \Gamma}&(x^2_{10},x^2_{21},z's) = {\widetilde G}^{(0)}(x^2_{10},z's) + \frac{\alpha_s N_c}{2\pi}\int_{\max\{\Lambda^2,1/x^2_{10}\}/s}^{z'}\frac{dz''}{z''}\int_{1/z''s}^{\min\{x^2_{10},x^2_{21}z'/z''\}} \frac{dx^2_{32}}{x^2_{32}} \\
&\;\;\;\;\times  \left[3 \, {\widetilde G} (x^2_{32},z''s) + {\widetilde  \Gamma}(x^2_{10},x^2_{32},z''s) + 2 \, G_2(x^2_{32},z''s) +  2 \, \Gamma_2(x^2_{10},x^2_{32},z''s)\right] \notag \\
&- \frac{\alpha_sN_f}{8\pi}  \int_{\Lambda^2/s}^{z'} \frac{dz''}{z''}\int_{1/z''s}^{ x^2_{21}z'/z''} \frac{dx^2_{32}}{x^2_{32}}  \left[   \overline{\Gamma}^{\text{gen}}(x^2_{30},x^2_{32},z''s) +      2 \, \Gamma^{\text{gen}}_2(x^2_{30},x^2_{32},z''s)  \right]  .\notag 
\end{align}

Finally, we consider the adjoint dipole amplitude of the second kind. Since, as we mentioned above, the polarized Wilson line of this kind does not contain a sub-eikonal quark operator, the DLA evolution equation for $G_2(x^2_{10},zs)$ and $\Gamma_2(x^2_{10},x^2_{21},z's)$ can be taken directly from Eqs. \eqref{Gi_largeNc_DLA_evol} and \eqref{Gamma_i_largeNc_DLA_evol}, respectively, by replacing $G$ with ${\widetilde G}$ in them due to the difference in the definitions \eqref{G_dfn_NcNf} and \eqref{Ggdef}. This gives
\begin{subequations}\label{Gi_Gmi_evol_final}
\begin{align}
& G_2(x_{10}^2, z s)  =  G_2^{(0)} (x_{10}^2, z s) + \frac{\as \, N_c}{\pi} \, \int\limits_{\frac{\Lambda^2}{s}}^z \frac{d z'}{z'} \, \int\limits_{\max \left[ x_{10}^2 , \frac{1}{z' s} \right]}^{\frac{z}{z'} x_{10}^2} \frac{d x^2_{21}}{x_{21}^2} \left[ {\widetilde G} (x^2_{21} , z' s) + 2 \, G_2 (x_{21}^2, z' s)  \right] , \\
& \Gamma_2 (x_{10}^2, x_{21}^2, z' s)  =  G_2^{(0)} (x_{10}^2, z' s) + \frac{\as \, N_c}{\pi} \, \int\limits_{\frac{\Lambda^2}{s}}^{z' \frac{x_{21}^2}{x_{10}^2}} \frac{d z''}{z''} \, \int\limits_{\max \left[ x_{10}^2 , \frac{1}{z'' s} \right]}^{ \frac{z'}{z''} x_{21}^2} \frac{d x^2_{32}}{x_{32}^2} \left[ {\widetilde G} (x^2_{32} , z'' s) + 2 \, G_2(x_{32}^2, z'' s)  \right] .
\end{align}
\end{subequations}
A caveat in arriving at \eq{Gi_Gmi_evol_final} is that all the terms involving polarized Wilson lines of the first kind in Eqs.~\eqref{Gi_evol_main} and \eqref{Gi_adj_evol_main} got absorbed into the adjoint dipole amplitudes ${\widetilde G}$, that is, the amplitude $Q$ does not appear. Diagrammatically, this is due to the fact that there is no sub-eikonal emission of a polarized soft quark in any of the diagrams in \fig{FIG:Gi_evol}. Another difference between the large-$N_c$ counterparts, Eqs. \eqref{Gi_evol_main} and \eqref{Gi_adj_evol_main}, and \eq{Gi_Gmi_evol_final} is in the upper limit of the transverse integrals, where the constraints imposed by the infrared cutoff, $\Lambda^2$, in Eqs. \eqref{Gi_evol_main} and \eqref{Gi_adj_evol_main} were removed because $\Lambda^2$ no longer acts as the infrared cutoff in the large-$N_c\& N_f$ limit.

Equations \eqref{Q_evol_final}, \eqref{Q_evol_nb}, \eqref{G_adj_evol_final}, \eqref{G_adj_evol_nb} and \eqref{Gi_Gmi_evol_final} form a closed system of DLA evolution equations involving six polarized (neighbor) dipole amplitudes in the large-$N_c\& N_f$ limit. To summarize, we rewrite all the equations below, utilizing Eqs.~\eqref{Gamma_gen_NcNf} to separate all integrals into the UV and IR regions.
\begin{subequations}\label{eq_LargeNcNf}
\begin{tcolorbox}[ams align]
& Q(x^2_{10},zs) = Q^{(0)}(x^2_{10},zs) + \frac{\alpha_sN_c}{2\pi} \int_{\max\{\Lambda^2,1/x^2_{10}\}/s}^{z} \frac{dz'}{z'}   \int_{1/z's}^{x^2_{10}}  \frac{dx^2_{21}}{x_{21}^2}    \left[ 2 \, {\widetilde G}(x^2_{21},z's) + 2 \, {\widetilde \Gamma}(x^2_{10},x^2_{21},z's) \right. \\
&\hspace*{5cm}\left.+ \; Q(x^2_{21},z's) -  \overline{\Gamma}(x^2_{10},x^2_{21},z's) + 2 \, \Gamma_2(x^2_{10},x^2_{21},z's) + 2 \, G_2(x^2_{21},z's)   \right] \notag \\
&\hspace*{3cm}+ \frac{\alpha_sN_c}{4\pi} \int_{\Lambda^2/s}^{z} \frac{dz'}{z'}   \int_{1/z's}^{x^2_{10}z/z'}  \frac{dx^2_{21}}{x_{21}^2} \left[Q(x^2_{21},z's) + 2 \, G_2(x^2_{21},z's) \right] ,  \notag  \\
&\overline{\Gamma}(x^2_{10},x^2_{21},z's) = Q^{(0)}(x^2_{10},z's) + \frac{\alpha_sN_c}{2\pi} \int_{\max\{\Lambda^2,1/x^2_{10}\}/s}^{z'} \frac{dz''}{z''}   \int_{1/z''s}^{\min\{x^2_{10}, x^2_{21}z'/z''\}}  \frac{dx^2_{32}}{x_{32}^2}    \left[ 2\, {\widetilde G} (x^2_{32},z''s)  \right. \\
&\hspace*{2.5cm}\left.+ \; 2\, {\widetilde \Gamma} (x^2_{10},x^2_{32},z''s) +  Q(x^2_{32},z''s) -  \overline{\Gamma}(x^2_{10},x^2_{32},z''s) + 2 \, \Gamma_2(x^2_{10},x^2_{32},z''s) + 2 \, G_2(x^2_{32},z''s) \right] \notag \\
&\hspace*{3cm}+ \frac{\alpha_sN_c}{4\pi} \int_{\Lambda^2/s}^{z'} \frac{dz''}{z''}   \int_{1/z''s}^{x^2_{21}z'/z''}  \frac{dx^2_{32}}{x_{32}^2} \left[Q(x^2_{32},z''s) + 2 \, G_2(x^2_{32},z''s) \right] , \notag \\
& {\widetilde G}(x^2_{10},zs) = {\widetilde G}^{(0)}(x^2_{10},zs) + \frac{\alpha_s N_c}{2\pi}\int_{\max\{\Lambda^2,1/x^2_{10}\}/s}^z\frac{dz'}{z'}\int_{1/z's}^{x^2_{10}} \frac{dx^2_{21}}{x^2_{21}} \left[3 \, {\widetilde G}(x^2_{21},z's) + {\widetilde \Gamma}(x^2_{10},x^2_{21},z's) \right. \\
&\hspace*{4cm}\left.  + \; 2\,G_2(x^2_{21},z's)  +  \left(2 - \frac{N_f}{2N_c}\right) \Gamma_2(x^2_{10},x^2_{21},z's) - \frac{N_f}{4N_c}\,\overline{\Gamma}(x^2_{10},x^2_{21},z's)\right] \notag \\
&\hspace*{3cm}- \frac{\alpha_sN_f}{8\pi}  \int_{\Lambda^2/s}^z \frac{dz'}{z'}\int_{\max\{x^2_{10},\,1/z's\}}^{x^2_{10}z/z'} \frac{dx^2_{21}}{x^2_{21}}  \left[   Q(x^2_{21},z's) +     2 \, G_2(x^2_{21},z's)  \right] , \notag \\
& {\widetilde \Gamma} (x^2_{10},x^2_{21},z's) = {\widetilde G}^{(0)}(x^2_{10},z's) + \frac{\alpha_s N_c}{2\pi}\int_{\max\{\Lambda^2,1/x^2_{10}\}/s}^{z'}\frac{dz''}{z''}\int_{1/z''s}^{\min\{x^2_{10},x^2_{21}z'/z''\}} \frac{dx^2_{32}}{x^2_{32}} \left[3 \, {\widetilde G} (x^2_{32},z''s) \right. \\
&\hspace*{2.5cm}\left. + \; {\widetilde \Gamma}(x^2_{10},x^2_{32},z''s) + 2 \, G_2(x^2_{32},z''s)  +  \left(2 - \frac{N_f}{2N_c}\right) \Gamma_2(x^2_{10},x^2_{32},z''s) - \frac{N_f}{4N_c} \,\overline{\Gamma}(x^2_{10},x^2_{32},z''s)  \right] \notag \\
&\hspace*{3cm}- \frac{\alpha_sN_f}{8\pi}  \int_{\Lambda^2/s}^{z'x^2_{21}/x^2_{10}} \frac{dz''}{z''}\int_{\max\{x^2_{10},\,1/z''s\}}^{ x^2_{21}z'/z''} \frac{dx^2_{32}}{x^2_{32}}  \left[   Q(x^2_{32},z''s) +  2  \,  G_2(x^2_{32},z''s)  \right] , \notag \\
& G_2(x_{10}^2, z s)  =  G_2^{(0)} (x_{10}^2, z s) + \frac{\as N_c}{\pi} \, \int\limits_{\frac{\Lambda^2}{s}}^z \frac{d z'}{z'} \, \int\limits_{\max \left[ x_{10}^2 , \frac{1}{z' s} \right]}^{\frac{z}{z'} x_{10}^2} \frac{d x^2_{21}}{x_{21}^2} \left[ {\widetilde G} (x^2_{21} , z' s) + 2 \, G_2 (x_{21}^2, z' s)  \right] , \\
& \Gamma_2 (x_{10}^2, x_{21}^2, z' s)  =  G_2^{(0)} (x_{10}^2, z' s) + \frac{\as N_c}{\pi}  \int\limits_{\frac{\Lambda^2}{s}}^{z' \frac{x_{21}^2}{x_{10}^2}} \frac{d z''}{z''}  \int\limits_{\max \left[ x_{10}^2 , \frac{1}{z'' s} \right]}^{ \frac{z'}{z''} x_{21}^2} \frac{d x^2_{32}}{x_{32}^2} \left[ {\widetilde G} (x^2_{32} , z'' s) + 2 \, G_2(x_{32}^2, z'' s)  \right] .  
\end{tcolorbox}
\end{subequations}
Similar to \eq{non-homo} for the large-$N_c$ limit, the inhomogeneous terms of Eqs.~\eqref{eq_LargeNcNf} are given by the following expressions at Born level \cite{Kovchegov:2016zex,Kovchegov:2017lsr}:
\begin{align}\label{non-homo_NcNf}
{\widetilde G}^{(0)} (x_{10}^2, zs) = Q^{(0)} (x_{10}^2, zs) = \frac{\as^2 C_F}{2N_c} \pi \, \left[ C_F \, \ln \frac{zs}{\Lambda^2} - 2 \, \ln (z s x_{10}^2) \right],  \ \ \ G_2^{(0)} (x_{10}^2, zs) = \frac{\as^2 C_F}{N_c} \pi \, \ln \frac{1}{x_{10} \Lambda} \, .
\end{align}
These initial conditions assume that the projectile is much smaller than the target, $x_{10} \ll 1/\Lambda$.
To be used in Eqs.~\eqref{eq_LargeNcNf}, the expressions \eqref{non-homo_NcNf} may need to be generalized to also describe the large-projectile case, $x_{10} \gg 1/\Lambda$.


\subsection{Cross-check against the spin-dependent DGLAP evolution}
\label{sec:cross-check}

Let us cross-check our results against the spin-dependent DGLAP evolution equation \cite{Gribov:1972ri,Altarelli:1977zs,Dokshitzer:1977sg}. We are interested in the gluon sector only, since this is where the previous works' \cite{Kovchegov:2015pbl,Kovchegov:2016zex} agreement with DGLAP evolution was not completely clear. To this end we put the flavor-singlet quark helicity PDF to zero, $\Delta \Sigma (x, Q^2) =0$, (for instance, by putting $N_f =0$) and write the DGLAP equation for the gluon helicity PDF only
\begin{align}\label{DGLAP_DG}
    \frac{\pd \Delta G (x, Q^2)}{\pd \ln Q^2} = \int\limits_x^1 \frac{dz}{z} \, \Delta P_{GG} (z) \, \Delta G \left( \frac{x}{z}, Q^2 \right). 
\end{align}
We would like to stress that discarding $\Delta \Sigma$ is not a physical approximation. Rather, it is a mathematical step to verify that our evolution agrees with that driven by the splitting function $\Delta P_{GG} (z)$. The latter is known up to three loops \cite{Mertig:1995ny,Moch:2014sna} (see also \cite{Blumlein:2021ryt}). At small $z$ and large $N_c$ it reduces to
\begin{align}\label{P_GG}
    \Delta P_{GG} (z) = \frac{\as}{2 \pi} \, 4 N_c + \left( \frac{\as}{2 \pi}  \right)^2 \, 4 N_c^2 \, \ln^2 z + \left( \frac{\as}{2 \pi}  \right)^3 \, \frac{7}{3} N_c^3 \, \ln^4 z + \ldots . 
\end{align}

Since our goal is to check that our evolution in the gluon sector agrees with DGLAP, we will consider the large-$N_c$ evolution in Eqs.~\eqref{eq_LargeNc}. We choose the initial conditions to be 
\begin{align}\label{init_DGLAP}
    G^{(0)} (x^2_{10} , zs) =0, \ \ \ G_2^{(0)} (x_{10}^2, z' s) = 1. 
\end{align}
Employing \eq{JM_DeltaG}, we see that this choice of the initial conditions corresponds to the initial PDF $\Delta G^{(0)} (x, Q^2) =$const, where the value of the constant is not important for us. 

Inserting \eq{init_DGLAP} into the right-hand sides of Eqs.~\eqref{Geq} and \eqref{G2eq} yields the result of one iteration of our evolution
\begin{subequations}\label{G1eqs}
\begin{align}
&    G^{(1)} (x_{10}^2, z s) = \frac{\as N_c}{\pi} \, \ln^2 (z s x_{10}^2), \\
&    G_2^{(1)} (x_{10}^2, z s) = 2 \frac{\as N_c}{\pi} \, \ln (z s x_{10}^2) \, \ln \left( \frac{1}{x_{10}^2 \Lambda^2} \right). \label{G2eq1}
\end{align}
\end{subequations}
In arriving at Eqs.~\eqref{G1eqs} it is convenient to rewrite the kernel of \eq{G2eq} as
\begin{align}\label{kernels}
    \int\limits_{\frac{\Lambda^2}{s}}^z \frac{d z'}{z'} \, \int\limits_{\max \left[ x_{10}^2 , \frac{1}{z' s} \right]}^{\min \left[ \frac{z}{z'} x_{10}^2, \frac{1}{\Lambda^2} \right]} \frac{d x^2_{21}}{x_{21}^2} = \int\limits_{x_{10}^2}^\frac{1}{\Lambda^2} \frac{d x^2_{21}}{x_{21}^2}  \int\limits_\frac{1}{s x_{21}^2}^{z \, \frac{x_{10}^2}{x_{21}^2}} \frac{d z'}{z'} .
\end{align}

Identifying 
\begin{align}\label{identify}
    \frac{1}{x_{10}^2} \to Q^2, \ \ \ z s x_{10}^2 \to \frac{zs}{Q^2} \to \frac{1}{x}
\end{align}
we see that \eq{G2eq1}, via \eq{JM_DeltaG}, gives
\begin{align}\label{DG1}
    \Delta G^{(1)} (x, Q^2) = 2 \frac{\as N_c}{\pi} \, \ln \left(\frac{1}{x} \right) \, \ln \left( \frac{Q^2}{\Lambda^2} \right) \, \mbox{const}.
\end{align}
This is in complete agreement with one iteration of leading-order (LO) spin-dependent DGLAP equation: indeed, using $\Delta G^{(0)} (x, Q^2) = \text{const}$ on the right of \eq{DGLAP_DG} with the order-$\as$ part of the splitting function \eqref{P_GG} gives us \eq{DG1}. We see that we are in complete agreement with the one-loop DGLAP equation. 

To check the result at two loops, we substitute Eqs.~\eqref{G1eqs} into the right-hand side of \eq{G2eq}. Employing \eq{kernels} to simplify the integration we get
\begin{align}\label{G22}
    G_2^{(2)} (x_{10}^2, z s) = \left( \frac{\as N_c}{\pi} \right)^2 \,  \left[ \frac{1}{3} \, \ln^3 (z s x_{10}^2) \, \ln \left( \frac{1}{x_{10}^2 \Lambda^2} \right)  + \ln^2 (z s x_{10}^2) \, \ln^2 \left( \frac{1}{x_{10}^2 \Lambda^2} \right) \right],
\end{align}
which, with the help of \eq{identify}, corresponds to 
\begin{align}\label{DG2}
    \Delta G^{(2)} (x, Q^2) = \left( \frac{\as N_c}{\pi} \right)^2 \,  \left[ \frac{1}{3} \, \ln^3 \left(\frac{1}{x} \right)\, \ln \left( \frac{Q^2}{\Lambda^2} \right)  + \ln^2 \left(\frac{1}{x} \right) \, \ln^2 \left( \frac{Q^2}{\Lambda^2} \right) \right] \, \mbox{const}.
\end{align}
Inserting $\Delta G^{(0)} (x, Q^2) = \text{const}$ into the right side of \eq{DGLAP_DG} and employing the order-$\as^2$ part of the splitting function \eqref{P_GG} we arrive at the first term on the right of \eq{DG2}: hence, we agree with the next-to-leading order (NLO) spin-dependent DGLAP evolution (at large-$N_c$ and small-$x$) as well. The last term on the right of \eq{DG2} results from two iterations of the LO DGLAP, as can be verified explicitly as well.

Let us push the comparison one step further. To compare our evolution with the next-to-next-to-leading order (NNLO) DGLAP equation, we need to find $G_2^{(3)}$. To construct it, we first employ Eqs.~\eqref{init_DGLAP} in Eqs.~\eqref{Gamma_eq} and \eqref{Gamma2_eq} to obtain
\begin{subequations}\label{Gamma1_eqs}
\begin{align}
    & \Gamma^{(1)} (x_{10}^2, x_{21}^2,  z's) = 2 \frac{\as N_c}{\pi} \left[ \frac{1}{2} \, \ln^2 (z' s x_{21}^2) + \ln (z' s x_{21}^2) \, \ln \frac{x_{10}^2}{x_{21}^2} \right], \\
    & \Gamma_2^{(1)} (x_{10}^2, x_{21}^2,  z's) = 2 \frac{\as N_c}{\pi} \, \ln (z' s x_{21}^2) \, \ln \frac{1}{x_{10}^2 \Lambda^2}.
\end{align}
\end{subequations}
The calculation is simplified if one notices that the kernel of \eq{Gamma2_eq} can be rewritten as
\begin{align}\label{kernels2}
    \int\limits_{\frac{\Lambda^2}{s}}^{z' \frac{x_{21}^2}{x_{10}^2}} \frac{d z''}{z''} \, \int\limits_{\max \left[ x_{10}^2 , \frac{1}{z'' s} \right]}^{\min \left[ \frac{z'}{z''} x_{21}^2 , \frac{1}{\Lambda^2} \right]} \frac{d x^2_{32}}{x_{32}^2} = \int\limits_{x_{10}^2}^\frac{1}{\Lambda^2} \frac{d x^2_{32}}{x_{32}^2} \, \int\limits_\frac{1}{s \, x_{32}^2}^{z' \frac{x_{21}^2}{x_{32}^2}} \frac{d z''}{z''} .
\end{align}
Employing Eqs.~\eqref{Gamma1_eqs} and \eqref{G1eqs} in \eq{Geq} we arrive at
\begin{align}\label{G(2)}
    G^{(2)} (x_{10}^2, z s) = \left( \frac{\as N_c}{\pi} \right)^2 \,  \left[ \frac{7}{24} \, \ln^4 (z s x_{10}^2)  + \frac{2}{3} \, \ln^3 (z s x_{10}^2) \, \ln \left( \frac{1}{x_{10}^2 \Lambda^2} \right) \right].
\end{align}
Finally, inserting Eqs.~\eqref{G22} and \eqref{G(2)} into the right-hand side of \eq{G2eq} yields
\begin{align}\label{G23}
    G_2^{(3)} (x_{10}^2, z s) = \left( \frac{\as N_c}{\pi} \right)^3 \,  \bigg[ \frac{7}{120} \, \ln^5 (z s x_{10}^2) \, \ln \left( \frac{1}{x_{10}^2 \Lambda^2} \right)  + & \, \frac{1}{6} \, \ln^4 (z s x_{10}^2) \, \ln^2 \left( \frac{1}{x_{10}^2 \Lambda^2} \right)  \\ 
    & + \frac{2}{9} \, \ln^3 (z s x_{10}^2) \, \ln^3 \left( \frac{1}{x_{10}^2 \Lambda^2} \right) \bigg], \notag
\end{align}
which, using \eq{identify}, translates into
\begin{align}\label{DG3}
    \Delta G^{(3)} (x, Q^2) = \left( \frac{\as N_c}{\pi} \right)^3 \,  \left[ \frac{7}{120} \, \ln^5 \left(\frac{1}{x} \right)\, \ln \left( \frac{Q^2}{\Lambda^2} \right)  + \frac{1}{6} \, \ln^4 \left(\frac{1}{x} \right)\, \ln^2 \left( \frac{Q^2}{\Lambda^2} \right)  + \frac{2}{9} \, \ln^3 \left(\frac{1}{x} \right) \, \ln^3 \left( \frac{Q^2}{\Lambda^2} \right) \right] \, \mbox{const}.
\end{align}
The first term on the right of \eq{DG3} exactly corresponds to the contribution of the order-$\as^3$ part of the splitting function \eqref{P_GG} to \eq{DGLAP_DG}: our evolution \eqref{eq_LargeNc} thus agrees with the NNLO DGLAP gluon-gluon splitting function (at large-$N_c$ and small-$x$). One can also readily verify that the last term on the right of \eq{DG3} corresponds to three iterations of the LO DGLAP kernel, LO$^3$, while the second term on the right of \eq{DG3} is a sum of applying LO and NLO DGLAP in different orders, that is, LO$\times$NLO + NLO$\times$LO. 

Therefore, the agreement between our evolution and the small-$x$ limit of spin-dependent DGLAP equation in the gluon sector has been verified to three loops, the same order as the IREE of \cite{Bartels:1996wc,Blumlein:1995jp,Blumlein:1996hb}. Further iterations in the solution of our Eqs.~\eqref{eq_LargeNc} can be used to generate new higher-order corrections to the small-$x$ anomalous dimension \eqref{P_GG}, which have not been derived yet (but can also be extracted using the technique of \cite{Bartels:1996wc,Blumlein:1995jp,Blumlein:1996hb}). In addition, let us note here that the amplitude $G_2 (x_{10}^2, z s)$ obtained here in Eqs.~\eqref{DG1}, \eqref{DG2}, and \eqref{DG3} appears to only contain the solution of the spin-dependent DGLAP equation \eqref{DGLAP_DG} at small $x$: if an exact analytic solution of Eqs.~\eqref{eq_LargeNc} is constructed in the future work, it would contain the {\sl exact} expression for the small-$x$ large-$N_c$ spin-dependent gluon-gluon splitting function, generalizing \eq{P_GG} to all orders in the coupling.


\section{Helicity Evolution at Small $x$: the Background Field Method}
\label{sec:hel_evo_bfm}

In the previous Section we derived the helicity evolution equations at small $x$ in the LCOT approach. The key element of the calculation was the observation that in the helicity evolution quarks and gluons couple to the background shock-wave fields through the polarized Wilson lines (\ref{VqG_decomp}) and (\ref{UqG_decomp}). This is a non-trivial statement which requires further explanation. The most powerful framework which allows to unambiguously determine the form of the operators which define the coupling of ``quantum" quarks and gluons to the background field is the background field method \cite{Abbott:1980hw,Abbott:1981ke}. In this approach the separation of ``quantum" and background fields is done at the level of the QCD Lagrangian which allows to obtain the most general form of the propagator in the external background.

In this Section we will show how the polarized Wilson lines (\ref{VqG_decomp}) and (\ref{UqG_decomp}) appear in this approach and present an alternative  derivation of the helicity evolution equations \eqref{Q_evol_main} and \eqref{Gi_evol_main}. We will thus show that the helicity evolution equations obtained in the background field method are in full agreement with the above results obtained in the LCOT approach.


\subsection{The background field method\label{sec:bfm}}
To introduce the background field method, let us start with a matrix element of an arbitrary operator $\mathcal{O}(A, \psi, \bar{\psi})$ (corresponding to some observable) which is constructed out of quark and gluon fields. The matrix element can be represented as a functional integral over those fields,\footnote{For brevity we do not explicitly show the dependence on (and integrals over) the anti-quark fields $\bar \psi$.}
\begin{eqnarray}
&&\langle P_1|\mathcal{O}|P_2\rangle = \int \mathcal{D}A\,\int\mathcal{D}\psi \,\Psi^\ast_{P_1}(\vec{A}(t_f), \psi(t_f))\,\mathcal{O}(A,\psi)\,\Psi_{P_2}(\vec{A}(t_i), \psi(t_i)) e^{iS_{QCD}(A, \psi)}\,,
\label{FintO}
\end{eqnarray}
where $\Psi_{P_2}$ is the initial state wave function at the initial time $t_i\to-\infty$ and, similarly, $\Psi_{P_1}$ is the final state wave function at the final time $t_f\to\infty$.

The main idea of the background field method is that the fields in (\ref{FintO}) can be separated into the ``quantum" and background parts,
\begin{eqnarray}
&&A_\mu \to A^\textrm{q}_\mu + A^{\textrm{bg}}_\mu,\,\,\,\,\, \psi \to \psi^\textrm{q} + \psi^{\textrm{bg}}\,.
\end{eqnarray}
The way we separate the fields is completely arbitrary, see for example \cite{Balitsky:1987bk,Balitsky:1990ck,Beneke:2002ph,Bauer:2000yr,Bauer:2001yt}. However, in the context of small-$x$ physics the most efficient approach is to separate the fields based on their longitudinal momentum fraction (or, equivalently, rapidity). This is the rapidity factorization approach \cite{Balitsky:1995ub,Balitsky:1998ya}. In this approach the ``quantum" fields are defined to have momenta $p^- > \sigma$, and background fields are characterized by $p^- < \sigma$, where $\sigma$ is some rapidity factorization scale.\footnote{In the rapidity factorization approach the ``quantum" fields are usually called ``slow" fields, and the background fields are called ``fast" fields \cite{Balitsky:2015qba}.} Note that in the small-$x$ limit, due to Lorentz contraction, the background fields have a shock-wave form with a limited support in the $x^-$ direction (for the plus-direction moving proton).

Assuming that the wave functions depend only on the background fields, we rewrite the matrix element as
\begin{eqnarray}
&&\langle P_1|\mathcal{O}|P_2\rangle = \int \mathcal{D}A^{\textrm{bg}}\,\int\mathcal{D}\psi^{\textrm{bg}}\, \Psi^\ast_{P_1}(\vec{A}^{\textrm{bg}}(t_f), \psi^{\textrm{bg}}(t_f))\, \tilde{\mathcal{O}}(A^{\textrm{bg}},\psi^{\textrm{bg}},\sigma) \,\Psi_{P_2}(\vec{A}^{\textrm{bg}}(t_i), \psi^{\textrm{bg}}(t_i)) e^{iS_{QCD}(A^{\textrm{bg}}, \psi^{\textrm{bg}})}\,, 
\label{matrfac}
\end{eqnarray}
where
\begin{eqnarray}
&&\tilde{\mathcal{O}}(A^{\textrm{bg}},\psi^{\textrm{bg}},\sigma) = \int \mathcal{D}A^{\textrm{q}}\,\int\mathcal{D}\psi^{\textrm{q}} \,\mathcal{O}(A^{\textrm{q}} + A^{\textrm{bg}},\psi^{\textrm{q}} + \psi^{\textrm{bg}}) e^{iS_{bQCD}(A^{\textrm{q}}, \psi^{\textrm{q}}; A^{\textrm{bg}},  \psi^{\textrm{bg}})}
\end{eqnarray}
and the QCD action in the background fields is
\begin{eqnarray}
S_{bQCD}(A^{\textrm{q}}, \psi^{\textrm{q}}; A^{\textrm{bg}},  \psi^{\textrm{bg}}) = S_{QCD}(A^{\textrm{q}} + A^{\textrm{bg}}, \psi^{\textrm{q}} + \psi^{\textrm{bg}}) - S_{QCD}(A^{\textrm{bg}}, \psi^{\textrm{bg}})\,.
\end{eqnarray}

Now we can fix the background fields and evaluate the functional integral over the ``quantum" fields perturbatively to a certain order in the number of loops. This perturbative calculation in the background field is the essence of the background field method. In general, the result of calculating the functional integrals has a form of a product of the coefficient functions (``impact" factors) and the Wilson-line operators constructed from background fields which describe interaction of ``quantum" fields with the background,
\begin{align}\label{Otilde}
\tilde{\mathcal{O}}(A^{\textrm{bg}},\psi^{\textrm{bg}},\sigma) = \sum_i C_i(\sigma)\otimes \mathcal{V}_i(A^{\textrm{bg}},\psi^{\textrm{bg}}, \sigma)\,.
\end{align}
The sum goes over the different operators.
Equation \eqref{Otilde} should be substituted back into Eq.~(\ref{matrfac}). In particular, as we will see in our calculation below, the helicity-dependent interaction of quarks and gluons with the shock-wave background is described by polarized Wilson lines  (\ref{VqG_decomp}) and (\ref{UqG_decomp}).

To study the dependence of the Wilson-line operators on the rapidity factorization scale $\sigma$ one can repeat the procedure described above. We introduce a new scale $\sigma'$ and redefine the background fields as
\begin{eqnarray}
&&A^{\textrm{bg}}_\mu \to \hat{A}^\textrm{q}_\mu + \hat{A}^{\textrm{bg}}_\mu,\,\,\,\,\, \psi^{\textrm{bg}} \to \hat{\psi}^\textrm{q} + \hat{\psi}^{\textrm{bg}}\,,
\end{eqnarray}
where the ``quantum" fields now have momenta $\sigma > p^- > \sigma'$ and the background fields have $p^- < \sigma'$. After this we can perform the integration over new ``quantum" fields $\hat{A}^\textrm{q}_\mu$, $\hat{\psi}^\textrm{q}$ (keeping $\hat{A}^\textrm{bg}_\mu$ and $\hat{\psi}^\textrm{bg}$ fixed) in Eq. (\ref{matrfac}) which corresponds to the functional integral
\begin{eqnarray}
&&\textrm{T}\,[\mathcal{V}_i(A^{\textrm{bg}},\psi^{\textrm{bg}}, \sigma)] \equiv \int \mathcal{D}\hat{A}^{\textrm{q}}\,\int\mathcal{D}\hat{\psi}^{\textrm{q}}\,  \mathcal{V}_i(\hat{A}^{\textrm{q}} + \hat{A}^{\textrm{bg}},\hat{\psi}^{\textrm{q}} + \hat{\psi}^{\textrm{bg}}, \sigma) e^{iS_{bQCD}(\hat{A}^{\textrm{q}}, \hat{\psi}^{\textrm{q}}; \hat{A}^{\textrm{bg}},  \hat{\psi}^{\textrm{bg}})}\,.
\label{funcintbqcd}
\end{eqnarray}
This integral can be evaluated by a perturbative calculation in the background field which yields an evolution equation of the following form
\begin{eqnarray}
\textrm{T}\,[\mathcal{V}_i(A^{\textrm{bg}},\psi^{\textrm{bg}}, \sigma)] = \int\limits^\sigma_{\sigma'} \frac{dp^-}{p^-} \sum_j \mathcal{K}_{ij} \otimes \mathcal{V}_j(\hat{A}^{\textrm{bg}},\hat{\psi}^{\textrm{bg}}, \sigma')\, ,
\label{evbfm}
\end{eqnarray}
with some kernels $\mathcal{K}_{ij}$. In particular, in this paper we derive the evolution equation for the polarized Wilson lines  (\ref{VqG_decomp}) and (\ref{UqG_decomp}).

For perturbative calculations of the functional integral (\ref{funcintbqcd}) we need to know the propagators of ``quantum" particles in the background field. Note that the form of such propagators unambiguously fixes the set of the Wilson-line operators on the right-hand side of the evolution equation (\ref{evbfm}). In the next Section we will derive the quark and gluon propagators in the shock-wave background and later use them to construct helicity evolution equations for operators (\ref{VqG_decomp}) and (\ref{UqG_decomp}).

  
 \subsection{Quark and gluon propagators in the shock-wave background\label{sec:prbfm}}
 
 In this Section we will construct quark and gluon propagators in the external background field by direct resummation of the corresponding Feynman diagrams. While to solve this problem in full generality is a formidable task, see  \cite{Altinoluk:2014oxa,Balitsky:2015qba,Altinoluk:2015gia,Balitsky:2016dgz,Balitsky:2017flc,Balitsky:2017gis,Chirilli:2018kkw,Altinoluk:2020oyd,Altinoluk:2021lvu, Chirilli:2021lif}, it is still possible to separate a contribution which dominates at small $x$. To find this contribution we construct an expansion of the propagators in inverse powers of $p^-$ and find the first few terms in this expansion. Indeed, at small $x$, the $p^-$ component of the ``quantum" field is assumed to be large. As a result, the leading terms of the expansion in inverse powers of $p^-$ dominate at small $x$ yielding a large logarithm $\int \frac{dp^-}{p^-}$. In general, the expansion in the inverse powers of $p^-$ corresponds to the expansion in the powers of $x$ or in eikonality we employed above.
 
 The technique we use is similar to the one developed in Refs. \cite{Balitsky:2015qba,Balitsky:2016dgz} for the unpolarized evolution. However, for the helicity evolution we need to extend the approach and assume the most general form of the background field. In particular, we take into account the transverse component $A_i$ of the field, which was neglected in \cite{Balitsky:2015qba,Balitsky:2016dgz}. In our calculation we fix the gauge of the background field as $A_+ = A^- = 0$ and assume that the fields are independent of $x^+$, $A_\mu = A_\mu(x^-, {\ul x})$.
 
 
\subsubsection{Scalar propagator in the shock-wave background\label{scalarinbw}}
Before we consider quark and gluon propagators in the background field, let us start with a simpler problem and calculate the scalar propagator in the background field. In the Schwinger's notation, see Appendix \ref{Ap:Schnot}, we write the scalar propagator in the background field $A_\mu$ as\footnote{From here on we do not explicitly show the ``q" and ``bg" labels for ``quantum" and background fields.}
\begin{eqnarray}
 &&(x|\frac{1}{\hat{P}^2+i\epsilon}|y)=(x|\frac{1}{\hat{p}^2+g\{\hat{p}^\mu, A_\mu(\hat{x})\} + g^2A^\mu(\hat{x}) A_\mu(\hat{x}) + i\epsilon}|y)\,,
 \label{scpropinit}
 \end{eqnarray}
where $\hat{P}_\mu = \hat{p}_\mu + g A_\mu(\hat{x})$. Note that in the Schwinger's notation $\hat{p}$ and $A(\hat{x})$ are operators so one should take into account their ordering. In particular, one can immediately recognize in $\{\hat{p}^\mu, A_\mu(\hat{x})\}$ and $A^\mu(\hat{x}) A_\mu(\hat{x})$ the two vertices of scalar QED, where the latter is the ``seagull" vertex. Of course \eq{scpropinit} can be obtained by resummation of an infinite number of interactions of the background field $A_\mu$ with the propagating scalar particle.

Indeed, expanding the propagator we write\footnote{For brevity we are going to omit the hat sign over momentum and coordinate operators.}
\begin{eqnarray}
(x|\frac{1}{P^2+i\epsilon}|y)=(x|&&\frac{1}{p^2 + i\epsilon}|y) - (x|\frac{1}{p^2 + i\epsilon}(g\{p^\mu, A_\mu\} + g^2A^\mu A_\mu)\frac{1}{p^2 + i\epsilon}|y) 
 \\
 &&+(x|\frac{1}{p^2 + i\epsilon}(g\{p^\mu, A_\mu\} + g^2A^\mu A_\mu)\frac{1}{p^2 + i\epsilon}(\{p^\mu, A_\mu\} + A^\mu A_\mu)\frac{1}{p^2 + i\epsilon}|y)  +\dots\,. \notag
 \label{scexp}
 \end{eqnarray}

Let us start with the first term of this expansion which is a free propagator of the scalar particle. Using Eq. (\ref{Schsc}) and performing the integration over $p^+$ we find
\begin{eqnarray}
(x|\frac{1}{p^2 + i\epsilon}|y) &&= \Bigg(-\frac{i}{2\pi}\theta(x^- - y^-) \int\limits^\infty_0 \frac{dp^-}{2p^-} + \frac{i}{2\pi} \theta(y^- - x^-) \int\limits^0_{-\infty} \frac{dp^-}{2p^-} \Bigg) e^{-ip^-(x-y)^+}
 \\
&&\times ({\ul x}| e^{-i\frac{p^2_\perp}{2p^-}x^-} e^{i\frac{p^2_\perp}{2p^-} y^-} |{\ul y})\,. \notag
\label{scfree}
\end{eqnarray}

Substituting this result for each free propagator\footnote{One should also use Eq. (\ref{shstcompl}) to introduce the integration over intermediate coordinates.} in Eq. (\ref{scexp}) one finds the following form of the scalar propagator in the background field
\begin{eqnarray}
(x|\frac{1}{P^2 + i\epsilon}|y) &&= \Bigg(-\frac{i}{2\pi}\theta(x^- - y^-) \int\limits^\infty_0 \frac{dp^-}{2p^-} + \frac{i}{2\pi} \theta(y^- - x^-) \int\limits^0_{-\infty} \frac{dp^-}{2p^-} \Bigg) e^{-ip^-(x-y)^+}
 \\
&&\times ({\ul x}| e^{-i\frac{p^2_\perp}{2p^-}x^-}\mathcal{S}(x^-, y^-) e^{i\frac{p^2_\perp}{2p^-} y^-} |{\ul y})\,, \notag
\end{eqnarray}
where the operator $\mathcal{S}$ is constructed out of the background fields and describes the interaction of the ``quantum" scalar field with the background gluons. In general, this operator has a form of an expansion in inverse powers of $p^-$,
\begin{eqnarray}
&&\mathcal{S}(x^-, y^-) = \mathcal{S}_0(x^-, y^-) + \frac{1}{p^-}\mathcal{S}_1(x^-, y^-) + \frac{1}{(p^-)^2}\mathcal{S}_2(x^-, y^-) + \dots\,.
\label{scexppmin}
\end{eqnarray}

As we discussed above, the dominant contribution at small $x$ corresponds to the first few orders of expansion (\ref{scexppmin}). Fortunately, it is possible to obtain the exact form of those terms by considering the first few orders of the expansion in the coupling constant (\ref{scexp}).

 To show this let us go back to \eq{scexp}. Using Eq.~(\ref{scfree}) for the second term of the expansion and taking into account that $A_+=0$ we obtain
  \begin{eqnarray}
(x|\frac{1}{P^2+i\epsilon}|y) &&= \Bigg( -\frac{i}{2\pi} \theta(x^- - y^-) \int\limits^\infty_0 \frac{dp^-}{2p^-} + \frac{i}{2\pi} \theta(y^- - x^-) \int\limits^0_{-\infty} \frac{dp^-}{2p^-} \Bigg) e^{-ip^- (x-y)^+}
 \\
&&\times ({\ul x}| e^{-i\frac{p^2_\perp}{2p^-}x^-} \Bigg\{1 +  ig\int\limits^{x^-}_{y^-} dz^- e^{i\frac{p^2_\perp}{2p^-}z^-} \Big( A_-(z^-) + \frac{p^k}{2p^-}  A_k(z^-) + A_k(z^-) \frac{p^k}{2p^-}
\nonumber\\
&&+ \frac{g}{2p^-}A^k(z^-) A_k(z^-) \Big) e^{-i\frac{p^2_\perp}{2p^-}z^-} + \dots\Bigg\} e^{i\frac{p^2_\perp}{2p^-}y^-} |{\ul y})\,, \notag
\label{scpropbfexp}
 \end{eqnarray}
where the ellipsis stand for the higher-order terms of the expansion (\ref{scexp}).
 
Now let us use the following identity for an arbitrary operator $O$:
 \begin{eqnarray}
&&e^{i\frac{p^2_\perp}{2p^-}z^-}O e^{-i\frac{p^2_\perp}{2p^-}z^-}= O + i\frac{z^-}{2p^-}[p^2_\perp, O]-\frac{1}{2}\Big(\frac{z^-}{2p^-}\Big)^2[p^2_\perp, [p^2_\perp, O]] + \dots\,,
\end{eqnarray}
which, taking into account that
 \begin{eqnarray}
 &&[p^2_\perp, O] = -i \{p^s, \partial_s O\}\,,
 \end{eqnarray}
can be rewritten as
 \begin{eqnarray}
&&e^{i\frac{p^2_\perp}{2p^-}z^-}O e^{-i\frac{p^2_\perp}{2p^-}z^-}=O + \frac{z^-}{2p^-} \{p^s, \partial_s O\} + \frac{1}{2}\Big(\frac{z^-}{2p^-}\Big)^2 \{p^s, \{p^m, \partial_s \partial_m O\} \} + \dots\,.
\label{phaseexp}
\end{eqnarray}
Note that $p^s$ is an operator acting on everything to its right, while the partial derivatives in $\partial_s O$ and $\partial_s \partial_m O$ act only on $O$.

Employing this result in Eq.~(\ref{scpropbfexp}) we obtain
\begin{eqnarray}
(x|\frac{1}{P^2+i\epsilon}|y)^{ab} &&= \Bigg( -\frac{i}{2\pi} \theta(x^- - y^-) \int\limits^\infty_0 \frac{dp^-}{2p^-} + \frac{i}{2\pi} \theta(y^- - x^-) \int\limits^0_{-\infty} \frac{dp^-}{2p^-} \Bigg) e^{-ip^- (x-y)^+} \\
&&\times ({\ul x}| e^{-i\frac{p^2_\perp}{2p^-}x^-} \Bigg\{1 +  ig\int\limits^{x^-}_{y^-} dz^- A_-(z^-) + \frac{ig}{2p^-} \int\limits^{x^-}_{y^-} dz^- \Big( \{p^k, A_k(z^-)\} + gA^k(z^-) A_k(z^-) \Big)
\nonumber\\
&& + \frac{ig}{2p^-} \int\limits^{x^-}_{y^-} dz^- z^- \{p^s, \partial_s A_-(z^-) \} + \dots\Bigg\}^{ab} e^{i\frac{p^2_\perp}{2p^-}y^-} |{\ul y})\,, \notag
\label{scpropaftop}
\end{eqnarray}
where we explicitly keep only the first two terms of (\ref{phaseexp}).

A similar calculation can be done for the other terms in the expansion (\ref{scexp}). Eventually, each insertion of $\{p^\mu, A_\mu\} + A^\mu A_\mu$ generates a structure similar to (\ref{scpropaftop}). As a result we see that each coupling to the background field brings an extra inverse power of $p^-$. The only exception is the eikonal coupling $\{p^\mu, A_\mu\}\to p^- A^+$ which does not change the counting in inverse powers of $p^-$. However, these terms can be resummed into Wilson-line factors, which in the operator form are given by \eq{Vhat} in Appendix~\ref{Ap:Schnot}. After this resummation the scalar propagator takes the form
  \begin{eqnarray}\label{scalar5}
(x|\frac{1}{P^2+i\epsilon}|y) &&= \Bigg( -\frac{i}{2\pi} \theta(x^- - y^-) \int\limits^\infty_0 \frac{dp^-}{2p^-} + \frac{i}{2\pi} \theta(y^- - x^-) \int\limits^0_{-\infty} \frac{dp^-}{2p^-} \Bigg) e^{-ip^- (x-y)^+}   \\
&&\times ({\ul x}| e^{-i\frac{p^2_\perp}{2p^-}x^-} \Bigg\{V[x^-, y^-] + \frac{ig}{2p^-} \int\limits^{x^-}_{y^-} dz^- V[x^-, z^-]\Big( \{p^k, A_k(z^-)\} + gA^k(z^-) A_k(z^-) \Big)V[z^-, y^-]
\nonumber\\
&& + \frac{ig}{2p^-} \int\limits^{x^-}_{y^-} dz^- z^- V[x^-, z^-]\{p^s, \partial_s A_-(z^-) \}V[z^-, y^-] + O\Big(\frac{1}{(p^-)^2}\Big)\Bigg\} e^{i\frac{p^2_\perp}{2p^-}y^-} |{\ul y})\,. \notag
 \end{eqnarray}
Here $V[x^-, y^-]$ are the light-cone Wilson-line operators akin to those in \eq{Vhat}, but defined with finite integration limits. 
 
Now let us rewrite this result in a gauge-covariant form. Introducing $\frac{d}{dz^-}(z^-) = 1$ in the second term in the curly brackets of \eq{scalar5} and integrating by parts we can recombine the resulting terms to get the following form of the propagator,
  \begin{eqnarray}
(x|\frac{1}{P^2+i\epsilon}|y) &&= \Bigg( -\frac{i}{2\pi} \theta(x^- - y^-) \int\limits^\infty_0 \frac{dp^-}{2p^-} + \frac{i}{2\pi} \theta(y^- - x^-) \int\limits^0_{-\infty} \frac{dp^-}{2p^-} \Bigg) e^{-ip^- (x-y)^+} ({\ul x}| e^{-i\frac{p^2_\perp}{2p^-}x^-} \\
&&\times \Bigg\{V[x^-, y^-] + \frac{igx^-}{2p^-}( \{p^k, A_k\} + gA^k A_k )(x^-)V[x^-, y^-] - \frac{igy^-}{2p^-} V[x^-, y^-]( \{p^k, A_k\} + gA^k A_k )(y^-)
\nonumber\\
&&- \frac{ig}{2p^-} \int\limits^{x^-}_{y^-} dz^- z^- V[x^-, z^-] \{P^k, F_{-k} \} V[z^-, y^-] + O\Big(\frac{1}{(p^-)^2}\Big)\Bigg\} e^{i\frac{p^2_\perp}{2p^-}y^-} |{\ul y})\,, \notag
 \end{eqnarray}
 where the second and third terms in the curly brackets are the boundary terms which we obtained in the integration by parts.
 
 Note that up to this point our calculation has been completely general. Now let us consider the scalar propagator in the shock-wave approximation with the shock wave localized near $x^-=0$. Since there are no fields outside the shock-wave we can neglect the boundary terms and simplify the gauge factors  $V[x^-, y^-]\to V$ for $x^- > 0 > y^-$ (with $V$ the infinite light-cone Wilson line operator \eqref{Vhat}), which yields
   \begin{eqnarray}\label{Spropagator}
&&(x|\frac{1}{P^2+i\epsilon}|y) =  -\frac{i}{2\pi} \int\limits^\infty_0 \frac{dp^-}{2p^-}e^{-ip^- (x-y)^+} \\
&&\times ({\ul x} | e^{-i \frac{p^2_\perp}{2p^-}x^-} \Bigg\{V - \frac{ig}{2p^-} \int\limits^\infty_{-\infty} dz^- z^- V[\infty, z^-] \{P^k, F_{-k} \} V[z^-, -\infty] + O\Big(\frac{1}{(p^-)^2}\Big)\Bigg\} e^{i\frac{p^2_\perp}{2p^-}y^-} |{\ul y} )\,, \notag
\label{scpropfin}
 \end{eqnarray}
which agrees with Ref. \cite{Balitsky:2015qba}. Here we assume that $x^- > 0 > y^-$.
 
We find that at the leading order of the $1/p^-$ expansion the interaction of the scalar particle with the background field is defined by the eikonal Wilson line $V$, while at the next order the interaction is described by the sub-eikonal operator
\begin{eqnarray}\label{F-k}
&&\int\limits^\infty_{-\infty} dz^- z^- V[\infty, z^-] \{P^k, F_{-k} \} V[z^-, -\infty]\,.
\label{subiekf}
\end{eqnarray}
In the subsequent Sections we will relate this operator to the polarized Wilson lines (\ref{VxyG2}) and (\ref{Vi}).

Note that this type of sub-eikonal operators is neglected in the unpolarized evolution. Indeed the $z^-$ factor under the integral in \eq{F-k} leads to the suppression of the operator in the shock-wave approximation by a factor of $1/P^+$, which, when combined with the $1/p^-$ prefactor of this operator in \eq{Spropagator} gives a suppression by a factor of $1/s \sim x$. Therefore, the unpolarized evolution is driven by the eikonal gauge factors (light-cone Wilson lines). However, as we will see, the interactions via eikonal factors do not contribute to helicity evolution, which starts with sub-eikonal operators like the one in Eq.~(\ref{subiekf}).
 
 
\subsubsection{Gluon propagator in the shock-wave background} 

In this Section we are going to derive the gluon propagator in the background field in the axial gauge. Using the approach developed in the previous Section we will consider the expansion of the propagator in inverse powers of $p^-$ and reconstruct the first several terms of this expansion by analyzing the first few orders of the perturbative expansion in the background field,
\begin{eqnarray}
&&\textrm{T}\, [C^a_\mu(x)C^b_\nu(y)] = (x|\frac{-id_{\mu\nu}(p)\delta^{ab}}{p^2+i\epsilon} |y) \\
&&- i g (x| \frac{-id_{\mu\rho}(p)}{p^2+i\epsilon} \Big[ g^{\rho\sigma} \{ p_\alpha,  {\cal A}^{\alpha} \} + 2 i (\partial^\rho {\cal A}^\sigma - \partial^\sigma {\cal A}^\rho) - p^\rho {\cal A}^\sigma - {\cal A}^\rho p^\sigma \Big] \frac{-id_{\sigma\nu}(p)}{p^2+i\epsilon} |y)^{ab}
\nonumber\\
&&- g^2 (x|\frac{-id_{\mu\rho}(p)}{p^2} \bar{\psi}\gamma^\rho t^a \frac{i\slashed{p}}{\slashed{p}^2+i\epsilon} \gamma^\sigma t^b \psi \frac{-id_{\sigma\nu}(p)}{p^2} |y)
- g^2 (y| \frac{-id_{\nu\sigma}(p)}{p^2+i\epsilon} \bar{\psi}\gamma^\sigma t^b \frac{i\slashed{p}}{\slashed{p}^2+i\epsilon} \gamma^\rho t^a \psi \frac{-id_{\rho\mu}(p)}{p^2+i\epsilon} |x) + \dots\,, \notag
\label{glprop1}
\end{eqnarray}
where $\psi$, $\bar{\psi}$ are background quark and anti-quark fields and the expression in the square brackets is the QCD three-gluon vertex in the background field. For the free gluon propagator in the axial gauge we have
\begin{eqnarray}
&&d_{\mu\nu}(p)\equiv g_{\mu\nu} - \frac{n_\mu p_\nu + p_\mu n_\nu}{n\cdot p}\,,
\end{eqnarray}
where $n$ is a light-like vector with $n^+ =1$, $n^- = 0$, and ${\un n} =0$.

The first term in Eq. (\ref{glprop1}) is the free gluon propagator. Integrating over $p^+$ we can rewrite it as
\begin{eqnarray}\label{glue2}
(x|\frac{-id_{\mu\nu}(p)}{p^2 + i\epsilon}|y) &&= \Bigg(-\frac{1}{2\pi}\theta(x^- - y^-) \int\limits^\infty_0 \frac{dp^-}{2p^-} + \frac{1}{2\pi} \theta(y^- - x^-) \int\limits^0_{-\infty} \frac{dp^-}{2p^-} \Bigg) e^{-ip^-(x-y)^+} \\
&&\times ({\ul x}|(g_{\mu i}-\frac{n_\mu}{n\cdot p}p_i) e^{-i\frac{p^2_\perp}{2p^-}x^-} e^{i\frac{p^2_\perp}{2p^-} y^-} (\delta^i_\nu-p^i\frac{n_\nu}{n\cdot p}) |{\ul y}) + i(x| \frac{n_\mu n_\nu}{(n\cdot p)^2} |y)\,. \notag
\label{axialprop}
\end{eqnarray}
The last term here is the instantaneous term in the LCPT terminology which we neglect in our calculation. Substituting the right-hand side of \eq{glue2} for each free propagator in the expansion (\ref{glprop1}) one finds the following general structure of the gluon propagator in the background field,
\begin{eqnarray}
\textrm{T}\, [C^a_\mu(x)C^b_\nu(y)] &&= \Bigg(-\frac{1}{2\pi}\theta(x^- - y^-) \int\limits^\infty_0 \frac{dp^-}{2p^-} + \frac{1}{2\pi} \theta(y^- - x^-) \int\limits^0_{-\infty} \frac{dp^-}{2p^-} \Bigg) e^{-ip^-(x-y)^+} \\
&&\times ({\ul x}|(g_{\mu i}-\frac{n_\mu}{n\cdot p}\mathscr{P}_i) e^{-i\frac{p^2_\perp}{2p^-}x^-}\mathcal{G}^{ij}(x^-, y^-)  e^{i\frac{p^2_\perp}{2p^-} y^-} (g_{j\nu}-\mathscr{P}_j\frac{n_\nu}{n\cdot p}) |{\ul y})^{ab}\,, \notag
\end{eqnarray}
where $\mathscr{P}_\mu = p_\mu + g{\cal A}_\mu$ and the operator $\mathcal{G}(x^-, y^-)$ is constructed out of the background quark and gluon fields and describes the interaction of the ``quantum" gluon with the background. Similarly to the scalar case, the operator $\mathcal{G}$ can be expanded in the inverse powers of $p^-$,
\begin{eqnarray}\label{glue3}
&&\mathcal{G}^{ij}(x^-, y^-) = \mathcal{G}^{ij}_0(x^-, y^-) + \frac{1}{p^-}\mathcal{G}^{ij}_1(x^-, y^-) + \frac{1}{(p^-)^2}\mathcal{G}^{ij}_2(x^-, y^-) + \dots\,.
\end{eqnarray}

Following the approach developed for the scalar propagator we are going to construct the first few terms in the series \eqref{glue3} using the perturbative expansion (\ref{glprop1}). It is easy to observe that each intermediate propagator in (\ref{glprop1}) leads to suppression by an extra inverse power of $p^-$, see Eq. (\ref{axialprop}). This suppression can only be compensated by the eikonal term $p^- A^+$ of the three-gluon vertex. However, such terms can be easily resummed to all orders in the perturbation theory into Wilson-line gauge factors.

As a result, substituting \eq{axialprop} into \eq{glprop1} and performing manipulations similar to those done in Sec.~\ref{scalarinbw} we find
\begin{eqnarray}
&&\textrm{T}\, [C^a_\mu(x)C^b_\nu(y)] = \Bigg( -\frac{1}{2\pi}\theta(x^- - y^-) \int\limits^\infty_0 \frac{dp^-}{2p^-} + \frac{1}{2\pi} \theta(y^- - x^-) \int\limits^0_{-\infty} \frac{dp^-}{2p^-} \Bigg) e^{-ip^- (x-y)^+} \\
&&\times ({\ul x}| (g_{\mu i}-\frac{n_\mu}{p^-}\mathscr{P}_i)^{ac} e^{-i\frac{p^2_\perp}{2p^-}x^-} \Bigg[ g^{ij}U^{cd}[x^-, y^-]
\nonumber\\
&&- \frac{i g g^{ij}}{2p^-} \int^{x^-}_{y^-} dz^- z^- (U[x^-, z^-] \{\mathscr{P}^k, \mathcal{F}_{-k} \} U[z^-, y^-])^{cd}  - \frac{g\epsilon^{ij}}{p^-} \int^{x^-}_{y^-} dz^- ( U[x^-, z^-] \mathcal{F}^{12}(z^-) U[z^-, y^-])^{cd}
\nonumber\\
&& -  \frac{g^2g^{ij}}{4p^-}\int^{x^-}_{y^-} dz^- \int^{z^-}_{y^-} dz'^- \Big(U^{cc'}[x^-, z^-] \bar{\psi}(z^-)t^{c'} V[z^-, z'^-] \gamma^+ t^{d'} \psi(z'^-) U^{d'd}[z'^-, y^-] + \mbox{c.c.} \Big) 
\nonumber\\
&&+ \frac{ig^2\epsilon^{ij}}{4p^-}\int^{x^-}_{y^-} dz^- \int^{z^-}_{y^-} dz'^- \Big(U^{cc'}[x^-, z^-] \bar{\psi}(z^-)t^{c'} V[z^-, z'^-] \gamma^+ \gamma_5 t^{d'} \psi(z'^-) U^{d'd}[z'^-, y^-]  + \mbox{c.c.} \Big)
+ O\Big(\frac{1}{(p^-)^2}\Big)\Bigg] 
\nonumber\\
&&\times e^{i\frac{p^2_\perp}{2p^-} y^-} (g_{j\nu}-\mathscr{P}_j\frac{n_\nu}{p^-})^{db} |{\ul y}) + \dots\,. \notag
\end{eqnarray}
One can see that the structure of the operators in the gluon propagator in the background field, which is in agreement with Refs. \cite{Balitsky:2015qba,Chirilli:2018kkw,Kovchegov:2017lsr}, is richer than the one in the scalar propagator. But what is more important is that now we explicitly see that the interaction of the ``quantum" gluon with the shock-wave background fields is described be the polarized Wilson lines (\ref{UqG_decomp}). 
Indeed, taking into account that
\begin{eqnarray}
&&\int\limits^{x^-}_{y^-} dz^- z^- U[x^-, z^-] \{\mathscr{P}^k, \mathcal{F}_{-k} \} U[z^-, y^-] = \int\limits^{x^-}_{y^-} dz^- z^- \mathscr{P}^k U[x^-, z^-] \mathcal{F}_{-k} U[z^-, y^-]
\label{acomtozF}\\
&& + \int\limits^{x^-}_{y^-} dz^- z^- U[x^-, z^-] \mathcal{F}_{-k} U[z^-, y^-]\mathscr{P}^k - g\int\limits^{x^-}_{y^-} dz^-_1 \int\limits^{z^-_1}_{y^-} dz^-_2 (z^-_1 - z^-_2) U[x^-, z^-_1] \mathcal{F}_{-k} U[z^-_1, z^-_2] \mathcal{F}_{- k} U[z^-_2, y^-]
\nonumber
\end{eqnarray}
we can finally write the gluon propagator in the shock-wave background as 
\begin{eqnarray}
\textrm{T}\, [C^a_\mu(x)C^b_\nu(y)] &&= -\frac{1}{2\pi} \int\limits^\infty_0 \frac{dp^-}{2p^-} e^{-ip^- (x-y)^+}
\label{glinbackgf}\\
&&\times ({\ul x}| (g_{\mu i}-\frac{n_\mu}{p^-}p_i)^{ac} e^{-i\frac{p^2_\perp}{2p^-}x^-} \mathcal{G}^{ij}(\infty, -\infty) e^{i\frac{p^2_\perp}{2p^-} y^-} (g_{j\nu}-p_j\frac{n_\nu}{p^-})^{db} |{\ul y}) + \dots
\nonumber
\end{eqnarray}
where we assume that $x^- > 0 > y^-$ and
\begin{eqnarray}
\label{GopGl}
\mathcal{G}^{ij}(\infty, -\infty) &&= g^{ij}U + \frac{g^{ij}s}{2P^+p^-} U^{\textrm{q}[2]} + \frac{i\epsilon^{ij}s}{2P^+p^-}U^{\textrm{pol} [1]}
\label{Gforgl}\\
&& - \frac{igg^{ij}}{2p^-} p^k \int\limits^{\infty}_{-\infty} dz^- z^- U[\infty, z^-] \mathcal{F}_{-k} U[z^-, -\infty] - \frac{igg^{ij}}{2p^-}  \int\limits^{\infty}_{-\infty} dz^- z^- U[\infty, z^-] \mathcal{F}_{-k} U[z^-, -\infty]p^k
\nonumber\\
&&  + \frac{ig^2g^{ij}}{2p^-}\int\limits^{\infty}_{-\infty} dz^-_1 \int\limits^{z^-_1}_{-\infty} dz^-_2 (z^-_1 - z^-_2) U[\infty, z^-_1] \mathcal{F}_{-k} U[z^-_1, z^-_2] \mathcal{F}_{- k} U[z^-_2, -\infty]  + O\Big(\frac{1}{(p^-)^2}\Big).
\nonumber
\end{eqnarray}

As we will show in the next Section, see Eqs. (\ref{Pperpop}) and (\ref{zFtoDi}),  the operator
\begin{eqnarray}
&&\int\limits^{\infty}_{-\infty} dz^- z^- U[\infty, z^-] \mathcal{F}_{-k} U[z^-, -\infty]
\nonumber
\end{eqnarray}
can be further related to the polarized Wilson line (\ref{UG2}) and the adjoint version of the operator (\ref{Vi}) given in \eq{Ui}.\footnote{Note that the fundametal-representation version of this operator, given by \eq{Vi}, appears in the dipole gluon helicity TMD and the Jaffe-Manohar (JM) gluon helicity
PDF at small-$x$ after the expansion of the exponential phases, as shown above in Sec.~\ref{sec:glue_helicity}.} We will also see that operators $U$, $U^{\textrm{q}[2]}$ and the operator in the last line of (\ref{Gforgl}) do not contribute to the helicity evolution.


\subsubsection{Quark propagator in the shock-wave background field}

In this Section we will consider quark propagator in the background of quark and gluon fields. To simplify the problem we will start the derivation taking into account only the background gluon field and later extend it to include the contribution of background quarks.

The most general form of the quark propagator, which can be obtained by resummation of an infinite number of couplings to background gluons, is
\begin{eqnarray}
&&\textrm{T}\,[\psi (x) \bar{\psi}(y)]_A = (x|\frac{i}{\slashed{P} + i\epsilon}|y) = (x|\slashed{P}\frac{i}{P^2 + \frac{g}{2} \sigma^{\mu\nu} F_{\mu\nu} + i\epsilon}|y)\,,
\end{eqnarray}
where we use the identity
\begin{eqnarray}
&&\slashed{P}^2 = P^2 + \frac{g}{2} \sigma^{\mu\nu} F_{\mu\nu}\,.
\end{eqnarray}
Here the subscript $A$ denotes the gluon-only background field.

To construct the expansion in inverse powers of $p^-$ we write the propagator as an infinite series
\begin{eqnarray}
\textrm{T}\,[\psi (x) \bar{\psi}(y)]_A && = i(x|\slashed{P}\frac{1}{P^2 + i\epsilon}|y) - ig(x|\slashed{P}\frac{1}{P^2 + i\epsilon}\frac{1}{2} \sigma^{\mu\nu} F_{\mu\nu} \frac{1}{P^2 + i\epsilon}|y)
 \\
&&  + ig^2(x|\slashed{P}\frac{1}{P^2 + i\epsilon}\frac{1}{2} \sigma^{\mu\nu} F_{\mu\nu} \frac{1}{P^2 + i\epsilon} \frac{1}{2} \sigma^{\rho\sigma} F_{\rho\sigma} \frac{1}{P^2 + i\epsilon}|y) + \dots \notag
\label{qwexF}
\end{eqnarray}
and substitute Eq.~(\ref{scpropfin}) for each scalar propagator. Since each scalar propagator is proportional to $1/p^-$, it is easy to see that to find the leading contribution at small $x$, it is sufficient to consider only the first few orders of the expansion (\ref{qwexF}).

For brevity, let us also simplify the problem and instead of calculating the full quark propagator consider only its contraction with $\gamma^+\gamma^5$. Indeed, as shown above in the LCOT approach, it is the only contraction we need in order to derive the helicity evolution  equations (cf. Eq. (\ref{Vq1})).\footnote{For a more general derivation see Refs.~\cite{Chirilli:2018kkw,Altinoluk:2020oyd,Altinoluk:2021lvu}.}

Starting with Eq. (\ref{qwexF}) we obtain\footnote{Note that in the scalar propagators one should take care of the exponential factors which, after expansion, can be combined with boundary terms yielding contributions proportional to $P^2_\perp$.}
\begin{eqnarray}
&&\textrm{T}\,[\psi_\beta (x) \bar{\psi}_\alpha(y)]_A [\gamma^+ \gamma_5]_{\alpha\beta}
\\
&&= \Bigg( - \frac{1}{2\pi} \theta(x^- - y^-) \int\limits^\infty_0 \frac{dp^-}{(2p^-)^2} + \frac{1}{2\pi} \theta(y^- - x^-) \int\limits^0_{-\infty} \frac{dp^-}{(2p^-)^2} \Bigg) e^{-ip^- (x - y)^+} ({\ul x}| \tr\, \Bigg\{\left(\left(\frac{p^2_\perp}{2p^-} + A^+\right) \gamma^- - P^i \gamma^i\right) 
\nonumber\\
&&\times e^{-i\frac{p^2_\perp}{2p^-}x^-} \Bigg[ - \frac{ig}{2} \int\limits^{x^-}_{y^-} dz^- \Big(V[x^-, z^-] + \frac{igx^-}{2p^-}( \{p^k, A_k\} + gA^k A_k )(x^-)V[x^-, z^-] + \frac{iz^-}{2p^-} V[x^-, z^-] P^2_\perp 
\nonumber\\
&&- \frac{ig}{2p^-} \int\limits^{x^-}_{z^-} dz^-_1 z^-_1 V[x^-, z^-_1] \{P^k, F_{-k} \} V[z^-_1, z^-] \Big)
\sigma^{\mu\nu} F_{\mu\nu}(z^-) \Big(V[z^-, y^-] - \frac{iz^-}{2p^-}P^2_\perp V[z^-, y^-]
\nonumber\\
&& - \frac{igy^-}{2p^-} V[z^-, y^-]( \{p^k, A_k\} + gA^k A_k )(y^-) - \frac{ig}{2p^-} \int\limits^{z^-}_{y^-} dz^-_2 z^-_2 V[z^-, z^-_2] \{P^k, F_{-k} \} V[z^-_2, y^-] \Big)
\nonumber\\
&&+\frac{g^2}{8p^-} \int\limits^{x^-}_{y^-} dz^-_1 \int\limits^{z^-_1}_{y^-} dz^-_2 V[x^-, z^-_1] \sigma^{\mu\nu} F_{\mu\nu}(z^-_1) V[z^-_1, z^-_2]
 \sigma^{\rho\sigma} F_{\rho\sigma}(z^-_2) V[z^-_2, y^-] \Bigg] e^{i\frac{p^2_\perp}{2p^-}y^-} \gamma^+ \gamma_5\Bigg\} |{\ul y}) + O\Big(\frac{1}{(p^-)^4}\Big)\, .
 \nonumber
\end{eqnarray}

Now we calculate the trace of gamma matrices and simplify the structure of operators. To do the latter, we utilize the following relations:
\begin{eqnarray}
&&[P^2, P_\mu]=i\{P^\alpha, F_{\alpha\mu}\}
\end{eqnarray}
and
\begin{eqnarray}
&&g\int^{x^-}_{y^-}dz^- [x^-, z^-]F_{- m}[z^-, y^-] = P_m[x^-, y^-] - [x^-, y^-]P_m\,.
\end{eqnarray}

After a somewhat lengthy algebra we obtain
\begin{eqnarray}
\label{qvpropingl}
&&\textrm{T}\,[\psi_\beta (x) \bar{\psi}_\alpha(y)]_A [\gamma^+ \gamma_5]_{\alpha\beta}
\\
&&= \Bigg( - \frac{i}{2\pi} \theta(x^- - y^-) \int\limits^\infty_0 \frac{dp^-}{(2p^-)^2} + \frac{i}{2\pi} \theta(y^- - x^-) \int\limits^0_{-\infty} \frac{dp^-}{(2p^-)^2} \Bigg) e^{-ip^- (x - y)^+}
\nonumber\\
&&\times ({\ul x}| \Bigg\{ 4 \epsilon^{im} P^i e^{-i\frac{p^2_\perp}{2p^-}x^-} \Big[ P^m V[x^-, y^-] - V[x^-, y^-] P^m + \frac{ix^-}{2p^-} (p^2_\perp P^m - P^m P^2_\perp)V[x^-, y^-]
\nonumber\\
&&  + \frac{ix^-}{2p^-} (P^2_\perp - p^2_\perp) V[x^-, y^-] P^m + \frac{iy^-}{2p^-} V[x^-, y^-] (P^m p^2_\perp - P^2_\perp P^m)  +  \frac{iy^-}{2p^-} P^m V[x^-, y^-] (P^2_\perp - p^2_\perp)
\nonumber\\
&& - \frac{ig}{2p^-} P^m \int\limits^{x^-}_{y^-} dz^- z^- V[x^-, z^-] \{P^k, F_{-k}\} V[z^-, y^-] + \frac{ig}{2p^-} \int\limits^{x^-}_{y^-} dz^- z^- V[x^-, z^-] \{P^k, F_{-k}\} V[z^-, y^-] P^m  \Big]
\nonumber\\
&&+ g\epsilon^{mn} P^i e^{-i\frac{p^2_\perp}{2p^-}x^-} \frac{1}{p^-} \Big[ \int\limits^{x^-}_{y^-} dz^- V[x^-, z^-]  F_{mn}(z^-) V[z^-, y^-] P^i + P^i \int\limits^{x^-}_{y^-} dz^- V[x^-, z^-] F_{mn}(z^-) V[z^-, y^-]\Big]
\nonumber\\
&&- 2g\epsilon^{mn}  \left(\frac{p^2_\perp}{2p^-} + A^+\right) e^{-i\frac{p^2_\perp}{2p^-}x^-} \int\limits^{x^-}_{y^-} dz^-_1  V[x^-, z^-_1] F_{mn}(z^-_1) V[z^-_1, y^-] \Bigg\} e^{i\frac{p^2_\perp}{2p^-}y^-} |{\ul y}) + O\Big(\frac{1}{(p^-)^4}\Big)\,.
\nonumber
\end{eqnarray}

This result contains three types of operators. The first is the eikonal coupling of the quark to the background field via Wilson lines $V[x^-, y^-]$. As we will see in explicit calculation below, this operator does not contribute to helicity evolution. The helicity evolution is defined by the sub-eikonal coupling via operators
\begin{eqnarray}
&&\int\limits^{x^-}_{y^-} dz^- V[x^-, z^-] F_{mn}(z^-) V[z^-, y^-]
\label{PkFmn}
\end{eqnarray}
and
\begin{eqnarray}
\int\limits^{x^-}_{y^-} dz^- z^- V[x^-, z^-] \{P^k, F_{-k}\} V[z^-, y^-]\,.
\label{PkF}
\end{eqnarray}
While the former operator is obviously related to the small-$x$ polarized Wilson line (\ref{VG1}), the relation of the latter to Eqs.~(\ref{VxyG2}) and (\ref{Vi}) can be observed from the identity
\begin{eqnarray}
&&g\int\limits^{x^-}_{y^-} dz^- z^- V[x^-, z^-] \{P^k, F_{-k}\} V[z^-, y^-] = i \int\limits^{x^-}_{y^-} dz^- z^- V[x^-, z^-] [P_-, P^2_\perp] V[z^-, y^-]
 \\
&&= - x^- P^2_\perp V[x^-, y^-] + y^- V[x^-, y^-] P^2_\perp + \int\limits^{x^-}_{y^-} dz^- V[x^-, z^-] P^2_\perp V[z^-, y^-]\,, \notag
\label{Pperpop}
\end{eqnarray}
where the last operator is nothing else but the polarized Wilson line (\ref{VxyG2}).

Alternatively, one can use
\begin{eqnarray}\label{glue4}
&&g\int\limits^{x^-}_{y^-} dz^- z^- [x^-, z^-] \{P^k, F_{-k} \} [z^-, y^-] = P^k\,g \int\limits^{x^-}_{y^-} dz^- z^- [x^-, z^-] F_{-k} [z^-, y^-] + g\int\limits^{x^-}_{y^-} dz^- z^- [x^-, z^-] F_{-k} [z^-, y^-]P^k\notag
 \\
&& - g^2\int\limits^{x^-}_{y^-} dz^-_1 \int^{z^-_1}_{y^-} dz^-_2 (z^-_1 - z^-_2) [x^-, z^-_1] F_{-k} [z^-_1, z^-_2] F_{- k}[z^-_2, y^-] 
\label{trickpf}
\end{eqnarray}
and (\ref{zFtoDi}) at small-$x$ to relate the operator (\ref{PkF}) to the polarized Wilson line (\ref{Vi}). Note that the helicity-independent operators like the one in the last line of \eq{glue4} or the eikonal Wilson lines never contribute to helicity evolution which we will explicitly show in our calculation below.

For now, let us keep the form of \eq{qvpropingl} and calculate the coupling of the propagator to the background quark field. It is easy to see that each such coupling comes along with $1/p^-$, so that at the leading order of the expansion in the inverse powers of $p^-$ it is sufficient to add just a single quark insertion. As a result, for the full quark propagator in the background field (contracted with $\gamma^+ \gamma_5$) we have
\begin{eqnarray}
&&\textrm{T}\,[\psi_\beta (x) \bar{\psi}_\alpha(y)] [\gamma^+ \gamma_5]_{\alpha\beta} = \textrm{T}\,[\psi_\beta (x) \bar{\psi}_\alpha(y)]_A [\gamma^+ \gamma_5]_{\alpha\beta}
\\
&& - g^2\int d^4z_1 \int d^4z_2 \tr\,\Big\{ (x|\frac{i\slashed{P}}{P^2+i\epsilon}|z_1) \gamma^\rho t^a \psi(z_1) (z_1|\frac{-i}{P^2+i\epsilon}|z_2)^{ab}  \bar{\psi}(z_2)\gamma_\rho t^b (z_2|\frac{i\slashed{P}}{P^2+i\epsilon}|y)\gamma^+ \gamma_5\Big\} + O\Big(\frac{1}{(p^-)^4}\Big)\,.
\nonumber
\end{eqnarray}

Substituting the scalar propagators we obtain
\begin{eqnarray}
\label{qwaftscal}
&&\textrm{T}\,[\psi_\beta (x) \bar{\psi}_\alpha(y)] [\gamma^+ \gamma_5]_{\alpha\beta} = \textrm{T}\,[\psi_\beta (x) \bar{\psi}_\alpha(y)]_A [\gamma^+ \gamma_5]_{\alpha\beta}
\\
&&+ \Bigg( - \frac{i}{2\pi} \theta(x^- - y^-) \int\limits^\infty_0 \frac{dp^-}{(2p^-)^2}  +  \frac{i}{2\pi} \theta(y^- - x^-) \int\limits^0_{-\infty} \frac{dp^-}{(2p^-)^2} \Bigg) e^{-ip^- (x-y)^+} \tr\, \Bigg\{ ({\ul x}| P^i\gamma_i e^{-i\frac{p^2_\perp}{2p^-}x^-}
\nonumber\\
&&\times  \frac{ig^2}{2p^-} \int^{x^-}_{y^-} dz^-_1 \int^{z^-_1}_{y^-} dz^-_2 V[x^-, z^-_1] \gamma^\rho t^a \psi(z^-_1)  U[z^-_1, z^-_2]^{ab}  \bar{\psi}(z^-_2)\gamma_\rho t^b V[z^-_2, y^-] e^{i\frac{p^2_\perp}{2p^-}y^-} \gamma_jP^j |{\ul y}) \gamma^+ \gamma_5\Bigg\} + O\Big(\frac{1}{(p^-)^4}\Big)\,.
\nonumber
\end{eqnarray}

Next we use the Fierz identity to decompose the product of background quark fields in terms of the Dirac matrices, i.e., we apply
\begin{eqnarray}
\Gamma = \frac{1}{4}\tr\,[\Gamma] \,\textrm{I} + \frac{1}{4}\tr\,[\gamma_\mu\Gamma]\gamma^\mu + \frac{1}{8}\tr\,[\sigma_{\mu\nu}\Gamma] \sigma^{\mu\nu}-\frac{1}{4}\tr\,[\gamma_\mu\gamma^5\Gamma]\gamma^\mu\gamma^5 + \frac{1}{4}\tr\,[\gamma^5\Gamma]\gamma^5\,,
\label{Fierz}
\end{eqnarray}
which is valid for an arbitrary gamma-matrix $\Gamma$. Employing \eq{Fierz} we can calculate the trace in (\ref{qwaftscal}) getting
\begin{eqnarray}\label{Qpropagator_46}
&&\textrm{T}\,[\psi_\beta (x) \bar{\psi}_\alpha(y)] [\gamma^+ \gamma_5]_{\alpha\beta} = \textrm{T}\,[\psi_\beta (x) \bar{\psi}_\alpha(y)]_A [\gamma^+ \gamma_5]_{\alpha\beta} 
\label{qwprwithqwb}\\
&&+ \Bigg( - \frac{i}{2\pi} \theta(x^- - y^-) \int\limits^\infty_0 \frac{dp^-}{(2p^-)^2}  +  \frac{i}{2\pi} \theta(y^- - x^-) \int\limits^0_{-\infty} \frac{dp^-}{(2p^-)^2} \Bigg) e^{-ip^- (x-y)^+} ({\ul x} | P_i e^{-i\frac{p^2_\perp}{2p^-}x^-}
\nonumber\\
&&\times \frac{ig^2}{p^-} \int^{x^-}_{y^-} dz^- \int^{z^-}_{y^-} dz'^- V[x^-, z^-] t^a \Big( - i\epsilon^{ij} \psi_\beta(z^-)\bar{\psi}_\alpha(z'^-)[\gamma^+]_{\alpha\beta}
\nonumber\\
&& + g^{ij} \psi_\beta(z^-)\bar{\psi}_\alpha(z'^-) [\gamma^+\gamma^5]_{\alpha\beta} \Big) U[z^-, z'^-]^{ab} t^b V[z'^-, y^-] e^{i\frac{p^2_\perp}{2p^-}y^-} P_j |{\ul y} ) + O\Big(\frac{1}{(p^-)^4}\Big)\,.
\nonumber
\end{eqnarray}
At this point we clearly see that the coupling of the propagator to the background quark field is defined by the polarized Wilson line (\ref{Vq1}). We will also see that the coupling via (\ref{Vq2}) does not survive in helicity evolution.

Though the equation \eqref{Qpropagator_46} we obtained is quite lengthy, it can be significantly simplified in the case of the shock-wave background when there are no fields outside the shock-wave. The result (\ref{qwprwithqwb}) can be simplified even further if we integrate it over the longitudinal coordinates and consider a particular case of ${\ul x} = {\ul y} = {\ul x}_1$ which we will use later in the derivation of helicity evolution, see the diagram III in Fig.~\ref{FIG:Q_evol}.

Indeed, after changing the sign of $p^-$ and taking into account that
\begin{eqnarray}
&&({\ul x}_1|\frac{p_i}{p^2_\perp}\mathcal{O}\frac{p_j}{p^2_\perp}|{\ul x}_1) = \int d^2{\ul z} ({\ul x}_1|\frac{p_i}{p^2_\perp}|{\ul z})\mathcal{O}({\ul z})({\ul z}|\frac{p_j}{p^2_\perp}|{\ul x}_1) = \int d^2{\ul z} ({\ul x}_1|\frac{p_j}{p^2_\perp}|{\ul z})\mathcal{O}({\ul z})({\ul z}|\frac{p_i}{p^2_\perp}|{\ul x}_1) = ({\ul x}_1|\frac{p_j}{p^2_\perp}\mathcal{O}\frac{p_i}{p^2_\perp}|{\ul x}_1)
\end{eqnarray}
is symmetric under $i\leftrightarrow j$ for an arbitrary operator $\mathcal{O}({\ul z})$, we obtain
\begin{eqnarray}
&&\int\limits^0_{-\infty} dx^- \int\limits^\infty_0 dy^-\, \textrm{T}\,[\psi_\beta (x^-, {\ul x}_1) \bar{\psi}_\alpha(y^-, {\ul x}_1)] [\gamma^+ \gamma_5]_{\alpha\beta} \\
&&= - \frac{1}{\pi} \int\limits^\infty_0 \frac{dp^-}{p^-} ({\ul x}_1| \frac{\epsilon^{im} p^i}{p^2_\perp}  g\int\limits^{\infty}_{-\infty} dz^- z^- V[-\infty, z^-] \{P^k, F_{-k}\} V[z^-, \infty] \frac{p_m}{p^2_\perp} |{\ul x}_1)
\nonumber\\
&&- \frac{1}{2\pi} \int\limits^\infty_0 \frac{dp^-}{p^- } ({\ul x}_1| \frac{p_i  }{p^2_\perp} \Big( \epsilon^{mn}\,ig \int\limits_{-\infty}^{\infty} dz^- V[-\infty, z^-]  F_{mn} V[z^-, \infty]
\nonumber\\
&&- g^2\int\limits_{-\infty}^{\infty} dz^-_1 \int\limits_{-\infty}^{z^-_1} dz^-_2 V[-\infty, z^-_2] t^a \psi_\beta(z^-_2)U[z^-_2, z^-_1]^{ab}\bar{\psi}_\alpha(z^-_1) [\gamma^+\gamma^5]_{\alpha\beta} t^b V[z^-_1, \infty] \Big) \frac{p_i}{p^2_\perp} |{\ul x}_1) + O\Big(\frac{1}{(p^-)^2}\Big)\,. \notag
\end{eqnarray}
Note that the higher-order terms of the expansion in inverse powers of $p^-$ do not contain a logarithm $\int \frac{dp^-}{p^-}$ which dominates at small $x$. The reader should also note that the large logarithm arises in the terms with sub-eikonal operators (\ref{PkFmn}) and (\ref{PkF}), while the eikonal Wilson lines do not contribute.

Now we use the identity (\ref{trickpf}), introduce the integration over the intermediate transverse coordinate ${\ul x}_2$ and perform the Fourier transformations over transverse momenta using
\footnote{Note that in the last equation we neglect the instantaneous term, see the discussion after Eq.~(\ref{+perp_sub_eik_1}) above.}
\begin{eqnarray}\label{Fr2}
&&({\un x}_1|\frac{p^i}{p^2_\perp} |{\un x}_2) =  \frac{i}{2\pi}  \frac{{\ul x}^i_{12}}{{\ul x}^2_{12}}, \, \, \, \, \, \, \, \, ({\un x}_1| \frac{p^i p^k}{p^2_\perp} |{\un x}_2) = \frac{1}{2\pi} \frac{\delta^{ik}{\ul x}^2_{12} - 2{\ul x}^i_{12} {\ul x}^k_{12}}{{\ul x}^4_{12}} \; .
\end{eqnarray}
We obtain
\begin{eqnarray}\label{Qpropagator_47}
&&\int\limits^0_{-\infty} dx^- \int\limits^\infty_0 dy^-\, \textrm{T}\, [\psi_\beta (x^-, {\ul x}_1) \bar{\psi}_\alpha(y^-, {\ul x}_1)] [\gamma^+ \gamma_5]_{\alpha\beta} \\
&&= \frac{ig}{2\pi^3} \int\limits^\infty_0 \frac{dp^-}{p^-} \int d^2{\ul x}_2 \frac{\epsilon^{mk} {\ul x}^m_{12}}{{\ul x}^4_{12}} \left[ \int\limits^{\infty}_{-\infty} dz^- z^- V_{{\ul x}_2}[-\infty, z^-] F_{-k}(z^-, {\ul x}_2) V_{{\ul x}_2}[z^-, \infty] \right]
\nonumber\\
&&- \frac{i}{8\pi^3} \int\limits^\infty_0 \frac{dp^-}{p^- } \int d^2{\ul x}_2   \frac{1}{{\ul x}^2_{12}} \left[ 2g \int\limits_{-\infty}^{\infty} dz^- V_{{\ul x}_2}[-\infty, z^-]  F_{12}(z^-, {\ul x}_2) V_{{\ul x}_2}[z^-, \infty] \right.
\nonumber\\
&&+ \left. i g^2\int\limits_{-\infty}^{\infty} dz^-_1 \int\limits_{-\infty}^{z^-_1} dz^-_2 V_{{\ul x}_2}[-\infty, z^-_2] t^a \psi_\beta(z^-_2, {\ul x}_2)U_{{\ul x}_2}[z^-_2, z^-_1]^{ab}\bar{\psi}_\alpha(z^-_1, {\ul x}_2) [\gamma^+\gamma^5]_{\alpha\beta} t^b V_{{\ul x}_2}[z^-_1, \infty] \right]\,. \notag
\end{eqnarray}

We can finally relate the operator in the first term of \eqref{Qpropagator_47} to the polarized Wilson line \eqref{Vi}, 
\begin{align}
\label{zFtoDi}
&ig\int\limits_{-\infty}^\infty d z^-  \, z^- V_{{\un x}_2} [\infty, z^-] \,
    F_{-k} \, V_{{\un x}_2} [z^-, - \infty] = - i g\int\limits_{-\infty}^\infty d z^- V_{{\un x}_2} [\infty, z^-] \,
    ( \, z^- \, \pd_k A_- + A_k) \, V_{{\un x}_2} [z^-, - \infty] \\
    &= - \frac{ig}{2}\lim_{L^-\to\infty}\left[  \int\limits_{-\infty}^\infty d z^- V_{{\un x}_2} [\infty, z^-] \,
    \left( \, \int\limits^{z^-}_{-L^-} d\xi^- \, \pd_k A_-  + A_k \right) \, V_{{\un x}_2} [z^-, - \infty] \right. \notag 
    \\
    & \left. + \int\limits_{-\infty}^\infty d z^- V_{{\un x}_2} [\infty, z^-] \,
    \left( - \int\limits^{L^-}_{z^-} d\xi^- \, \pd_k A_-  + A_k \right) \, V_{{\un x}_2} [z^-, - \infty] \right] = \frac{1}{2}  \int\limits^{\infty}_{-\infty} dz^- V_{{\un x}_2} [\infty, z^-] \left[ D_k - \cev{D}_k \right] V_{{\un x}_2} [z^-, -\infty ] \; , \notag
\end{align}
and rewrite our result in a compact form
\begin{eqnarray}
&&\frac{P^+}{2s}\int\limits^0_{-\infty} dx^- \int\limits^\infty_0 dy^-\, \textrm{T}\, [\psi_\beta (x^-, {\ul x}_1) \bar{\psi}_\alpha(y^-, {\ul x}_1)] [\gamma^+ \gamma_5]_{\alpha\beta}
= \frac{1}{8\pi^3} \int\limits^\infty_0 \frac{dp^-}{p^-} \int \frac{d^2{\ul x}_2}{{\ul x}^2_{21}} \left[ 2 \frac{\epsilon^{km} {\ul x}^m_{21}}{{\ul x}^2_{21}} V^{k\textrm{G}[2]\dag}_{{\un x}_2} + V^{\textrm{pol}[1]\dag}_{{\ul x}_2} \right]\,.
\label{qpinb}
\end{eqnarray}
This is our final result for the quark propagator in the background field which we will use in the calculation of the helicity evolution equations. While we consider a particular projection of the propagator, we should mention that our method of derivation is completely general and can be used beyond the problem of helicity evolution.


\subsection{Evolution equation for $Q_{10}$ in the background field method}
\label{EvQ10}

In this Section we will use the results we obtained in the previous Section for the gluon and quark propagators in the background field to derive the evolution equation for the polarized dipole amplitude $Q_{10}$. Following the logic of the background field method and the rapidity factorization approach we define the amplitude as
\begin{align}\label{Q10rap}
Q_{10} (\sigma) \equiv \frac{1}{2 N_c} \, \llangle \mbox{T} \, \tr \left[ V_{\ul 0} \, V_{\un 1}^{\textrm{pol} [1] \, \dagger} \right] + \mbox{T} \, \tr \left[ V_{\ul 1}^{\textrm{pol} [1]} \, V_{\ul 0}^\dagger \right]  \rrangle (\sigma),
\end{align}
where the operators are constructed from fields with longitudinal momentum fraction $p^-$ restricted from above by a cutoff scale  $\sigma$. As we discussed in Sec. \ref{sec:bfm}, to construct the evolution equation for the amplitude we shift the scale to a lower value $\sigma'$ and integrate the matrix element in (\ref{Q10rap}) over the fields with $\sigma > p^- > \sigma'$, see Eq. (\ref{funcintbqcd}), while keeping the fields with momenta $p^- < \sigma'$ fixed. We will perform this integration at the one-loop level which is represented by the diagrams in Fig. \ref{FIG:Q_evol}.

Let us start with the calculation of the diagram I. Expanding Wilson lines of the operators in Eq. (\ref{Q10rap}) one can readily obtain
\begin{eqnarray}
\left(\mbox{T} \, \tr \left[ V_{\ul 0} \, V_{\un 1}^{\textrm{pol} [1] \, \dagger} \right] + \textrm{c.c.} \right)_\textrm{I} &&= \frac{g^2 \, P^+}{s} \, \int\limits^\infty_0 dx^-_0 \int\limits^0_{-\infty} d{x}^-_1 \tr \big[ V_{\ul 0}[\infty, x^-_0] t^a V_{\ul 0}[x^-_0, -\infty] \\
&&\times \, V_{\un{1}} [ -\infty, x^-_1] \, t^b \, \, V_{\un{1}} [ x^-_1, \infty] \big] \, \mbox{T}\, [A^{a+}(x^-_0, {\ul x}_0) F^{b12} (x^-_1, {\un x}_1)] + \textrm{c.c.} 
\nonumber\\
&&= \frac{g^2 \, P^+ \epsilon^{ij}}{s} \, \tr \left[ t^a V_{\ul 0} \, t^b \, \, V^\dag_{\un{1}} \right] \int\limits^\infty_0 dx^-_0 \int\limits^0_{-\infty} d{x}^-_1 \mbox{T}\, [A^{a+}(x^-_0, {\ul x}_0) \partial_i A^b_j (x^-_1, {\un x}_1)] + \textrm{c.c.}\,, \notag
\end{eqnarray}
where in the last line we use the shock-wave approximation to simplify the gauge factors as $V_{\ul 0}[\infty, x^-_0]\to 1$, $V_{\ul 0}[x^-_0, -\infty]\to V_{\ul 0}$, both for $x_0^- >0$, etc.

The subsequent steps of the calculation are straightforward. Substituting the gluon propagator in the shock-wave background field (\ref{glinbackgf})
\begin{eqnarray}
&&\mbox{T}\, [A^{a+}(x^-_0, {\ul x}_0) \partial^i A^{bj}(x^-_1, {\un x}_1)] \Big|_{x^-_0 > x^-_1} = \frac{i}{2\pi} \int\limits^\infty_0 \frac{dp^-}{2p^-} ({\ul x}_0| e^{-i\frac{p^2_\perp}{2p^-}x^-_0} \frac{p_m}{p^-} \mathcal{G}^{mn}(\infty, -\infty) p^i \delta^j_n e^{i\frac{p^2_\perp}{2p^-} x^-_1} |{\un x}_1)^{ab},
\end{eqnarray}
where operator $\mathcal{G}^{mn}$ describes the interaction of the ``quantum" gluon with the shock-wave background, and integrating over the longitudinal coordinates we obtain
\begin{eqnarray}
&&\left(\mbox{T} \, \tr \left[ V_{\ul 0} \, V_{\un 1}^{\textrm{pol} [1] \, \dagger} \right] + \textrm{c.c.} \right)_\textrm{I} = - \frac{ i g^2 \, P^+ \epsilon^{ij}}{\pi s} \, \tr \left[ t^a V_{\ul 0} \, t^b \, \, V^\dag_{\un{1}} \right] \int\limits^\infty_0 dp^- ({\ul x}_0| \frac{p_m}{p^2_\perp} \mathcal{G}^{mn}(\infty, -\infty) \frac{p^i \delta^j_n}{p^2_\perp} |{\un x}_1)^{ab} + \textrm{c.c.}\,.
\label{QevrapI}
\end{eqnarray}
Note that until this point we have not explicitly restricted the integration over the longitudinal momentum $p^-$. However, one should take into account that the matrix element \ref{Q10rap} is integrated over the fields with $\sigma > p^- >  \sigma'$. As a result, the integration over $p^-$ in Eq. \ref{QevrapI} should be restricted to
\begin{eqnarray}
\int\limits^\infty_0 dp^- \to \int\limits^\sigma_{\sigma'} dp^-\,.
\end{eqnarray}
For brevity, we will perform this substitution at the very end of our calculation.

In an analogous way one can perform the calculation of diagrams II, I$'$ and II$'$ in Fig. \ref{FIG:Q_evol}. Adding all terms together we obtain
\begin{eqnarray}
&&\left(\mbox{T} \, \tr \left[ V_{\ul 0} \, V_{\un 1}^{\textrm{pol} [1] \, \dagger} \right] + \textrm{c.c.} \right)_{\textrm{I} + \textrm{II} + \textrm{I}' + \textrm{II}'} = - \frac{ i g^2 \, P^+ \epsilon^{ij}}{\pi s} \, \tr \left[ t^a V_{\ul 0} \, t^b \, \, V^\dag_{\un{1}} \right] \int\limits^\infty_0 dp^- \\
&&\times\left[  ({\ul x}_0| \frac{p_m}{p^2_\perp} \mathcal{G}^{mn}(\infty, -\infty) \frac{p^i \delta^j_n}{p^2_\perp} |{\un x}_1) - ({\un x}_1| \frac{p^i \delta^j_m}{p^2_\perp} \mathcal{G}^{mn}(\infty, -\infty) \frac{p_n}{p^2_\perp} | {\ul x}_0) - ({\ul x}_0\to {\ul x}_1) \right]^{ab} + \textrm{c.c.}\,, \notag
\label{Qevolrapford}
\end{eqnarray}
where the second term in the last line corresponds to the diagram II, and the last two terms are the sum of the diagrams I$'$ and II$'$.

Now we need to substitute the explicit form of the operator $\mathcal{G}^{mn}$ from Eq. (\ref{Gforgl}). Let us show that the first term of the operator, i.e., the interaction described by the Wilson line $U$ does not provide any contribution to the evolution equation. Indeed, substituting this term into Eq.~(\ref{Qevolrapford}) we obtain a trivial combination
\begin{eqnarray}
&&({\ul x}_0| \frac{p_m}{p^2_\perp} \mathcal{G}^{mn}(\infty, -\infty) \frac{p^i \delta^j_n}{p^2_\perp} |{\un x}_1)^{ab} - ({\un x}_1| \frac{p^i \delta^j_m}{p^2_\perp} \mathcal{G}^{mn}(\infty, -\infty) \frac{p_n}{p^2_\perp} | {\ul x}_0)^{ab} \\
&&\to ({\ul x}_0| \frac{p^j}{p^2_\perp} U^{ab} \frac{p^i }{p^2_\perp} |{\un x}_1) - ({\un x}_1| \frac{p^i }{p^2_\perp} U^{ab} \frac{p^j}{p^2_\perp} | {\ul x}_0) = ({\un x}_1| \frac{p^i }{p^2_\perp} U^{ab} \frac{p^j}{p^2_\perp}|{\ul x}_0) - ({\un x}_1| \frac{p^i }{p^2_\perp} U^{ab} \frac{p^j}{p^2_\perp} | {\ul x}_0) = 0\,. \notag
\end{eqnarray}
Moreover, for the same reason, the interaction of the gluon with the shock wave via operators $U^{\textrm{pol} [1]}$ and
\begin{eqnarray}\label{helindop5}
&&\int\limits^{\infty}_{-\infty} dz^-_1 \int\limits^{z^-_1}_{-\infty} dz^-_2 (z^-_1 - z^-_2) U[\infty, z^-_1] \mathcal{F}_{-k} U[z^-_1, z^-_2] \mathcal{F}_{- k}U[z^-_2, -\infty]
\end{eqnarray}
does not contribute to helicity evolution as well.

Substituting the remaining three terms of Eq. (\ref{Gforgl}) into Eq. (\ref{Qevolrapford}) and introducing the integration over the intermediate coordinate ${\un x}_2$ (see Eq. (\ref{shstcompl})), after some straightforward algebra we obtain
\begin{eqnarray}\label{Qevol22}
&&\left(\mbox{T} \, \tr \left[ V_{\ul 0} \, V_{\un 1}^{\textrm{pol} [1] \, \dagger} \right] + \textrm{c.c.} \right)_{\textrm{I} + \textrm{II} + \textrm{I}' + \textrm{II}'} = \frac{ g^2 \, \epsilon^{ij}}{\pi} \, \tr \left[ t^a V_{\ul 0} \, t^b \, \, V^\dag_{\un{1}} \right] \int\limits^\infty_0 \frac{dp^-}{p^-} \int d^2{\un x}_2 \\
&&\times\left[ \left( ({\un x}_1|\frac{p^i }{p^2_\perp} |{\un x}_2) ({\un x}_2| \frac{p^j p^k}{p^2_\perp}|{\ul x}_0) + ({\un x}_1| \frac{p^i p^k}{p^2_\perp}|{\un x}_2) ({\un x}_2|\frac{p^j}{p^2_\perp} | {\ul x}_0) \right) \frac{gP^+}{s}\int\limits^{\infty}_{-\infty} dz^- z^- U_{{\un 2}} [\infty, z^-] \mathcal{F}_{-k}(z^-, {\un x}_2) U_{{\un 2}}[z^-, -\infty]\right.
\nonumber\\
&&- \left.\epsilon^{jn}  ({\un x}_1|\frac{p^i }{p^2_\perp} |{\un x}_2)  ({\un x}_2| \frac{p_n}{p^2_\perp}|{\ul x}_0)  U_{\un 2}^{\textrm{pol} [1]}
- ({\ul x}_0\to {\ul x}_1) \right]^{ab} + \textrm{c.c.}\,. \notag
\end{eqnarray}

Finally, we need to substitute the Fourier transformations \footnote{Note again that in the last equation we neglect the instantaneous contribution, see the discussion after Eq. (\ref{+perp_sub_eik_1}).}
\begin{eqnarray}
&&({\un x}_1|\frac{p^i}{p^2_\perp} |{\un x}_2) =  \frac{i}{2\pi}  \frac{{\ul x}^i_{12}}{{\ul x}^2_{12}}, \, \, \, \, \, \, \, \, ({\un x}_1| \frac{p^i p^k}{p^2_\perp} |{\un x}_2) = \frac{1}{2\pi} \frac{\delta^{ik}{\ul x}^2_{12} - 2{\ul x}^i_{12} {\ul x}^k_{12}}{{\ul x}^4_{12}}
\end{eqnarray}
into \eq{Qevol22}, which yields 
\begin{eqnarray}\label{V10_237}
&&\left(\mbox{T} \, \tr \left[ V_{\ul 0} \, V_{\un 1}^{\textrm{pol} [1] \, \dagger} \right] + \textrm{c.c.} \right)_{\textrm{I} + \textrm{II} + \textrm{I}' + \textrm{II}'} = - \frac{g^2}{4\pi^3} \, \tr \left[ t^a V_{\ul 0} \, t^b \, \, V^\dag_{\un{1}} \right] \int\limits^\infty_0 \frac{dp^-}{p^-} \int d^2{\un x}_2 \\
&&\times \left\{ \left[ 2 \epsilon^{kj} \frac{ {\ul x}^j_{21}}{{\ul x}^4_{21}} - \frac{ \epsilon^{kj} ({\ul x}^j_{21} + {\ul x}^j_{20} )}{{\ul x}^2_{21}{\ul x}^2_{20}} - \frac{2 {\ul x}_{20}\times {\ul x}_{21} }{{\ul x}^2_{21}{\ul x}^2_{20}} \left( \frac{ {\ul x}^k_{21}}{{\ul x}^2_{21} } - \frac{ {\ul x}^k_{20}}{{\ul x}^2_{20}} \right) \right] \frac{ igP^+}{s}\int\limits^{\infty}_{-\infty} dz^- z^- U_{{\un 2}} [\infty, z^-] \mathcal{F}_{-k}(z^-, {\un x}_2) U_{{\un 2}}[z^-, -\infty]\right.
\nonumber\\
&&- \left.\left[ \frac{ 1 }{{\ul x}^2_{21}} - \frac{{\ul x}_{21}\cdot{\ul x}_{20}}{{\ul x}^2_{21}{\ul x}^2_{20}} \right] U_{\un 2}^{\textrm{pol} [1]} \right\}^{ab} + \textrm{c.c.}\,. \notag
\label{Qpolevsumofall1}
\end{eqnarray}

Now let us discuss the operator in the second line of \eq{V10_237}. It is easy to see that this operator is nothing else but a small-$x$ version of the operator in the dipole gluon helicity TMD and the Jaffe-Manohar (JM) gluon helicity PDF (see also the discussion in Sec.~\ref{sec:glue_helicity}). This operator can be obtained by expanding the exponential factor and keeping only the term linear in $x$, c.f. Eq. (\ref{JMexp1}),
\begin{align}\label{phexp}
& \int\limits_{-\infty}^\infty d z^- e^{i x P^+ \, z^-} U_{{\un 2}} [\infty, z^-] \,
    \mathcal{F}^{+k} (z^- , {\un x}_2) \, U_{{\un 2}} [z^-, - \infty]  \\ 
    &= \int\limits_{-\infty}^\infty d z^-  U_{{\un 2}} [\infty, z^-] \,
    \mathcal{F}^{+k} \, U_{{\un 2}} [z^-, - \infty]
    + i x P^+ \, \int\limits_{-\infty}^\infty d z^-  \, z^- U_{{\un 2}} [\infty, z^-] \,
    \mathcal{F}^{+k} \, U_{{\un 2}} [z^-, - \infty]\notag + \dots\\ 
&= - \int\limits_{-\infty}^\infty d z^-  U_{{\un 2}} [\infty, z^-] \,
    \partial^k {\cal A}^+ \, U_{{\un 2}} [z^-, - \infty]
    + i x P^+ \, \int\limits_{-\infty}^\infty d z^-  \, z^- U_{{\un 2}} [\infty, z^-] \,
    \mathcal{F}^{+k} \, U_{{\un 2}} [z^-, - \infty] + \dots\notag\,.
\end{align}
Here the first term of the last line can be rewritten as a derivative of the Wilson line and for this reason describes the eikonal helicity-independent coupling of the ``quantum" gluon to the shock-wave background. The helicity-dependent coupling in the small-$x$ limit is described by the second term which explicitly appears in Eqs.~(\ref{Qpolevsumofall1}) and  (\ref{glinbackgf}) for the gluon propagator.

Using Eq.~(\ref{zFtoDi}) one can rewrite this operator in terms of the adjoint polarized Wilson line of the second kind defined in \eq{Ui}. We obtain
\begin{eqnarray}
&&\left(\mbox{T} \, \tr \left[ V_{\ul 0} \, V_{\un 1}^{\textrm{pol} [1] \, \dagger} \right] + \textrm{c.c.} \right)_{\textrm{I} + \textrm{II} + \textrm{I}' + \textrm{II}'} = \frac{ g^2 }{4\pi^3} \, \tr \left[ t^b V_{\ul 0} \, t^a \, \, V^\dag_{\un{1}} \right] \int\limits^\infty_0 \frac{dp^-}{p^-} \int d^2{\un x}_2
\\
&&\times\left\{ \left[ 2 \epsilon^{kj} \frac{ {\ul x}^j_{21}}{{\ul x}^4_{21}} - \frac{ \epsilon^{kj} ({\ul x}^j_{21} + {\ul x}^j_{20} )}{{\ul x}^2_{21}{\ul x}^2_{20}} - \frac{2 {\ul x}_{20}\times {\ul x}_{21} }{{\ul x}^2_{21}{\ul x}^2_{20}} \left( \frac{ {\ul x}^k_{21}}{{\ul x}^2_{21} } - \frac{ {\ul x}^k_{20}}{{\ul x}^2_{20}} \right) \right]  U^{k\textrm{G}[2]}_{{\un 2}}
+ \left[ \frac{ 1 }{{\ul x}^2_{21}} - \frac{{\ul x}_{21}\cdot{\ul x}_{20}}{{\ul x}^2_{21}{\ul x}^2_{20}} \right] U^{\textrm{pol}[1]}_{\un 2} \right\}^{ba} + \textrm{c.c.}\,.
\nonumber
\end{eqnarray}

Let us now calculate the contribution of the diagram III. Using the shock-wave approximation we can write the following expression for this diagram
\begin{eqnarray}
&&\left(\mbox{T} \, \tr \left[ V_{\ul 0} \, V_{\un 1}^{\textrm{pol} [1] \, \dagger} \right] + \textrm{c.c.} \right)_\textrm{III}   \\
&&= \frac{ g^2 P^+}{2 \, s} \int\limits_{-\infty}^0 \!\! d{x}_1^- \! \int\limits_0^\infty d x_2^- \, \tr \left(\, V_{\ul 0}\, t^a\, \textrm{T} [\psi_\beta (x_1^-,\un{x}_1) \bar{\psi}_\alpha (x_2^-,\un{x}_1) ]\, [\gamma^+ \gamma^5]_{\alpha\beta} \,t^b \,\right) U_{\un{1}}^{ba} + \textrm{c.c.}\,. \notag
\end{eqnarray}

Substituting the expression for the quark propagator in the background field (\ref{qpinb}) and integrating over longitudinal coordinates we obtain
\begin{eqnarray}
&&\left(\mbox{T} \, \tr \left[ V_{\ul 0} \, V_{\un 1}^{\textrm{pol} [1] \, \dagger} \right] + \textrm{c.c.} \right)_\textrm{III}  \\
&&= \frac{\alpha_s}{2\pi^2} \int\limits^\infty_0 \frac{dp^-}{p^-} \int d^2{\ul x}_2 \left( 2 \frac{\epsilon^{ij} {\ul x}^j_{21}}{{\ul x}^4_{21}} \tr [ t^b V_{\ul 0}\, t^a\,  V^{i\textrm{G}[2]\dag}_{\un 2} ] U_{\un{1}}^{ba} + \frac{1}{{\ul x}^2_{21}} \tr [ V_{\ul 0} t^a V^{\textrm{pol}[1]\dag}_{\ul 2} t^b ] U_{\un{1}}^{ba}\right) + \textrm{c.c.}\,. \notag
\label{dIIIfin}
\end{eqnarray}

Finally, one needs to calculate the eikonal diagrams. Since this calculation is similar to the derivation of the Balitsky-Kovchegov (BK) evolution equation \cite{Balitsky:1995ub,Balitsky:1998ya,Kovchegov:1999yj,Kovchegov:1999ua} with the well-known kernel $\frac{x_{10}^2}{x_{21}^2 \, x_{20}^2}$, we will not present the details of the calculation here. Let us just mention that this calculation can be done similar to the calculation presented above. However, for the eikonal diagrams, the dominant contribution with the logarithmic integral $\int \frac{d p^-}{p^-}$ comes from the interaction through the eikonal Wilson line of the gluon propagator in the background field, see the first term in Eq.~(\ref{GopGl}).

Let us now assemble all the terms together. We have
\begin{eqnarray}\label{V10evol}
&&\frac{1}{2 N_c} \, \llangle \mbox{T} \, \tr \left[ V_{\ul 0} \, V_{\un 1}^{\textrm{pol} [1] \, \dagger} \right] + \textrm{c.c.} \rrangle(\sigma) = \frac{1}{2 N_c} \, \llangle \mbox{T} \, \tr \left[ V_{\ul 0} \, V_{\un 1}^{\textrm{pol} [1] \, \dagger} \right] + \textrm{c.c.} \rrangle_0(\sigma)
\\
&&+\frac{ \alpha_sN_c }{2\pi^2} \, \int\limits^\sigma_{\sigma'} \frac{dp^-}{p^-} \int d^2{\un x}_2 \Bigg\{ \left[ \frac{ 1 }{{\ul x}^2_{21}} - \frac{{\ul x}_{21}\cdot{\ul x}_{20}}{{\ul x}^2_{21}{\ul x}^2_{20}} \right] \frac{1}{N^2_c} \llangle \tr \left[ t^b V_{\ul 0} \, t^a \, \, V^\dag_{\un{1}} \right] (U^{\textrm{pol}[1]}_{\un 2})^{ba} + \textrm{c.c.} \rrangle(\sigma')
\nonumber\\
&&+ \left[ 2 \epsilon^{ij} \frac{ {\ul x}^j_{21}}{{\ul x}^4_{21}} - \frac{ \epsilon^{ij} ({\ul x}^j_{21} + {\ul x}^j_{20} )}{{\ul x}^2_{21}{\ul x}^2_{20}} - \frac{2 {\ul x}_{20}\times {\ul x}_{21} }{{\ul x}^2_{21}{\ul x}^2_{20}} \left( \frac{ {\ul x}^i_{21}}{{\ul x}^2_{21} } - \frac{ {\ul x}^i_{20}}{{\ul x}^2_{20}} \right) \right] \frac{1}{N^2_c} \llangle \tr \left[ t^b V_{\ul 0} \, t^a \, \, V^\dag_{\un{1}} \right] (U^{i\textrm{G}[2]}_{\un 2})^{ba} + \textrm{c.c.} \rrangle(\sigma') \Bigg\}
\nonumber\\
&&+ \frac{\alpha_sN_c}{4\pi^2} \int\limits^\sigma_{\sigma'} \frac{dp^-}{p^-} \int  \frac{d^2{\ul x}_2}{{\ul x}^2_{21}} \left\{ \frac{1}{N^2_c}\llangle \tr [ V_{\ul 0} t^a V^{\textrm{pol}[1]\dag}_{\ul 2} t^b ] U_{\un{1}}^{ba}\rrangle(\sigma') + 2 \frac{\epsilon^{ij} {\ul x}^j_{21}}{{\ul x}^2_{21}}\frac{1}{N^2_c} \llangle \tr [ t^b V_{\ul 0}\, t^a\,  V^{i\textrm{G}[2]\dag}_{\un 2} ] U_{\un{1}}^{ba}\rrangle(\sigma') + \textrm{c.c.}\right\}
\nonumber\\
&&+ \frac{\as \, N_c}{2 \pi^2} \, \int\limits^\sigma_{\sigma'} \frac{d p^-}{p^-} \, \int d^2 {\ul x}_2 \, \frac{x_{10}^2}{x_{21}^2 \, x_{20}^2} \,  \left\{ \frac{1}{N_c^2} \, \llangle \tr \left[ t^b \, V_{\un 0} \, t^a \, V_{\un 1}^{\textrm{pol} [1] \, \dagger} \right] \, U_{\un 2}^{ba} \rrangle (\sigma')  - \frac{C_F}{N_c^2} \, \llangle \tr \left[ V_{\un 0} \, V_{\un 1}^{\textrm{pol} [1] \, \dagger} \right] \rrangle (\sigma') + \textrm{c.c.}  \right\}\,,
\nonumber
\end{eqnarray}
where, following the logic of the background field method, we insert the limits of the integral over $p^-$ and identify the Wilson lines as constructed out of the background fields with $p^- < \sigma'$. Now we can see that up to a trivial change of variables we are in a full agreement with the result obtained in the LCOT approach above, given in \eq{Q_evol_main}. The result of a similar calculation employing the background field method in the adjoint representation, which is not shown here, is also in agreement with Eq.~(\ref{G_adj_evol}).


\subsection{Evolution equation for $G^i_{10}$ in the background field method}

The operator definition of the polarized dipole amplitude $G^i_{10}$ is given by Eq. (\ref{Gj2}). Of course, to derive the evolution equation for the amplitude one can directly start with that definition. However, we would like to remind the reader that the corresponding operator \eqref{Vi} is a small-$x$ version of the operator in the definition of the dipole gluon helicity TMD and the Jaffe-Manohar (JM) gluon helicity PDF in Eqs. (\ref{glue_hel_TMD}) and (\ref{e:coll}). To emphasise this relation let us start with an alternative definition of the amplitude $G^i_{10}$ which is more obviously related to the aforementioned distributions.

Indeed, using Eq. (\ref{zFtoDi}) one can rewrite the definition (\ref{Gj2}) as
\begin{align}\label{GiasF}
G^i_{10} (\sigma) \equiv \frac{ig P^+}{2 s N_c} \, \llangle  \mbox{T} \, \tr \left[  V^\dagger_{\un 0} \, \int\limits_{-\infty}^\infty d z^-  \, z^- V_{{\un 1}} [\infty, z^-] \,
    F^{+i} \, V_{{\un 1}} [z^-, - \infty]\right] + \textrm{c.c.} \rrangle (\sigma)\,.
\end{align}

Following the background field method, to derive the evolution equation for $G^i_{10}$ let us start with the operator definition (\ref{GiasF}) and integrate the matrix element over fields with $\sigma > p^- > \sigma'$ while keeping the background fields with $p^- < \sigma'$ fixed. At the one-loop level this corresponds to the calculation of diagrams in Fig.~\ref{FIG:Gi_evol}.

Let us start with the calculation of the diagram IV. Rewriting the initial operator as
\begin{eqnarray}
&&\mbox{T} \, \tr \left[  V^\dagger_{\un 0} \, \int\limits_{-\infty}^\infty d z^-  \, z^- V_{{\un 1}} [\infty, z^-] \,
    F^{+i} \, V_{{\un 1}} [z^-, - \infty]\right] + \textrm{c.c.} \\
&&= - \mbox{T} \, \tr \left[  V^\dagger_{\un 0} \, \int\limits_{-\infty}^\infty d z^- V_{{\un 1}} [\infty, z^-] \,
    ( \, z^- \, \pd^i A^+ + A^i) \, V_{{\un 1}} [z^-, - \infty]\right] + \textrm{c.c.}\,, \notag
\end{eqnarray}
expanding the Wilson lines and simplifying the gauge factors using the shock-wave approximation we obtain
\begin{eqnarray}
&&\Bigg(\mbox{T} \, \tr \Bigg[  V^\dagger_{\un 0} \, \int\limits_{-\infty}^\infty d z^-  \, z^- V_{{\un 1}} [\infty, z^-] \,
    F^{+i} \, V_{{\un 1}} [z^-, - \infty]\Bigg] + \textrm{c.c.}\Bigg)_\textrm{IV}
\\
&&= \tr \left[ V^\dag_{\un 0} t^a V_{\un 1}  t^b \right]\, ig \int\limits^{\infty}_0 dx^- \int\limits_{-\infty}^0 dz^- \left( z^- \mbox{T} \, [A^{a+}(x^-, {\un x}_0) \partial^i A^{b+}(z^-, {\un x}_1)]  + \mbox{T} \, [A^{a+}(x^-, {\un x}_0) A^{bi}(z^-, {\un x}_1)] \right) + \textrm{c.c.}\,.
\nonumber
\end{eqnarray}

Now we need to substitute the gluon propagators in the shock-wave background. Using Eq. (\ref{glinbackgf}) we find
\begin{eqnarray}
&&\mbox{T}[ A^{a+}(x^-, {\un x}_0) \partial^i A^{b+}(z^-, {\un x}_1) ]\Big|_{x^- > z^-} = -\frac{i}{2\pi} \int\limits^\infty_0 \frac{dp^-}{2p^-} ({\un x}_0| e^{-i\frac{p^2_\perp}{2p^-}x^-} \frac{p_m}{p^-} \mathcal{G}^{mn}(\infty, -\infty) \frac{p^i p_n}{p^-} e^{i\frac{p^2_\perp}{2p^-} z^-} |{\un x}_1)^{ab}
\end{eqnarray}
and
\begin{eqnarray}
&&\mbox{T} [ A^{a+}(x^-, {\un x}_0) A^{bi}(z^-, {\un x}_1) \rangle \Big|_{x^- > z^-} = \frac{1}{2\pi} \int\limits^\infty_0 \frac{dp^-}{2p^-} ({\un x}_0| e^{-i\frac{p^2_\perp}{2p^-}x^-} \frac{p_m}{p^-} \mathcal{G}^{mn}(\infty, -\infty) \delta^i_n e^{i\frac{p^2_\perp}{2p^-} z^-} |{\un x}_1)^{ab}\,.
\end{eqnarray}
With this result it is straightforward to integrate over the longitudinal coordinates $x^-$ and $z^-$, obtaining
\begin{eqnarray}
&&\Bigg(\mbox{T} \, \tr \Bigg[  V^\dagger_{\un 0} \, \int\limits_{-\infty}^\infty d z^-  \, z^- V_{{\un 1}} [\infty, z^-] \,
    F^{+i} \, V_{{\un 1}} [z^-, - \infty]\Bigg] + \textrm{c.c.}\Bigg)_\textrm{IV} \\
&&= - \tr \left[ V^\dag_{\un 0} t^a V_{\un 1}  t^b \right] \frac{ig}{\pi} \int\limits^\infty_0 dp^- \left\{ 2 ({\un x}_0| \frac{p_m}{p^2_\perp} \mathcal{G}^{mn}(\infty, -\infty) \frac{p^i p_n}{ p^4_\perp } |{\un x}_1)^{ab} + ({\un x}_0| \frac{p_m}{p^2_\perp} \mathcal{G}^{mi}(\infty, -\infty) \frac{1}{p^2_\perp} |{\un x}_1)^{ab} \right\}\,. \notag
\end{eqnarray}

Employing a similar technique one can calculate the diagrams V, IV$'$ and V$'$ in Fig.~(\ref{FIG:Gi_evol}). For the sum of the diagrams we have
\begin{eqnarray}\label{Gievol77}
&&\Bigg(\mbox{T} \, \tr \Bigg[  V^\dagger_{\un 0} \, \int\limits_{-\infty}^\infty d z^-  \, z^- V_{{\un 1}} [\infty, z^-] \,
    F^{+i} \, V_{{\un 1}} [z^-, - \infty]\Bigg] + \textrm{c.c.}\Bigg)_{\textrm{IV}+\textrm{V}+\textrm{IV}'+\textrm{V}'} \\
&&= - \tr \left[ V^\dag_{\un 0} t^a V_{\un 1}  t^b \right] \frac{ig}{\pi} \int\limits^\infty_0 dp^- \left\{ 2 ({\un x}_0| \frac{p_m}{p^2_\perp} \mathcal{G}^{mn}(\infty, -\infty) \frac{p^i p_n}{ p^4_\perp } |{\un x}_1) + 2 ({\un x}_1| \frac{p^i p_m}{p^4_\perp} \mathcal{G}^{mn}(\infty, -\infty) \frac{p_n}{p^2_\perp} |{\un x}_0) \right.
\nonumber\\
&&+ \left.({\un x}_0| \frac{p_m}{p^2_\perp} \mathcal{G}^{mi}(\infty, -\infty) \frac{1}{p^2_\perp} |{\un x}_1)
+ ({\un x}_1| \frac{1}{p^2_\perp} \mathcal{G}^{i n}(\infty, -\infty) \frac{p_n}{ p^2_\perp } |{\un x}_0)  - ({\un x}_0\to{\un x}_1)\right\}^{ab}\,. \notag
\end{eqnarray}

After this we need to substitute the operator $\mathcal{G}^{mn}$ which describes the interaction of the ``quantum" gluon in Fig.~(\ref{FIG:Gi_evol}) with the shock-wave background field. Similar to the case of the dipole amplitude $Q_{10}$, the operators $U$, $U^{\textrm{pol} [1]}$, and (\ref{helindop5}) do not contribute to the evolution of the dipole amplitude $G^i_{10}$. For example, substituting
\begin{eqnarray}
&&\mathcal{G}^{mn}(\infty, -\infty)\to g^{mn}U
\end{eqnarray}
we obtain
\begin{eqnarray}
&& 2 ({\un x}_0| \frac{p^m}{p^2_\perp} U \frac{p^i p_m}{ p^4_\perp } |{\un x}_!) + 2 ({\un x}_1| \frac{p^i p^m}{p^4_\perp} U \frac{p_m}{p^2_\perp} |{\un x}_0) 
+ ({\un x}_0| \frac{p^i}{p^2_\perp} U \frac{1}{p^2_\perp} |{\un x}_1)
+ ({\un x}_1| \frac{1}{p^2_\perp} U \frac{p^i}{ p^2_\perp } |{\un x}_0)  - ({\un x}_0\to{\un x}_1) \\
&&= -2 ({\un x}_1|\frac{p^i p^m}{ p^4_\perp } U \frac{p_m}{p^2_\perp}|{\un x}_0) + 2 ({\un x}_1| \frac{p^i p^m}{p^4_\perp} U \frac{p_m}{p^2_\perp} |{\un x}_0) 
- ({\un x}_1|\frac{1}{p^2_\perp} U  \frac{p^i}{p^2_\perp}|{\un x}_0)
+ ({\un x}_1| \frac{1}{p^2_\perp} U \frac{p^i}{ p^2_\perp } |{\un x}_0)  - ({\un x}_0\to{\un x}_1) = 0\,. \notag
\end{eqnarray}

Substituting the remaining terms of $\mathcal{G}^{mn}$ and introducing the integration over the intermediate variable ${\un x}_2$ we rewrite \eq{Gievol77} as
\begin{eqnarray}
&&\Bigg(\mbox{T} \, \tr \Bigg[  V^\dagger_{\un 0} \, \int\limits_{-\infty}^\infty d z^-  \, z^- V_{{\un 1}} [\infty, z^-] \,
    F^{+i} \, V_{{\un 1}} [z^-, - \infty]\Bigg] + \textrm{c.c.}\Bigg)_{\textrm{IV}+\textrm{V}+\textrm{IV}'+\textrm{V}'} \\
&&= - \tr \left[ V^\dag_{\un 0} t^a V_{\un 1}  t^b \right] \frac{g}{\pi} \int\limits^\infty_0 \frac{dp^-}{p^-} \int d^2{\un x}_2 \left\{ \left[ ({\un x}_1| \frac{1}{p^2_\perp}\left( \delta^{im} - \frac{2 p^i p^m}{ p^2_\perp } \right) |{\un x}_2) ({\un x}_2| \frac{p^m p^k}{p^2_\perp} |{\un x}_0)\right.\right.
\nonumber\\
&&\left. + 2 ({\un x}_1| \frac{p^i p^m p^k}{p^4_\perp} |{\un x}_2) ({\un x}_2| \frac{p_m}{p^2_\perp} |{\un x}_0)
+ ({\un x}_1| \frac{p^k}{p^2_\perp} |{\un x}_2) ({{\un x}_2}| \frac{p^i}{ p^2_\perp } |{\un x}_0) \right] g \int\limits^{\infty}_{-\infty} dz^- z^- U_{\un 2}[\infty, z^-]_z \mathcal{F}_{-k} U_{{\un 2}}[z^-, -\infty]_z
\nonumber\\
&&\left. - \epsilon^{mn} ({\un x}_1 | \frac{1}{p^2_\perp}\left(\delta^{im} - \frac{2 p^i p^m}{ p^2_\perp }\right) |{\un x}_2) ({\un x}_2| \frac{p_n}{p^2_\perp} |{\un x}_0) \frac{s}{P^+}U_{\un 2}^{\textrm{pol} [1]} - ({\un x}_0\to{\un x}_1) \right\}^{ab}\,. \notag
\end{eqnarray}

The Fourier transformations in this equation can be calculated using Eqs. (\ref{Fr1}), (\ref{Fr2}), and
\begin{eqnarray}
&&({\un x}_1 | \frac{p^i p^m p^k}{p^4_\perp} |{\un x}_2)
= \frac{i}{4\pi} \Bigg[ \frac{\delta^{im} {\un x}_{12}^k}{{\un x}_{12}^2} + \frac{\delta^{ik} {\un x}_{12}^m }{{\un x}_{12}^2 } + \frac{\delta^{mk} {\un x}_{12}^i }{{\un x}_{12}^2 } - 2  \frac{ {\un x}_{12}^i {\un x}_{12}^m {\un x}_{12}^k }{{\un x}_{12}^4 } \Bigg] ,
\end{eqnarray}
which gives
\begin{eqnarray}
&&\Bigg(\mbox{T} \, \tr \Bigg[  V^\dagger_{\un 0} \, \int\limits_{-\infty}^\infty d z^-  \, z^- V_{{\un 1}} [\infty, z^-] \,
    F^{+i} \, V_{{\un 1}} [z^-, - \infty]\Bigg] + \textrm{c.c.}\Bigg)_{\textrm{IV}+\textrm{V}+\textrm{IV}'+\textrm{V}'} = \tr \left[ V^\dag_{\un 0} t^a V_{\un 1}  t^b \right] \frac{g}{8\pi^3} \int\limits^\infty_0 \frac{dp^-}{p^-} \int d^2{\un x}_2 \\
&&\times \Bigg\{ \left[ \delta^{ik} \left( - 2\frac{ {\un x}_{12}\cdot {\un x}_{20} }{{\un x}_{12}^2 {\un x}_{20}^2 } + \frac{ 1 }{{\un x}_{20}^2} \right) + 2\frac{ {\un x}_{12}^i {\un x}_{20}^k }{{\un x}_{12}^2 {\un x}_{20}^2 } \left( 2\frac{ {\un x}_{12} \cdot {\un x}_{20}}{ {\un x}_{20}^2 } - 1 \right) + 2\frac{{\un x}_{12}^i {\un x}_{12}^k}{ {\un x}_{12}^2 {\un x}_{20}^2 } \left( 2  \frac{ {\un x}_{12} \cdot {\un x}_{20} }{{\un x}_{12}^2 } - 1 \right)\right.
\nonumber\\
&&\left.- 2\frac{ {\un x}_{20}^i {\un x}_{20}^k}{{\un x}_{20}^4} \right] g\int\limits^{\infty}_{-\infty} dz^- z^- U_{\un 2}[\infty, z^-] \mathcal{F}_{-k} U_{\un 2}[z^-, -\infty] 
 + \left[ \epsilon^{in} \frac{ {\un x}_{20}^n }{{\un x}_{20}^2} - 2 {\un x}_{12}^i \frac{ {\un x}_{12}\times {\un x}_{20}}{ {\un x}_{12}^2 {\un x}_{20}^2 } \right]\frac{is}{P^+} U_{\un 2}^{\textrm{pol} [1]} - ({\un x}_0\to{\un x}_1) \Bigg\}^{ab}. \notag
\end{eqnarray}
From this result we see that at small $x$ the helicity evolution operator, 
\begin{eqnarray}
&&\int\limits^{\infty}_{-\infty} dz^- z^- V_{\ul x}[\infty, z^-] F_{-k} V_{\ul x}[z^-, -\infty] \,,
\end{eqnarray}
mixes with the adjoint version of the same operator and with $U_{\un 2}^{\textrm{pol} [1]}$. We can finally use Eq.~(\ref{zFtoDi}) and write the sum of all the diagrams in \fig{FIG:Gi_evol} as a single evolution equation\footnote{Here we also explicitly introduce the limits for the integral over $p^-$.}
\begin{eqnarray}\label{V10i_evol}
&&\frac{1}{2 N_c} \, \llangle  \tr \Big[  V^\dagger_{\un 0} \, V_{\un{1}}^{i \, \textrm{G} [2]} \Big] + \textrm{c.c.}  \rrangle (\sigma) = \frac{1}{2 N_c} \, \llangle  \tr \Big[  V^\dagger_{\un 0} \, V_{\un{1}}^{i \, \textrm{G} [2]} \Big] + \textrm{c.c.}  \rrangle_0 (\sigma)
\\
&&+ \frac{\alpha_s N_c}{4\pi^2} \int\limits^\sigma_{\sigma'} \frac{dp^-}{p^-} \int d^2{\un x}_2 \left\{ \left[ \epsilon^{ij} \frac{ {\un x}_{21}^j }{{\un x}_{21}^2} -  \epsilon^{ij} \frac{ {\un x}_{20}^j }{{\un x}_{20}^2} + 2 {\un x}_{21}^i \frac{ {\un x}_{21}\times {\un x}_{20}}{ {\un x}_{12}^2 {\un x}_{20}^2 } \right] \frac{1}{ N^2_c} \llangle \tr \Big[ t^b V^\dag_{\un 0} t^a V_{\un 1} \Big] (U_{\un 2}^{\textrm{pol} [1]})^{ab} + \textrm{c.c.}\rrangle(\sigma')\right.
\nonumber\\
&&+ \left[ \delta^{ij} \left( 3\frac{ 1 }{ {\un x}_{21}^2 } - 2\frac{ {\un x}_{21}\cdot {\un x}_{20} }{{\un x}_{12}^2 {\un x}_{20}^2 } - \frac{ 1 }{{\un x}_{20}^2} \right) - 2\frac{ {\un x}_{21}^i {\un x}_{20}^j }{{\un x}_{12}^2 {\un x}_{20}^2 } \left( 2\frac{ {\un x}_{21} \cdot {\un x}_{20}}{ {\un x}_{20}^2 } + 1 \right) 
+ 2\frac{{\un x}_{21}^i {\un x}_{21}^j}{ {\un x}_{21}^2 {\un x}_{20}^2 } \left( 2  \frac{ {\un x}_{21} \cdot {\un x}_{20} }{{\un x}_{21}^2 } + 1 \right) \right.
\nonumber\\
&&\left.\left.\left.+ 2\frac{ {\un x}_{20}^i {\un x}_{20}^j}{{\un x}_{20}^4} - 2\frac{ {\un x}_{21}^i {\un x}_{21}^j}{{\un x}_{21}^4} \right. \right]\frac{1}{ N^2_c} \llangle \tr \Big[ t^b V^\dag_{\un 0} t^a V_{\un 1} \Big] (U_{\un{2}}^{j \, \textrm{G} [2]})^{ab} + \textrm{c.c.}\rrangle(\sigma')
 - ({\un x}_0\to{\un x}_1) \right\}
 \nonumber\\
 &&+ \frac{\as \, N_c}{2 \pi^2} \, \int\limits_{\sigma'}^\sigma \frac{d p^-}{p^-} \, \int d^2 {\ul x}_2 \, \frac{{\ul x}_{10}^2}{{\ul x}_{21}^2 \, {\ul x}_{20}^2} \,  \Bigg\{ \frac{1}{N_c^2} \, \llangle \tr \left[ t^b \, V^\dag_{\un 0} \, t^a \, V_{\un 1}^{i \, \textrm{G} [2] } \right] \, \left( U_{\un 2} \right)^{ab} \rrangle (\sigma')  - \frac{C_F}{N_c^2} \, \llangle \tr \left[ V^\dag_{\un 0} \, V_{\un 1}^{i \, \textrm{G} [2] \, } \right] \rrangle (\sigma') + \textrm{c.c.}  \Bigg\}\,,
 \nonumber
\end{eqnarray}
where the last line is the sum of the eikonal diagrams, see the discussion after Eq.~(\ref{dIIIfin}). After a trivial change of variables we find a complete agreement with Eq.~(\ref{Gi_evol_main}) above. A similar calculation employing the background field method in the adjoint representation yields the helicity evolution equation (\ref{Gi_adj_evol_main}).


\section{Small-$x$ Asymptotics of the Quark and Gluon Helicity Distributions and $g_1$ Structure Function in the Large-$N_c$ Limit}

\label{sec:numerics}

As can be seen in Eqs.~\eqref{glue_hel_TMD57}, \eqref{JM_DeltaG}, \eqref{DeltaSigma}, \eqref{TMD19} and \eqref{g1_final}, gluon and quark helicity TMD and PDF, together with the $g_1$ structure function, can be determined for small $x$ using the polarized dipole amplitudes. In particular, the small-$x$ asymptotics of the former will have the same intercepts  as the large-$zs$ asymptotics of the latter. In the large-$N_c$ limit considered in this Section, the polarized dipole amplitudes, $G(x^2_{10},zs)$ and $G_2(x^2_{10},zs)$, can be specified by solving Eqs.~\eqref{eq_LargeNc}. Since, at large $N_c$, $Q(x^2_{10},zs) \approx G(x^2_{10},zs)$, knowing $G$ and $G_2$ gives us all the flavor-singlet helicity PDFs and TMDs, along with the $g_1$ structure function. Owing to the complicated form of Eqs.~\eqref{eq_LargeNc}, we solve the system numerically.

As mentioned above, we begin by examining the asymptotic forms of $G(x^2_{10},zs)$ and $G_2(x^2_{10},zs)$ as $zs$ grows large. As discussed in \cite{Kovchegov:2016weo, Kovchegov:2020hgb}, it is more convenient to express Eqs.~\eqref{eq_LargeNc} in terms of
\begin{align}
\begin{split}
    \eta &= \sqrt{\frac{\alpha_sN_c}{2\pi}} \, \ln\frac{zs}{\Lambda^2} \;\;\text{,}\;\;\;\;\eta' = \sqrt{\frac{\alpha_sN_c}{2\pi}} \,\ln\frac{z's}{\Lambda^2}\;\;\;\;\text{and}\;\;\;\;\eta'' = \sqrt{\frac{\alpha_sN_c}{2\pi}} \,\ln\frac{z''s}{\Lambda^2} \; , \\
    s_{10} &= \sqrt{\frac{\alpha_sN_c}{2\pi}} \,\ln\frac{1}{x^2_{10}\Lambda^2} \;\;\text{,}\;\;\;\;s_{21} = \sqrt{\frac{\alpha_sN_c}{2\pi}} \,\ln\frac{1}{x^2_{21}\Lambda^2}\;\;\;\;\text{and}\;\;\;\;s_{32} = \sqrt{\frac{\alpha_sN_c}{2\pi}} \,\ln\frac{1}{x^2_{32}\Lambda^2}  \; .
\label{nume1}
\end{split}
\end{align}
In terms of these parameters, Eqs.~\eqref{eq_LargeNc}, with the help of \eq{kernels}, can be written as
\begin{subequations}\label{nume2}
\begin{align}
& G (s_{10} , \eta) = G^{(0)} (s_{10} , \eta) +  \int\limits_{s_{10}}^{\eta} d\eta' \, \int\limits_{s_{10}}^{\eta'} ds_{21} \,  \Bigg[  \Gamma (s_{10}, s_{21},  \eta') + 3 \, G (s_{21}, \eta')  \label{nume2a} \\ 
& \hspace*{9cm} + 2 \, G_2 (s_{21}, \eta') + 2 \, \Gamma_2 (s_{10}, s_{21},  \eta') \Bigg]  , \notag  \\
& \Gamma (s_{10} , s_{21}, \eta') = G^{(0)} (s_{10} , \eta') +  \int\limits_{s_{10}}^{\eta'} d\eta'' \, \int\limits_{\max \left[ s_{10} , \,  s_{21}+\eta''-\eta' \right]}^{\eta''} ds_{32} \,  \Bigg[  \Gamma (s_{10},s_{32},\eta'') + 3 \, G (s_{32}, \eta'') \label{nume2b}  \\ 
& \hspace*{9cm} + 2 \, G_2 (s_{32}, \eta'') + 2 \, \Gamma_2 (s_{10},s_{32},\eta'') \Bigg] , \notag \\
& G_2 (s_{10} , \eta)  =  G_2^{(0)} (s_{10} , \eta) + 2 \, \int\limits_0^{s_{10}} d s_{21} \int\limits_{s_{21}}^{\eta - s_{10} + s_{21}} d \eta' \, \left[ G (s_{21} , \eta') + 2 \, G_2 (s_{21}, \eta')  \right] , \label{nume2c} \\
& \Gamma_2 (s_{10} , s_{21}, \eta')  =  G_2^{(0)} (s_{10} , \eta') + 2 \, \int\limits_{0}^{s_{10}} ds_{32}  \, \int\limits^{\eta' - s_{21} + s_{32}}_{s_{32}} d\eta'' \left[ G (s_{32} , \eta'') + 2 \, G_2 (s_{32},\eta'')  \right] , \label{nume2d}
\end{align}
\end{subequations}
where the ordering $0\leq s_{10}\leq s_{21}\leq\eta'$ is assumed in Eqs. \eqref{nume2b} and \eqref{nume2d}. This is the only region where $\Gamma$ and $\Gamma_2$ appear in any large-$N_c$ evolution kernel.

Now, we discretize the integrals in Eqs.~\eqref{nume2} with step size $\delta$ both in $\eta$ and $s_{10}$ directions. We express the discretized version of the dipole amplitudes such that 
\begin{align}
    \begin{split}
        G_{ij} &= G\left(i\delta, j\delta\right) \;\;\;\,, \;\;\;\;\;\;\;\Gamma_{ikj} = \Gamma\left(i\delta, k\delta, j\delta\right) \; , \\
        G_{2,ij} &= G_2\left(i\delta, j\delta\right) \;\;,\;\;\;\;\;
        \Gamma_{2,ikj} = \Gamma_2\left(i\delta, k\delta, j\delta\right) .
        \label{nume3}
    \end{split}
\end{align}
With all the definitions outlined above, we obtain the following discretized evolution equations.
\begin{subequations}\label{nume4}
\begin{align}
 G_{ij} &= G^{(0)}_{ij} + \delta^2 \, \sum_{j'=i}^{j-1} \, \sum_{i'=i}^{j'} \, \left[\Gamma_{ii'j'} + 3 \, G_{i'j'} + 2 \, G_{2,i'j'} + 2 \, \Gamma_{2,ii'j'}\right] , \label{nume4a} \\
 \Gamma_{ikj} &= G^{(0)}_{ij} + \delta^2 \, \sum_{j'=i}^{j-1} \, \sum_{i'=\max\left[i,\,k+j'-j\right]}^{j'}  \,  \left[  \Gamma_{ii'j'} + 3 \, G_{i'j'} + 2 \, G_{2,i'j'} + 2 \, \Gamma_{2,ii'j'} \right] , \label{nume4b} \\
 G_{2,ij}  &=  G_{2,ij}^{(0)} + 2 \, \delta^2 \, \sum\limits_{i'=0}^{i-1} \, \sum\limits_{j'=i'}^{j-i+i'} \,  \left[  G_{i'j'} + 2 \, G_{2,i'j'}  \right] , \label{nume4c} \\
 \Gamma_{2,ikj} &=  G_{2,ij}^{(0)} + 2 \, \delta^2 \, \sum_{i'=0}^{i-1} \, \sum_{j'=i'}^{j-k+i'}  \,  \left[ G_{i'j'} + 2 \, G_{2,i'j'} \right] .
\end{align}
\end{subequations}
To obtain the values of $G_{ij}$ and $G_{2,ij}$ for $0\leq i \leq i_{\max}$ and $0\leq j\leq j_{\max}$, we only need to know the following dipole amplitudes:
\begin{itemize}
    \item $G_{ij}$ and $G_{2,ij}$ such that $0\leq i < j$, with $i\leq i_{\max}$ and $j\leq j_{\max}$. Note that if $i\geq j$, then we have $G_{ij}=G^{(0)}_{ij}$ and $G_{2,ij}=G^{(0)}_{2,ij}$, as can be seen from \eq{nume4a} and \eq{nume4c}.
    \item $\Gamma_{ikj}$ and $\Gamma_{2,ikj}$ such that $0\leq i\leq k\leq j$, with $k\leq i_{\max}$ and $j\leq j_{\max}$. This is because the neighbor dipole amplitudes only appear in Eqs. \eqref{nume4a} and \eqref{nume4b}.
\end{itemize}
In a fashion similar to \cite{Kovchegov:2020hgb}, the numerical computation becomes more efficient once we realize the following recursive relations that follow directly from Eqs.~\eqref{nume4} for $j>0$: 
\begin{subequations}\label{nume5}
\begin{align}
 G_{ij} &= \begin{cases} 
 G^{(0)}_{ij} - G^{(0)}_{i(j-1)} + G_{i(j-1)} + \delta^2  \, \sum\limits_{i'=i}^{j-1} \, \left[\Gamma_{ii'(j-1)} + 3 \, G_{i'(j-1)} + 2 \, G_{2,i'(j-1)} + 2 \, \Gamma_{2,ii'(j-1)}\right] &, \;\;\; i < j \\
 G^{(0)}_{ij} &, \;\;\; i = j
 \end{cases} \, , \label{nume5a} \\
 \Gamma_{ikj} &= \begin{cases}
 G^{(0)}_{ij} - G^{(0)}_{i(j-1)} + \Gamma_{i(k-1)(j-1)} + \delta^2  \sum\limits_{i'=k-1}^{j-1}    \left[  \Gamma_{ii'(j-1)} + 3 \, G_{i'(j-1)} + 2 \, G_{2,i'(j-1)} + 2 \, \Gamma_{2,ii'(j-1)} \right] &, \;\;\; i < k \\
 G_{ij} &, \;\;\; i=k 
 \end{cases} \, , \label{nume5b} \\
  G_{2,ij}  &= \begin{cases} 
G^{(0)}_{2,ij} - G^{(0)}_{2,i(j-1)} + G_{2,i(j-1)} + 2\, \delta^2  \, \sum\limits_{i'=0}^{i-1} \, \left[G_{i'(i'+j-i)} + 2 \, G_{2,i'(i'+j-i)}\right] &, \;\;\; i < j \\
 G^{(0)}_{2,ij} &, \;\;\; i = j
 \end{cases} \, , \label{nume5c} \\
 \Gamma_{2,ikj} &=  \begin{cases}
 G^{(0)}_{2,ij} - G^{(0)}_{2,i(j-1)} + \Gamma_{2,i(k-1)(j-1)} &, \;\;\; i < k \\
 G_{2,ij} &, \;\;\; i=k 
  \end{cases} \, .
\end{align}
\end{subequations}
In the case where $j=0$, each of the dipole amplitudes simply equals its corresponding inhomogeneous term, as can be seen from Eqs.~\eqref{nume4}.

In order to perform the numerical computation, we also need to rewrite the non-homogeneous terms, \eq{non-homo}, in terms of the new variables, $s_{10}$ and $\eta$. This gives
\begin{subequations}\label{nume_ic_general}
\begin{align}
G^{(0)}(s_{10},\eta) &= \frac{\as^2 C_F}{2 N_c} \pi  \sqrt{\frac{2\pi}{\alpha_sN_c}} \, \left[ C_F  \, \eta - 2 \, (\eta-s_{10}) \right]\;\;\;\;\Rightarrow\;\;\;\;G^{(0)}_{ij} =  \frac{\as^2 C_F}{2 N_c} \pi  \sqrt{\frac{2\pi}{\alpha_sN_c}} \, \left[ (C_F-2)  \, j   + 2 \, i   \right] \delta   \, , \\ 
G^{(0)}_{2}(s_{10},\eta) &= \frac{\as^2 C_F}{2N_c} \pi \sqrt{\frac{2\pi}{\alpha_sN_c}} \, s_{10}\;\;\;\;\Rightarrow\;\;\;\;G^{(0)}_{2,ij} = \frac{\as^2 C_F}{2N_c} \pi \sqrt{\frac{2\pi}{\alpha_sN_c}} \, i  \delta \, .
\end{align}
\end{subequations}
In particular, in terms of the discrete variables, $i$ and $j$, the one-step differences of the non-homogeneous terms are
\begin{align}\label{nume_ic}
G^{(0)}_{ij} - G^{(0)}_{i(j-1)}  &= \frac{\as^2 C_F}{2 N_c} \pi  \sqrt{\frac{2\pi}{\alpha_sN_c}} \, (C_F-2) \, \delta \, , \ \ \ G^{(0)}_{2,ij} - G^{(0)}_{2,i(j-1)} = 0 \, .
\end{align}

\begin{figure}
     \centering
     \begin{subfigure}[b]{0.45\textwidth}
         \centering
         \includegraphics[width=\textwidth]{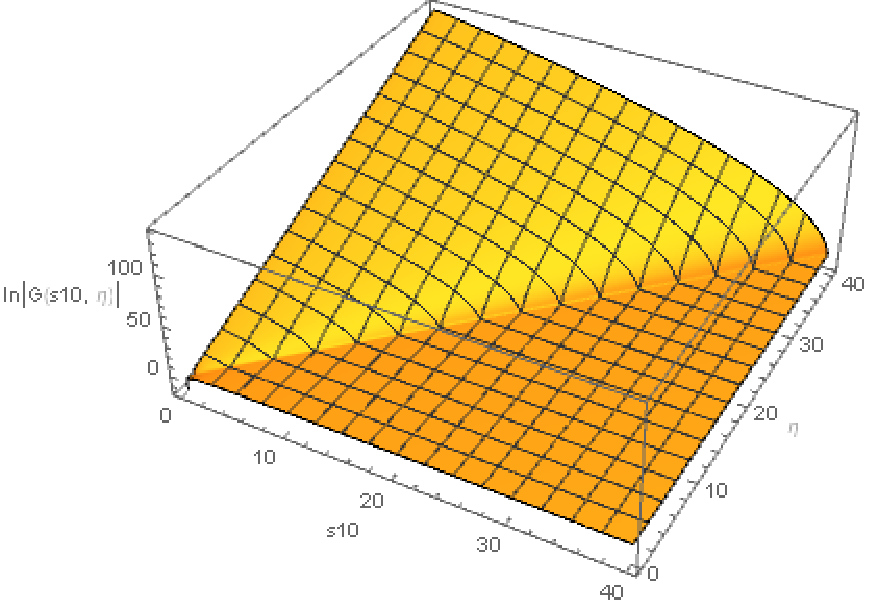}
         \caption{$\ln\left|G(s_{10},\eta)\right|$}
         \label{fig:lnG_3d}
     \end{subfigure} 
     \;\;\;\;\;
     \begin{subfigure}[b]{0.45\textwidth}
         \centering
         \includegraphics[width=\textwidth]{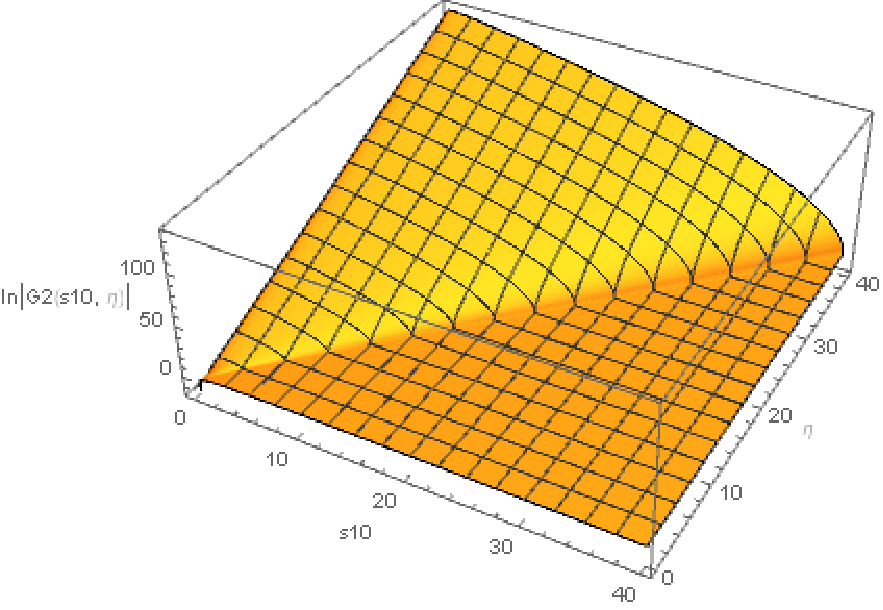}
         \caption{$\ln\left|G_2(s_{10},\eta)\right|$}
         \label{fig:lnG2_3d}
     \end{subfigure}
     \caption{The plots of logarithms of the absolute values of the two polarized dipole amplitudes $G$ and $G_2$ versus $s_{10}$ and $\eta$, for the $0\leq s_{10},\eta\leq\eta_{\max}=40$ range. The amplitudes are computed numerically using step size $\delta = 0.05$. The inhomogeneities near the $\eta = s_{10}$ line result from the Born initial conditions and the discretization error.}
     \label{fig:ln_3d}
\end{figure}

Now, we numerically compute all the dipole amplitudes in Eqs.~\eqref{nume5} with the help of \eq{nume_ic}, using the step size of $\delta = 0.05$. In the range where $0\leq \eta, s_{10}\leq \eta_{\max} = 40$, the logarithms of $G(x^2_{10},zs)$ and $G_2(x^2_{10},zs)$ are plotted in Fig. \ref{fig:ln_3d}.
From the plots, we see that both amplitudes grow roughly linearly with $\eta-s_{10}$, which corresponds to an exponential growth in $zsx^2_{10}$. Mild deviations from the aforementioned pattern, including the inhomogeneities along $\eta=s_{10}$ line, likely  result from discretization errors. However, their actual cause must be determined with certainty through an analytic solution. 

\begin{figure}
     \centering
     \begin{subfigure}[b]{0.4\textwidth}
         \centering
         \includegraphics[width=\textwidth]{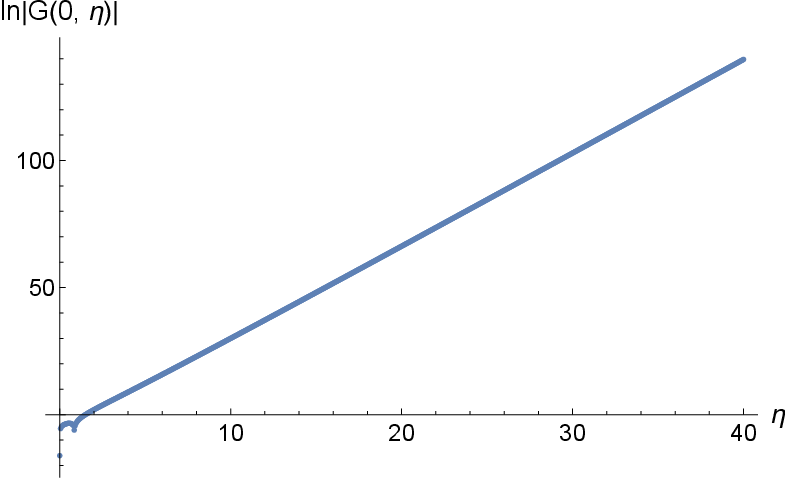}
         \caption{$\ln\left|G(0,\eta)\right|$}
         \label{fig:lnG_2d}
     \end{subfigure} 
     \;\;\;\;\;
     \begin{subfigure}[b]{0.4\textwidth}
         \centering
         \includegraphics[width=\textwidth]{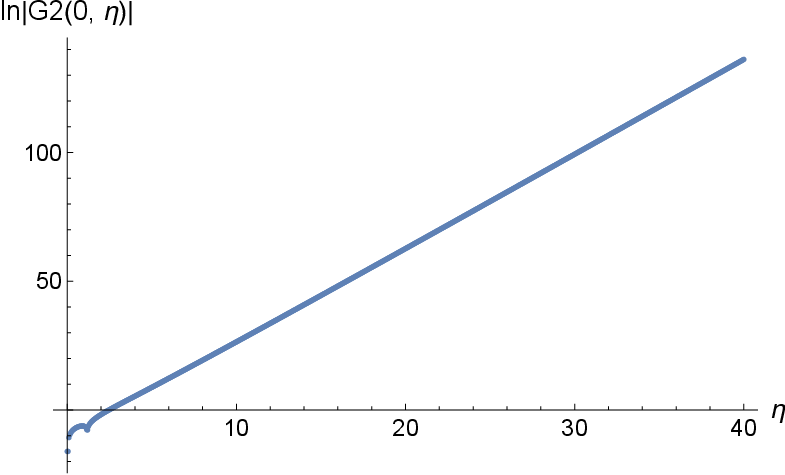}
         \caption{$\ln\left|G_2(0,\eta)\right|$}
         \label{fig:lnG2_2d}
     \end{subfigure}
     \caption{The plots of logarithms of the absolute values of the two polarized dipole amplitudes at $s_{10}=0$ versus $\eta$, for the $0\leq\eta\leq\eta_{\max}=40$ range. The amplitudes are computed numerically using step size $\delta = 0.05$. The kinks near $\eta=0$ occur due to sign flips in $G(0,\eta)$ and $G_2(0,\eta)$. By Eqs.~\eqref{nume_ic_general}, the Born initial condition leads to $G^{(0)}_{0j} < 0$ for $G(0,\eta)$ at any $j>0$.}
     \label{fig:ln_2d}
\end{figure}

As mentioned previously, for the purpose of this Section, it is sufficient for us to determine the asymptotic form of $G(s_{10}=0,\eta)$ and $G_2(s_{10}=0,\eta)$ as $\eta\to\infty$. To do so, we plot the logarithm of each amplitude at $s_{10}=0$ against $\eta$. These plots are shown in Fig. \ref{fig:ln_2d}.
As expected, both functions increase linearly once we get sufficiently far away from $\eta=0$, where the non-homogeneous term and the discretization error remain relatively significant. This justifies the following {\sl  ansatze} as $\eta\to\infty$,
\begin{align}\label{ansatz}
G(s_{10}=0,\eta) \sim e^{\alpha_h\eta\sqrt{\frac{2\pi}{\as N_c}}} \, , \ \ \ G_2(s_{10}=0,\eta) \sim e^{\alpha_{h,2}\eta\sqrt{\frac{2\pi}{\as N_c}}} \, ,
\end{align}
where $\alpha_h$ and $\alpha_{h,2}$ are given by the slopes of the functions in Figs. \ref{fig:lnG_2d} and \ref{fig:lnG2_2d}, respectively. Since the exponential growth is more dominant at larger $\eta$'s, we deduce the approximation of $\alpha_h$ and $\alpha_{h,2}$ for this step size, $\delta$, and maximum rapidity, $\eta_{\max}$, by regressing $\ln\left[G(0,\eta)\right]$ and $\ln\left[G_2(0,\eta)\right]$, respectively, on $\eta$ over the range where $0.75\,\eta_{\max}\leq\eta\leq\eta_{\max}$. For example, at $\delta=0.05$ and $\eta_{\max}=40$, corresponding to Fig. \ref{fig:ln_2d}, we obtain $\alpha_h = (3.6825\pm 0.0002)\sqrt{\frac{\as N_c}{2\pi}}$ and $\alpha_{h,2}=(3.6821\pm 0.0002)\sqrt{\frac{\as N_c}{2\pi}}$. The uncertainty is estimated from the residual of linear regression performed on $\ln\left[G(0,\eta)\right]$ or $\ln\left[G_2(0,\eta)\right]$ at 95\% confidence level.

Having estimated the intercepts, $\alpha_h$ and $\alpha_{h,2}$, at $\delta=0.05$ and $\eta_{\max}=40$, we then repeat the steps for other choices of $\delta$ and $\eta_{\max}$. In particular, for each step size, $\delta$, we numerically compute the intercepts for $\eta_{\max}\in\{10,20,\ldots,M(\delta)\}$, where $M(\delta)$ is given in Table \ref{tab:M_delta} for each $\delta$ employed in this work.
\begin{table}[h]
\begin{tabular}{|c|c|c|c|c|c|c|c|c|c|c|}
\hline
$\delta$ 
& \,0.0125\,
& \,\,0.016\,\,
& \,\,0.025\,\,
& \,\,0.032\,\,
& \,0.0375\,
& \,\,\,0.05\,\,\,
& \,0.0625\,
& \,\,0.075\,\,
& \,\,\,0.08\,\,\,
& \,\,\,\,0.1\,\,\,\,
\\ \hline 
\,$M(\delta)$ \,
& 10
& 10
& 20
& 20
& 30
& 40
& 50
& 60
& 60
& 70
\\ \hline
\end{tabular}
\caption{The maximum, $M(\delta)$, of $\eta_{\max}$ computed for each step size, $\delta$.}
\label{tab:M_delta}
\end{table}

Now, we obtain the estimated intercepts and their uncertainties for all 37 combinations of $\delta$ and $\eta_{\max}$. Since the continuum limit corresponds to $\delta\to 0$ and $\eta_{\max}\to\infty$, we attempt to model the intercepts using $\delta$ and $1/\eta_{\max}$ as independent variables. Afterward, with the correct model at hand, we will be able to predict the intercepts at $\delta=1/\eta_{\max}=0$ and use them as our best estimate for the actual intercepts in the continuum limit. 

In what follows, we will detail our process to determine the intercept, $\alpha_h$, in the continuum limit. The process for $\alpha_{h,2}$ will be similar. Inspired by the success of \cite{Kovchegov:2016weo} in numerically estimating the correct intercept as verified by the analytic solution \cite{Kovchegov:2017jxc}, we employ polynomial regression models of various degrees, with interaction terms included, weighted by the uncertainties of the estimated intercepts. In particular, we consider four following nested models with increasing maximum polynomial degrees:
\begin{itemize}
    \item Model 1: $\alpha_h = a_1\,$,
    \item Model 2: $\alpha_h = a_1 + a_2\delta + \frac{a_3}{\eta_{\max}}\,$,
    \item Model 3: $\alpha_h = a_1 + a_2\delta + \frac{a_3}{\eta_{\max}} + a_4\delta^2 + \frac{a_5\delta}{\eta_{\max}} + \frac{a_6}{\eta^2_{\max}}\,$,
    \item Model 4: $\alpha_h = a_1 + a_2\delta + \frac{a_3}{\eta_{\max}} + a_4\delta^2 + \frac{a_5\delta}{\eta_{\max}} + \frac{a_6}{\eta^2_{\max}} + a_7\delta^3 + \frac{a_8\delta^2}{\eta_{\max}} + \frac{a_9\delta}{\eta^2_{\max}} + \frac{a_{10}}{\eta^3_{\max}}\,$.
\end{itemize}
Once we fit and evaluate all four models to our numerical estimates for $\alpha_h$, the Akaike information criterion (AIC) \cite{Akaike:1974} decreases significantly from model 1 to model 2 and from model 2 to model 3. However, the AIC is roughly equal for models 3 and 4. Furthermore, the parameters $a_7$, $a_8$, $a_9$ and $a_{10}$ are all insignificant when the $t$-test is performed at 10\% significance level for each of them. This implies that all degree-3 terms in model 4 are not significantly different from zero, that is, model 4 would not account for our intercept results any better than model 3. Together with the fact that all parameters for model 3 are significant, we decide to use model 3, the quadratic model, to fit the values of $\alpha_h$. The process and, more importantly, the conclusion about the final model choice are exactly the same for $\alpha_{h,2}$, although the resulting parameter values are slightly different.

\begin{figure}
     \centering
     \begin{subfigure}[b]{0.45\textwidth}
         \centering
         \includegraphics[width=\textwidth]{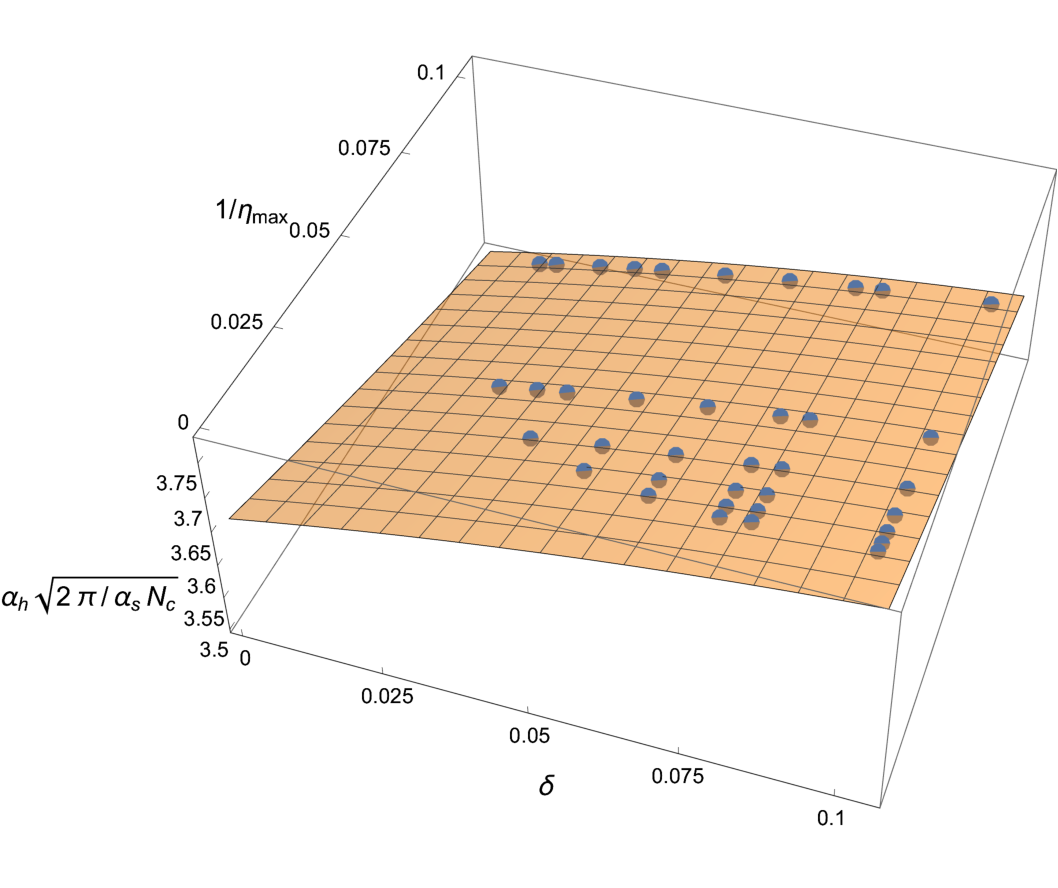}
         \caption{$\alpha_h$}
         \label{fig:quad_surf_ah}
     \end{subfigure} 
     \;\;\;\;\;
     \begin{subfigure}[b]{0.45\textwidth}
         \centering
         \includegraphics[width=\textwidth]{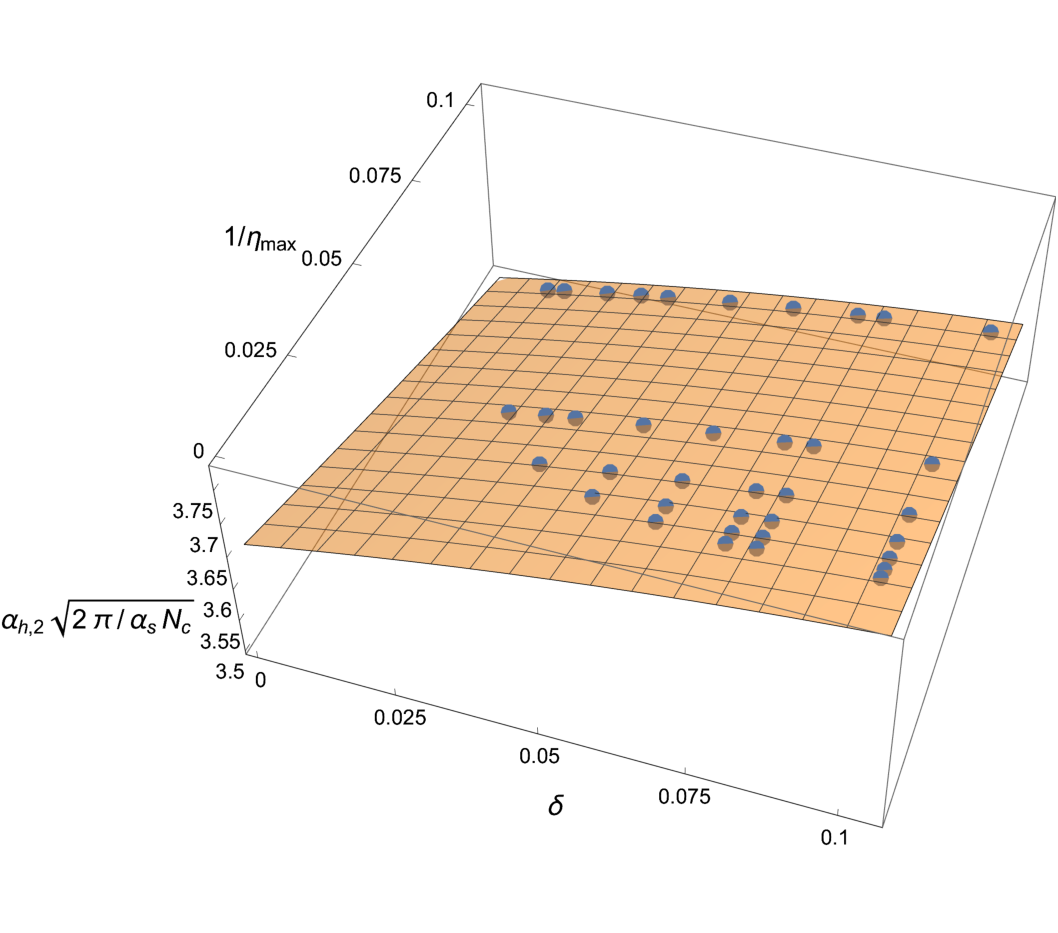}
         \caption{$\alpha_{h,2}$}
         \label{fig:quad_surf_ah2}
     \end{subfigure}
     \caption{The plots of estimated intercepts, $\alpha_h$ and $\alpha_{h,2}$, at each $\delta$ and $1/\eta_{\max}$ (blue dots), together with the best-fitted quadratic surface given by \eq{quadratic_relation_intercepts} (yellow surfaces). The continuum limit, $\delta=1/\eta_{\max}=0$, corresponds to the lower-left corner of each plot.}
     \label{fig:quad_surf}
\end{figure}

With model 3, the estimated relation between each intercept and $\delta$ and $1/\eta_{\max}$ are given by
\begin{subequations}\label{quadratic_relation_intercepts}
\begin{align}
    \alpha_h &= \left[3.661 + 1.503\,\delta - 1.740\,(1/\eta_{\max}) - 4.414\,\delta^2 + 0.116\,\delta\,(1/\eta_{\max}) + 1.429\,(1/\eta_{\max})^2\right] \sqrt{\frac{\as N_c}{2\pi}} \, , \label{quad_ah} \\
    \alpha_{h,2} &= \left[3.660 + 1.509\,\delta - 1.734\,(1/\eta_{\max}) - 4.438\,\delta^2 - 0.034\,\delta\,(1/\eta_{\max}) + 0.873\,(1/\eta_{\max})^2\right] \sqrt{\frac{\as N_c}{2\pi}} \, . \label{quad_ah2}
\end{align}
\end{subequations}
The estimated quadratic surfaces are plotted together with the intercepts we computed previously for various combinations of $\delta$ and $1/\eta_{\max}$ in Fig. \ref{fig:quad_surf}.

Next, we compute the continuum-limit intercepts, whose estimated values are the first terms in the right-hand sides of \eq{quadratic_relation_intercepts}. The uncertainties are estimated while taking into account both the residuals of the quadratic model and the uncertainties of each data point, i.e., intercept estimated at each $\delta$ and $1/\eta_{\max}$. This gives 
\begin{align}\label{intercept_results}
\alpha_h  &= (3.661\pm 0.006)\sqrt{\frac{\as N_c}{2\pi}} \, , \ \ \ \alpha_{h,2} = (3.660 \pm 0.009)\sqrt{\frac{\as N_c}{2\pi}} \, .
\end{align}
Recall that the uncertainties in \eq{intercept_results} come from (i) the residual of linear regression performed on $\ln\left|G(0,\eta)\right|$ and $\ln\left|G_2(0,\eta)\right|$ at each $\delta$ and $\eta_{\max}$, and (ii) the residual of polynomial regression performed on $\alpha_h$ and $\alpha_{h,2}$.

Now, the pure-glue BER intercept can be shown to be \begin{align}\label{BERintercept}
    \alpha_h = \sqrt{\frac{17 + \sqrt{97}}{2}} \, \sqrt{\frac{\as \, N_c}{2 \pi}} \approx 3.664 \, \sqrt{\frac{\as \, N_c}{2 \pi}} 
\end{align}
by solving the corresponding IREE from \cite{Bartels:1996wc} analytically \cite{Kovchegov:2016zex}. \eq{BERintercept} agrees with both $\alpha_h$ and $\alpha_{h,2}$ from \eq{intercept_results}, within the uncertainties. While the construction of an analytic solution for Eqs.~\eqref{eq_LargeNc} is left for future work, \eq{BERintercept} already provides us with the analytic expression for the intercept. 

Finally, empoying Eqs. \eqref{ansatz} and \eqref{intercept_results} in Eqs. \eqref{JM_DeltaG}, \eqref{DeltaSigma} and \eqref{g1_DLA}, we obtain the following small-$x$ asymptotics for the quark and gluon helicity PDF, together with the $g_1$ structure function:
\begin{align}\label{small_x_asymp}
    \Delta \Sigma (x, Q^2) \sim \Delta G (x, Q^2) 
    \sim g_1 (x, Q^2) \sim \left( \frac{1}{x} \right)^{3.66 \, \sqrt{\frac{\as \, N_c}{2 \pi}}} .
\end{align}


\section{Conclusions and Outlook}
\label{sec:conclusion}

Let us summarize what we have accomplished here. We have extended the helicity evolution formalism of \cite{Kovchegov:2015pbl, Kovchegov:2016zex, Kovchegov:2016weo, Kovchegov:2017jxc, Kovchegov:2017lsr, Kovchegov:2018znm, Cougoulic:2019aja, Kovchegov:2020hgb, Kovchegov:2021lvz}
to include the sub-eikonal operator $\cev{D}^i \, D^i$ (or, equivalently, $D^i - \cev{D}^i$). This generalized the small-$x$ evolution equations for the relevant sub-eikonal operators $D^i - \cev{D}^i$, $F^{12}$, and ${\bar \psi} \gamma^+ \gamma^5 \psi$ to those in Eqs.~\eqref{Q_evol_main}, \eqref{G_adj_evol}, \eqref{Gi_evol_main}, and \eqref{Gi_adj_evol_main}. The corresponding DLA evolution equations are given by Eqs.~\eqref{eq_LargeNc}
 and \eqref{eq_LargeNcNf} in the large-$N_c$ and large-$N_c \& N_f$ limits, respectively. We demonstrated that the large-$N_c$ equations agree with the spin-dependent DGLAP evolution at small $x$ including up to three loops in the splitting function. We solved these equations numerically showing that the resulting asymptotics of the gluon and flavor-singlet quark helicity distributions, along with the $g_1$ structure function, are given by \eq{small_x_asymp} and agree with that found by BER \cite{Bartels:1996wc} in the pure-glue case. We have thus completed the construction of the DLA helicity evolution equations at small-$x$ in the $s$-channel/shock wave formalism, which we also refer to as the light-cone operator treatment (LCOT). We have also cross-checked the LCOT calculation using the background field method and found a full agreement between the two. 
 
 The future steps of working with this now-complete LCOT formalism include solving the large-$N_c \& N_f$ equations \eqref{eq_LargeNcNf} and comparing the solution to those found in \cite{Bartels:1996wc} and in \cite{Kovchegov:2020hgb}. The two solutions in \cite{Bartels:1996wc} and \cite{Kovchegov:2020hgb} have a qualitatively different dependence on $x$: the latter exhibits sign-changing oscillations with $\ln (1/x)$, while the former changes sign only once with decreasing $x$. It would be important to identify which, if any, of those behaviours are exhibited by the solution of Eqs.~\eqref{eq_LargeNcNf}. 
 
 In the effort to go beyond the large-$N_c$ and large-$N_c \& N_f$ limits, a helicity version of the Jalilian-Marian--Iancu--McLerran--Weigert--Leonidov--Kovner
(JIMWLK)
\cite{Jalilian-Marian:1997dw,Jalilian-Marian:1997gr,Weigert:2000gi,Iancu:2001ad,Iancu:2000hn,Ferreiro:2001qy} evolution was constructed in \cite{Cougoulic:2019aja}, also without taking the operator $\cev{D}^i \, D^i$ into account. The helicity JIMWLK kernel from \cite{Cougoulic:2019aja} also needs to be extended to include the effects of this operator. The initial conditions for the helicity JIMWLK evolution are given by the helicity-dependent version of the McLerran--Venugopalan (MV) model \cite{McLerran:1993ni,McLerran:1993ka,McLerran:1994vd} derived in \cite{Cougoulic:2020tbc}, which may also have to be extended to include the terms into the weight functional needed for the calculations of the expectation value of the $D^i - \cev{D}^i$ operator. 

To further improve the precision of helicity evolution one should go beyond the DLA limit. This was attempted in \cite{Kovchegov:2021lvz} using the earlier $s$-channel helicity formalism of 
 \cite{Kovchegov:2015pbl, Kovchegov:2016zex, Kovchegov:2016weo, Kovchegov:2017jxc, Kovchegov:2017lsr, Kovchegov:2018znm, Cougoulic:2019aja, Kovchegov:2020hgb}. In addition to resumming all the DLA and SLA$_L$ terms, the evolution equations constructed in \cite{Kovchegov:2021lvz} sum up all the single logarithmic corrections coming from the UV transverse integrals. These corrections were labeled SLA$_T$ in \cite{Kovchegov:2021lvz}. It remains to be seen whether the results of \cite{Kovchegov:2021lvz} can simply be added to the equations obtained in this work for a complete DLA+SLA helicity evolution at small $x$. It appears likely that the IR transverse logarithms need to be resummed as well, such that an interfacing of our evolution found above with the full spin-dependent DGLAP equation may also be needed for the DLA+SLA helicity evolution. 
 
 Last but not least, the helicity formalism of \cite{Kovchegov:2015pbl, Kovchegov:2016zex, Kovchegov:2016weo, Kovchegov:2018znm} has recently been used to successfully describe the world data on the proton and neutron $g_1$ structure functions at small $x$ \cite{Adamiak:2021ppq}. This was the first-ever helicity phenomenology work based on small-$x$ evolution only, not taken as an improvement of the DGLAP anomalous dimension \cite{Blumlein:1995jp,Blumlein:1996hb}. It would be interesting and important to see how much the conclusions of \cite{Adamiak:2021ppq} would be affected by the corrections included in this work. At the very least, the formalism presented here would allow for a natural inclusion of the gluon helicity PDF \eqref{JM_DeltaG} into the calculation. The fact that the intercept/power of $x$ in \eq{small_x_asymp} is larger than that given by the evolution in \cite{Kovchegov:2015pbl, Kovchegov:2016zex, Kovchegov:2016weo, Kovchegov:2017jxc, Kovchegov:2017lsr, Kovchegov:2018znm, Cougoulic:2019aja, Kovchegov:2020hgb} may generate more quark and gluon spin at small $x$, while simultaneously challenging the convergence of the integrals in Eqs.~\eqref{eqn:SqSG} at small $x$. The latter problem may be addressed by including saturation corrections (non-trivial unpolarized dipole $S$-matrices) and/or running of the coupling constant in the kernels of our helicity evolution.


\section*{Acknowledgments}

\label{sec:acknowledgement}

One of the authors (Y.K.) would like to thank Bob Jaffe and Cedric Lorc\'{e} for discussions of the Jaffe-Manohar distribution and its role in small-$x$ evolution. A.T. would like to thank Ian Balitsky and Raju Venugopalan for useful and inspiring conversations. Y.T. would like to thank Daniel Adamiak for discussions leading to important corrections to the results and great insights to the next steps.

This material is based upon work supported by the U.S. Department of
Energy, Office of Science, Office of Nuclear Physics under Award
Number DE-SC0004286. 

The work is performed within the framework of the TMD Topical
Collaboration.

The work of F.C. has been supported by the Academy of Finland, by the Centre of Excellence in Quark Matter and project 321840, and under the European Union’s Horizon 2020 research and innovation programme by the STRONG-2020 project (grant agreement No 824093) and by the European Research Council, grant agreement ERC-2015-CoG-681707. The content of this article does not reflect the official opinion of the European Union and responsibility for the information and views expressed therein lies entirely with the authors.


 \appendix
 \section{Schwinger's notation\label{Ap:Schnot}}
In this Appendix, we introduce the Schwinger's notation that we use for the quark and gluon propagators in Sec.~\ref{sec:hel_evo_bfm}. Following Schwinger \cite{Schwinger:1951nm}, we consider the coherent states $|x)$ and $|p)$ which are eigenvectors of the position and momentum operators,
\begin{eqnarray}
&&\hat{x}_\mu|x) = x_\mu |x) , \,\,\,\, \hat{p}_\mu|p) = p_\mu|p)\,.
\end{eqnarray}
The states define a particle with position $x$ and momentum $p$, respectively, and satisfy the completeness
\begin{eqnarray}
&&\int d^4x |x)(x| = 1, \,\,\,\, \int \frac{d^4p}{(2\pi)^4} |p)(p| = 1,
\label{shstcompl}
\end{eqnarray}
and orthogonality 
\begin{eqnarray}
(x|p) = e^{-ipx}, \,\,\,\,(p|x) = e^{ipx}, \,\,\,\, (x|y) =\delta^4(x-y), \,\,\,\, (p|q) = (2\pi)^4 \delta^4(p-q)
\end{eqnarray}
relations.

For an arbitrary function of the momentum operator, we have
\begin{eqnarray}
&&(x|f(\hat{p})|y) = \int \frac{d^4q}{(2\pi)^4}\, (x|f(\hat{p})|q)\,(q|y) = \int \frac{d^4q}{(2\pi)^4}\, f(q)\,e^{-iq(x-y)}\,.
\end{eqnarray}
In particular, this motivates the following representation for the scalar propagator:
\begin{eqnarray}
&&(x|\frac{1}{\hat{p}^2 + i\epsilon}|y) = \int \frac{d^4p}{(2\pi)^4}\, \frac{1}{p^2 + i\epsilon}\,e^{-ip(x-y)}\,.
\label{Schsc}
\end{eqnarray}

Similarly, for an arbitrary function of the position operator $f(\hat{x})|x) = f(x)|x)$. As a result, for the Wilson line operator
\begin{eqnarray}\label{Vhat}
&&\hat{V} \equiv \mathcal{P} \exp\left[ig\int\limits^\infty_{-\infty} dx^- A^+(x^-, \hat{\ul x})\right],
\end{eqnarray}
we write
\begin{eqnarray}
&&\hat{V}|{\ul x}) = V_{\ul x}|{\ul x})\,.
\end{eqnarray}
Note that, for brevity, in the main text of the paper we omit the hat symbol over the position and momentum operators.


\bibliographystyle{JHEP}

\providecommand{\href}[2]{#2}\begingroup\raggedright\endgroup

\end{document}